\documentclass{article}
\usepackage{graphicx} 

\usepackage{amsxtra}
\usepackage{mathtools}
\usepackage{float}
\usepackage{xcolor}
\usepackage{tikz}
\usepackage{macros}
\usepackage{hyperref}

\usepackage{rotating}  

\newcommand{\Figref}[1]{Figure~\ref{#1}}
\newcommand{\Quiver}[1]{$\mathcal Q_{\ref{#1}}$}

\newcommand{\hs}{\mathrm{HS}}
\newcommand{\pe}{\mathrm{PE}}
\newcommand{\hwg}{\mathrm{HWG}}
\usepackage{placeins}
\usetikzlibrary{calc}
\usetikzlibrary{arrows.meta, bending, patterns}
\usetikzlibrary{snakes}

\newcommand{\convexpath}[2]{
  [   
  create hullcoords/.code={
    \global\edef\namelist{#1}
    \foreach [count=\counter] \nodename in \namelist {
      \global\edef\numberofnodes{\counter}
      \coordinate (hullcoord\counter) at (\nodename);
    }
    \coordinate (hullcoord0) at (hullcoord\numberofnodes);
    \pgfmathtruncatemacro\lastnumber{\numberofnodes+1}
    \coordinate (hullcoord\lastnumber) at (hullcoord1);
  },
  create hullcoords
  ]
  ($(hullcoord1)!#2!-90:(hullcoord0)$)
  \foreach [
  evaluate=\currentnode as \previousnode using \currentnode-1,
  evaluate=\currentnode as \nextnode using \currentnode+1
  ] \currentnode in {1,...,\numberofnodes} {
    let \p1 = ($(hullcoord\currentnode) - (hullcoord\previousnode)$),
    \n1 = {atan2(\y1,\x1) + 90},
    \p2 = ($(hullcoord\nextnode) - (hullcoord\currentnode)$),
    \n2 = {atan2(\y2,\x2) + 90},
    \n{delta} = {Mod(\n2-\n1,360) - 360}
    in 
    {arc [start angle=\n1, delta angle=\n{delta}, radius=#2]}
    -- ($(hullcoord\nextnode)!#2!-90:(hullcoord\currentnode)$) 
  }
}

\tikzset{gaugebl/.style={circle,draw=black,fill=black,inner sep=1.5pt}}
\tikzset{gaugeblnormal/.style={circle,draw=black,fill=black,inner sep=2.5pt}}
\tikzset{hasse/.style={circle, fill,inner sep=2pt}}
\tikzset{->-/.style={decoration={
  markings,
  mark=at position #1 with {\arrow{>}}},postaction={decorate}}}

\definecolor{greed}{HTML}{2DBE06}
\definecolor{purpld}{HTML}{C14CE0}

\allowdisplaybreaks

\usepackage{todonotes}

\preprint{Imperial/TP/24/AH/03, UWThPh 2024-21}

\title{Quiver Subtraction on the Higgs Branch}

\author[a]{Sam Bennett, }
\author[a]{Amihay Hanany, }
\author[a]{Guhesh Kumaran, }
\author[a]{Chunhao Li, }
\author[a]{Deshuo Liu, and }
\author[b]{Marcus Sperling}

\affiliation[a]{Theoretical Physics Group, The Blackett Laboratory, Imperial College London,\\ Prince Consort Road
London, SW7 2AZ, UK}

\affiliation[b]{Fakultät für Physik, Universität Wien,\\
Boltzmanngasse 5, 1090 Wien, Austria}
\emailAdd{samuel.bennett18@imperial.ac.uk}
\emailAdd{a.hanany@imperial.ac.uk}
\emailAdd{guhesh.kumaran18@imperial.ac.uk}
\emailAdd{chunhao.li21@imperial.ac.uk}
\emailAdd{deshuo.liu21@imperial.ac.uk}
\emailAdd{marcus.sperling@univie.ac.at}

\abstract{This paper classifies all Higgs branch Higgsing patterns for simply-laced unitary quiver gauge theories with eight supercharges (including multiple loops) and introduces a Higgs branch subtraction algorithm. All possible minimal transitions are given, identifying differences between slices that emerge on the Higgs and Coulomb branches.
In particular, the algorithm is sensitive to global information including monodromies and Namikawa-Weyl groups.
Guided by symplectic duality, the algorithm further determines the global symmetry on the Coulomb branch, and verifies the exclusion of $C$ type or $F_4$ global symmetry for (simply-laced) unitary quiver gauge theories. The Higgs branches of some unitary quivers are verified to give slices in the nilpotent cones of exceptional simple Lie algebras.
}
\begin{document}
\maketitle

\section{Introduction}
A hall-mark of supersymmetric theories is the existence of a large continuous space of supersymmetric vacua. For theories with 8 supercharges in space-time dimension $d=3,4,5,6$, this moduli space of vacua contains maximal branches like the Higgs and Coulomb branch. By virtue of an $\surm(2)_R$ symmetry these Higgs branches exhibit strongly restricted geometric structures - known as singular hyper-K\"ahler spaces or symplectic singularities. In contrast, for $d=3,4,5$ theories, the Coulomb branches of theories with 8 supercharges do not exhibit uniform behaviour across dimensions. Higgs branches $\Higgs$ are therefore exceedingly useful objects in the study of the vast landscape of supersymmetric theories with 8 supercharges across various dimensions.

Owing to the high amount of supersymmetry, the Higgs branches $\Higgs$ of theories with eight supercharges are understood as a stratification of leaves related under a partial order and transverse slices.
Physically, each leaf represents a set of massless states, where moving within a leaf changes the mass of massive states but preserves the massless states. Hence, it is justified to call each leaf a phase of the theory. A slice, which in many instances can be represented as a moduli space of a quiver (either Higgs or Coulomb branch), is a set of moduli that, if tuned, changes the set of massless states and gives rise to a (second order) phase transition. In most cases, the mechanism in which the states acquire mass is the Brout-Englert-Guralnik-Higgs-Hagen-Kibble mechanism  \cite{englert1964broken,higgs1964broken,guralnik1964global,kibble1967symmetry} (hereafter Higgs mechanism for short), but more general cases include, for example, small instanton transitions.
The structure of the Higgs branch may be summarised in a Hasse diagram which displays the partial order of the leaves together with the slices between them. For Lagrangian theories, the Higgs branch Hasse diagram can be simply derived from the Higgs mechanism. Starting with a Lagrangian theory with gauge group $G$, each leaf in the Higgs branch Hasse diagram corresponds to a now partially Higgsed theory with gauge group $H\subset G$. The slice between them corresponds to a gauge theory with gauge group $C$, the commutant of $H$ inside $G$, coupled to matter that is derived from the embedding of $H\subset G$. In practice, the gauge group $C$ turns out to be of classical type.

While the Higgs mechanism is intuitive and familiar for Lagrangian theories, the derivation of the Higgs branch stratification for strongly-coupled and/or non-Lagrangian theories is beyond classical analysis. 
The characterisation of a leaf as a set of massless states remains, but does not necessarily enjoy a Lagrangian interpretation. The slices in turn are of non-classical nature, associated for example with the exceptional groups $E_{6,7,8}$, $F_4$, and $G_2$.

The magnetic quiver program (see \cite{Cremonesi:2015lsa,Ferlito:2017xdq,Cabrera:2018jxt,Cabrera:2019izd,Bourget:2019rtl,Cabrera:2019dob} and subsequent works) has opened a window to the systematic exploration of such Higgs branches. This includes the understanding of discrete and continuous actions on the moduli space as well as probing its structure. Based on progress involving 3d $\Ncal=4$ Coulomb branches (see \cite{Aharony:1997bx,Borokhov:2002ib,Borokhov:2002cg,Gaiotto:2008ak,Bashkirov:2010hj,Cremonesi:2013lqa,Cremonesi:2014kwa,Cremonesi:2014xha,Cremonesi:2014uva,Bullimore:2015lsa,Nakajima:2015txa,Braverman:2016wma} and later works), the magnetic quiver toolbox has been expanded by several key techniques — among them: quiver subtraction \cite{Cabrera:2018ann,Bourget:2019aer} and the decay and fission algorithm \cite{Bourget:2023dkj,Bourget:2024mgn}. Although both techniques produce the same stratification, they arise from entirely different principles.

The same ideas can be realised purely mathematically by viewing the Higgs branch from the algebro-geometric perspective. The coordinate ring of $\Higgs$ has several Poisson ideals, partially ordered by inclusion, which correspond to singular loci in the geometry \cite{kaledin2006symplectic}. From the point of view of representation theory, each point on $\Higgs$ gives an equivalence class of representations of the quiver algebra. These representations are preserved by some symmetry group, thus partially ordered by inclusion relations of the groups. 

Based on intuition from brane systems, quiver subtraction postulates subtraction rules that allow for the derivation of the Coulomb branch Hasse diagram of the magnetic quiver.\footnote{This is, of course, equivalent to the Higgs branch Hasse diagram of the electric theory in the magnetic quiver context.} It has been understood only recently that the Coulomb branch stratification can be derived without relying on subtraction rules — this is the essence of the decay and fission algorithm, which is the natural setting for the Higgs mechanism along the Coulomb branch. 

However, a basic question remains: how does one derive the Higgs branch stratification directly from the (electric) theory? For concreteness, focus on unitary 3d $\Ncal=4$ quiver gauge theories\footnote{i.e.\ the gauge group factors are only of the type $\urm(n_i)$ and the matter content is restricted to bifundamental and adjoint hypermultiplets. The former are called edges, while the latter are loops. Both of which may have multiplicity.} --- this is precisely the setting where the Coulomb branch techniques are the most developed. For such quivers a variety of results exist in the literature. In mathematics, these Higgs branches are known as Nakajima quiver varieties \cite{Nakajima:1994nid,Nakajima:1998} and their stratification data has been partially established in \cite{Crawley-Boevey1,Crawley-Boevey2}. Specifically, the latter computes the quivers after the partial Higgs mechanism along the Higgs branch. In other words, this algorithm establishes the stratification, but does not provide the geometric information on the minimal degeneration (the minimal Higgs mechanism).\footnote{This missing information on transition types is subject of an upcoming work \cite{Travis_toappear}.}
In physics, minimal transitions along the Higgs branch have been studied in D5-D3-NS5 brane systems in \cite{Gaiotto:2013bwa,Cabrera:2016vvv,Gu:2022dac} and sometimes go under the name Kraft-Procesi transitions.\footnote{Similar types of Higgsing have appeared in other Type II brane systems \cite{Hayashi:2018bkd,Hayashi:2018lyv}.} Therein, hints appeared that one might be able to recast the Higgs branch transitions as subtractions according to a set of rules --- 
such rules are the main results presented in this work.
This is more than just a rewriting for a variety of reasons:
\begin{compactenum}
    \item  The Higgs branch subtraction algorithm offers the transition data at every step, allowing for a complete construction of the Higgs branch Hasse diagram and
    extending the algorithm of \cite{Crawley-Boevey1,Crawley-Boevey2} via physical reasoning.
    \item The Higgs branch subtraction algorithm naturally displays the partial order of the Hasse diagram, which is otherwise not always transparent.
    \item The global structure (monodromies, etc.) can be naturally extracted from the Higgs branch subtraction algorithm by utilising intuition from moduli spaces of instantons, symmetric products, and decorated magnetic quivers.
\end{compactenum}

The algorithm introduced in this paper further illuminates the role in gauge theory of various concepts from the mathematics literature. 
For instance, monodromies \cite{Generic_singularities,losev2024unipotent} in symplectic singularities are translated into \emph{decorations} on the quivers appearing in the Higgs branch subtraction.
Similarly, the so-called Namikawa-Weyl group \cite{Namikawa_2011,Namikawa_2010} is a discrete group acting on deformation/resolution parameters, such as FI-terms for the Higgs branch or masses for the Coulomb branch. The Higgs branch subtraction algorithm offers a simple way to extract the Namikawa-Weyl group.

The rules given in this paper lead to several new examples of theories whose Higgs and Coulomb branches arise as S{\l}odowy intersections in the nilpotent cones (nilcones) of simple Lie algebras. Nilcones are among the most studied symplectic singularities in the mathematics literature -- in some sense, their associated minimal nilpotent orbits act as units from which most other symplectic singularities encountered in physics can be constructed. Moreover, their Hasse diagrams, encoding precisely this singular stratification (alongside additional data such as the canonical quotient of Lusztig) are well known. Attempts to identify singular-symplectic properties of the moduli spaces of gauge theories with eight supercharges thus often begin with a variation on nilcone divination, and it is for this reason that examples of theories with Higgs/Coulomb branches as S{\l}odowy intersections are so useful. 

It is worth noting that the Nakajima quiver varieties studied in the mathematics literature are often constructed so as to avoid loops, which are known to introduce further complexities. From the physics perspective, loops on gauge nodes simply correspond to hypermultiplets in the adjoint representation of the gauge group, and as such the Higgs branch subtraction algorithm introduced in this work is equally as effective when applied to quivers with loops as to those without.

The remainder is organised as follows: in Section~\ref{sec_RHBQS} the rules for Higgs branch subtraction are introduced; these determine the local structure. The global structure of the Higgs branch is then subject of Section~\ref{sec_DQDS}. For the convenience of the reader, Section~\ref{sec:cheat_sheet} provides a summary of the rules for the Higgs branch subtraction algorithm.
Equipped with all relevant techniques, several examples are discussed in Section~\ref{sec_INSI} and serve as non-trivial consistency checks of the rules proposed here. Lastly, a summary and outlook is provided in Section~\ref{sec_outlook}.
The main body is supplemented by a number of appendices that cover background material.

%
\section{Incomplete Local Rules for Higgs Branch Quiver Subtraction}
\label{sec_RHBQS}
This section presents a set of \emph{local} rules delineating all possible quiver subtractions on the Higgs branch of the 3d $\mathcal{N}=4$ gauge theories considered here, along with justifications via explicit calculations using the classical Higgs mechanism. The subtraction slices themselves are the Higgs branches of minimal degenerate transverse theories; by sequential application of the relevant rules, the full local information of all minimal degenerate slices in the Higgs branch Hasse diagram can be specified. It is important to note here the distinction between the \emph{local} and \emph{global} structure of the Higgs branch, with the latter given a more exhaustive treatment in Section~\ref{sec_DQDS}; \emph{global} structure is sensitive to monodromies, which generically change the subtraction slice via the action of a discrete group. The local rules presented in this section are for this reason incomplete, and are included only as part of the derivation of the full global rules of Section~\ref{sec_DQDS}. A proof of completeness of the local rules is offered in the work of Gwyn Bellamy and Travis Schedler \cite{Travis_toappear}.

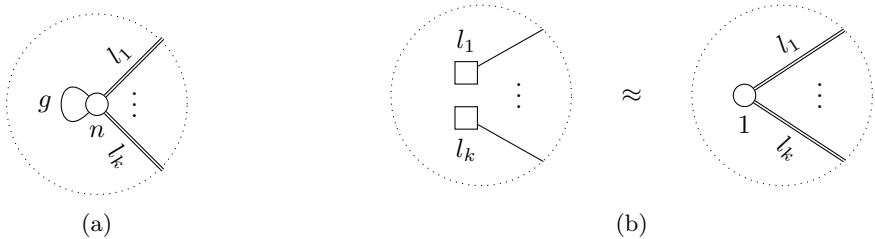
\begin{figure}
    \centering
   \begin{subfigure}[t]{0.30\textwidth}
   \centering
    \begin{tikzpicture}
        \node[gauge,label=below:{$n$}] (0) at (0,0) {};
        \draw (0) to [in=225,out=135,looseness=8]node[pos=0.5,left]{$g$} (0);
        \draw[dotted] (0) circle (1.2cm);
        \node (1) at (1,1) {};
        \node (2) at (1,-1) {};
        \draw[double] (0) to node[pos=0.5,above,sloped]{$l_1$} (1);
        \draw[double] (0) to node[pos=0.5,below,sloped]{$l_k$} (2);
        \node (3) at (0.5,0.1) {$\vdots$};
    \end{tikzpicture}
    \caption{}
    \label{fig_node}
   \end{subfigure}
      \begin{subfigure}[t]{0.60\textwidth}
      \centering
    \begin{tikzpicture}
    \node (a) at (0,0) {\begin{tikzpicture}
        \node[flavour,label=above:{$l_1$}] (0u) at (-.2,0.3) {};
        \node[flavour,label=below:{$l_k$}] (0d) at (-.2,-0.3) {};
        \draw[dotted] (0,0) circle (1.2cm);
        \node (1) at (1-.05,1-.05) {};
        \node (2) at (1-.05,-1+.05) {};
        \draw (0u)--(1);
        \draw (0d)--(2);
        \node (3) at (0.5,0.1) {$\vdots$};
    \end{tikzpicture}};
    \node (b) at (2,0) {$\approx$};
    \node (c) at (4,0) {\begin{tikzpicture}
        \node[gauge,label=below:{$1$}] (0) at (-.5,0) {};
        \draw[dotted] (0,0) circle (1.2cm);
        \node (1) at (1-.05,1-.05) {};
        \node (2) at (1-.05,-1+.05) {};
        \draw[double] (0) to node[pos=0.5,above,sloped]{$l_1$} (1);
        \draw[double] (0) to node[pos=0.5,below,sloped]{$l_k$} (2);
        \node (3) at (0.5,0.1) {$\vdots$};
    \end{tikzpicture}};
    \end{tikzpicture}
    \caption{}
    \label{fig_unframe}
   \end{subfigure}
    \caption{\subref{fig_node}: A $\urm(n)$ gauge node with $g$ loops ($g$ hypermultiplets in the adjoint representation of $\urm(n)$). This gauge node is in general connected to $k$ other gauge nodes with simply laced edges of multiplicity $l_1,\cdots,l_k$, respectively.
    \subref{fig_unframe}: The ``equivalence'' between framed and unframed quivers. The difference is a $T^*\C^\times$ factor on the Coulomb branch, which is invisible to the Hasse diagram as it is non-singular. The Higgs branch is unchanged.
    }
    \label{}
\end{figure}

 All 3d $\mathcal{N}=4$ quiver gauge theories considered in this work are unframed simply-laced quivers with gauge group $G=\prod_i \urm(n_i)/\urm(1)$; where each node in the quiver contributes a 3d $\mathcal{N}=4$ $\urm(n_i)$ vector multiplet. A simply-laced quiver is one in which all 3d $\mathcal{N}=4$ hypermultiplets (i.e.\ the quiver's edges) are in either the bi-fundamental or adjoint representation of the gauge group(s). Two gauge nodes connected by $k$ edges are said to be connected by an edge of multiplicity $k$. 
Graphically, each node takes the form as shown in \Figref{fig_node}.

Without loss of generality, framed quivers can be treated as unframed quivers by absorption of all flavour nodes in a single $\urm(1)$ gauge node, as in \Figref{fig_unframe}. This is called the ``Crawley-Boevey move'' in the mathematics literature \cite{Crawley-Boevey1}, and results in the Coulomb branch picking up a non-singular factor of $T^*\C^\times$ while leaving the Higgs branch unchanged.

\subsection{Local Rules for Minimal Higgsing}
\label{subsec_RMH}
Along the Higgs branch, the Higgs mechanism breaks the gauge group of the theory to a subgroup by giving non-zero VEVs to a gauge-invariant combination of the hypermultiplet scalars \cite{Englert:1964,Higgs:1964,Guralnik:1964,Kibble:1967}. All vector multiplets not in the adjoint representation of the unbroken residual gauge group become massive, absorbing hypermultiplet scalars in the same representation. In this paper all allowed unbroken subgroups are continuous. In general, the Higgsing pattern for a given theory can be quite complicated and so it is convenient to organise this into a Hasse diagram. The vertices of the Hasse diagram are labelled by the residual (massless) theories after Higgsing and the edges are labelled by the minimal transitions between the theories.

Suppose the original gauge group is $G$, the unbroken subgroup is $H$, and the (reduced) normaliser is $T=N_G(H)/H$. Then the transverse slice is the Higgs branch (up to normalisation) of the transverse theory with gauge group $T$ and hypermultiplets charged trivially under $H$ and non-trivially under $T$.

There are three types of local minimal transition, which are encoded in the following rules:
\begin{figure}[t]
\centering
\begin{subfigure}[t]{0.28\textwidth}
  \centering
    \begin{tikzpicture}
        \node (a) at (0,0) {$\raisebox{-.5 \height}{ \begin{tikzpicture}
            \node (1u) at (-0.5,0.1) {};
            \node (1d) at (-0.5,-0.1) {};
            \draw[dotted] (1,0) circle (1.5cm);
            \node[gauge,label=below:{$n-m$}] (2) at (1,-0.8) {};
            \node[gauge,label=left:{$m$}] (3) at (1,0.8) {};
            \draw (2) to [in=45,out=-45,looseness=8]node[pos=0.5,right]{$g$} (2);
            \draw (3) to [in=45,out=-45,looseness=8]node[pos=0.5,right]{$g$} (3);
            \draw[double] (3) to node[pos=0.5,above,sloped]{$2g\!-\!2$} (2);
            \draw[dashed] (3)--(1u) (2)--(1d);
        \end{tikzpicture}}$};
        \node at (-1,-2.5) {$c_g/m_g$};
        \node (b) at (0,-5) {$\raisebox{-.5 \height}{ \begin{tikzpicture}
            \node (1) at (-0.5,0) {};
            \draw[dotted] (1,0) circle (1.5cm);
            \node[gauge,label=below:{$n$}] (2) at (1,0) {};
            \draw (2) to [in=45,out=-45,looseness=8]node[pos=0.5,right]{$g$} (2);
            \draw[dashed] (1)--(2);
        \end{tikzpicture}}$};
        \draw[->] (b)--(a);
    \end{tikzpicture}
    \caption{Rule 1}
    \label{fig_cgmg}
\end{subfigure}
%
\begin{subfigure}[t]{0.40\textwidth}
\centering
\begin{tikzpicture}
        \node (a) at (0,0) {$\raisebox{-.5 \height}{ \begin{tikzpicture}
            \node (1l) at (0,0) {};
            \node (1r) at (4,0) {};
            \node (1lu) at (0,0.5) {};
            \node (1ru) at (4,0.5) {};
            \draw[dotted] (2,0.7) circle (2cm);
            \node[gauge,label=below:{$n_1\!-\!m$}] (2) at (1,0) {};
            \node[gauge,label=below:{$n_2\!-\!m$}] (3) at (3,0) {};
            \node[gauge,label=left:{$m$}] (4) at (2,1.5) {};
            \draw[double] (2)--(3)node[pos=0.5,above,sloped]{$g\!+\!1$};
            \draw[double] (2)--(4)node[pos=0.5,above,sloped]{$g\!-\!1$};
            \draw[double] (3)--(4)node[pos=0.5,above,sloped]{$g\!-\!1$};
            \draw (4) to [in=45,out=135,looseness=8]node[pos=0.5,above,sloped]{$g$}(4);
            \draw[dashed] (2)--(1l) (3)--(1r) (1lu)--(4)--(1ru);
        \end{tikzpicture}}$};
        \node at (-0.8,-3) {$a_g$};
        \node at (2,-3) {\begin{tikzpicture}
            \node[gauge,label=below:{$m$}] (2) at (1,0) {};
            \node[gauge,label=below:{$m$}] (3) at (3,0) {};
            \node[] (1) at (0.5,0) {$-$};
            \draw[double] (2)--(3)node[pos=0.5,above,sloped]{$g\!+\!1$};
        \end{tikzpicture}};
        \node (b) at (0,-6) {$\raisebox{-.5 \height}{ \begin{tikzpicture}
            \node (1l) at (0,0) {};
            \node (1r) at (4,0) {};
            \draw[dotted] (2,0) circle (2cm);
            \node[gauge,label=below:{$n_1$}] (2) at (1,0) {};
            \node[gauge,label=below:{$n_2$}] (3) at (3,0) {};
            \draw[dashed] (2)--(1l) (3)--(1r);
            \draw[double] (2)--(3)node[pos=0.5,above,sloped]{$g\!+\!1$};
        \end{tikzpicture}}$};
        \draw[->] (b)--(a);
\end{tikzpicture}
 \caption{Rule 2}
    \label{fig_ag}
\end{subfigure}
%
\begin{subfigure}[t]{0.28\textwidth}
 \centering
    \begin{tikzpicture}
        \node (a) at (0,0) {$\raisebox{-.5 \height}{ \begin{tikzpicture}
            \node (1u) at (-2,0.6) {};
            \node (1d) at (-2,0.4) {};
            \draw[dotted] (-0.2,0.5) circle (1.7cm);
            \node (2) at (0,0) {$(n\!-\!m)\times\mathsf{Q}_{ADE}$};
            \node[gauge,label=right:{$m$}] (3) at (-0.2,1) {};
            \draw (3) to [out=45,in=135,looseness=8] (3);
            \draw[dashed] (2)--(1d) (3)--(1u);
        \end{tikzpicture}}$};
        \node at (-1,-2.5) {$ADE$};
        \node at (1.5,-2.5) {\begin{tikzpicture}
            \node (1) at (1,0) {$-$};
            \node (2) at (2,0) {$m\times\mathsf{Q}_{ADE}$};
        \end{tikzpicture}};
        \node (b) at (0,-5) {$\raisebox{-.5 \height}{ \begin{tikzpicture}
            \node (1) at (-0.2,0) {};
            \draw[dotted] (1.5,0) circle (1.7cm);
            \node (2) at (2,0) {$n\times\mathsf{Q}_{ADE}$};
            \draw[dashed] (1)--(2);
        \end{tikzpicture}}$};
        \draw[->] (b)--(a);
    \end{tikzpicture}
     \caption{Rule 3}
    \label{fig_ADE}
\end{subfigure}    
    \caption{\subref{fig_cgmg}:
    \hyperref[fig_cgmg]{Rule 1} --- The $\urm(n)$ node with $g$ loops splits into two nodes of $\urm(m)$ and $\urm(n-m)$ with $g$ loops. Between them, there are $2g\!-\!2$ edges connecting the two nodes. The transverse slice is $c_g\cong\C^{2g}/\Z_2$ if $m=n-m$, and is $m_g$ (whose normalisation is $\C^{2g}$) if $m\neq n-m$. The dashed lines denote the possible edges connecting to the rest of the quiver.
    \subref{fig_ag}:
    \hyperref[fig_ag]{Rule 2} --- $\urm(n_1)$ and $\urm(n_2)$ nodes connected by $g+1$ edges splits into $\urm(n_1-m)$, $\urm(n_2-m)$, and $\urm(m)$ gauge nodes. The $\urm(n_1-m)$ and $\urm(n_2-m)$ gauge nodes have $g+1$ edges between them and $g-1$ edges extending to the $\urm(m)$ node, which has $g$ loops on it. This transition can be visualised as a subtraction of a quiver of two $\urm(m)$ nodes connected by $g\!+\!1$ edges contracting into a node of $\urm(m)$ with $g$ loops. The transverse slice is $a_g$.
    \subref{fig_ADE}:
    \hyperref[fig_ADE]{Rule 3} --- It can be visualised as the subtraction of $m$ copies of an $\mathsf{Q}_{ADE}$ sub-quiver, leaving a $\urm(m)$ node with a loop. The transverse slice is the corresponding $ADE$ Klein singularity.
    Note that the slice here denoted $m_1$ is more often in the quiver literature given the name $m$.
    }
    \label{fig:local_rules}
\end{figure}
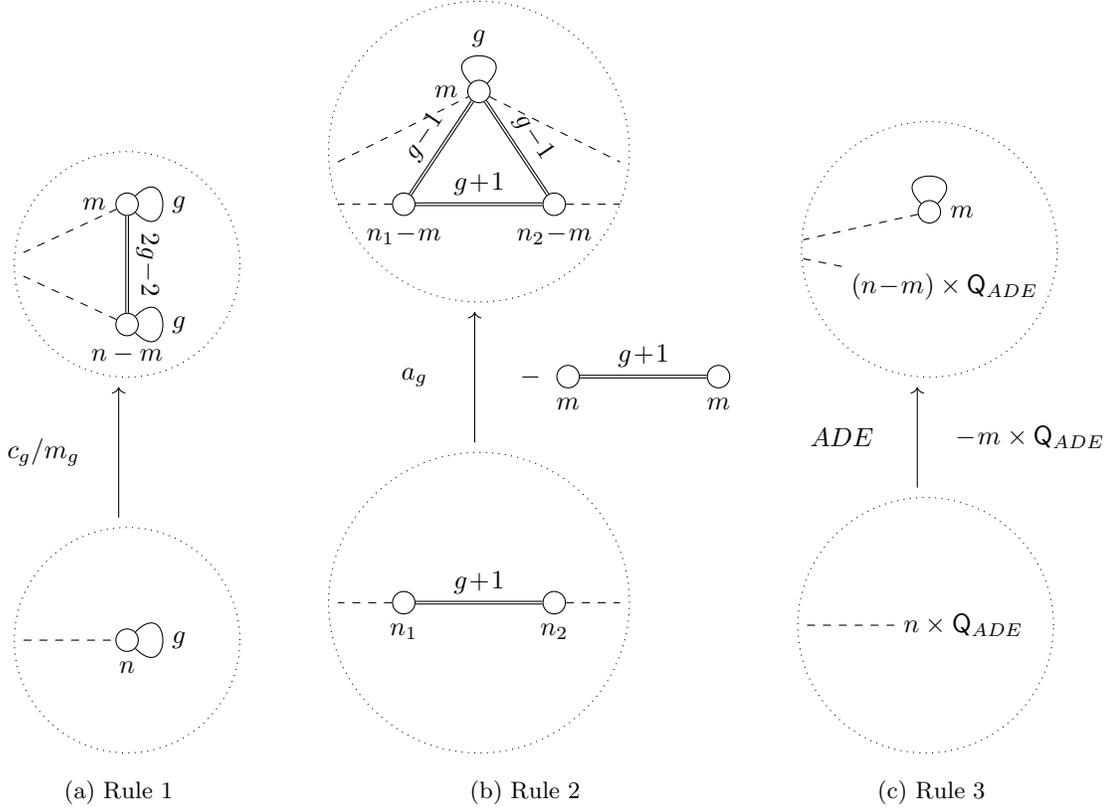

\subsubsection{\texorpdfstring{Local Rule 1: Adjoint Higgsing
}{Rule 1: Adjoint Higgsing}}
\label{sec_cg}

The local \hyperref[fig_cgmg]{Rule 1}, which is a consequence of the adjoint Higgs mechanism,
applies to a $\urm(n)$ gauge node with $g$ adjoint loops. As shown in \Figref{fig_cgmg}, the $\urm(n)$ node splits into a $\urm(n-m)$ and a $\urm(m)$ node (with integer $m$ such that $1\leq m\leq \frac{n}{2}$), each with $g$ adjoint loops and connected via $2g\!-\!2$ bi-fundamental hypermultiplets. For $n=2m$ the reduced normaliser is $T=S_2 \subset S_n$, where $S_n$ is the Weyl group of $\urm(n)$. Hence the transverse theory is an $S_2$ gauge theory with $g$ hypermultiplets in the sign representation of $S_2$. The Higgs branch of this transverse theory is $c_g\cong\C^{2g}/\Z_2$, which is the slice associated to the transition. For $n\neq m$, although the reduced normaliser is trivial the corresponding transverse gauge theory is not --- the slice is actually the non-normal variety $m_g$, with normalisation $\C^{2g}$. More details on the non-normal variety $m_g$ are provided below and in Appendix \ref{sec_mg}.

Note that repeated application of this rule to the node of rank $n$ in \Figref{fig_cgmg} recovers the complete graph quiver $\CG_{n,2g-2}$ \cite{Hanany:2023uzn} describing $n$ gauge nodes of $\urm(1)$ each with $g$ adjoint loops on each node and $2g\!-\!2$ edges between each pair.

The following describes a computation of this rule via the Higgs mechanism. The gauge group breaks as \begin{equation}
    \urm(n)\to\urm(n-m)\times\urm(m).
\end{equation}
After breaking, all vector multiplets not in the adjoint representation of the unbroken residual group become massive, by absorbing hypermultiplets in the same representation.
The adjoint representation decomposes as:
\begin{equation}
   \adj_{\urm(n)}\to \adj_{\urm(n-m)}+\adj_{\urm(m)}+
   \F_{\urm(n-m)}\times\bar{\F}_{\urm(m)}+\bar{\F}_{\urm(n-m)}\times\F_{\urm(m)}
   \label{eq:Rule1AdjDecomp}.
\end{equation}

This Higgsing pattern is incomplete, i.e.\ the moment map (in the adjoint representation of $\urm(n)$) has a trivial trace contribution. Each adjoint of $\urm(n)$ thus contributes a free sector $\C^2$, which is the trace of the $\urm(n)$ adjoint matrix together with its conjugate. After branching into $\urm(n-m)\times\urm(m)$ the free sector becomes $\C^2\times\C^2$ as the trace and conjugate of the $\urm(n-m)$ and $\urm(m)$ adjoint matrices. The free sectors are removed by requiring the traces to be zero. This is equivalent to using $\urm(n)\simeq\surm(n) \times \urm(1)$.
Hence, the transverse theory has a $2g$-dimensional Higgs branch, which is the difference between the singular sectors before and after breaking.

\paragraph{Transverse Slice $c_g/m_g$.}
Now consider the chiral ring of the slice. Note that the singlets live in the Cartan subalgebra and its conjugate, and hence the action of the gauge group $\urm(n)$ can be reduced to the action of the Weyl group $\Sorigin_n$, which permutes the diagonal elements. The singlets are invariant under the residual group $\urm(m)\times\urm(n-m)$, which is reduced to $\Sorigin_{m}\times\Sorigin_{n-m}$. The invariant $2g$ complex-dimensional sub vector space spanned by the singlets can be written as $V=\prod_i^g V_i$, where each $V_i$ is a 2 complex-dimensional subspace parameterised by the coordinates $(x_i, y_i)$ and takes the form (tracelessness is required to remove the free sector): 
\begin{equation}
V_i=(\underbrace{m x_i, m y_i, \cdots, m x_i, m y_i}_{n-m}, \underbrace{-(n-m) x_i, -(n-m) y_i, \cdots, -(n-m) x_i, -(n-m) y_i}_m).
\end{equation}
The invariant ring of $V$ under the action of $S_n$ gives exactly the ring of the non-normal slice $m_g$ for $m\neq n-m$ and the $c_g$ singularity for $m=n-m$:
\begin{equation}
V/\Sorigin_n=
\begin{cases}
m_g &\quad\text{when }m\neq n-m \,,\\
c_g &\quad\text{when }m=n-m \,.
\end{cases}
\label{eq_mgcg}
\end{equation}
Note that for $m=n-m$, the action of $S_n$ is reduced to the action of the normaliser $T=S_2$.

\subsubsection{\texorpdfstring{Local Rule 2: Bi-fundamental Higgsing 
}{Local Rule 2: Bi-fundamental Higgsing }}
\label{sec_ag}
\hyperref[fig_ag]{Rule 2}, which is a consequence of the bi-fundamental Higgs mechanism, applies when two nodes $\urm(n_1)$ and $\urm(n_2)$ are connected by $g+1$ edges. Here the edge parametrisation is such that after the transition  $g$ loops remain, as shown in \Figref{fig_ag}. 
%
The transition involves adding an additional $\urm(m)$ node for any integer $m$ in the range $1\leq m\leq \min(n_1,n_2)$. The ranks of the $\urm(n_1)$ and $\urm(n_2)$ gauge nodes are reduced by $m$ each with $g\!-\!1$ hypermultiplets connecting them to the $\urm(m)$ gauge node. The requirement that the multiplicities are non-negative implies that $g\geq 1$ for a transition to happen. The transverse slice is the $a_g$ singularity.

To demonstrate this, the Higgsing pattern is considered. The gauge group breaks as:
\begin{equation}
    \urm(n_1)\times\urm(n_2)\to\urm(n_1-m)\times\urm(n_2-m)\times\urm(m)
\end{equation}
and the relevant representations decompose as follows:
\begin{align}
\adj_{\urm(n_i)} \to & \adj_{\urm(n_i-m)}+\adj_{\urm(m)} 
+\F_{\urm(n_i-m)}\times\bar{\F}_{\urm(m)}+\bar{\F}_{\urm(n_i-m)}\times\F_{\urm(m)} \,,
\label{eq:Rule2AdjDecomp}
\\
\F_{\urm(n_1)}\times\bar{\F}_{\urm(n_2)} \to & \F_{\urm(n_1-m)}\times\bar{\F}_{\urm(n_2-m)} 
+\F_{\urm(n_1-m)}\times\bar{\F}_{\urm(m)} 
+\bar{\F}_{\urm(n_2-m)}\times\F_{\urm(m)} +\adj_{\urm(m)}
\,.
\label{eq:Rule2FundDecomp} 
\end{align}

\paragraph{Transverse Slice $a_g$.}
The transverse slice is determined as follows. For the same reason as in Section \ref{sec_cg}, each adjoint of the residual $\urm(m)$ contributes a singlet which enters the transverse theory as matter content. The singlets have charge $1$ under $T=\urm(1)$.
Hence, the transverse theory is SQED with $g\!+\!1$ flavours, whose Higgs branch is $a_g$.

\paragraph{Local Rule 2 with Adjoint Loops.}
\label{Par:Local_Rule_2_with_Adjoint_Loops}
There exists a subtlety in the application of \hyperref[fig_ag]{Rule 2} to subquivers with adjoint loops. 
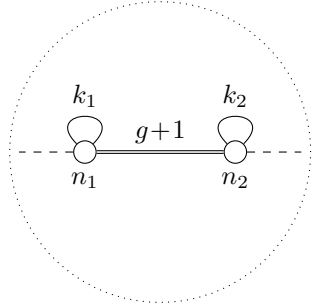
\begin{figure}[h!]
    \centering
    \begin{tikzpicture}
        \node (1l) at (0,0) {};
        \node (1r) at (4,0) {};
        \draw[dotted] (2,0) circle (2cm);
        \node[gauge,label=below:{$n_1$}] (2) at (1,0) {};
        \node[gauge,label=below:{$n_2$}] (3) at (3,0) {};
        \draw[dashed] (2)--(1l) (3)--(1r);
        \draw[double] (2)--(3)node[pos=0.5,above,sloped]{$g\!+\!1$};
        \draw (2) to [out=45,in=135,looseness=8]node[pos=0.5,above,sloped]{$k_1$} (2);
        \draw (3) to [out=45,in=135,looseness=8]node[pos=0.5,above,sloped]{$k_2$} (3);
    \end{tikzpicture}
    \caption{For this quiver, either \hyperref[fig_cgmg]{Rule 1} or \hyperref[fig_ag]{2} are minimal transitions depending on the values of $n_{1,2}$, $k_{1,2}$. For $n_1 >n_2$ and $k_1\neq 0$, $k_2 =0$, \hyperref[fig_cgmg]{Rule 1} is minimal, while \hyperref[fig_ag]{Rule 2} is not.
    For $n_1 >n_2$ and $k_1 =0$, $k_2\neq 0$,  \hyperref[fig_ag]{Rule 2} is minimal for $m=\min(n_1,n_2)$.}
    \label{fig:Rule2_Adjoint_Loops}
\end{figure}
If the node with higher rank has adjoint loops (e.g.\ take $n_1 > n_2$ in \Figref{fig:Rule2_Adjoint_Loops} with the requirement $k_1 \neq 0$, $k_2 = 0$) then \hyperref[fig_ag]{Rule 2} is \emph{not} a minimal transition, and \hyperref[fig_cgmg]{Rule 1} is.
If the node with lower (or equal) rank has adjoint loops (e.g.\ take $n_1 > n_2$ in \Figref{fig:Rule2_Adjoint_Loops} with  $k_1 = 0$, $k_2 \neq 0$), then \hyperref[fig_ag]{Rule 2} is minimal \emph{only} with precisely $m=\min(n_1,n_2)$, instead of any $1\leq m\leq \min(n_1,n_2)$, i.e, only a subtraction of the maximum number of subquivers is minimal. The number of adjoint loops on the $\urm(m)$ nodes are also inherited by the new $\urm(m)$ node after contracting --- this is the non-Abelian analog of bullet point~\ref{para_loopu1} below. See \Figref{fig_23withadjoints} for an example. Otherwise, if $m<\min(n_1,n_2)$, the transition can be treated as a combination of \hyperref[fig_cgmg]{Rule 1} and \hyperref[fig_ag]{Rule 2}, which is not a minimal transition. For this non-minimal transition, an example is shown in \Figref{fig_nonminimala}.

\subsubsection{\texorpdfstring{Local Rule 3: $ADE$ Higgsing 
}{Rule 3: ADE Higgsing}}
\label{sec_ADE}
\hyperref[fig_ADE]{Rule 3} applies when there is an affine $ADE$-shaped sub-quiver with each node's rank no less than $n$ times the dual Coxeter label of the affine $ADE$ Dynkin diagram, i.e, it contains $n$ copies of $\mathsf{Q}_{ADE}$.\footnote{Here the label $\mathsf{Q}_{ADE}$ is used to denote a minimal balanced affine $ADE$ quiver. The $ADE$ affine Dynkin quivers with their dual Coxeter labels are shown in Table \ref{tab_ADEaffinedynkins}.} The transition is shown in \Figref{fig_ADE}.

 This rule can be derived from the Higgs mechanism as follows. Given a quiver $\mathcal{Q}$, suppose that it contains a sub-quiver of the form of some rank $k$ ADE finite quiver. Let the gauge group coming from this sub-quiver be a gauge subgroup $G=\urm(n_0)\times\cdots\times\urm(n_{k})$ with $n\leq\min(n_0/h_0^\vee,\cdots,n_{k}/h_k^\vee) < n+1$, where $h^\vee_i$ are the dual Coxeter labels for the rank $k$ ADE group.\footnote{In other words, the maximal balanced rank $k$ affine $ADE$ quiver contained in $\mathcal{Q}$ has $n$ times the rank of the minimal balanced $\mathsf{Q}_{ADE}$.} The aim is to Higgs $1\leq m \leq n$ copies of this $ADE$ sub-quiver in the sense of \Figref{fig_ADE}.  If the subquiver to be subtracted has ranks $h^\vee_i$ at the $i$-th node, then the gauge groups breaks as 
\begin{equation}
\urm(n_0)\times\cdots\times\urm(n_{k})\to\urm\left(n_0-mh^\vee_1\right)\times\cdots\times\urm\left(n_{k}-mh^\vee_{k}\right)\times\urm(m)
\end{equation}
and the adjoint representations decompose as follows:
\begin{align}
\adj_{\urm\left(n_i\right)}&\to  \adj_{\urm\left(n_i-mh^\vee_i\right)} \label{higgsADEv}
+(h^\vee_i)^2 \cdot \adj_{\urm(m)}  \\ 
&+ h^\vee_i\left(\F_{\urm\left(n_i-mh^\vee_i\right)}\times \overline{\F}_{\urm(m)} + \F_{\urm(m)}\times\overline{\F}_{\urm\left(n_i-mh^\vee_i\right)}\right)
\notag
\end{align}
and the bifundamental hypermultiplets (whenever there exits one between node $i$ and $j$) decomposes
\begin{align}
\F_{\urm(n_i)}\times \overline{\F}_{\urm(n_j)}&\to 
\F_{\urm(n_i-mh^\vee_i)}\times \overline{\F}_{\urm(n_j-mh^\vee_{j})} +
h^\vee_ih^\vee_j \cdot \adj_{\urm(m)}
\notag\\&
+
h^\vee_j\cdot\F_{\urm\left(n_i-mh^\vee_i\right)}\times \overline{\F}_{\urm(m)} + h^\vee_i\cdot\F_{\urm(m)}\times\overline{\F}_{\urm\left(n_j-mh^\vee_j\right)} \,.
\label{higgsADEh}
\end{align}
Let $c_{ij}$ be the Cartan matrix for the affine Dynkin diagram. It has a zero eigenvector given by $h^\vee$ such that $ \sum_{j=0}^k c_{ij} h_j^\vee =0$. The conventions are such that the index $j=0$ denotes the affine node for which $h_0^\vee =1$. 
To verify the consistent Higgsing, count the multiplicities. To begin with, consider $\adj_{\urm(m)}$ for which one finds $
\frac{1}{2}\sum_{i,j=0}^k h^\vee_i (2\delta_{ij}- c_{ij}) h^\vee_j
-\sum_{i=0}^{k}(h_i^\vee)^2=0
$ from the hypermultiplets minus vector multiplets. As the Higgsing pattern makes one $\adj_{\urm(m)}$ survive, one vector multiplet and one adjoint hypermultiplet are present.
Next, consider the bifundamental $\F_{\urm\left(n_i-mh^\vee_i\right)}\times \overline{\F}_{\urm(m)}$ for which the multiplicity reads
$\frac{1}{2}\sum_{j=0}^{k} (2\delta_{ij}-c_{ij}) h_j^\vee -  h_i^\vee=0$; again from hypermultiplets minus vector multiplets.
In conclusion, the Higgsing reduces the quiver precisely to the one shown in \Figref{fig_ADE}.

\paragraph{Transverse Slice $ADE$.}
All of the singlets that are not charged under $H=\urm\left( n_{0} -mh_{0}^{\lor }\right) \times \cdots \times \urm\left( n_{k} -mh_{k}^{\lor }\right) \times \urm( m)/\urm(1)$ form the quiver data of $\mathsf{Q}_{ADE}$.
The vector multiplet decomposition \eqref{higgsADEv} yields 
$ \sum_{j=0}^k (h_j^\vee)^2-1$ singlets (after taking away the vectormultiplets of the residual theory), which is precisely the dimension of the $T=\prod_{i=0}^k \urm(h_i^\vee)/\urm(1)$ vector multiplets.
Analogously, the hypermultiplet decomposition \eqref{higgsADEh} yields $\frac{1}{2} h_i^\vee (2 \delta_{ij} - c_{ij})h_j^\vee $, which is in the form of the matter content of the $\mathsf{Q}_{ADE}$ quiver. 
Hence, the transverse theory is the $\mathsf{Q}_{ADE}$ quiver, whose Higgs branch is the $ADE$ singularity.

\begin{table}[t]
\ra{1.5}
    \centering
    \begin{tabular}{ccc}
    \toprule
         Label  &  Affine Dynkin Diagrams  &   HS\\ \midrule
         
         $A_n$ &
         \raisebox{-0.5\height}{\begin{tikzpicture}[x=1cm,y=.8cm]
            \node (g8) at (-1,0) [gauge,label=below:{$1$}] {};
            \node (g5) at (0,0) {$\cdots$};
            \node (g6) at (1,0) [gauge,label=below:{$1$}] {};
            \node (g7) at (0,1) [gauge,label=above:{$1$}] {};
            \draw (g8)--(g5)--(g6)--(g7)--(g8);
            \draw [decorate,decoration={brace,mirror,amplitude=6pt}] (-1,-0.8) --node[below=6pt] {$n$} (1,-0.8);
    \end{tikzpicture}} &  $\frac{1-t^{2n+2}}{(1-t^2)(1-t^{n+1})^2}$ \\ \midrule

        $\begin{array}{c}
	   D_n\\
	   n\geq 4
	   \end{array}$ &
        \raisebox{-0.5\height}{\begin{tikzpicture}[x=1cm,y=.8cm]
            \node (g2) at (-2,0) [gauge,label=below:{$1$}] {};
            \node (g3) at (-1,0) [gauge,label=below:{$2$}] {};
            \node (g4) at (0,0) {$\cdots$};
            \node (g5) at (1,0) [gauge,label=below:{$2$}] {};
            \node (g12) at (2,0) [gauge,label=below:{$1$}] {};
            \node (g13) at (1,1) [gauge,label=above:{$1$}] {};
            \node (g11) at (-1,1) [gauge,label=above:{$1$}] {};
            \draw (g2)--(g3)--(g4)--(g5)--(g12);
            \draw (g13)--(g5);
            \draw (g11)--(g3);
            \draw [decorate,decoration={brace,mirror,amplitude=6pt}] (-1,-0.8) --node[below=6pt] {$n-3$} (1,-0.8);
    \end{tikzpicture}} &  $\frac{1-t^{4n-4}}{(1-t^4)(1-t^{2n-4})(1-t^{2n-2})}$ \\ \midrule

         $E_6$ &  \raisebox{-0.5\height}{\begin{tikzpicture}[x=1cm,y=.8cm]
            \node (g2) at (-2,0) [gauge,label=below:{$1$}] {};
            \node (g3) at (-1,0) [gauge,label=below:{$2$}] {};
            \node (g4) at (0,0) [gauge,label=below:{$3$}] {};
            \node (g5) at (1,0) [gauge,label=below:{$2$}] {};
            \node (g12) at (2,0) [gauge,label=below:{$1$}] {};
            \node (g10) at (0,1) [gauge,label=right:{$2$}] {};
            \node (g11) at (0,2) [gauge,label=right:{$1$}] {};
            \draw (g2)--(g3)--(g4)--(g5)--(g12);
            \draw (g4)--(g10)--(g11);
    \end{tikzpicture}}&  $\frac{1-t^{24}}{(1-t^6)(1-t^8)(1-t^{12})}$ \\ \midrule
    
         $E_7$ & \raisebox{-0.5\height}{\begin{tikzpicture}[x=1cm,y=.8cm]
            \node (g3) at (-3,0) [gauge,label=below:{$1$}] {};
            \node (g4) at (-2,0) [gauge,label=below:{$2$}] {};
            \node (g13) at (-1,0) [gauge,label=below:{$3$}] {};
            \node (g5) at (0,0) [gauge,label=below:{$4$}] {};
            \node (g12) at (1,0) [gauge,label=below:{$3$}] {};
            \node (g6) at (2,0) [gauge,label=below:{$2$}] {};
            \node (g10) at (3,0) [gauge,label=below:{$1$}] {};
            \node (g7) at (0,1) [gauge,label=above:{$2$}] {};
            \draw (g3)--(g4)--(g13)--(g5)--(g12)--(g6)--(g10);
            \draw (g7)--(g5);
    \end{tikzpicture}}&  $\frac{1-t^{36}}{(1-t^8)(1-t^{12})(1-t^{18})}$ \\ \midrule
         $E_8$ & \raisebox{-0.5\height}{\begin{tikzpicture}[x=1cm,y=.8cm]
            \node (g3) at (-3.5,0) [gauge,label=below:{$1$}] {};
            \node (g4) at (-2.5,0) [gauge,label=below:{$2$}] {};
            \node (g5) at (-1.5,0) [gauge,label=below:{$3$}] {};
            \node (g6) at (-0.5,0) [gauge,label=below:{$4$}] {};
            \node (g7) at (0.5,0) [gauge,label=below:{$5$}] {};
            \node (g8) at (1.5,0) [gauge,label=below:{$6$}] {};
            \node (g9) at (2.5,0) [gauge,label=below:{$4$}] {};
            \node (g10) at (3.5,0) [gauge,label=below:{$2$}] {};
            \node (g11) at (1.5,1) [gauge,label=above:{$3$}] {};
            \draw (g3)--(g4)--(g5)--(g6)--(g7)--(g8)--(g9)--(g10);
            \draw (g8)--(g11);
    \end{tikzpicture}}&  $\frac{1-t^{60}}{(1-t^{12})(1-t^{20})(1-t^{30})}$ \\ \bottomrule
    \end{tabular}
    \caption{Affine Dynkin diagrams with dual Coxeter labels $h_i^\vee$. Note that $h_0^\vee =1$ for the  affine node. Subtraction of these quivers gives $ADE$ surface singularities listed in the first column, alongside their Higgs branch Hilbert series in the third.}
    \label{tab_ADEaffinedynkins}
\end{table}

\paragraph{Local Rule 3 with Adjoint Loops.}
Similar to the case of local \hyperref[fig_ag]{Rule 2}, local \hyperref[fig_ADE]{Rule 3} presents a subtlety when applied to affine $ADE$ Dynkin subquivers with adjoint loops. If there are adjoint loops on the node(s) of the affine $ADE$ Dynkin subquiver that have dual Coxeter number larger than $1$, then \hyperref[fig_ADE]{Rule 3} does \emph{not} give rise to minimal transition. In this case, \hyperref[fig_cgmg]{Rule 1} must be applied first.

If there are adjoint loops on the node(s) of the subquiver that have dual Coxeter number $1$, and all such node(s) have rank exactly $n$, then \hyperref[fig_ADE]{Rule 3} is minimal only for $m=n$. This means that only $m=n$ copies of the $\mathsf{Q}_{ADE}$ subquiver can be subtracted, instead of any $1\leq m\leq n$. The slice is of course the corresponding $ADE$ singularity. The number of adjoint loops on the $\urm(m)$ nodes are also inherited by the new $\urm(m)$ node after contracting --- this is the non-Abelian analog of Remark \ref{para_loopu1} below. See \Figref{fig_2d4withadj} for an example. Otherwise, if $m<n$, the transition can be treated as a combination of \hyperref[fig_cgmg]{Rule 1} and \hyperref[fig_ADE]{Rule 3}, which is not a minimal transition. For this non-minimal transition, an example is shown in \Figref{fig_nonminimalb}.

\subsubsection{Remarks}
A few comments are in order.
\begin{enumerate}
    \item \textbf{Branching of External Edges.}
The three rules above omit the discussion of the decomposition of the external edges connecting from the background to the subquiver. A simple principle to keep in mind is that all those edges are inherited by the new node after contracting, which is equivalent to the requirement of re-balancing.

\item \textbf{Number of Loops.}
\label{para_loopu1}
The adjoint loops on the $\urm(1)$ nodes are also inherited by the new node after contracting. These loops contribute to the free sector. At each leaf, the local geometry can be written as: $\Mcal_{\text{Sing.}}\times\mathbb{H}^{l}$, where $l$ is the number of $\urm(1)$ adjoint loops. The dimension of the singular and free sectors must sum to the Higgs branch dimension, i.e, $\mathrm{dim}_\mathbb{H}\Mcal_{\text{Sing.}}+l=\mathrm{dim}_\mathbb{H}\Hcal$.

\item \textbf{Degeneracy of Rules.}
\label{para_degenrules}
When $g=1$, \hyperref[fig_cgmg]{Rule 1} can be treated as the $A_0$ limit of \hyperref[fig_ADE]{Rule 3} and \hyperref[fig_ag]{Rule 2} can be treated as the $A_1$ limit of \hyperref[fig_ADE]{Rule 3}.

\item \textbf{Bad Theories and Incomplete Higgsing}
\label{subsec_BTIH}
The algorithm presented in this paper is equally effective at determining the singular structure of the Higgs branch of a theory regardless of whether it is \emph{good}, \emph{bad}, or \emph{ugly} in the sense of \cite{Gaiotto:2008ak}. In any case, repeated application of the algorithm reaches a quiver that cannot be operated on by the three above rules. This quiver is hence a top leaf on the Higgs branch, which is a non-singular phase.
\\Incomplete Higgsing corresponds to the case where the final quiver at the top leaf is \emph{bad}. Note that for an unframed quiver this condition should be relaxed to the following: if the final quiver of the top leaf is still bad after ungauging a $\urm(1)$, then there is incomplete Higgsing.
\\It is a simple exercise to construct a family of quivers with the same Higgs branch by adding a quiver with trivial Higgs branch on top of an original quiver. The Coulomb branches of the theories in this family also have the same singular structure, but differ in the smooth factor \cite{Assel:2017jgo}.
\end{enumerate}

\FloatBarrier
\subsection{Examples of Local Subtraction}
\label{subsec_Ex123}
This section presents simple examples of the rules introduced above. 
\paragraph{Disclaimer.} The examples in this section are only applications of the (incomplete) local rules, and not the full global rules introduced in Section~\ref{sec_DQDS} (in other words, they do not contain decorations); they are simply presented as instances of the local rules above, without consideration for global phenomena such as monodromies. The slices in the examples here have been chosen such that they are not altered by monodromies, however those considered later on in this work do in some cases significantly differ from the na\"ive local viewpoint, as explained in Section~\ref{sec_DQDS}.
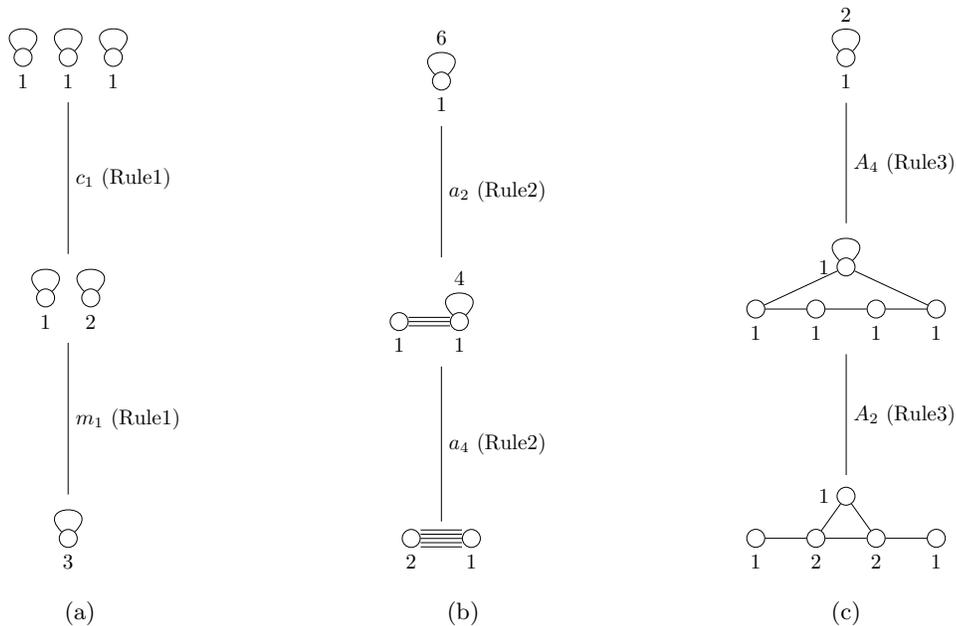
\begin{figure}[H]
    \centering
       \begin{subfigure}{0.32\textwidth}
       \centering
       \scalebox{.8}{
       \begin{tikzpicture}
        \node (a) at (0,0){$\begin{tikzpicture}

        \node[gauge, label=below:$3$] (2) at (1,0){};

        \draw[-] (2) to[out=45,in=135,looseness=8] (2);
        \end{tikzpicture}$};
        
        \node (b) at (0,4){$\begin{tikzpicture}
        \node[gauge, label=below:$1$] (1) at (0,0){};
        \node[gauge, label=below:$2$] (2) at (0.75,0){};
        \draw[-] (1) to[out=45,in=135,looseness=8](1){};
        \draw[-] (2) to[out=45,in=135,looseness=8](2){};
        
        \end{tikzpicture}$};
        \node (c) at (0,8){$\begin{tikzpicture}
        \node[gauge, label=below:$1$] (1) at (0,0){};
        \node[gauge, label=below:$1$] (2) at (0.75,0){};
        \node[gauge, label=below:$1$] (3) at (1.5,0){};
        \draw[-] (1) to[out=45,in=135,looseness=8](1){};
        \draw[-] (2) to[out=45,in=135,looseness=8](2){};
        \draw[-] (3) to[out=45,in=135,looseness=8](3){};
        \end{tikzpicture}$};
        \draw[-] (a)--(b) node[pos=0.5,midway, right]{$m_1\ (\text{Rule} 1)$}--(c) node[pos=0.5,midway,right]{$c_1\ (\text{Rule} 1)$};
        \end{tikzpicture}}
        \caption{}
        \label{fig_Hasseu3adj}
        \end{subfigure}
        \begin{subfigure}{0.32\textwidth}
        \centering
        \scalebox{.8}{
        \begin{tikzpicture}
            \node (a) at (0,0){$\begin{tikzpicture}
            \node[gauge, label=below:$2$] (1) at (0,0){};
            \node[gauge, label=below:$1$] (2) at (1,0){};
            \draw[] (1)--(2);
            \draw[transform canvas={yshift=-2pt}] (1)--(2);
            \draw[transform canvas={yshift=2pt}] (1)--(2);
            \draw[transform canvas={yshift=-4pt}] (1)--(2);
            \draw[transform canvas={yshift=4pt}] (1)--(2);
            \end{tikzpicture}$};
            \node (b) at (0,4){$\begin{tikzpicture}
            \node[gauge, label=below:$1$] (1) at (0,0){};
            \node[gauge, label=below:$1$] (2) at (1,0){};
            \draw[] (1)--(2);
            \draw[transform canvas={yshift=-2pt}] (1)--(2);
            \draw[transform canvas={yshift=2pt}] (1)--(2);
            \draw[-] (2) to[out=45,in=135,looseness=8] node[pos=0.5,above]{4}(2){};
            \end{tikzpicture}$};
            \node (c) at (0,8){$\begin{tikzpicture}
            \node[gauge, label=below:$1$] (1) at (0,0){};
            \draw[-] (1) to[out=45,in=135,looseness=8] node[pos=0.5,above]{6}(1){};
            \end{tikzpicture}$};
            \draw[-] (a)--(b) node[pos=0.5,midway, right]{$a_4\ (\text{Rule} 2)$}--(c) node[pos=0.5,midway,right]{$a_2\ (\text{Rule} 2)$};
    \end{tikzpicture}}
    \caption{}
    \label{fig_Hasseu25}
       \end{subfigure}
       \begin{subfigure}{0.32\textwidth}
       \centering
           \scalebox{.8}{
        \begin{tikzpicture}
        \node (a) at (0,0){$\begin{tikzpicture}

        \node[gauge, label=below:$1$] (1) at (-1,0){};
        \node[gauge, label=below:$2$] (2) at (0,0){};
        \node[gauge, label=below:$2$] (3) at (1,0){};
        \node[gauge, label=below:$1$] (4) at (2,0){};
        \node[gauge, label=left:$1$] (5) at (0.5,0.7){};

        \draw[] (1)--(2)--(3)--(4);
        \draw[] (2)--(5)--(3);
        
        \end{tikzpicture}$};
        
        \node (b) at (0,4){$\begin{tikzpicture}
        \node[gauge, label=below:$1$] (1) at (-1,0){};
        \node[gauge, label=below:$1$] (2) at (0,0){};
        \node[gauge, label=below:$1$] (3) at (1,0){};
        \node[gauge, label=below:$1$] (4) at (2,0){};
        \node[gauge, label=left:$1$] (5) at (0.5,0.7){};

        \draw[] (1)--(2)--(3)--(4)--(5)--(1);
        \draw[-] (5) to[out=45,in=135,looseness=8] (5){};
        
        \end{tikzpicture}$};
        \node (c) at (0,8){$\begin{tikzpicture}
        \node[gauge, label=below:$1$] (1) at (0,0){};
        
        \draw[-] (1) to[out=45,in=135,looseness=8] node[pos=0.5,above]{2}(1){};
        \end{tikzpicture}$};
        \draw[-] (a)--(b) node[pos=0.5,midway, right]{$A_2\ (\text{Rule} 3)$}--(c) node[pos=0.5,midway,right]{$A_4\ (\text{Rule} 3)$};
    \end{tikzpicture}}
    \caption{}
    \label{fig_Hassea4a2}
       \end{subfigure}
\caption{(\subref{fig_Hasseu3adj}): Hasse diagram of $\urm(3)$ with an adjoint loop, recovered by applying \hyperref[fig_cgmg]{Rule 1} twice. (\subref{fig_Hasseu25}): Hasse diagram of $\urm(2)$ with $5$ flavours (unframed), recovered by applying \hyperref[fig_ag]{Rule 2} twice. (\subref{fig_Hassea4a2}): Hasse diagram of the (unframed) electric quiver for the S{\l}odowy slice $\Scal^{A_4}_{(5),(3,1^2)}$ in the $A_4$ nilcone, recovered by applying \hyperref[fig_ADE]{Rule 3} twice. 
Note the addition of adjoint loops on $\urm(1)$ nodes as explained in Paragraph \ref{para_loopu1}.}
\label{fig_Hasseex1}
\end{figure}

In \Figref{fig_Hasseex1} the examples each involve one rule only. 
The theory in \Figref{fig_Hasseu3adj} is Higgsed to its Levi subgroup $\urm(1)^3$, which is an incomplete Higgsing. The theories in \Figref{fig_Hasseu25} and \Figref{fig_Hassea4a2} can be treated as complete Higgsing after factoring out the centre-of-mass $\urm(1)$.
\begin{figure}[H]
    \centering
    
       \begin{subfigure}{0.4\textwidth}
       \centering
       \scalebox{.8}{
        \begin{tikzpicture}
            \node (a) at (0,0){$\begin{tikzpicture}
            \node[gauge, label=below:$3$] (1) at (0,0){};
            \node[gauge, label=below:$1$] (2) at (1,0){};
            \draw[] (1)--(2);
            \draw[transform canvas={yshift=-2pt}] (1)--(2);
            \draw[transform canvas={yshift=2pt}] (1)--(2);
            \draw[transform canvas={yshift=-4pt}] (1)--(2);
            \draw[transform canvas={yshift=4pt}] (1)--(2);
            \end{tikzpicture}$};
            \node (b) at (0,4){$\begin{tikzpicture}
            \node[gauge, label=below:$2$] (1) at (0,0){};
            \node[gauge, label=below:$1$] (2) at (1,0){};
            \draw[] (1)--(2);
            \draw[transform canvas={yshift=-2pt}] (1)--(2);
            \draw[transform canvas={yshift=2pt}] (1)--(2);
            \draw[-] (2) to[out=45,in=135,looseness=8] node[pos=0.5,above]{4}(2){};
            \end{tikzpicture}$};
            \node (c) at (0,8){$\begin{tikzpicture}
            \node[gauge, label=below:$1$] (1) at (0,0){};
            \node[gauge, label=below:$1$] (2) at (-1,0){};
            \draw[-] (1) to[out=45,in=135,looseness=8] node[pos=0.5,above]{6}(1){};
            \draw[-] (1)--(2) ;
            \end{tikzpicture}$};
            \draw[-] (a)--(b) node[pos=0.5,midway, right]{$a_4\ (\text{Rule} 2)$}--(c) node[pos=0.5,midway,right]{$a_2\ (\text{Rule} 2)$};
    \end{tikzpicture}}
        \caption{}
        \label{fig_Hasseu35}
        \end{subfigure}
        \begin{subfigure}{0.4\textwidth}
        \centering
        \scalebox{.8}{
        \begin{tikzpicture}
            \node (a) at (0,0){$\begin{tikzpicture}
            \node[gauge, label=below:$n$] (1) at (0,0){};
            \node[gauge, label=below:$1$] (2) at (1,0){};
            \draw[] (1)--(2);
            \draw[transform canvas={yshift=-2pt}] (1)--(2);
            \draw[transform canvas={yshift=2pt}] (1)--(2);
            \draw[transform canvas={yshift=-4pt}] (1)--(2);
            \draw[transform canvas={yshift=4pt}] (1)--(2);
            \end{tikzpicture}$};
            \node (b) at (0,4){$\begin{tikzpicture}
            \node[gauge, label=below:$n-1$] (1) at (0,0){};
            \node[gauge, label=below:$1$] (2) at (1,0){};
            \draw[] (1)--(2);
            \draw[transform canvas={yshift=-2pt}] (1)--(2);
            \draw[transform canvas={yshift=2pt}] (1)--(2);
            \draw[-] (2) to[out=45,in=135,looseness=8] node[pos=0.5,above]{4}(2){};
            \end{tikzpicture}$};
            \node (c) at (0,8){$\begin{tikzpicture}
            \node[gauge, label=below:$1$] (1) at (0,0){};
            \node[gauge, label=below:$n-2$] (2) at (-1,0){};
            \draw[-] (1) to[out=45,in=135,looseness=8] node[pos=0.5,above]{6}(1){};
            \draw[-] (1)--(2);
            \end{tikzpicture}$};
            \draw[-] (a)--(b) node[pos=0.5,midway, right]{$a_4\ (\text{Rule} 2)$}--(c) node[pos=0.5,midway,right]{$a_2\ (\text{Rule} 2)$};
    \end{tikzpicture}}
    \caption{}
    \label{fig_Hasseun5}
       \end{subfigure}
\caption{
(\subref{fig_Hasseu35}): Hasse diagram of $\urm(3)$ with $5$ flavours (unframed) showing incomplete Higgsing. (\subref{fig_Hasseun5}): Hasse diagram of $\urm(n)$ with $5$ flavours (unframed), where $n>3$ gives incomplete Higgsing.}
\label{fig_Hasseex2}
\end{figure}
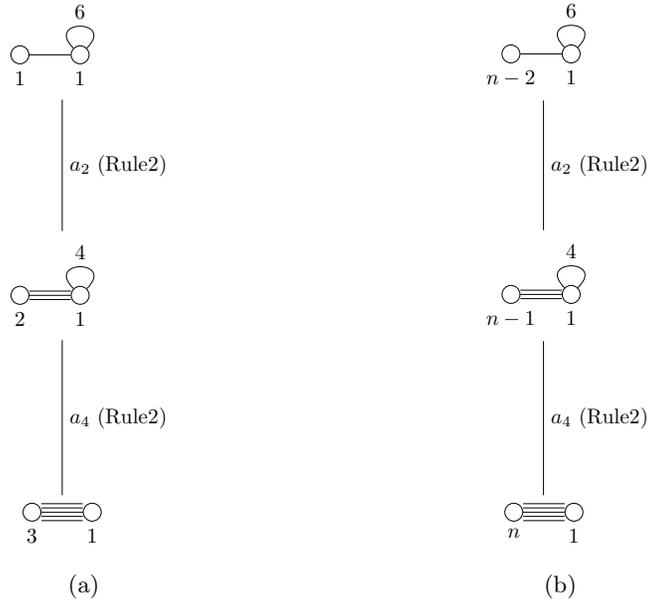

\Figref{fig_Hasseu35} and \Figref{fig_Hasseun5} are extensions of \Figref{fig_Hasseu25} --- they have the same Higgs branch, shown from repeated application of Rule~\hyperref[fig_ag]{2}. Each leaf in \Figref{fig_Hasseu35} is ugly, while each leaf in \Figref{fig_Hasseun5} is bad for $n>3$, implying incomplete Higgsing.

\begin{figure}[H]
    \centering
        \begin{subfigure}{0.45\textwidth}
       \centering
       \scalebox{.8}{
        \begin{tikzpicture}
        \node (a) at (0,0){$\begin{tikzpicture}

        \node[gauge, label=below:$2$] (1) at (0,0){};
        \node[gauge, label=below:$2$] (2) at (1,0){};

        \draw[transform canvas={yshift=-2pt}] (1)--(2);
        \draw[transform canvas={yshift=2pt}] (1)--(2);
        \end{tikzpicture}$};
        
        \node (b) at (-3,4){$\begin{tikzpicture}
        \node[gauge, label=below:$1$] (1) at (0,0){};
        \node[gauge, label=below:$1$] (2) at (1,0){};
        \node[gauge, label=below:$1$] (3) at (1,1){};

        \draw[-] (3) to[out=45,in=135,looseness=8] (3);
        \draw[transform canvas={yshift=-2pt}] (1)--(2);
        \draw[transform canvas={yshift=2pt}] (1)--(2);
        \end{tikzpicture}$};
        \node (c) at (3,4){$\begin{tikzpicture}
            \node[gauge,label=below:$2$] (1) at (0,0){};
            \draw[-] (1) to[out=45,in=135,looseness=8] (1);
        \end{tikzpicture}$};
        \node (d) at (0,8){$\begin{tikzpicture}
            \node[gauge,label=below:$1$] (1) at (0,0){};
            \node[gauge,label=below:$1$] (2) at (1,0){};
            \draw[-] (1) to[out=45,in=135,looseness=8] (1);
            \draw[-] (2) to[out=45,in=135,looseness=8] (2);
        \end{tikzpicture}$};
        \draw[-] (a)--(c) node[pos=0.5,midway, right]{$A_1\ (\text{Rule}\;2/\text{Rule}\;3)$}--(d) node[pos=0.5,midway, right]{$A_1\ (\text{Rule} 1)$};
        \draw[-] (a)--(b) node[pos=0.5,midway, left]{$A_1\ (\text{Rule}\;2/\text{Rule}\;3)$}--(d) node[pos=0.5,midway, left]{$A_1\ (\text{Rule}\;2/\text{Rule}\;3)$};
    \end{tikzpicture}}
        \caption{}
        \label{fig_Hasse22}
        \end{subfigure}
        \begin{subfigure}{0.45\textwidth}
       \centering
       \scalebox{.8}{
        \begin{tikzpicture}
        \node (a) at (0,0){$\begin{tikzpicture}

        \node[gauge, label=below:$2$] (1) at (0,0){};
        \node[gauge, label=below:$2$] (2) at (1,0){};
        \node[gauge, label=below:$1$] (0) at (-1,0){};

        \draw[transform canvas={yshift=-2pt}] (1)--(2);
        \draw[transform canvas={yshift=2pt}] (1)--(2);
        \draw[] (0)--(1);
        \end{tikzpicture}$};
        
        \node (b) at (-3,4){$\begin{tikzpicture}
        \node[gauge, label=below:$1$] (1) at (0,0){};
        \node[gauge, label=below:$1$] (2) at (1,0){};
        \node[gauge, label=below:$1$] (3) at (0,1){};
        \node[gauge, label=below:$1$] (0) at (-1,0.5){};

        \draw[-] (3) to[out=45,in=135,looseness=8] (3);
        \draw[transform canvas={yshift=-2pt}] (1)--(2);
        \draw[transform canvas={yshift=2pt}] (1)--(2);
        \draw[] (0)--(1) (0)--(3);
        \end{tikzpicture}$};
        \node (c) at (3,4){$\begin{tikzpicture}
            \node[gauge,label=below:$2$] (1) at (0,0){};
            \node[gauge,label=below:$1$] (0) at (-1,0){};
            \draw[-] (1) to[out=45,in=135,looseness=8] (1);
            \draw[] (0)--(1);
        \end{tikzpicture}$};
        \node (d) at (0,8){$\begin{tikzpicture}
            \node[gauge,label=below:$1$] (1) at (0,0){};
            \node[gauge,label=right:$1$] (2) at (0,1){};
            \node[gauge,label=below:$1$] (0) at (-1,0.5){};
            \draw[-] (1) to[out=45,in=135,looseness=8] (1);
            \draw[-] (2) to[out=45,in=135,looseness=8] (2);
            \draw[] (0)--(1) (0)--(2);
        \end{tikzpicture}$};
        \draw[-] (a)--(c) node[pos=0.5,midway, right]{$A_1\ (\text{Rule}\;2/\text{Rule}\;3)$}--(d) node[pos=0.5,midway, right]{$A_1\ (\text{Rule} 1)$};
        \draw[-] (a)--(b) node[pos=0.5,midway, left]{$A_1\ (\text{Rule}\;2/\text{Rule}\;3)$}--(d) node[pos=0.5,midway, left]{$A_1\ (\text{Rule}\;2/\text{Rule}\;3)$};
    \end{tikzpicture}}
        \caption{}
        \label{fig_Hasse122}
        \end{subfigure}
    \caption{(\subref{fig_Hasse22}): The affine $A_1$ quiver with doubled node numbers. The Higgs branch is the $2^{\text{nd}}$ symmetric product of the $A_1$ singularity. (\subref{fig_Hasse122}): This quiver is provided in \cite{de_Boer_1997,Cremonesi_2014} as the $3d$ mirror of the ADHM quiver. Both quivers in (\subref{fig_Hasse22}) and (\subref{fig_Hasse122}) have the same Higgs branch and the same singular part of the Coulomb branch. Note that (\subref{fig_Hasse22}) has incomplete Higgsing, while (\subref{fig_Hasse122}) can be completely Higgsed.}
    \label{fig_Hasse2A1}
\end{figure}

\Figref{fig_Hasse2A1} gives two examples of quivers with the Higgs branch as the $2^{\text{nd}}$ symmetric product of the $A_1$ singularity, which is an orbifold of $\C^4$ by the dihedral group $Dih_4$ of order $8$. Especially \Figref{fig_Hasse22} is the case that all the nodes are balanced but the theory has incomplete Higgsing.

\begin{figure}[H]
    \centering
    \begin{subfigure}{0.45\textwidth}
        \centering
        \scalebox{.85}{
\begin{tikzpicture}
        \node (a) at (0,0) {$\begin{tikzpicture}
            \node (1) [gauge, label=below:$3$] at (0,0) {};
            \node (2) [gauge, label=below:$2$] at (1,0) {};
            \draw[-] (2) to[out=45,in=315,looseness=8] (2);
            \draw[transform canvas={yshift=-2pt}] (1)--(2);
            \draw[transform canvas={yshift=0pt}] (1)--(2);
            \draw[transform canvas={yshift=2pt}] (1)--(2);
        \end{tikzpicture}$};
        \node (b) at (-2,4) {$\begin{tikzpicture}
            \node (1) [gauge, label=below:$1$] at (0,0) {};
            \node (2) [gauge, label=below:$2$] at (1,0) {};
            \draw[-] (2) to[out=315,in=45,looseness=8] node[pos=0.5,right]{$3$} (2);
            \draw[-] (1)--(2);
        \end{tikzpicture}$};
        \draw[-] (a)--(b) node[pos=0.5,midway, left]{$a_2\ (\text{Rule}\ 2)$};
            \node (c) at (2,4) {$\begin{tikzpicture}
            \node (1) [gauge, label=below:$3$] at (0,0) {};
            \node (2) [gauge, label=below:$1$] at (1,0) {};
            \draw[-] (2) to[out=45,in=315,looseness=8] (2);
            \node (3) [gauge, label=left:$1$] at (0,1) {};
            \draw[-] (3) to[out=45,in=135,looseness=8] (3);
            \draw[transform canvas={yshift=-2pt}] (1)--(2);
            \draw[transform canvas={yshift=0pt}] (1)--(2);
            \draw[transform canvas={yshift=2pt}] (1)--(2);
            \draw[transform canvas={xshift=2pt}] (1)--(3);
            \draw[transform canvas={xshift=0pt}] (1)--(3);
            \draw[transform canvas={xshift=-2pt}] (1)--(3);
        \end{tikzpicture}$};
        \draw[-] (a)--(c) node[pos=0.5,midway, right]{$c_1\ (\text{Rule}\ 1)$};
        \draw[loosely dotted,ultra thick] (0,5)--(0,6.1);
        \end{tikzpicture}}
        \caption{}
        \label{fig_23withadjointsc}
    \end{subfigure}
    \begin{subfigure}{0.45\textwidth}
        \centering
        \scalebox{.85}{
\begin{tikzpicture}
        \node (a) at (0,0) {$\begin{tikzpicture}
            \node (1) [gauge, label=below:$3$] at (0,0) {};
            \node (2) [gauge, label=below:$2$] at (1,0) {};
            \draw[-] (1) to[out=135,in=225,looseness=8] (1);
            \draw[-] (2) to[out=45,in=315,looseness=8] (2);
            \draw[transform canvas={yshift=-2pt}] (1)--(2);
            \draw[transform canvas={yshift=0pt}] (1)--(2);
            \draw[transform canvas={yshift=2pt}] (1)--(2);
        \end{tikzpicture}$};
        \node (b) at (-2,4) {$\begin{tikzpicture}
            \node (1) [gauge, label=below:$2$] at (0,0) {};
            \node (2) [gauge, label=below:$2$] at (-1,0) {};
            \draw[-] (2) to[out=135,in=225,looseness=8] (2);
            \node (3) [gauge, label=left:$1$] at (0,1) {};
            \draw[-] (3) to[out=45,in=135,looseness=8] (3);
            \draw[transform canvas={yshift=-2pt}] (1)--(2);
            \draw[transform canvas={yshift=0pt}] (1)--(2);
            \draw[transform canvas={yshift=2pt}] (1)--(2);
            \draw[transform canvas={xshift=2pt}] (1)--(3);
            \draw[transform canvas={xshift=0pt}] (1)--(3);
            \draw[transform canvas={xshift=-2pt}] (1)--(3);
        \end{tikzpicture}$};
        \node (c) at (2,4) {$\begin{tikzpicture}
            \node (1) [gauge, label=below:$3$] at (0,0) {};
            \draw[-] (1) to[out=135,in=225,looseness=8] (1);
            \node (2) [gauge, label=below:$1$] at (1,0) {};
            \draw[-] (2) to[out=45,in=315,looseness=8] (2);
            \node (3) [gauge, label=left:$1$] at (0,1) {};
            \draw[-] (3) to[out=45,in=135,looseness=8] (3);
            \draw[transform canvas={yshift=-2pt}] (1)--(2);
            \draw[transform canvas={yshift=0pt}] (1)--(2);
            \draw[transform canvas={yshift=2pt}] (1)--(2);
            \draw[transform canvas={xshift=2pt}] (1)--(3);
            \draw[transform canvas={xshift=0pt}] (1)--(3);
            \draw[transform canvas={xshift=-2pt}] (1)--(3);
        \end{tikzpicture}$};
        \draw[-] (a)--(b) node[pos=0.5,midway, left]{$m_1\ (\text{Rule}\ 1)$};
        \draw[-] (a)--(c) node[pos=0.5,midway, right]{$c_1\ (\text{Rule}\ 1)$};
        \draw[loosely dotted,ultra thick] (0,5)--(0,6.1);
        \end{tikzpicture}}
        \caption{}
        \label{fig_23withadjointsd}
    \end{subfigure}
     \caption{Examples of \hyperref[fig_ag]{Rule 2} involving loops. In (\subref{fig_23withadjointsc}), the loop is on the node with lower rank - \hyperref[fig_ag]{Rule 2} can be applied with $m=2$. In (\subref{fig_23withadjointsd}), the loop on the higher rank node means that \hyperref[fig_ag]{Rule 2} no longer gives a minimal transition.}
    \label{fig_23withadjoints}
\end{figure}

\Figref{fig_23withadjoints} gives examples of \hyperref[fig_ag]{Rule 2} involving one loop. For brevity, only bottom slices are considered. In \Figref{fig_23withadjointsc}, \hyperref[fig_ag]{Rule 2} can be applied only with $m=2$, in contrast with \Figref{fig_23withadjointsd}, where the node with higher rank has an adjoint loop and so \hyperref[fig_ag]{Rule 2} cannot be minimal transition before \hyperref[fig_cgmg]{Rule 1} is applied.

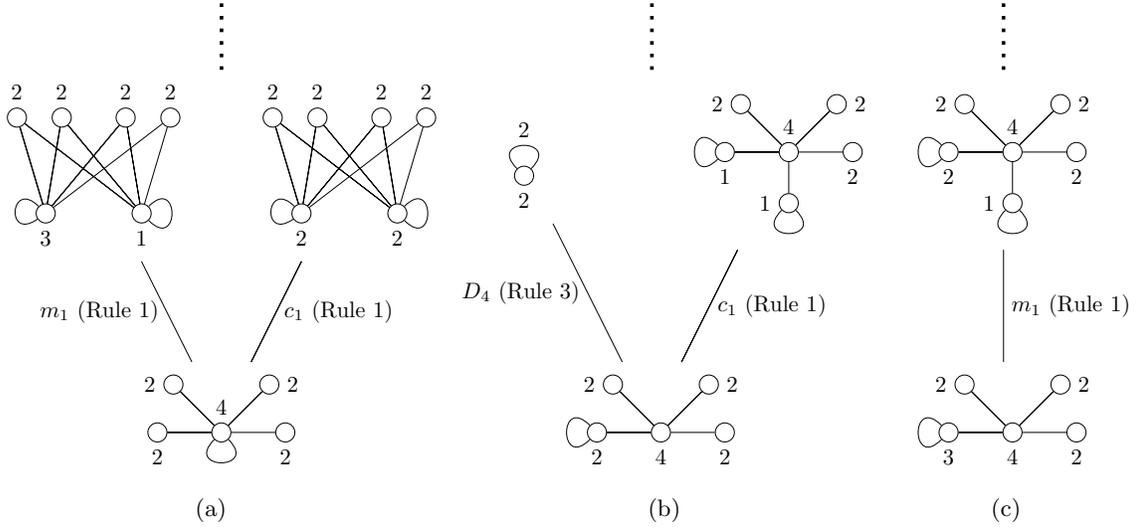
\begin{figure}[H]
    \centering
    \begin{subfigure}{0.38\textwidth}
        \centering
        \scalebox{.85}{
\begin{tikzpicture}
        \node (a) at (0,0) {$\begin{tikzpicture}
    \node (1) [gauge, label=above:$4$] at (0,0) {};
    \node (2) [gauge, label=below:$2$] at (1,0) {};
    \node (3) [gauge, label=below:$2$] at (-1,0) {};
    \node (4) [gauge, label=right:$2$] at (0.75,0.75) {};
    \node (5) [gauge, label=left:$2$] at (-0.75,0.75) {};
    \draw[-] (2)--(1)--(3)--(1)--(4)--(1)--(5)--(1);
    \draw[-] (1) to[out=-45,in=-135,looseness=8] (1);
    \end{tikzpicture}$};
    \node (b) at (2,4) {$\begin{tikzpicture}
    \node (0) [gauge, label=below:$2$] at (-0.75,0) {};
    \node (1) [gauge, label=below:$2$] at (0.75,0) {};
    \node (2) [gauge, label=above:$2$] at (1.2,1.5) {};
    \node (3) [gauge, label=above:$2$] at (0.5,1.5) {};
    \node (4) [gauge, label=above:$2$] at (-0.5,1.5) {};
    \node (5) [gauge, label=above:$2$] at (-1.2,1.5) {};
    \draw[-] (2)--(1)--(3)--(1)--(4)--(1)--(5)--(1);
    \draw[-] (2)--(0)--(3)--(0)--(4)--(0)--(5)--(0);
    \draw[-] (0) to[out=135,in=215,looseness=8] (0);
    \draw[-] (1) to[out=45,in=-45,looseness=8] (1);
    \end{tikzpicture}$};
    \node (c) at (-2,4) {$\begin{tikzpicture}
    \node (0) [gauge, label=below:$3$] at (-0.75,0) {};
    \node (1) [gauge, label=below:$1$] at (0.75,0) {};
    \node (2) [gauge, label=above:$2$] at (1.2,1.5) {};
    \node (3) [gauge, label=above:$2$] at (0.5,1.5) {};
    \node (4) [gauge, label=above:$2$] at (-0.5,1.5) {};
    \node (5) [gauge, label=above:$2$] at (-1.2,1.5) {};
    \draw[-] (2)--(1)--(3)--(1)--(4)--(1)--(5)--(1);
    \draw[-] (2)--(0)--(3)--(0)--(4)--(0)--(5)--(0);
    \draw[-] (0) to[out=135,in=215,looseness=8] (0);
    \draw[-] (1) to[out=45,in=-45,looseness=8] (1);
    \end{tikzpicture}$};
    \draw[-] (a) -- (b) node[pos=0.5,midway, right]{$c_1\ (\text{Rule}\ 1)$} -- (a) -- (c) node[pos=0.5,midway, left]{$m_1\ (\text{Rule}\ 1)$};
    \draw[loosely dotted,ultra thick] (0,5.5)--(0,6.6);
        \end{tikzpicture}}
        \caption{}
        \label{fig_2d4withadja}
    \end{subfigure}
    \begin{subfigure}{0.38\textwidth}
        \centering
        \scalebox{.85}{
\begin{tikzpicture}
        \node (a) at (0,0) {$\begin{tikzpicture}
    \node (1) [gauge, label=below:$4$] at (0,0) {};
    \node (2) [gauge, label=below:$2$] at (1,0) {};
    \node (3) [gauge, label=below:$2$] at (-1,0) {};
    \node (4) [gauge, label=right:$2$] at (0.75,0.75) {};
    \node (5) [gauge, label=left:$2$] at (-0.75,0.75) {};
    \draw[-] (2)--(1)--(3)--(1)--(4)--(1)--(5)--(1);
    \draw[-] (3) to[out=225,in=135,looseness=8] (3);
    \end{tikzpicture}$};
    \node (b) at (2,4) {$\begin{tikzpicture}
    \node (1) [gauge, label=above:$4$] at (0,0) {};
    \node (2) [gauge, label=below:$2$] at (1,0) {};
    \node (3) [gauge, label=below:$1$] at (-1,0) {};
    \node (4) [gauge, label=right:$2$] at (0.75,0.75) {};
    \node (5) [gauge, label=left:$2$] at (-0.75,0.75) {};
    \node (6) [gauge, label=left:$1$] at (0,-0.8) {};
    \draw[-] (2)--(1)--(3)--(1)--(4)--(1)--(5)--(1)--(6);
    \draw[-] (3) to[out=225,in=135,looseness=8] (3);
    \draw[-] (6) to[out=225,in=315,looseness=8] (6);
    \end{tikzpicture}$};
    \node (c) at (-2,4) {$\begin{tikzpicture}
    \node (0) [gauge, label=below:$2$] at (0,0) {};
    \draw[-] (0) to[out=135,in=45,looseness=8] node[pos=0.5,above]{$2$} (0);
    \end{tikzpicture}$};
    \draw[-] (a) -- (b) node[pos=0.5,midway, right]{$c_1\ (\text{Rule}\ 1)$} -- (a) -- (c) node[pos=0.5,midway, left]{$D_4\ (\text{Rule}\ 3)$};
    \draw[loosely dotted,ultra thick] (0,5.5)--(0,6.6);
        \end{tikzpicture}}
        \caption{}
        \label{fig_2d4withadjb}
    \end{subfigure}
        \begin{subfigure}{0.19\textwidth}
        \centering
        \scalebox{.85}{
\begin{tikzpicture}
        \node (a) at (0,0) {$\begin{tikzpicture}
    \node (1) [gauge, label=below:$4$] at (0,0) {};
    \node (2) [gauge, label=below:$2$] at (1,0) {};
    \node (3) [gauge, label=below:$3$] at (-1,0) {};
    \node (4) [gauge, label=right:$2$] at (0.75,0.75) {};
    \node (5) [gauge, label=left:$2$] at (-0.75,0.75) {};
    \draw[-] (2)--(1)--(3)--(1)--(4)--(1)--(5)--(1);
    \draw[-] (3) to[out=225,in=135,looseness=8] (3);
    \end{tikzpicture}$};
    \node (b) at (0,4) {$\begin{tikzpicture}
    \node (1) [gauge, label=above:$4$] at (0,0) {};
    \node (2) [gauge, label=below:$2$] at (1,0) {};
    \node (3) [gauge, label=below:$2$] at (-1,0) {};
    \node (4) [gauge, label=right:$2$] at (0.75,0.75) {};
    \node (5) [gauge, label=left:$2$] at (-0.75,0.75) {};
    \node (6) [gauge, label=left:$1$] at (0,-0.8) {};
    \draw[-] (2)--(1)--(3)--(1)--(4)--(1)--(5)--(1)--(6);
    \draw[-] (3) to[out=225,in=135,looseness=8] (3);
    \draw[-] (6) to[out=225,in=315,looseness=8] (6);
    \end{tikzpicture}$};
    \draw[-] (a) -- (b) node[pos=0.5,midway, right]{$m_1\ (\text{Rule}\ 1)$};
    \draw[loosely dotted,ultra thick] (0,5.5)--(0,6.6);
        \end{tikzpicture}}
        \caption{}
        \label{fig_2d4withadjc}
    \end{subfigure}
    \caption{Examples of quivers containing at least the affine $D_4$ quiver with node numbers doubled, i.e, at least $2$ copies of $\mathsf{Q}_{D_4}$. An additional loop is added. Recall that for $\mathsf{Q}_{D_4}$, the dual Coxeter labels are 2 for the middle node and 1 for each of the other nodes. In (\subref{fig_2d4withadja}), the loop is on the node with label $2$ and \hyperref[fig_ADE]{Rule 3} is not minimal transition. In (\subref{fig_2d4withadjb}), the loop is on one of the nodes with label $1$, also the rank of the node is exactly $2$ and \hyperref[fig_ADE]{Rule 3} can be applied by subtracting exactly $2$ copies of $\mathsf{Q}_{D_4}$. In (\subref{fig_2d4withadjc}), although the loop is on a node with dual Coxeter label $1$ in the $\mathsf{Q}_{D_4}$ quiver, the node is of rank-3 which does not coincide with the number of copies (2) of $\mathsf{Q}_{D_4}$ present. \hyperref[fig_ADE]{Rule 3} is thus not a minimal transition.}
    \label{fig_2d4withadj}
\end{figure}

In \Figref{fig_2d4withadj}, examples of \hyperref[fig_ADE]{Rule 3} involving a loop are shown. In \Figref{fig_2d4withadja} and \Figref{fig_2d4withadjc}, \hyperref[fig_ADE]{Rule 3} is not a minimal transition. In \Figref{fig_2d4withadjb}, \hyperref[fig_ADE]{Rule 3} with $m=2$ is a minimal transition.

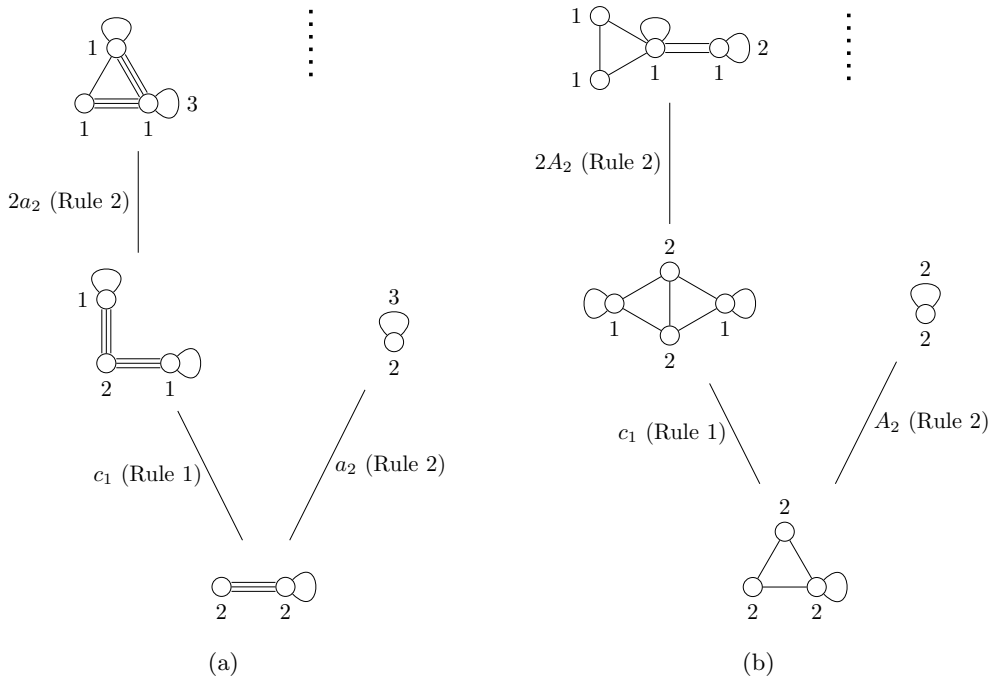
\begin{figure}[H]
    \centering
    \begin{subfigure}{0.45\textwidth}
        \centering
        \scalebox{.85}{
\begin{tikzpicture}
        \node (a) at (0,0) {$\begin{tikzpicture}
            \node (1) [gauge, label=below:$2$] at (0,0) {};
            \node (2) [gauge, label=below:$2$] at (1,0) {};
            \draw[-] (2) to[out=45,in=315,looseness=8] (2);
            \draw[transform canvas={yshift=-2pt}] (1)--(2);
            \draw[transform canvas={yshift=0pt}] (1)--(2);
            \draw[transform canvas={yshift=2pt}] (1)--(2);
        \end{tikzpicture}$};
        \node (b) at (-2,4) {$\begin{tikzpicture}
            \node (1) [gauge, label=below:$2$] at (0,0) {};
            \node (2) [gauge, label=below:$1$] at (1,0) {};
            \draw[-] (2) to[out=45,in=315,looseness=8] (2);
            \node (3) [gauge, label=left:$1$] at (0,1) {};
            \draw[-] (3) to[out=45,in=135,looseness=8] (3);
            \draw[transform canvas={yshift=-2pt}] (1)--(2);
            \draw[transform canvas={yshift=0pt}] (1)--(2);
            \draw[transform canvas={yshift=2pt}] (1)--(2);
            \draw[transform canvas={xshift=2pt}] (1)--(3);
            \draw[transform canvas={xshift=0pt}] (1)--(3);
            \draw[transform canvas={xshift=-2pt}] (1)--(3);
        \end{tikzpicture}$};
        \node (c) at (-2,8) {$\begin{tikzpicture}
            \node (1) [gauge, label=below:$1$] at (0,0) {};
            \node (2) [gauge, label=below:$1$] at (1,0) {};
            \node (3) [gauge, label=left:$1$] at (0.5,0.866) {};
            \draw[-] (2) to[out=315,in=45,looseness=8] node[pos=0.5,right]{$3$} (2);
            \draw[-] (3) to[out=135,in=45,looseness=8] (3);
            \draw[-] (1)--(3);
            \draw[transform canvas={yshift=-2pt}] (1)--(2);
            \draw[transform canvas={yshift=0pt}] (1)--(2);
            \draw[transform canvas={yshift=2pt}] (1)--(2);
            \draw[transform canvas={xshift=1.732pt,yshift=1pt}] (3)--(2);
            \draw[transform canvas={yshift=0pt}] (3)--(2);
            \draw[transform canvas={xshift=-1.732pt,yshift=-1pt}] (3)--(2);
        \end{tikzpicture}$};
        \node (d) at (2,4) {$\begin{tikzpicture}
            \node (1) [gauge, label=below:$2$] at (0,0) {};
            \draw[-] (1) to[out=135,in=45,looseness=8] node[pos=0.5,above]{$3$} (1);
        \end{tikzpicture}$};
        \draw[-] (a)--(b) node[pos=0.5,midway, left]{$c_1\ (\text{Rule}\ 1)$};
        \draw[-] (a)--(d) node[pos=0.5,midway, right]{$a_2\ (\text{Rule}\ 2)$};
        \draw[-] (b)--(c) node[pos=0.5,midway, left]{$2a_2\ (\text{Rule}\ 2)$};
        \draw[loosely dotted,ultra thick] (0.7,8)--(0.7,9.1);
        \end{tikzpicture}}
        \caption{}
        \label{fig_nonminimala}
    \end{subfigure}
    \begin{subfigure}{0.45\textwidth}
        \centering
        \scalebox{.85}{
\begin{tikzpicture}
        \node (a) at (0,0) {$\begin{tikzpicture}
            \node (1) [gauge, label=below:$2$] at (0,0) {};
            \node (2) [gauge, label=below:$2$] at (1,0) {};
            \node (3) [gauge, label=above:$2$] at (0.5,0.866) {};
            \draw[-] (2) to[out=45,in=315,looseness=8] (2);
            \draw [-] (1)--(2)--(3)--(1);
        \end{tikzpicture}$};
        \node (b) at (-2,4) {$\begin{tikzpicture}
            \node (1) [gauge, label=below:$1$] at (0,0) {};
            \node (2) [gauge, label=below:$1$] at (1.732,0) {};
            \node (3) [gauge, label=above:$2$] at (0.866,0.5) {};
            \node (4) [gauge, label=below:$2$] at (0.866,-0.5) {};
            \draw[-] (1)--(3)--(4)--(1);
            \draw[-] (3)--(2)--(4);
            \draw[-] (2) to[out=45,in=315,looseness=8] (2);
            \draw[-] (1) to[out=135,in=225,looseness=8] (1);
        \end{tikzpicture}$};
        \node (c) at (-2,8) {$\begin{tikzpicture}
            \node (1) [gauge, label=below:$1$] at (0,0) {};
            \node (2) [gauge, label=below:$1$] at (1,0) {};
            \node (3) [gauge, label=left:$1$] at (-0.866,0.5) {};
            \node (4) [gauge, label=left:$1$] at (-0.866,-0.5) {};
            \draw[-] (1)--(3)--(4)--(1);
            \draw[transform canvas={yshift=-2pt}] (1)--(2);
            \draw[transform canvas={yshift=2pt}] (1)--(2);
            \draw[-] (2) to[out=315,in=45,looseness=8] node[pos=0.5,right]{$2$} (2);
            \draw[-] (1) to[out=135,in=45,looseness=8] (1);
        \end{tikzpicture}$};
        \node (d) at (2,4) {$\begin{tikzpicture}
            \node (1) [gauge, label=below:$2$] at (0,0) {};
            \draw[-] (1) to[out=135,in=45,looseness=8] node[pos=0.5,above]{$2$} (1);
        \end{tikzpicture}$};
        \draw[-] (a)--(b) node[pos=0.5,midway, left]{$c_1\ (\text{Rule}\ 1)$};
        \draw[-] (a)--(d) node[pos=0.5,midway, right]{$A_2\ (\text{Rule}\ 2)$};
        \draw[-] (b)--(c) node[pos=0.5,midway, left]{$2A_2\ (\text{Rule}\ 2)$};
        \draw[loosely dotted,ultra thick] (0.8,7.5)--(0.8,8.6);
        \end{tikzpicture}}
        \caption{}
        \label{fig_nonminimalb}
    \end{subfigure}
    \caption{Examples of Rules \hyperref[fig_ag]{2} and \hyperref[fig_ADE]{3} involving a loop. If Rules \hyperref[fig_ag]{2} and \hyperref[fig_ADE]{3} are applied, only the maximum number of subtractions give rise to a minimal transition. Subtracting a non-maximal number of sub-diagrams is a non-minimal transition, and arises as the use of \hyperref[fig_cgmg]{Rule 1} followed by Rules \hyperref[fig_ag]{2}/\hyperref[fig_ADE]{3}. The appearance of two unions of cones in the transition is explained in Section \ref{sec_MMU}. In the figure, only the relevant part of the Hasse diagram is shown for brevity.}
    \label{fig_nonminimal}
\end{figure}

\Figref{fig_nonminimal}, shows why only the maximal number of subtractions is possible for minimal Higgsing; otherwise, it is equivalent to using \hyperref[fig_cgmg]{Rule 1} to split the node and then using Rules \hyperref[fig_ag]{2} or \hyperref[fig_ADE]{3} to do the subtraction (as shown in the left direction in the Hasse diagrams).

\section{Full Global Rules and Decoration}
\label{sec_DQDS}

In Section~\ref{sec_RHBQS} a set of rules were introduced for computing the slices associated with subtracted theories on the Higgs branch locally. However, the stratification of a symplectic singularity in general admits non-trivial global structure which is not captured by this algorithm; the following section demonstrates that the global structure plays a crucial role in determining the Hasse diagram and promotes the local rules introduced in Section~\ref{sec_RHBQS} to a full global argument.
\\\par Recall that a slice in a symplectic singularity is transverse to a local base leaf in some other higher dimensional leaf. In general, the local structure of a slice can be seen from any given point of the base leaf. If the global structure is trivial, moving around the base leaf does not change the appearance of the slice. However, non-trivial global structure means that moving on a closed path on the base leaf results in a non-trivial action, called a monodromy map, on the chiral ring of the slice \cite{Slodowy1980SimpleSA,OALiegroup,Generic_singularities}.\footnote{Moreover, the Weyl symmetry of the deformation parameters is also reduced under this action.} 
\\ This global information can be recorded via a decoration in analogy to those used in top-down Coulomb branch quiver subtraction \cite{Bourget_2022instanton,Bourget_2022dim6}.

Decorated quivers were introduced to capture quotients of $S_n$ outer-automorphisms on the Coulomb branch. In brane systems, Coulomb branch decoration corresponds to the indistinguishability of multiple branes or sub-brane systems. The Higgs branch decoration introduced here replaces the quotient that arises on the Coulomb branch with the action of a discrete symmetry on the Higgs branch. There is currently no brane-system interpretation of Higgs branch decoration.
\subsection{Global Rules}
\label{subsec:create_merg_deco}
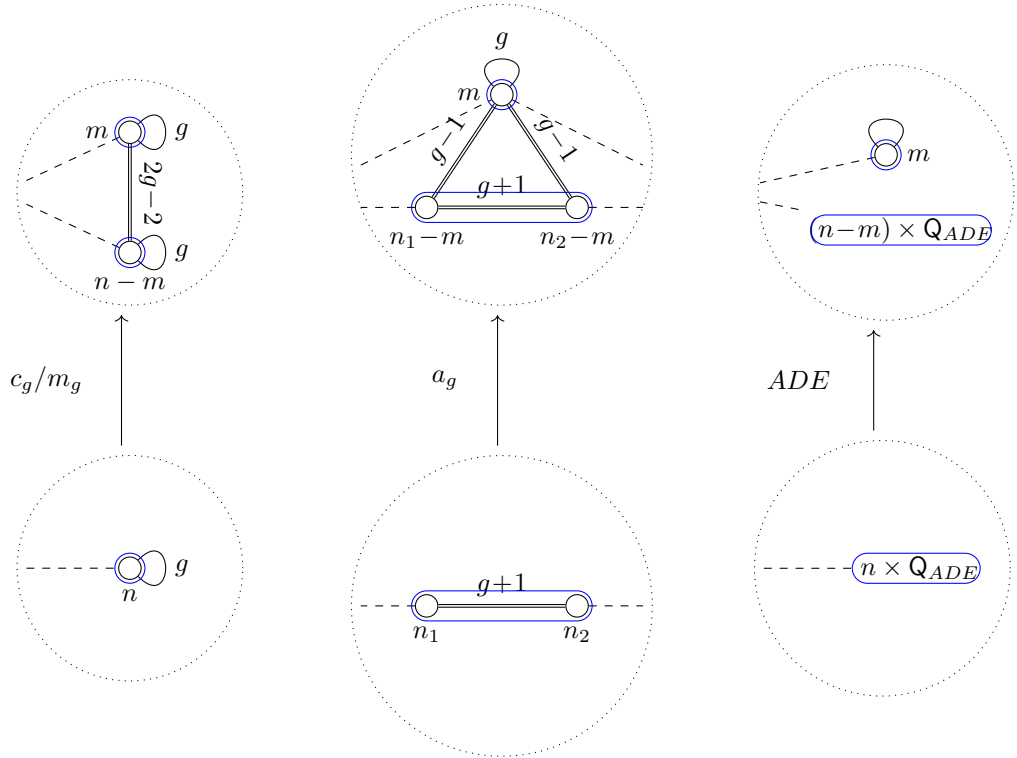
\begin{figure}[t]
    \centering
    \begin{tikzpicture}
        \node (aa) at (-5,0) {      \begin{tikzpicture}
        \node (a) at (0,0) {$\raisebox{-.5 \height}{ \begin{tikzpicture}
            \node (1u) at (-0.5,0.1) {};
            \node (1d) at (-0.5,-0.1) {};
            \draw[dotted] (1,0) circle (1.5cm);
            \node[gauge,label=below:{$n-m$}] (2) at (1,-0.8) {};
            \node[gauge,label=left:{$m$}] (3) at (1,0.8) {};
            \draw (2) to [in=45,out=-45,looseness=8]node[pos=0.5,right]{$g$} (2);
            \draw (3) to [in=45,out=-45,looseness=8]node[pos=0.5,right]{$g$} (3);
            \draw[double] (3) to node[pos=0.5,above,sloped]{$2g\!-\!2$} (2);
            \draw[dashed] (3)--(1u) (2)--(1d);
            \draw[blue] (2) circle (0.2cm);
            \draw[blue] (3) circle (0.2cm);
        \end{tikzpicture}}$};
        \node at (-1,-2.5) {$c_g/m_g$};
        \node (b) at (0,-5) {$\raisebox{-.5 \height}{ \begin{tikzpicture}
            \node (1) at (-0.5,0) {};
            \draw[dotted] (1,0) circle (1.5cm);
            \node[gauge,label=below:{$n$}] (2) at (1,0) {};
            \draw (2) to [in=45,out=-45,looseness=8]node[pos=0.5,right]{$g$} (2);
            \draw[dashed] (1)--(2);
            \draw[blue] (2) circle (0.2cm);
        \end{tikzpicture}}$};
        \draw[->] (b)--(a);
    \end{tikzpicture}  };
        \node (bb) at (0,0) {  \begin{tikzpicture}
        \node (a) at (0,0) {$\raisebox{-.5 \height}{ \begin{tikzpicture}
            \node (1l) at (0,0) {};
            \node (1r) at (4,0) {};
            \node (1lu) at (0,0.5) {};
            \node (1ru) at (4,0.5) {};
            \draw[dotted] (2,0.7) circle (2cm);
            \node[gauge,label=below:{$n_1\!-\!m$}] (2) at (1,0) {};
            \node[gauge,label=below:{$n_2\!-\!m$}] (3) at (3,0) {};
            \node[gauge,label=left:{$m$}] (4) at (2,1.5) {};
            \draw[double] (2)--(3)node[pos=0.5,above,sloped]{$g\!+\!1$};
            \draw[double] (2)--(4)node[pos=0.5,above,sloped]{$g\!-\!1$};
            \draw[double] (3)--(4)node[pos=0.5,above,sloped]{$g\!-\!1$};
            \draw (4) to [in=45,out=135,looseness=8]node[pos=0.5,above,sloped]{$g$}(4);
            \draw[dashed] (2)--(1l) (3)--(1r) (1lu)--(4)--(1ru);
            \draw[blue] (4) circle (0.2cm);
            \draw[blue] \convexpath{2,3} {0.2cm};
        \end{tikzpicture}}$};
        \node at (-0.7,-3) {$a_g$};
        \node (b) at (0,-6) {$\raisebox{-.5 \height}{ \begin{tikzpicture}
            \node (1l) at (0,0) {};
            \node (1r) at (4,0) {};
            \draw[dotted] (2,0) circle (2cm);
            \node[gauge,label=below:{$n_1$}] (2) at (1,0) {};
            \node[gauge,label=below:{$n_2$}] (3) at (3,0) {};
            \draw[dashed] (2)--(1l) (3)--(1r);
            \draw[double] (2)--(3)node[pos=0.5,above,sloped]{$g\!+\!1$};
            \draw[blue] \convexpath{2,3} {0.2cm};
        \end{tikzpicture}}$};
        \draw[->] (b)--(a);
\end{tikzpicture} };
        \node (cc) at (5,0) {      \begin{tikzpicture}
        \node (a) at (0,0) {$\raisebox{-.5 \height}{ \begin{tikzpicture}
            \node (1u) at (-2,0.6) {};
            \node (1d) at (-2,0.4) {};
            \draw[dotted] (-0.2,0.5) circle (1.7cm);
            \node (2) at (0,0) {$(n\!-\!m)\times\mathsf{Q}_{ADE}$};
            \node[gauge,label=right:{$m$}] (3) at (-0.2,1) {};
            \draw (3) to [out=45,in=135,looseness=8] (3);
            \draw[dashed] (2)--(1d) (3)--(1u);
            \draw[blue] (3) circle (0.2cm);
            \node (aux1) at (-1,0) {};
            \node (aux2) at (1,0) {};
            \draw[blue] \convexpath{aux1,aux2} {0.2cm};
        \end{tikzpicture}}$};
        \node at (-1,-2.5) {$ADE$};
        \node (b) at (0,-5) {$\raisebox{-.5 \height}{ \begin{tikzpicture}
            \node (1) at (-0.2,0) {};
            \draw[dotted] (1.5,0) circle (1.7cm);
            \node (2) at (2,0) {$n\times\mathsf{Q}_{ADE}$};
            \draw[dashed] (1)--(2);
            \node (aux1) at (1.3,0) {};
            \node (aux2) at (2.6,0) {};
            \draw[blue] \convexpath{aux1,aux2} {0.2cm};
        \end{tikzpicture}}$};
        \draw[->] (b)--(a);
    \end{tikzpicture}  };
    \end{tikzpicture}
    \caption{Decorations are inherited from the original theory before subtraction. The decorations in the bottom row are trivial, but become non-trivial in the rows above upon application of the subtraction rules.}
    \label{fig_decorationinhert}
\end{figure}

As mentioned above, the local rules introduced in Section~\ref{sec_RHBQS} are incomplete since they lack information relating to the global structure of the symplectic singularity. The introduction of decorations ameliorates this --- giving the final form of the rules in \Figref{fig_decorationinhert} --- although they in general add further complexity to the subtraction process and mandate extra care when a single quiver (or even a single node) contains multiple decorations. The decoration rules introduced below are supported via non-trivial examples involving nilpotent cones as given in Section \ref{sec_INSI}.

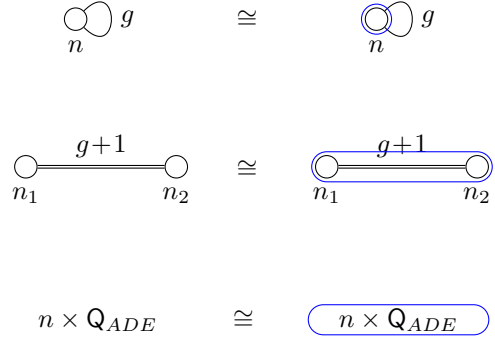
\begin{figure}[t]
    \centering
    \begin{tikzpicture}
        \node (aa) at (0,0) {    \begin{tikzpicture}
        \node (a) at (0,0) {$\raisebox{-.5 \height}{ \begin{tikzpicture}
            \node[gauge,label=below:{$n$}] (2) at (1,0) {};
            \draw (2) to [in=45,out=-45,looseness=8]node[pos=0.5,right]{$g$} (2);
        \end{tikzpicture}}$};
        \node at (2,0) {$\cong$};
        \node (b) at (4,0) {$\raisebox{-.5 \height}{ \begin{tikzpicture}
            \node[gauge,label=below:{$n$}] (2) at (1,0) {};
            \draw (2) to [in=45,out=-45,looseness=8]node[pos=0.5,right]{$g$} (2);
            \draw[blue] (2) circle (0.2cm);
        \end{tikzpicture}}$};
    \end{tikzpicture}};
        \node (bb) at (0,-2) {\begin{tikzpicture}
        \node (a) at (0,0) {$\raisebox{-.5 \height}{ \begin{tikzpicture}
            \node[gauge,label=below:{$n_1$}] (2) at (1,0) {};
            \node[gauge,label=below:{$n_2$}] (3) at (3,0) {};
            \draw[double] (2)--(3)node[pos=0.5,above,sloped]{$g\!+\!1$};
        \end{tikzpicture}}$};
        \node at (2,0) {$\cong$};
        \node (b) at (4,0) {$\raisebox{-.5 \height}{ \begin{tikzpicture}
            \node[gauge,label=below:{$n_1$}] (2) at (1,0) {};
            \node[gauge,label=below:{$n_2$}] (3) at (3,0) {};
            \draw[double] (2)--(3)node[pos=0.5,above,sloped]{$g\!+\!1$};
            \draw[blue] \convexpath{2,3} {0.2cm};
        \end{tikzpicture}}$};
\end{tikzpicture}};
        \node (cc) at (0,-4) {    \begin{tikzpicture}
        \node (a) at (0,0) {$\raisebox{-.5 \height}{ \begin{tikzpicture}
            \node (2) at (2,0) {$n\times\mathsf{Q}_{ADE}$};
            \node (aux1) at (1,0) {};
            \node (aux2) at (3,0) {};
        \end{tikzpicture}}$};
        \node at (2,0) {$\cong$};
        \node (b) at (4,0) {$\raisebox{-.5 \height}{ \begin{tikzpicture}
            \node (2) at (2,0) {$n\times\mathsf{Q}_{ADE}$};
            \node (aux1) at (1,0) {};
            \node (aux2) at (3,0) {};
            \draw[blue] \convexpath{aux1,aux2} {0.2cm};
        \end{tikzpicture}}$};
    \end{tikzpicture}};
    \end{tikzpicture}
    \caption{A (formally trivial) decoration is assigned to each sub-quiver that is not already entirely decorated.}
    \label{fig_decorationcreation}
\end{figure}

\paragraph{Higgs Branch Decoration} Firstly, recall that the rules of Section~\ref{sec_RHBQS} identify, for a given quiver $\mathcal{Q}$, several possible sub-quivers whose subtraction yields a transition on the Higgs branch. The most rigorous way in which to think about the decoration associated to this transition is the following. 
\begin{enumerate}
    \item Take all sub-quivers of $\mathcal{Q}$ that admit a subtraction and decorate them in the sense shown in \Figref{fig_decorationcreation} (do not add another decoration to a sub-quiver if it is already decorated in its entirety\footnote{That is to say, if the sub-quivers given in the bottom row of \Figref{fig:local_rules} are decorated as in \Figref{fig_decorationinhert}, do not add any further decorations. (The decorations in the lower row of \Figref{fig_decorationinhert} are in fact formally trivial, as shown in \Figref{fig_decorationcreation}.)}).
    \item The application of the subtraction rules of Section~\ref{sec_RHBQS} causes the decoration to be ``inherited'' by the subtraction products, as shown in \Figref{fig_decorationinhert}.
    \item After a subtraction, the process of identifying new sub-quivers to be subtracted begins again, and new decorations need be applied.
\end{enumerate}  
\par Take as an example \hyperref[fig_ADE]{Rule 3}, shown rightmost in \Figref{fig_decorationinhert}. If $m \geq 2$ one can apply \hyperref[fig_cgmg]{Rule 1} (shown leftmost in \Figref{fig_decorationinhert}) to the node of rank $m$ and split it into two nodes of rank $m'$ and $m-m'$ for some $1 \leq m' < m$. In this case, \emph{no new decoration should be applied}, since all nodes undergoing the transition (in this case just the single node of rank $m$) are already decorated by the decoration introduced in the initial subtraction. Similarly, further $ADE$ subtractions to the same set of nodes as the initial subtraction also do not introduce a new decoration. A non-trivial example of this can be seen in \Figref{fig_HasseE72a}, in which only one decoration has been used throughout.

\par This rather formal approach to decoration is in practice not the most intuitive, and it is common to think of subtraction products as 'picking up' a decoration in the process of subtraction instead of inheriting the decorations applied to sub-quivers beforehand. Indeed, the process of allocating decorations to subquivers before performing a subtraction does not modify the quiver as it introduces no global information, in other words, only the decorations in the upper row of \Figref{fig_decorationinhert} are non-trivial\footnote{This is somewhat similar to the redundancy of a decoration on a node of rank 1 in a bouquet of a single node in Coulomb-branch quiver subtraction}. Of course, it is easiest to understand the process of decoration via illustrations, and the reader is encouraged to consider the numerous Hasse diagrams in this paper for examples of the rules in action.

\begin{figure}[t]
    \centering
    \begin{tikzpicture}
        \node (a) at (0,0) {$\raisebox{-.5 \height}{ \begin{tikzpicture}
            \node[gauge,label=below left:$1$] (1) at (0,0) {};
            \node[gauge,label=below:{$n_1-m_1$}] (2) at (1,0) {};
            \node[gauge,label=left:{$m_1$}] (3) at (0,1) {};
            \node[gauge,label=above:{$n_2-m_2$}] (4) at (-1,0) {};
            \node[gauge,label=right:{$m_2$}] (5) at (0,-1) {};
            \draw (2) to [in=45,out=-45,looseness=8]node[pos=0.5,right]{$g$} (2);
            \draw (3) to [out=45,in=135,looseness=8]node[pos=0.5,above,sloped]{$g$} (3);
            \draw[double] (2) to node[pos=0.5,above,sloped]{$2g\!-\!2$} (3);
            \draw (3)--(1)--(2);
            \draw[blue] (2) circle (0.2cm);
            \draw[blue] (3) circle (0.2cm);
            \draw (4) to [in=135,out=225,looseness=8]node[pos=0.5,left]{$g$} (4);
            \draw (5) to [out=225,in=315,looseness=8]node[pos=0.5,below,sloped]{$g$} (5);
            \draw[double] (4) to node[pos=0.5,below,sloped]{$2g\!-\!2$} (5);
            \draw (5)--(1)--(4);
            \draw[red] (4) circle (0.2cm);
            \draw[red] (5) circle (0.2cm);
        \end{tikzpicture}}$};
        \node at (-2.1,-3) {non-minimal transition};
        \node (b) at (0,-5) {$\raisebox{-.5 \height}{ \begin{tikzpicture}
            \node[gauge,label=below:{$1$}] (1) at (0,0) {};
            \node[gauge,label=below:{$n_1$}] (2) at (1,0) {};
            \node[gauge,label=below:{$n_2$}] (3) at (-1,0) {};
            \draw (2) to [in=45,out=-45,looseness=8]node[pos=0.5,right]{$g$} (2);
            \draw (3) to [in=135,out=225,looseness=8]node[pos=0.5,left]{$g$} (3);
            \draw (3)--(1)--(2);
            \draw[blue] (2) circle (0.2cm);
            \draw[red] (3) circle (0.2cm);
        \end{tikzpicture}}$};
        \node (c) at (-5,-5) {$\raisebox{-.5 \height}{ \begin{tikzpicture}
            \node[gauge,label=below:{$1$}] (1) at (0,0) {};
            \node[gauge,label=below:{$n_1$}] (2) at (1,0) {};
            \node[gauge,label=below:{$n_2$}] (3) at (-1,0) {};
            \draw (2) to [in=45,out=-45,looseness=8]node[pos=0.5,right]{$g$} (2);
            \draw (3) to [in=135,out=225,looseness=8]node[pos=0.5,left]{$g$} (3);
            \draw (3)--(1)--(2);
        \end{tikzpicture}}$};
                \node (d) at (-2.5,-5) {$\cong$};
        \draw[->] (b)--(a);
    \end{tikzpicture}
    \caption{An example of a single quiver with different decorations on different sub-quivers (nodes). Note that the transition shown here is non-minimal (in other words, the transition is the result of several applications of the rules).}
    \label{fig_decorationdifferent}
\end{figure}
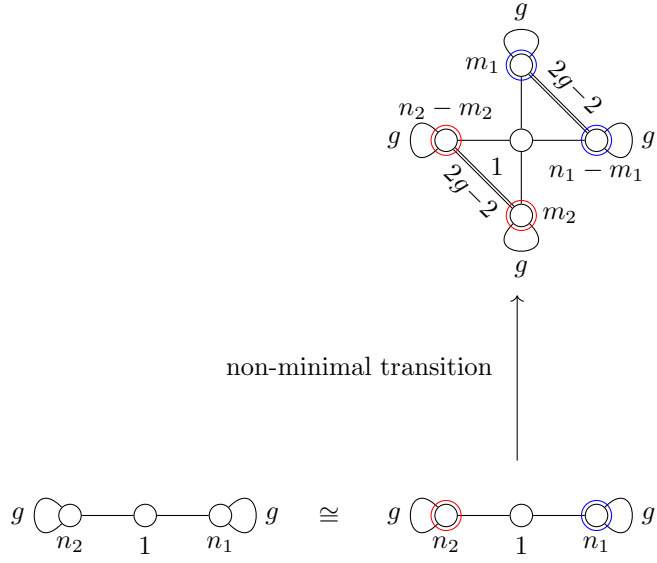

If a quiver $\mathcal{Q}$ contains different sub-quivers (each not already entirely decorated) then each sub-quiver picks up a new decoration in the process of subtraction. Such an example is given in  \Figref{fig_decorationdifferent}, which shows the formal process of first assigning decorations to each of the nodes of ranks $n_1$ and $n_2$ before performing both subtractions. Note that the transition in \Figref{fig_decorationdifferent} is not minimal.

\begin{figure}[t]
    \centering
    \begin{tikzpicture}
            \node (a) at (0,0) {$\raisebox{-.5 \height}{ \begin{tikzpicture}
            \node[gauge,label=left:$1$] (1) at (0,0) {};
            \node[gauge,label=right:{$n$}] (2) at (2,1) {};
            \node[gauge,label=right:{$n$}] (3) at (2,-1) {};
            \draw[double] (2)--(3)node[pos=0.5,above,sloped]{$g\!+\!1$};
            \draw (2) to [in=45,out=135,looseness=8]node[pos=0.5,above,sloped]{$g$}(2);
            \draw (3) to [in=225,out=315,looseness=8]node[pos=0.5,below,sloped]{$g$}(3);
            \draw (2)--(1)--(3);
            \draw[red] (2) circle (0.2cm);
            \draw[blue] (2) circle (0.25cm);
            \draw[red] (3) circle (0.2cm);
            \draw[blue] (3) circle (0.25cm);
        \end{tikzpicture}}$};
        \node at (2.5,0) {$\cong$};
        \node (c) at (5,0) {$\raisebox{-.5 \height}{ \begin{tikzpicture}
            \node[gauge,label=left:$1$] (1) at (0,0) {};
            \node[gauge,label=right:{$n$}] (2) at (2,1) {};
            \node[gauge,label=right:{$n$}] (3) at (2,-1) {};
            \draw[double] (2)--(3)node[pos=0.5,above,sloped]{$g\!+\!1$};
            \draw (2) to [in=45,out=135,looseness=8]node[pos=0.5,above,sloped]{$g$}(2);
            \draw (3) to [in=225,out=315,looseness=8]node[pos=0.5,below,sloped]{$g$}(3);
            \draw (2)--(1)--(3);
            \draw[greed] (2) circle (0.2cm);
            \draw[greed] (3) circle (0.2cm);
        \end{tikzpicture}}$};
        \node at (-2,-3) {non-minimal transition};
        \node (b) at (0,-5.5) {$\raisebox{-.5 \height}{ \begin{tikzpicture}
            \node[gauge,label=left:$1$] (1) at (0,1) {};
            \node[gauge,label=below:{$n$}] (2) at (0,0) {};
            \node[gauge,label=left:{$n$}] (3) at (0,2) {};
            \node[gauge,label=below:{$n$}] (4) at (2,0) {};
            \node[gauge,label=right:{$n$}] (5) at (2,2) {};
            \draw (3)--(1)--(2);
            \draw[double] (3)--(5)node[pos=0.5,above,sloped]{$g\!+\!1$}--(4)node[pos=0.5,above,sloped]{$g\!+\!1$}--(2)node[pos=0.5,below,sloped]{$g\!+\!1$};
            \draw[red] (2) circle (0.2cm);
            \draw[red] (3) circle (0.2cm);
            \draw[blue] (4) circle (0.2cm);
            \draw[blue] (5) circle (0.2cm);
        \end{tikzpicture}}$};
        \draw[->] (b)--(a);
    \end{tikzpicture}
    \caption{The procedure of decoration merging for individual nodes that are decorated twice. Such situations arise in the event that two individually-decorated nodes connected via $g+1$ links undergo an $a_g$ transition.}
    \label{fig_decorationannihilation}
\end{figure}
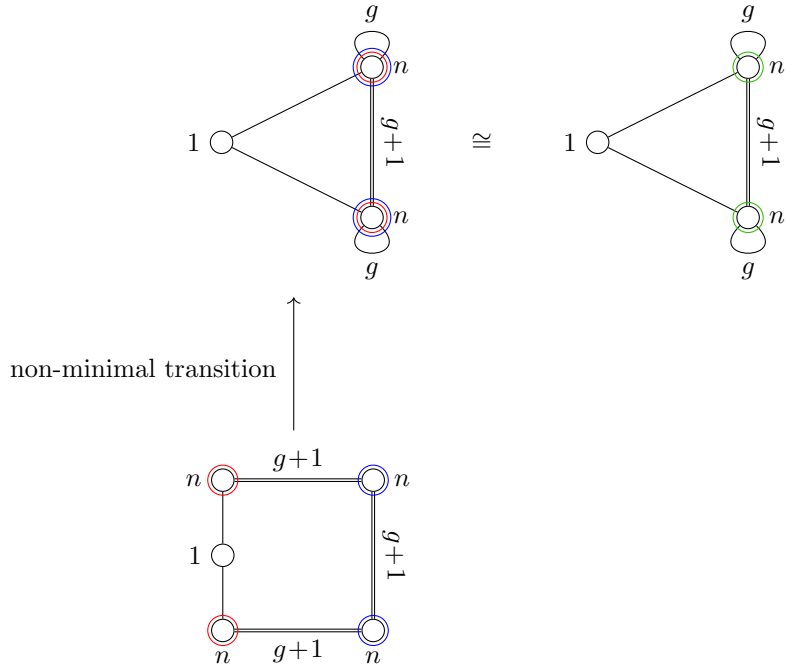

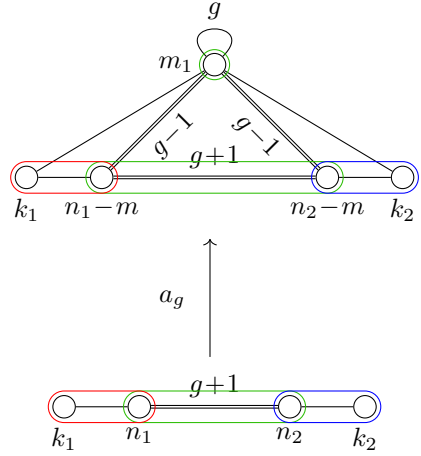
\begin{figure}[t]
    \centering
    \begin{tikzpicture}
        \node (a) at (0,0) {$\raisebox{-.5 \height}{ \begin{tikzpicture}
            \node[gauge,label=below:$k_1$] (1) at (0,0) {};
            \node[gauge,label=below:$k_2$] (5) at (5,0) {};
            \node[gauge,label=below:$n_1\!-\!m$] (2) at (1,0) {};
            \node[gauge,label=below:$n_2\!-\!m$] (3) at (4,0) {};
            \node[gauge,label=left:$m_1$] (4) at (2.5,1.5) {};
            \draw (1)--(2);
            \draw (3)--(5);
            \draw[double] (2)--(3)node[pos=0.5,above,sloped]{$g\!+\!1$};
            \draw[double] (2)--(4)node[pos=0.5,below,sloped]{$g\!-\!1$};
            \draw[double] (3)--(4)node[pos=0.5,below,sloped]{$g\!-\!1$};
            \draw (4) to [in=45,out=135,looseness=8]node[pos=0.5,above,sloped]{$g$}(4);
            \draw (1)--(4);
            \draw (5)--(4);
            \draw[greed] \convexpath{2,3} {0.21cm};
            \draw[red] \convexpath{1,2} {0.21cm};
            \draw[blue] \convexpath{3,5} {0.21cm};
            \draw[greed] (4) circle (0.2cm);
        \end{tikzpicture}}$};
        
        \node at (-0.5,-2.5) {$a_g$};
        \node (b) at (0,-4) {$\raisebox{-.5 \height}{ \begin{tikzpicture}
            \node[gauge,label=below:$k_1$] (1) at (0,0) {};
            \node[gauge,label=below:$k_2$] (4) at (4,0) {};
            \node[gauge,label=below:$n_1$] (2) at (1,0) {};
            \node[gauge,label=below:$n_2$] (3) at (3,0) {};
            \draw (1)--(2);
            \draw (3)--(4);
            \draw[double] (2)--(3)node[pos=0.5,above,sloped]{$g\!+\!1$};
            \draw[greed] \convexpath{2,3} {0.21cm};
            \draw[red] \convexpath{1,2} {0.21cm};
            \draw[blue] \convexpath{3,4} {0.21cm};
        \end{tikzpicture}}$};
        \draw[->] (b)--(a);
    \end{tikzpicture}
    \caption{An example in which the blue and red decorations are not inherited.}
    \label{fig_no_inheritance}
\end{figure}

\paragraph{Merging of Decoration.}
The next example of decoration consists of the case in which two decorations are applied to the same node, as shown in \Figref{fig_decorationannihilation}. In this case, the two decorations can be said to `merge', forming a new decorated node. Note that decoration merging is only applied in the situation in which it is a single node that holds two decorations. It is not uncommon to have quivers with overlapping decorations arising from $ADE$ affine Dynkin subquivers that share nodes. In these cases, the two decorations are not inherited, leading to no merging, but instead the creation of a third decoration overlapping the other two. An example of such a case is provided in \Figref{fig_no_inheritance}, contrasting with the case in \Figref{fig_decorationannihilation} in which merging is allowed.

\subsection{Monodromy Map and Union}
\label{sec_MMU}

The decoration determines the non-trivial global structure of the slice in one of the following two ways:
\begin{enumerate}
    \item The slice is locally $ADE$ or $a_g$ and the corresponding quiver contains $n$ identically decorated identical nodes. There is a monodromy map of (up to) $S_n$. The slices are shown in Table~\ref{tab_foldedslices}.
    \item The slice is locally $k$ copies of $ADE$ or $a_g$ and the corresponding quiver for each copy has the same decoration. The $n$-copies become a union that intersects at the singular point.
\end{enumerate}

The two cases can take place simultaneously. For the first case, if the slice is of type $ADE$ then the monodromy map is actually a subgroup of the automorphism of the corresponding Dynkin diagram \cite{Generic_singularities}, so it often happens that the actual monodromy map is smaller than the decoration indicates. Here there is a connection between the decoration and folding on the finite Dynkin diagram\footnote{The relation between quiver folding and Coulomb branch geometry was studied in \cite{Bourget:2020bxh}.} as in Table~\ref{tab_folding}.

The origin of the discrete symmetry is the action of Weyl group on the unbroken subgroup. There are in general different ways to embed the unbroken groups into the original group and these different embeddings are related by the action of Weyl group. For example, there are $n!$ different ways to embed $\urm(1)^n$ into $\urm(n)$: each $\urm(1)\subset\urm(1)^n$ is assigned to a certain $\urm(1)$ in the Cartan subgroup of $\urm(n)$. These embeddings are exchanged into each other under the action of $\Wcal_{\urm(n)}=S_n$. From the point of view of the stratification, going around a closed path on the base leaf corresponds to changing the embedding. This action also acts on the deformation space, as is discussed in the following subsection.

\subsection{Namikawa Weyl Group}
\label{sec_NWG}

Despite its use in the theory of symplectic singularities, the Namikawa-Weyl group \cite{Namikawa_2011,Namikawa_2010}, which arises as the reflection group acting on deformation parameters, has heretofore gone largely unmentioned in the physics literature. However, the Namikawa-Weyl group plays an important physical role in the moduli spaces of vacua of 3d $\mathcal{N}=4$ gauge theories: on the Higgs branch it is the reflection group acting on FI parameters, while on the Coulomb branch it is the reflection group acting on mass parameters. In other words, the Namikawa-Weyl group acts to exchange different chambers of deformation parameters associated to a generic symplectic singularity; under symplectic duality \cite{Braden:2014iea}, the Namikawa-Weyl group of one branch is exchanged with the Weyl group of the global symmetry of the other.
This translates to a D3-/D5-/NS5-brane system as follows. The Weyl group of the heavy D5-branes is the Weyl group of the Higgs branch global symmetry, which under symplectic duality corresponds to the Namikawa-Weyl group of mass parameters on the Coulomb branch. Conversely, the Weyl group of the NS5-branes corresponds to the Namikawa-Weyl group of FI parameters on the Higgs branch.

In the Hasse diagram, the Namikawa-Weyl group $\Wcal^X$ (of the symplectic singularity $X$) appears as the Weyl group of the Lie group associated to the slices from the top of the diagram to the quaternionic codimension-1 subvarieties \cite{wu2023namikawaweyl}
\footnote{Namikawa Weyl groups are not defined for non-normal varieties. However, the Higgs branch of a unitary quiver is always normal and the only (quaternionic) one-dimensional normal symplectic singularities are Kleinian singularities, so there is no ambiguity in this statement.}:
\begin{equation}
    \Wcal^X = \prod_i\Wcal_i,
    \label{eq_Weyl}
\end{equation}
where $i$ runs over all quaternionic 1-dimensional top slices, and $\Wcal_i$ is the Weyl group associated with the slice.

From the brane system picture, \eqref{eq_Weyl} is a simple consequence of the Kraft-Procesi transition. When there are more than one D5/NS5 branes with the same linking number in the same segment between two NS5/D5 branes, there is a valid transition to codimension-1 subvarieties. A natural Weyl symmetry acting on the D5/NS5 branes gives the Weyl symmetry of the deformation parameters. The direct product of all such Weyl symmetry gives the whole all Weyl symmetry of the deformation parameters.

Using the above formula, it is simple to calculate the Namikawa-Weyl group for a given Higgs branch from its Hasse diagram. To find each $\Wcal_i$, first take the decorated quiver whose Higgs branch corresponds to the given 1-dimensional slice --- $\Wcal_i$ is then the Weyl group of the Lie algebra given by the folded Dynkin diagram associated to the decorated quiver, according to the correspondence in Tables \ref{tab_foldedslices} and \ref{tab_folding}. For example, the $A_n$ slice gives the Namikawa-Weyl group $\Wcal_{A_n} = S_n$, and $C_n$ slice gives the Namikawa-Weyl group $\Wcal_{B_n}$. Given a Higgs branch Hasse diagram, the Coulomb branch global symmetry is determined up to an ambiguity between the $\bfrak$ and $\cfrak$ algebras. Thus it seems natural to conjecture that the algebra of the Coulomb branch global symmetry is precisely that of the Higgs branch's top slices.
For simply-laced unitary quivers, the continuous Coulomb branch symmetry takes the form:
\begin{equation}
    G_\Ccal=\prod_i G_i,
\end{equation}
where $G_i$ is one of the $\urm(1)$, $\surm(n)$, $\sorm(n)$, $G_2$, $E_6$, $E_7$, $E_8$ determined by $\Qcal_i$. The absence of $\sprm(n)$ and $F_4$ is due to the fact that the corresponding folded quivers cannot be constructed by adding decorations to single nodes, but to legs of multiple nodes.

\begin{table}[H]
\ra{1.5}
    \centering
    \begin{tabular}{cccc}
    \toprule
         Label  &   \multicolumn{2}{c}{Quivers}\\ \midrule
         $C_n \cong D_{n+1}$ with $S_2$ &
         \raisebox{-0.5\height}{\begin{tikzpicture}
            \node[gauge,label=left:{1}] (4) at (-1.7,0.7) {};
            \node[gauge,label=left:{1}] (6) at (-1.7,-0.7) {};
            \node[gauge,label=below:{2}] (1) at (-1,0) {};
            \node[] (5) at (-0.5,0) {$\cdots$};
            \node[gauge,label=below:{2}] (0) at (0,0) {};
            \node[gauge,label=right:{1}] (2) at (0.7,0.7) {};
            \node[gauge,label=right:{1}] (3) at (0.7,-0.7) {};
            \draw (2)--(0)--(3) (4)--(1)--(6);
            \draw[blue] (2) circle (0.2cm);
            \draw[blue] (3) circle (0.2cm);
        \end{tikzpicture}} & \raisebox{-0.5\height}{\begin{tikzpicture}
            \node[gauge,label=left:{1}] (4) at (-1.7,0.7) {};
            \node[gauge,label=left:{1}] (6) at (-1.7,-0.7) {};
            \node[gauge,label=below:{2}] (1) at (-1,0) {};
            \node[] (5) at (-0.5,0) {$\cdots$};
            \node[gauge,label=below:{2}] (0) at (0,0) {};
            \node[gauge,label=right:{1}] (2) at (0.7,0.7) {};
            \node[gauge,label=right:{1}] (3) at (0.7,-0.7) {};
            \draw (2)--(0)--(3) (4)--(1)--(6);
            \draw[red] (4) circle (0.2cm);
            \draw[red] (6) circle (0.2cm);
            \draw[blue] (2) circle (0.2cm);
            \draw[blue] (3) circle (0.2cm);
        \end{tikzpicture}} \\ 
        $C_2 \cong A_3$ with $S_2$ & \raisebox{-0.5\height}{\begin{tikzpicture}
            \node[gauge,label=left:{1}] (1) at (0,0.7) {};
            \node[gauge,label=below:{1}] (2) at (0,-0.7) {};
            \node[gauge,label=right:{1}] (3) at (0.7,0) {};
            \node[gauge,label=left:{1}] (4) at (-0.7,0) {};
            \draw (1)--(3)--(2)--(4)--(1);
            \draw[blue] (1) circle (0.2cm);
            \draw[blue] (2) circle (0.2cm);
        \end{tikzpicture}} & \raisebox{-0.5\height}{\begin{tikzpicture}
            \node[gauge,label=left:{1}] (1) at (0,0.7) {};
            \node[gauge,label=below:{1}] (2) at (0,-0.7) {};
            \node[gauge,label=right:{1}] (3) at (0.7,0) {};
            \node[gauge,label=left:{1}] (4) at (-0.7,0) {};
            \draw (1)--(3)--(2)--(4)--(1);
            \draw[blue] (1) circle (0.2cm);
            \draw[blue] (2) circle (0.2cm);
            \draw[red] (3) circle (0.2cm);
            \draw[red] (4) circle (0.2cm);
        \end{tikzpicture}} \\
         $G_2 \cong D_4$ with $S_3$ &  \raisebox{-0.5\height}{\begin{tikzpicture}
            \node[gauge,label=below:{2}] (0) at (0,0) {};
            \node[gauge,label=left:{1}] (1) at (-0.7,-0.7) {};
            \node[gauge,label=right:{1}] (2) at (0.7,0.7) {};
            \node[gauge,label=right:{1}] (3) at (0.7,-0.7) {};
            \node[gauge,label=left:{1}] (4) at (-0.7,0.7) {};
            \draw (1)--(0)--(2) (3)--(0)--(4);
            \draw[blue] (1) circle (0.2cm);
            \draw[blue] (2) circle (0.2cm);
            \draw[blue] (3) circle (0.2cm);
        \end{tikzpicture}} &  \raisebox{-0.5\height}{\begin{tikzpicture}
            \node[gauge,label=below:{2}] (0) at (0,0) {};
            \node[gauge,label=left:{1}] (1) at (-0.7,-0.7) {};
            \node[gauge,label=right:{1}] (2) at (0.7,0.7) {};
            \node[gauge,label=right:{1}] (3) at (0.7,-0.7) {};
            \node[gauge,label=left:{1}] (4) at (-0.7,0.7) {};
            \draw (1)--(0)--(2) (3)--(0)--(4);
            \draw[blue] (1) circle (0.2cm);
            \draw[blue] (2) circle (0.2cm);
            \draw[blue] (3) circle (0.2cm);
            \draw[blue] (4) circle (0.2cm);
        \end{tikzpicture}} \\ 
         $B_n \cong A_{2n-1}$ with $S_2$ & \raisebox{-0.5\height}{\begin{tikzpicture}
            \node[gauge,label=below:{1}] (4) at (-2,0) {};
            \node[gauge,label=below:{1}] (1) at (-1,0) {};
            \node[] (5) at (-1.5,0) {$\cdots$};
            \node[gauge,label=below:{1}] (0) at (0,0) {};
            \node[gauge,label=above:{1}] (7) at (0,1) {};
            \node[] (6) at (1.5,0) {$\cdots$};
            \node[gauge,label=below:{1}] (2) at (1,0) {};
            \node[gauge,label=below:{1}] (3) at (2,0) {};
            \draw (2)--(0)--(1) (4)--(7)--(3);
            \draw[blue] \convexpath{4,1} {0.2cm};
            \draw[blue] \convexpath{2,3} {0.2cm};
        \end{tikzpicture}} &  \\ 
         $F_4 \cong E_6$ with $S_2$ & \raisebox{-0.5\height}{\begin{tikzpicture}
            \node[gauge,label=below:{1}] (4) at (-2,0) {};
            \node[gauge,label=below:{2}] (1) at (-1,0) {};
            \node[gauge,label=below:{3}] (0) at (0,0) {};
            \node[gauge,label=below:{2}] (2) at (1,0) {};
            \node[gauge,label=below:{1}] (3) at (2,0) {};
            \node[gauge,label=above:{1}] (5) at (0,2) {};
            \node[gauge,label=right:{2}] (6) at (0,1) {};
            \draw (3)--(2)--(0)--(1)--(4) (0)--(6)--(5);
            \draw[blue] \convexpath{4,1} {0.2cm};
            \draw[blue] \convexpath{2,3} {0.2cm};
        \end{tikzpicture}} & \raisebox{-0.5\height}{\begin{tikzpicture}
            \node[gauge,label=below:{1}] (4) at (-2,0) {};
            \node[gauge,label=below:{2}] (1) at (-1,0) {};
            \node[gauge,label=below:{3}] (0) at (0,0) {};
            \node[gauge,label=below:{2}] (2) at (1,0) {};
            \node[gauge,label=below:{1}] (3) at (2,0) {};
            \node[gauge,label=above:{1}] (5) at (0,2) {};
            \node[gauge,label=right:{2}] (6) at (0,1) {};
            \draw (3)--(2)--(0)--(1)--(4) (0)--(6)--(5);
            \draw[blue] \convexpath{4,1} {0.2cm};
            \draw[blue] \convexpath{2,3} {0.2cm};
            \draw[blue] \convexpath{5,6} {0.2cm};
        \end{tikzpicture}} \\ 
                 $a_g^+ \cong a_g$ with $S_2$ & \raisebox{-0.5\height}{\begin{tikzpicture}
                 \node[gauge,label=below:{$1$}] (2) at (0,0) {};
            \node[gauge,label=below:{$1$}] (3) at (1.5,0) {};
            \draw[double] (2)--(3)node[pos=0.5,above,sloped]{$g\!+\!1$};
            \draw[blue] (2) circle (0.2cm);
            \draw[blue] (3) circle (0.2cm);
        \end{tikzpicture}} & \\ \bottomrule
    \end{tabular}
    \caption{ The Higgs branches of the quivers on the right are labelled by the first column. In each case, the decorations give a monodromy on the corresponding slice. The third column shows that in some cases adding further decorations does not alter the Higgs branch monodromy. As the decorations on $B_n$ and $F_4$ slices involve multiple nodes, they do not appear for a unitary quiver gauge theory. The last row shows the slice $a_g^+$ constructed as a Higgs branch.}
    
    \label{tab_foldedslices}
\end{table}

\begin{table}[H]
\ra{1.75}
    \centering
    \begin{tabular}{cccc}
    \toprule
         $G$ to $G'$  &  $G$ Dynkin Diagram & $G'$ Dynkin Diagram  \\ \midrule
         $D_{n+1}$ to $B_n$ & \raisebox{-0.5\height}{\begin{tikzpicture}
            \node[gauge] (4) at (-2,0) {};
            \node[gauge] (1) at (-1,0) {};
            \node[] (5) at (-0.5,0) {$\cdots$};
            \node[gauge] (0) at (0,0) {};
            \node[gauge] (2) at (0.7,0.7) {};
            \node[gauge] (3) at (0.7,-0.7) {};
            \draw (2)--(0)--(3) (4)--(1);
            \draw[blue] (2) circle (0.2cm);
            \draw[blue] (3) circle (0.2cm);
        \end{tikzpicture}} & \raisebox{-0.5\height}{\begin{tikzpicture}
            \node[gauge] (3) at (-2,0) {};
            \node[gauge] (2) at (-1,0) {};
            \node[] (5) at (-0.5,0) {$\cdots$};
            \node[gauge] (0) at (0,0) {};
            \node[gauge] (1) at (1,0) {};
            \draw (3)--(2);
            \draw[transform canvas={yshift=-1pt}] (1)--(0);
            \draw[transform canvas={yshift=1pt}] (1)--(0);
            \draw (.4,0.2)--(.6,0)--(.4,-.2);
        \end{tikzpicture}} \\ 
         $D_4$ to $G_2$ &  \raisebox{-0.5\height}{\begin{tikzpicture}
            \node[gauge] (0) at (0,0) {};
            \node[gauge] (1) at (-1,0) {};
            \node[gauge] (2) at (0.7,0.7) {};
            \node[gauge] (3) at (0.7,-0.7) {};
            \draw (1)--(0)--(2) (3)--(0);
            \draw[blue] (1) circle (0.2cm);
            \draw[blue] (2) circle (0.2cm);
            \draw[blue] (3) circle (0.2cm);
        \end{tikzpicture}} & \raisebox{-0.5\height}{\begin{tikzpicture}
            \node[gauge] (0) at (0,0) {};
            \node[gauge] (1) at (1,0) {};
            \draw[transform canvas={yshift=-2pt}] (1)--(0);
            \draw (1)--(0);
            \draw[transform canvas={yshift=2pt}] (1)--(0);
            \draw (.4,0.2)--(.6,0)--(.4,-.2);
        \end{tikzpicture}} \\
         $A_{2n-1}$ to $C_n$ & \raisebox{-0.5\height}{\begin{tikzpicture}
            \node[gauge] (4) at (-2,0) {};
            \node[gauge] (1) at (-1,0) {};
            \node[] (5) at (-1.5,0) {$\cdots$};
            \node[gauge] (0) at (0,0) {};
            \node[] (6) at (1.5,0) {$\cdots$};
            \node[gauge] (2) at (1,0) {};
            \node[gauge] (3) at (2,0) {};
            \draw (2)--(0)--(1);
            \draw[blue] \convexpath{4,1} {0.2cm};
            \draw[blue] \convexpath{2,3} {0.2cm};
        \end{tikzpicture}} & \raisebox{-0.5\height}{\begin{tikzpicture}
            \node[] (5) at (.5,0) {$\cdots$};
            \node[gauge] (0) at (-1,0) {};
            \node[gauge] (1) at (0,0) {};
            \node[gauge] (2) at (1,0) {};
            \draw[transform canvas={yshift=-1pt}] (1)--(0);
            \draw[transform canvas={yshift=1pt}] (1)--(0);
            \draw (-.6,0.2)--(-.4,0)--(-.6,-.2);
        \end{tikzpicture}} \\ 
         $E_6$ to $F_4$ & \raisebox{-0.5\height}{\begin{tikzpicture}
            \node[gauge] (4) at (-2,0) {};
            \node[gauge] (1) at (-1,0) {};
            \node[gauge] (0) at (0,0) {};
            \node[gauge] (2) at (1,0) {};
            \node[gauge] (3) at (2,0) {};
            \node[gauge] (5) at (0,1) {};
            \draw (3)--(2)--(0)--(1)--(4) (0)--(5);
            \draw[blue] \convexpath{4,1} {0.2cm};
            \draw[blue] \convexpath{2,3} {0.2cm};
        \end{tikzpicture}} & \raisebox{-0.5\height}{\begin{tikzpicture}
            \node[gauge] (0) at (-1,0) {};
            \node[gauge] (1) at (0,0) {};
            \node[gauge] (2) at (1,0) {};
            \node[gauge] (3) at (2,0) {};
            \draw (0)--(1) (2)--(3);
            \draw[transform canvas={yshift=-1pt}] (1)--(2);
            \draw[transform canvas={yshift=1pt}] (1)--(2);
            \draw (.4,0.2)--(.6,0)--(.4,-.2);
        \end{tikzpicture}} \\ \bottomrule
         
    \end{tabular}
    \caption{Foldings of Lie algebras and their finite Dynkin diagrams. The automorphism subgroups for the folding are indicated from decoration.}
    \label{tab_folding}
\end{table}

\begin{table}[H]
    \centering
\begin{subtable}[t]{0.485\textwidth}
\centering
\begin{tabular}{cc}
    \toprule
        $\mathfrak{g}$ & Weyl group \\\midrule
        $\urm(1)$ & Trivial\\
        $\surm(n)$ &$S_n$\\
        $\sorm(2n+1)$& $\mathbb Z_2^n\rtimes S_n$\\
        $\sprm(n)$ & $\mathbb Z_2^n\rtimes S_n$\\
        $\sorm(2n)$ & $\mathbb Z_2^{n-1}\rtimes S_n$\\
        \bottomrule
    \end{tabular}
\end{subtable}
\begin{subtable}[t]{0.485\textwidth}
\centering
\begin{tabular}{cc}
    \toprule
        $\mathfrak{g}$ & Weyl group \\\midrule
        $E_6$ & $\mathcal W_{E_6}$\\
        $E_7$ & $\mathcal W_{E_7}$\\
        $E_8$ & $\mathcal W_{E_8}$\\
        $F_4$ & $\mathcal W_{F_4}$\\
        $G_2$ & $\textrm{Dih}_{6}=\mathbb Z_6\rtimes \mathbb Z_2$\\\bottomrule
    \end{tabular}
\end{subtable}    
    \caption{Table of Weyl groups for semi-simple Lie algebras: left side lists classical algebras, while right side contains exceptional algebras.}
    \label{tab:WeylGroups}
\end{table}

\subsection{Global Structure in Practice}
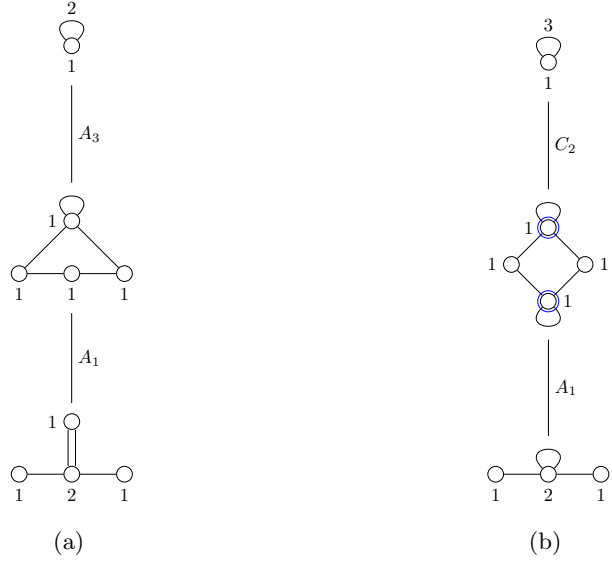
\begin{figure}[H]
    \centering
        \begin{subfigure}{0.4\textwidth}
        \centering
        \scalebox{.7}{
        \begin{tikzpicture}
        \node (a) at (0,0){$\begin{tikzpicture}

        \node[gauge, label=below:$1$] (1) at (0,0){};
        \node[gauge, label=below:$2$] (2) at (1,0){};
        \node[gauge, label=below:$1$] (3) at (2,0) {};
        \node[gauge, label=left:$1$] (4) at (1,1){};

        \draw[-] (1)--(2)--(3);
        \draw[transform canvas={xshift=-2pt}] (2)--(4);
        \draw[transform canvas={xshift=2pt}] (2)--(4);
        \end{tikzpicture}$};
        
        \node (b) at (0,4){$\begin{tikzpicture}

        \node[gauge, label=below:$1$] (1) at (0,0){};
        \node[gauge, label=below:$1$] (2) at (1,0){};
        \node[gauge, label=below:$1$] (3) at (2,0) {};
        \node[gauge, label=left:$1$] (4) at (1,1){};

        \draw[-] (1)--(2)--(3)--(4)--(1);
        \draw[-] (4) to[out=45,in=135,looseness=8] (4);
        \end{tikzpicture}$};
        \node (c) at (0,8){$\begin{tikzpicture}
            \node[gauge,label=below:$1$] (1) at (0,0){};
            \draw[-] (1) to[out=45,in=135,looseness=8] node[pos=0.5,above]{2} (1);
        \end{tikzpicture}$};
        \draw[-] (a)--(b) node[pos=0.5,midway, right]{$A_1$}--(c) node[pos=0.5,midway,right]{$A_3$};
    \end{tikzpicture}}
    \caption{}
    \label{fig_hassea3}
    \end{subfigure}
    \begin{subfigure}{0.4\textwidth}
    \centering
        \scalebox{.7}{
        \begin{tikzpicture}
        \node (a) at (0,0){$\begin{tikzpicture}

        \node[gauge, label=below:$1$] (1) at (0,0){};
        \node[gauge, label=below:$2$] (2) at (1,0){};
        \node[gauge, label=below:$1$] (3) at (2,0) {};

        \draw[-] (1)--(2)--(3);
        \draw[-] (2) to[out=45,in=135,looseness=8] (2);
        \end{tikzpicture}$};
        
        \node (b) at (0,4){$\begin{tikzpicture}
        \node[gauge, label=left:$1$] (1l) at (-0.7,0){};
        \node[gauge, label=right:$1$] (1r) at (0.7,0){};
        \node[gauge, label=right:$1$] (1d) at (0,-0.7){};
        \node[gauge, label=left:$1$] (1t) at (0,0.7){};
        \draw[-] (1t) to[out=45,in=135,looseness=8] (1t);
        \draw[-] (1l)--(1t)--(1r)--(1d)--(1l);
        \draw[-] (1d) to[out=-45,in=225,looseness=8](1d){};

        \draw[blue] (1t) circle (0.2cm);
        \draw[blue] (1d) circle (0.2cm);
        \end{tikzpicture}$};
        \node (c) at (0,8){$\begin{tikzpicture}
            \node[gauge,label=below:$1$] (1) at (0,0){};
            \draw[-] (1) to[out=45,in=135,looseness=8] node[pos=0.5,above]{3} (1);
        \end{tikzpicture}$};
        \draw[-] (a)--(b) node[pos=0.5,midway, right]{$A_1$}--(c) node[pos=0.5,midway,right]{$C_2$};
    \end{tikzpicture}}
    \caption{}
    \label{fig_hasseb2}
    \end{subfigure}
    \caption{(\subref{fig_hassea3}): The Higgs branch of the quiver at the bottom of the Hasse diagram is the S{\l}odowy Slice $\Scal^{[4]}_{[2,1^2]}$ in the $A_3$ nilcone. (\subref{fig_hasseb2}): The Higgs branch of the quiver at the bottom of the Hasse diagram is the S{\l}odowy Slice $\Scal^{[4]}_{[2,1^2]}$ in the $C_2$ nilcone.}
    \label{fig_hassea3b2}
\end{figure}

This section further studies the global structure of the Higgs branch and considers several examples.
\paragraph{From $A_3$ to $C_2$.}
The following is one of the simplest cases in which the monodromy map appears. The two examples in \Figref{fig_hassea3b2} manifestly have the same local slices; however, the top slice in \Figref{fig_hasseb2} has a non-trivial global structure determined by the decoration, changing the corresponding slice from $A_3$ to $C_2$.

\begin{figure}[H]
    \centering
        \begin{subfigure}{0.45\textwidth}
        \centering
        \scalebox{.7}{
        \begin{tikzpicture}
        \node[biggauge, label=left:$1$] (1l) at (-1.5,0){};
        \node[biggauge, label=right:$1$] (1r) at (1.5,0){};
        \node[biggauge, label=below:$1$] (1d) at (0,-1.5){};
        \node[biggauge, label=above:$1$] (1t) at (0,1.5){};
        \draw[->-=.5] (1l) to [out=45+20,in=225-20]node[pos=0.5,above,sloped]{$X_1$} (1t);
        \draw[->-=.5] (1t) to [out=225+20,in=45-20]node[pos=0.5,below,sloped]{$\tilde{X}_1$} (1l);
        \draw[->-=.5] (1t) to [out=-45+20,in=135-20]node[pos=0.5,above,sloped]{$X_2$} (1r);
        \draw[->-=.5] (1r) to [out=135+20,in=-45-20]node[pos=0.5,below,sloped]{$\tilde{X}_2$} (1t);
        \draw[->-=.5] (1r) to [out=-135+20,in=45-20]node[pos=0.5,below,sloped]{$X_3$} (1d);
        \draw[->-=.5] (1d) to [out=45+20,in=-135-20]node[pos=0.5,above,sloped]{$\tilde{X}_3$} (1r);
        \draw[->-=.5] (1d) to [out=135+20,in=-45-20]node[pos=0.5,below,sloped]{$X_4$} (1l);
        \draw[->-=.5] (1l) to [out=-45+20,in=135-20]node[pos=0.5,above,sloped]{$\tilde{X}_4$} (1d);
        \draw[blue] (1t) circle (0.25cm);
        \draw[blue] (1d) circle (0.25cm);
    \end{tikzpicture}}
    \caption{}
    \label{fig_midleaf121}
    \end{subfigure}
    \begin{subfigure}{0.45\textwidth}
    \centering
        \scalebox{.7}{
        \begin{tikzpicture}
        \node[biggauge, label=below:$1$] (1) at (0,0){};
        \node[biggauge, label=left:$2$] (2) at (1.5,0){};
        \node[biggauge, label=below:$1$] (3) at (3,0) {};
        \draw[->-=.5] (1) to [out=0+25,in=180-25]node[pos=0.5,above,sloped]{$Y_1^a$} (2);
        \draw[->-=.5] (2) to [out=180+25,in=0-25]node[pos=0.5,below,sloped]{$\tilde{Y}_{1\, a}$} (1);
        \draw[->-=.5] (2) to [out=0+25,in=180-25]node[pos=0.5,above,sloped]{$Y_2^a$} (3);
        \draw[->-=.5] (3) to [out=180+25,in=0-25]node[pos=0.5,below,sloped]{$\tilde{Y}_{2\, a}$} (2);
        \draw[->-=.5] (2) to[out=135,in=45,looseness=8]node[pos=0.5,above,sloped]{$\Phi^a_{1\, b}$} (2);
        \draw[->-=.5] (2) to[out=-45,in=-135,looseness=8]node[pos=0.5,below,sloped]{$\Phi^a_{2\, b}$} (2);
    \end{tikzpicture}}
    \caption{}
    \label{fig_bottomleaf121}
    \end{subfigure}
    \begin{subfigure}{0.45\textwidth}
        \centering
        \scalebox{.7}{
        \begin{tikzpicture}
        \node[biggauge, label=left:$1$] (1l) at (-1.5,0){};
        \node[biggauge, label=right:$1$] (1r) at (1.5,0){};
        \node[biggauge, label=below:$1$] (1d) at (0,-1.5){};
        \node[biggauge, label=above:$1$] (1t) at (0,1.5){};
        \draw[->-=.5] (1l) to [out=45+20,in=225-20]node[pos=0.5,above,sloped]{$X_1$} (1t);
        \draw[->-=.5] (1t) to [out=225+20,in=45-20]node[pos=0.5,below,sloped]{$\tilde{X}_1$} (1l);
        \draw[->-=.5] (1t) to [out=-45+20,in=135-20]node[pos=0.5,above,sloped]{$X_2$} (1r);
        \draw[->-=.5] (1r) to [out=135+20,in=-45-20]node[pos=0.5,below,sloped]{$\tilde{X}_2$} (1t);
        \draw[->-=.5] (1r) to [out=-135+20,in=45-20]node[pos=0.5,below,sloped]{$X_3$} (1d);
        \draw[->-=.5] (1d) to [out=45+20,in=-135-20]node[pos=0.5,above,sloped]{$\tilde{X}_3$} (1r);
        \draw[->-=.5] (1d) to [out=135+20,in=-45-20]node[pos=0.5,below,sloped]{$X_4$} (1l);
        \draw[->-=.5] (1l) to [out=-45+20,in=135-20]node[pos=0.5,above,sloped]{$\tilde{X}_4$} (1d);
        \draw[blue] (1t) circle (0.25cm);
        \draw[blue] (1d) circle (0.25cm);
        \draw[red] (1l) circle (0.25cm);
        \draw[red] (1r) circle (0.25cm);
    \end{tikzpicture}}
    \caption{}
    \label{fig_midleaf121_more_decorations}
    \end{subfigure}
    \caption{(\subref{fig_midleaf121}): the non-trivial hypermultiplets of the middle leaf of \Figref{fig_hasseb2} in $3d$ $\Ncal=2$ representation. (\subref{fig_bottomleaf121}): the hypermultiplets of bottom leaf of \Figref{fig_hasseb2} in $3d$ $\Ncal=2$ representation.(\subref{fig_midleaf121_more_decorations}): The same quiver as in (\subref{fig_midleaf121}), but with a further decoration added to the previously undecorated $\urm(1)$ nodes.}
    \label{fig_n4ton2}
\end{figure}
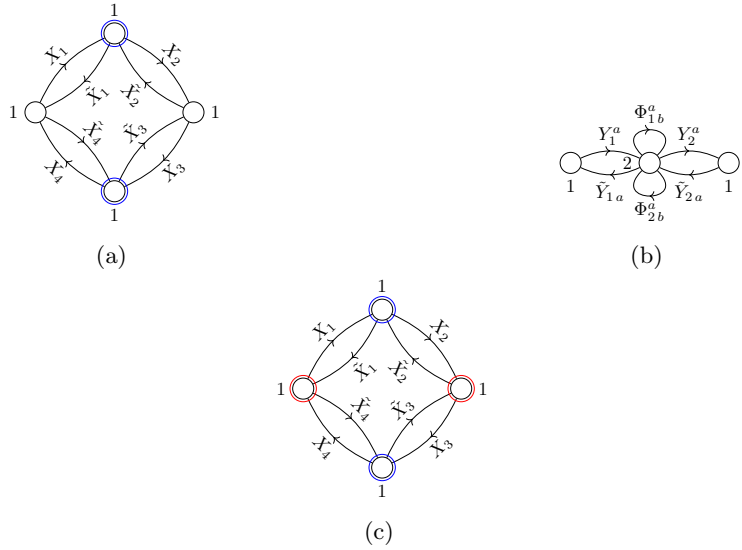

Consider how the decoration affects the operators in the chiral ring of \Figref{fig_hasseb2}; \Figref{fig_n4ton2} shows the hypermultiplets rewritten in terms of $3d$ $\Ncal=2$ multiplets (note that the free chiral hypermultiplets are omitted since they shall not appear in the F-term equations and GIOs).

The non-trivial Higgs branch operators in \Figref{fig_midleaf121} are given in Table \ref{tab_A3relation}.

\begin{table}[H]
\centering
\scalebox{1}{
\begin{tabular}{ccc}
\toprule
Generators & Chiral Polynomials & Degree\\ \midrule 
$x$ & $X_1X_2X_3X_4$ & $4$ \\  
$y$ & $\tilde{X}_1\tilde{X}_2\tilde{X}_3\tilde{X}_4$ & $4$ \\  
$z$ & $X_1\tilde{X}_1$ & $2$ \\  \addlinespace \midrule
 & Relations (F-term) & \\ \midrule
 & $X_1\tilde{X}_1=X_2\tilde{X}_2=X_3\tilde{X}_3=X_4\tilde{X}_4$ & $2$ \\ \addlinespace \midrule
Relations (GIOs) & & \\ \midrule
$xy=z^4$ & & $8$ \\ \bottomrule
\end{tabular}}
\caption{Chiral ring operators of \Figref{fig_midleaf121}, the middle leaf in \Figref{fig_hasseb2}.}
\label{tab_A3relation}
\end{table}
The Weyl group action exchanges the two decorated $\urm(1)$ nodes, which is inherited from the $S_2$ Weyl action of the original $\urm(2)$ exchanging the two $\urm(1)$s in its Cartan matrix. Under this action, the scalars transform as:
\begin{equation}
    X_1 \longleftrightarrow \tilde{X}_4,\ X_2 \longleftrightarrow \tilde{X}_3,\ X_3 \longleftrightarrow \tilde{X}_2,\ X_4 \longleftrightarrow \tilde{X}_1.
\label{eq_S2map}
\end{equation}
Note that this is not a gauge transformation but an equivalence between two choices of fields to describe the Higgs branch, so no quotient is taken under this action. The induced action on the GIOs is $x\longleftrightarrow y$, which thus gives a ring isomorphism (i.e. the Higgs branch is still the $A_3$ singularity under the change of embedding but the canonical sets of generators are different). This is a monodromy map as it `moves around' the singular locus of the base leaf, modifying the deformation space as described in \cite{Generic_singularities, wu2023namikawaweyl, Travis_toappear}. Roughly speaking, the two locally separated $\mathbb{P}^1$s in the fully resolved Higgs branch are connected globally, which means that going around the singular locus of the base leaf connects one $\mathbb{P}^1$ to another. Turning on FI parameters, the F-term equations are modified to the following:
\begin{subequations}
\begin{align}
    X_1\tilde{X}_1-X_2\tilde{X}_2 &=\zeta_{12}\,,\\
    X_2\tilde{X}_2-X_3\tilde{X}_3 &=\zeta_{23}\,,\\
    X_3\tilde{X}_3-X_4\tilde{X}_4 &=\zeta_{34}\,,\\
    X_4\tilde{X}_4-X_1\tilde{X}_1 &=\zeta_{41}\,.
\end{align}
\end{subequations}
It is useful to choose a new parameterization $\zeta_{ij}=h_i-h_j$, such that the constraint $\zeta_{12}+\zeta_{23}+\zeta_{34}+\zeta_{41}=0$ is manifest. The algebraic equations are thus deformed to:
\begin{equation}
    xy=z(z-\zeta_{12})(z-\zeta_{13})(z-\zeta_{14}).
\end{equation}
The action \eqref{eq_S2map} induces a map on the FI terms as:
\begin{equation}
    \zeta_{12}\to-\zeta_{34},\ \zeta_{23}\to-\zeta_{23},\ \zeta_{34}\to-\zeta_{12},\ \zeta_{41}\to-\zeta_{41}.
\label{eq_S2mapFI}
\end{equation}
This map can be treated as an orientifold action on the deformation space, reducing the Higgs branch Namikawa-Weyl group from $\Wcal_{A_3}\cong S_4$ to $\Wcal_{B_2}\cong \Z_2^2\rtimes S_2$.

\begin{figure}[H]
    \centering

\def\centerarc[#1](#2)(#3:#4:#5)
    { \draw[#1] ($(#2)+({#5*cos(#3)},{#5*sin(#3)})$) arc (#3:#4:#5); }

\begin{tikzpicture}[scale=0.8, transform shape]
  \def\smallradius{1cm}
  \def\bigradius{2cm}
  \def\labeldist{0.75cm}

  \coordinate (C1) at (0,0);
  \coordinate (C2) at ($(C1) + (2*\smallradius, 0)$);
  \coordinate (C3) at ($(C2) + (2*\smallradius, 0)$);
  
  \foreach \center in {C1, C2, C3}
    \centerarc[line width=1pt](\center)(0:360:\smallradius);
  
  \centerarc[thick]($(C1) - (\bigradius,0) - (\smallradius,0)$)(270:450:\bigradius);
  \centerarc[thick]($(C3) + (\bigradius,0) + (\smallradius,0)$)(90:270:\bigradius);

  \draw[dashed, red] ($(C2) - (0, 1.5*\bigradius)$) -- ++(0, 3*\bigradius);
  \draw[dotted] ($(C1) - (\smallradius,0)$) -- ++ (0, -\bigradius);
  \draw[dotted] ($(C2) - (\smallradius,0)$) -- ++ (0, -\bigradius);
  \draw[dotted] ($(C3) - (\smallradius,0)$) -- ++ (0, -\bigradius);
  \draw[dotted] ($(C3) + (\smallradius,0)$) -- ++ (0, -\bigradius);

  \centerarc[<->,thick,red]($(C2) + (0, 1.5*\smallradius + 10pt)$)(20:160:0.7*\smallradius);
  \node[red, above] at ($(C2) + (0, 1.5*\bigradius + 3pt)$) {$S_2$};

  \draw[<->] ($(C1) - (\smallradius, 1.8*\labeldist)$) -- ($(C2) - (\smallradius, 1.8*\labeldist)$) node[midway, fill=white] {$\zeta_{12}$};
  \draw[<->] ($(C2) - (\smallradius, 1.8*\labeldist)$) -- ($(C3) - (\smallradius, 1.8*\labeldist)$) node[midway, fill=white] {$\zeta_{23}$};
  \draw[<->] ($(C3) - (\smallradius, 1.8*\labeldist)$) -- ($(C3) - (-\smallradius, 1.8*\labeldist)$) node[midway, fill=white] {$\zeta_{34}$};

  \node[below] at ($(C1) - (\smallradius, 2.5*\labeldist)$) {$h_1$};
  \node[below] at ($(C2) - (\smallradius, 2.5*\labeldist)$) {$h_2$};
  \node[below] at ($(C3) - (\smallradius, 2.5*\labeldist)$) {$h_3$};
  \node[below] at ($(C3) + (\smallradius, -2.5*\labeldist)$) {$h_4$};

\end{tikzpicture}
    \caption{Schematic illustration of $A_3$ resolution under $S_2$ action. The FI parameters are the diameters of the $\mathbb{P}_1$ fibers.}
    \label{fig_A3FI}
\end{figure}
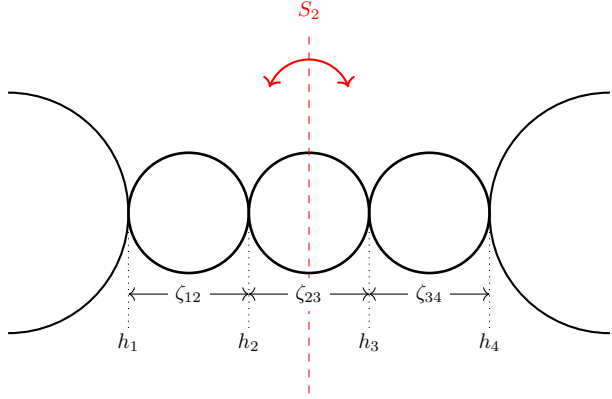
Now 
consider what happens if another set of decorations are placed on the two undercoated $\urm(1)$s in \Figref{fig_midleaf121}, as shown in \Figref{fig_midleaf121_more_decorations}. Interestingly, this new decoration does not give rise to a new $S_2$ monodromy map but in fact reproduces the same map as the first. This is of course expected from the fact that the $A_3$ finite Dynkin diagram has a single $S_2$ outer automorphism.
\\ This retention of the original $S_2$ action is also manifest from the deformation parameters of \Figref{fig_bottomleaf121}, the bottom leaf of \Figref{fig_hasseb2}.
The $F$-term equations read:
\begin{subequations}
\begin{align}
    Y_1^a \tilde{Y}_{1\, a}&=\zeta_1\,,\\
    Y_2^a \tilde{Y}_{2\, a}&=\zeta_2\,,\\
    Y_1^a \tilde{Y}_{1\, b} - Y_2^a \tilde{Y}_{2\, b} + \Phi_{1\, c}^a \Phi_{2\, b}^c - \Phi_{2\, c}^a \Phi_{1\, b}^c &=\zeta_{12} \boldone^a_b\,.
\end{align}
\end{subequations}
Which gives rise to the constraint $\zeta_1-\zeta_2-2\zeta_{12}=0$. The Weyl symmetry of the FI term is $\Wcal_{B_2}\cong \Z_2^2\rtimes S_2$, generated by the changes of sign of $\zeta_1$ and $\zeta_2$ and the permutation between them:
\begin{equation}
    \Z_2^2\rtimes S_2=\langle\zeta_1\to-\zeta_1\,,\zeta_2\to-\zeta_2\,,\zeta_1\longleftrightarrow\zeta_2\rangle\,.
\end{equation}

\begin{figure}[H]
    \centering
        \begin{subfigure}{0.45\textwidth}
        \centering
        \scalebox{.7}{
        \begin{tikzpicture}
        \node[gauge, label=below:$2$] (0) at (0,0){};
        \node[gauge, label=above:$1$] (1) at (1,1){};
        \node[gauge, label=right:$1$] (2) at (1,-1){};
        \node[gauge, label=above:$1$] (3) at (-1,1){};
        \node[gauge, label=left:$1$] (4) at (-1,-1){};
        \draw (1)--(0)--(2) (3)--(0)--(4);
        \draw[blue] (1) circle (0.2cm);
        \draw[blue] (2) circle (0.2cm);
    \end{tikzpicture}}
    \caption{}
    \label{fig_con1}
    \end{subfigure}
        \begin{subfigure}{0.45\textwidth}
        \centering
        \scalebox{.7}{
        \begin{tikzpicture}
        \node[gauge, label=below:$2$] (0) at (0,0){};
        \node[gauge, label=above:$1$] (1) at (1,1){};
        \node[gauge, label=right:$1$] (2) at (1,-1){};
        \node[gauge, label=above:$1$] (3) at (-1,1){};
        \node[gauge, label=left:$1$] (4) at (-1,-1){};
        \draw (1)--(0)--(2) (3)--(0)--(4);
        \draw[blue] (1) circle (0.2cm);
        \draw[blue] (2) circle (0.2cm);
        \draw[blue] (3) circle (0.2cm);
    \end{tikzpicture}}
    \caption{}
    \label{fig_con2}
    \end{subfigure}
            \begin{subfigure}{0.45\textwidth}
        \centering
        \scalebox{.7}{
        \begin{tikzpicture}
        \node[gauge, label=below:$2$] (0) at (0,0){};
        \node[gauge, label=above:$1$] (1) at (1,1){};
        \node[gauge, label=right:$1$] (2) at (1,-1){};
        \node[gauge, label=above:$1$] (3) at (-1,1){};
        \node[gauge, label=left:$1$] (4) at (-1,-1){};
        \draw (1)--(0)--(2) (3)--(0)--(4);
        \draw[blue] (1) circle (0.2cm);
        \draw[blue] (2) circle (0.2cm);
        \draw[blue] (3) circle (0.2cm);
        \draw[blue] (4) circle (0.2cm);
    \end{tikzpicture}}
    \caption{}
    \label{fig_con3}
    \end{subfigure}
            \begin{subfigure}{0.45\textwidth}
        \centering
        \scalebox{.7}{
        \begin{tikzpicture}
        \node[gauge, label=below:$2$] (0) at (0,0){};
        \node[gauge, label=above:$1$] (1) at (1,1){};
        \node[gauge, label=right:$1$] (2) at (1,-1){};
        \node[gauge, label=above:$1$] (3) at (-1,1){};
        \node[gauge, label=left:$1$] (4) at (-1,-1){};
        \draw (1)--(0)--(2) (3)--(0)--(4);
        \draw[blue] (1) circle (0.2cm);
        \draw[blue] (2) circle (0.2cm);
        \draw[red] (3) circle (0.2cm);
        \draw[red] (4) circle (0.2cm);
    \end{tikzpicture}}
    \caption{}
    \label{fig_con4}
    \end{subfigure}
    \caption{Affine $D_4$ under (\subref{fig_con1}): Two decorated nodes. (\subref{fig_con2}): Three decorated nodes. (\subref{fig_con3}): Four decorated nodes. (\subref{fig_con4}): Two nodes under one decoration and two two nodes under another.}
    \label{fig_D4decoration}
\end{figure}
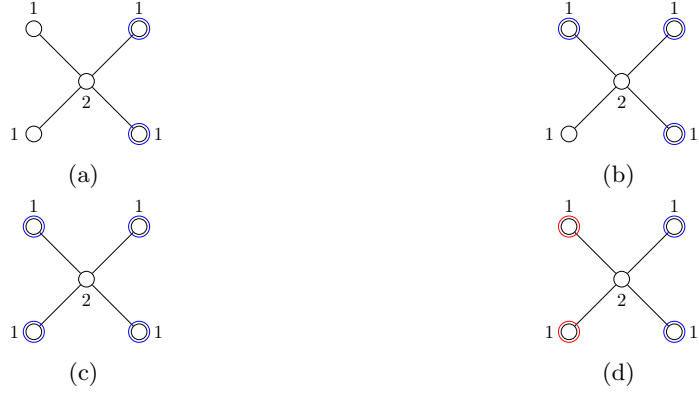

\paragraph{From $D_4$ to $B_3$, $G_2$.}
Consider the various examples of the affine $D_4$ quiver with decorations, as given in \Figref{fig_D4decoration}. There are four different cases: (\subref{fig_con1}) two decorations, (\subref{fig_con2}) three decorations, (\subref{fig_con3}) four decorations, (\subref{fig_con4}) two decorations of one type and two decorations of another type.

\begin{figure}[H]
\centering
\begin{tikzpicture}
        \node[biggauge, label=left:$2$] (1) at (-1.5,0){};
        \node[biggauge, label=below:$1$] (1dr) at (0,-1.5){};
        \node[biggauge, label=above:$1$] (1ur) at (0,1.5){};
        \node[biggauge, label=below:$1$] (1dl) at (-3,-1.5){};
        \node[biggauge, label=above:$1$] (1ul) at (-3,1.5){};
        \draw[->-=.5] (1) to [out=45+20,in=225-20]node[pos=0.5,above,sloped]{$X_1$} (1ur);
        \draw[->-=.5] (1ur) to [out=225+20,in=45-20]node[pos=0.5,below,sloped]{$\tilde{X}_1$} (1);
        \draw[->-=.5] (1) to [out=-45+20,in=135-20]node[pos=0.5,above,sloped]{$X_2$} (1dr);
        \draw[->-=.5] (1dr) to [out=135+20,in=-45-20]node[pos=0.5,below,sloped]{$\tilde{X}_2$} (1);
        \draw[->-=.5] (1) to [out=-135+20,in=45-20]node[pos=0.5,below,sloped]{$X_3$} (1dl);
        \draw[->-=.5] (1dl) to [out=45+20,in=-135-20]node[pos=0.5,above,sloped]{$\tilde{X}_3$} (1);
        \draw[->-=.5] (1) to [out=135+20,in=-45-20]node[pos=0.5,below,sloped]{$X_4$} (1ul);
        \draw[->-=.5] (1ul) to [out=-45+20,in=135-20]node[pos=0.5,above,sloped]{$\tilde{X}_4$} (1);
    \end{tikzpicture}
    \caption{$3d$ $\Ncal=2$ decomposition of hypermultiplets of $D_4$ quiver.}
    \label{fig_D43dn2}
\end{figure}
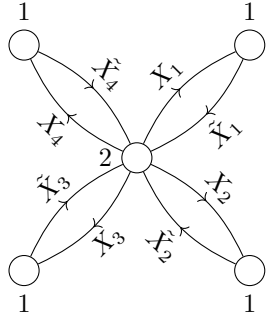

The $3d$ $\Ncal=2$ decomposition of hypermultiplets of $D_4$ quiver is shown in \Figref{fig_D43dn2}, and the non-trivial Higgs branch operators are given in in Table \ref{tab_A3relation}.

\begin{table}[H]
\centering
\scalebox{1}{
\begin{tabular}{ccc}
\toprule
Generators & Chiral Polynomials & Degree\\ \midrule 
$w$ & $X_3^a \tilde{X}_{3\, b} X_4^b \tilde{X}_{4\, a}$ & $4$ \\  
$v$ & $(X_1^a \tilde{X}_{1\, b}-X_2^a \tilde{X}_{2\, b})X_3^b \tilde{X}_{3\, a}$ & $4$ \\  
$u$ & $(X_1^a \tilde{X}_{1\, c}-X_2^a \tilde{X}_{2\, c}) X_3^c \tilde{X}_{3\, b} X_4^b \tilde{X}_{4\, a}$ & $6$ \\  \addlinespace \midrule
 & Relations (F-term) & \\ \midrule
 & $X_1^a \tilde{X}_{1\, a}=X_2^a\tilde{X}_{2\, a} = X_3^a \tilde{X}_{3\, a}=X_4^a \tilde{X}_{4\, a}=0$ & $2$ \\
 & $X_1^a \tilde{X}_{1\, b}+X_2^a \tilde{X}_{2\, b}+X_3^a\tilde{X}_{3\, b}+X_4^a \tilde{X}_{4\, b}=0$ & $2$ \\ \addlinespace \midrule
Relations(GIO) &  & \\ \midrule
$u^2+v^2w=w^3$ &  & $12$ \\  \bottomrule
\end{tabular}}
\caption{Chiral ring operators of the affine $D_4$ quiver.}
\label{tab_D4relation}
\end{table}

For case (\subref{fig_con1}), these two decorated $\urm(1)$s are exchanged by an $S_2$ - the scalars change as:
\begin{equation}
    X_1\longleftrightarrow X_2,\ \tilde{X}_1\longleftrightarrow\tilde{X}_2.
\end{equation}

The induced action on the GIOs is $v\to -v,\ u\to -u$, which gives a ring isomorphism. Adding a further set of decorations on the undecorated $\urm(1)$s generates the same $S_2$ action; hence, there is an identification of the quivers in \Figref{fig_con1} and \Figref{fig_con4}. The Namikawa Weyl group is $\Wcal_{B_3}$. The slice is called $C_3$ following the notation of \cite{Generic_singularities}, which will become clearer following the discussion of the symplectic dual of this slice.

The identification between \Figref{fig_con1} and \Figref{fig_con4} also implies the identification between \Figref{fig_con2} and \Figref{fig_con3}. It is easy to check that \Figref{fig_con2} and \Figref{fig_con3} have the same $S_3$ monodromy map; the Namikawa Weyl group is $\Wcal_{G_2}$ and the slice is $G_2$.

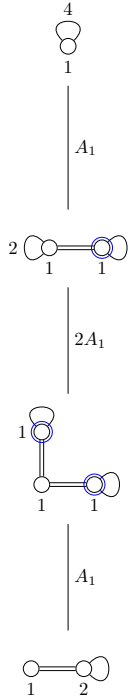
\begin{figure}[H]
    \centering
    \scalebox{.7}{
    \begin{tikzpicture}
        \node (a) at (0,0){$\begin{tikzpicture}

        \node[gauge, label=below:$1$] (1) at (0,0){};
        \node[gauge, label=below:$2$] (2) at (1,0){};

        \draw[transform canvas={yshift=-1pt}] (1)--(2);
        \draw[transform canvas={yshift=1pt}] (1)--(2);
        \draw[-] (2) to[out=-45,in=45,looseness=8] (2);
        \end{tikzpicture}$};
        
        \node (b) at (0,4){$\begin{tikzpicture}
        \node[gauge, label=below:$1$] (1) at (0,0){};
        \node[gauge, label=below:$1$] (2) at (1,0){};
        \node[gauge, label=left:$1$] (3) at (0,1){};

        \draw[transform canvas={yshift=-1pt}] (1)--(2);
        \draw[transform canvas={yshift=1pt}] (1)--(2);
        \draw[transform canvas={xshift=-1pt}] (1)--(3);
        \draw[transform canvas={xshift=1pt}] (1)--(3);
        \draw[-] (2) to[out=-45,in=45,looseness=8] (2);
        \draw[-] (3) to[out=45,in=135,looseness=8] (3);

        \draw[blue] (2) circle (0.2cm);
        \draw[blue] (3) circle (0.2cm);
        \end{tikzpicture}$};
        \node (c) at (0,8){$\begin{tikzpicture}
        \node[gauge, label=below:$1$] (1) at (0,0){};
        \node[gauge, label=below:$1$] (2) at (1,0){};

        \draw[transform canvas={yshift=-1pt}] (1)--(2);
        \draw[transform canvas={yshift=1pt}] (1)--(2);
        \draw[-] (2) to[out=-45,in=45,looseness=8] (2);
        \draw[-] (1) to[out=135,in=225,looseness=8]node[pos=0.5,left]{2} (1);
        \draw[blue] (2) circle (0.2cm);
        \end{tikzpicture}$};
        \node (d) at (0,12){$\begin{tikzpicture}
            \node[gauge,label=below:$1$] (1) at (0,0){};
            \draw[-] (1) to[out=45,in=135,looseness=8] node[pos=0.5,above]{4} (1);
        \end{tikzpicture}$};
        \draw[-] (a)--(b) node[pos=0.5,midway, right]{$A_1$};
        \draw[-] (c)--(d) node[pos=0.5,midway, right]{$A_1$};
        \draw (b)--(c) node[pos=0.5,midway, right]{$2A_1$};
    \end{tikzpicture}}
    \caption{The Higgs branch of this quiver is the $2^{\text{nd}}$ symmetric product of $A_1$.}
    \label{fig_HassenSym2A1}
\end{figure}

\paragraph{A union of $2$.}
A simple case containing a union of slices appears in the Higgs branch Hasse diagram in Figure \ref{fig_HassenSym2A1}. On the next-to-minimal leaf, there are two $A_1$ sub-quivers that contribute two $A_1$ components of the transverse slice. The two decorated $\urm(1)$s are exchanged under the action of $\Wcal_{\urm(2)} \cong S_2$, the Weyl group of the gauge group $\urm(2)$. Since this $S_2$ Weyl action belongs to the automorphism of the gauge transformation, these two components should not be connected to two different Higgsing phases/leaves, but form a union and are connected to a single phase/leaf.
\\In general, the union and monodromy maps can appear at the same time. Examples of this are given in section \ref{sec_INSI}.
\section{Examples from Nilpotent Cones}
\label{sec_INSI}
Nilpotent cones (nilcones) are a distinguished class of symplectic singularities that underpin much of the progress made in recent years concerning the moduli spaces of vacua of gauge theories with eight supercharges. Although generic Hasse diagrams can be constructed straightforwardly using minimal degenerations, nilpotent cones are one of the few examples wherein the Higgs branch of a gauge theory can be checked against a known Hasse diagram -- in this sense, graduating the nilpotent d\={o}j\={o} provides a non-trivial test of any subtraction algorithm.\footnote{Reciprocally, most recent additions to the combinatoric toolbox for quiver gauge theories have their origin in nilcones.}

The Higgs branch subtraction algorithm presented in this paper identifies several gauge theories whose Higgs branches are S{\l}odowy intersections inside nilcones of simple Lie algebras. Such theories are identified both by their Hasse diagrams and Hilbert series. Of course, Hasse diagrams are not unique to a moduli space, and thus explicit Hilbert series computations are required, alongside calculations involving the localisation formula \cite{Cremonesi:2014uva,Gadde:2011uv,Hanany:2017ooe,Cabrera:2018ldc,Hanany:2019tji}, to verify the identity of the moduli spaces presented in the following section.

Some quivers presented in this section contain hypermultiplets in the adjoint representation of a unitary gauge group -- each such hypermultiplet contributes an extra degree 1 generator of the Higgs branch and hence gives a free factor $\mathbb H$ to the Higgs branch. In the analysis of this section, such extra factors are ignored.

A further interesting property of nilcones concerns the action of \textit{Lusztig-Spaltenstein duality} \cite{lusztig1979class, lusztig1982class, spaltenstein2006classes, juteau2023minimal}, which acts as an involution and appears to relate the Higgs and Coulomb branches of theories when both lie in the same nilcone.

In the following, a quiver will be denoted by $\mathcal Q$ with either an equation number or figure number in the subscript. The Higgs branch of a quiver will be denoted with $\mathcal H$ and its Coulomb branch by $\mathcal C$.

\subsection{Examples of S{\l}odowy Intersections in $G_2$, $F_4$, and $E_8$ Nilcones}
The following one-parameter family of quivers $\mathcal Q_{\ref{eq_12n+1loop}}$ contains several examples of theories whose moduli spaces of vacua appear in the nilcones of exceptional algebras. 

They equally arise as magnetic quivers corresponding to intermediate leaves in the phase diagram of $n$ M5 branes probing an $A_1$ Klein singularity with either an M9 brane or a non-trivial boundary condition on one side.

This one-parameter family of quivers is also the $3d$ mirror quiver for the $4d\;\mathcal N=2$ class $\mathcal S$ theory \cite{Gaiotto:2009we, Benini:2010uu} specified by algebra $A_{n-1}$ on a torus with one puncture labelled by partition $(n-2,1^2)$.

The following examples consider the cases $n=3,4,5$, as these cases are the only ones where the Higgs branch and the Coulomb branch (by Lusztig-Spaltenstein duality \cite{Generic_singularities}) are slices in a nilcone. The moduli spaces of vacua for the cases $n=3,4,5$ are slices in the $G_2,\;F_4,$ and $E_8$ nilcones respectively. In fact, any nilcone with a Lusztig canonical quotient $S_{n}$, $n=3,4,5$ associated to a leaf will contain both the Higgs and Coulomb branches of \eqref{eq_12n+1loop} for $n=3,4,5$ respectively.

\begin{equation}
    \mathcal Q_{\ref{eq_12n+1loop}}=\raisebox{-.4\height}{\begin{tikzpicture}
        \node[gauge, label=below:$1$] (1) at (0,0){};
        \node[gauge, label=below:$2$] (2) at (1,0){};
        \node[gauge, label=below:$n$] (3) at (2,0) {};
        \draw[-] (1)--(2)--(3);
        \draw[-] (3) to[out=-45,in=45,looseness=8] (3);\label{eq_12n+1loop}
    \end{tikzpicture}}.
\end{equation}

Note that the following examples will make use of Coulomb branch quiver subtraction \cite{Bourget:2023dkj, Bourget:2024mgn, Cabrera:2018ann} to compute Hasse diagrams for the Coulomb branches of various unitary quivers.

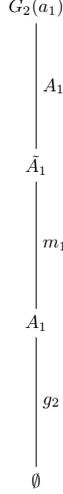
\begin{figure}[h!]
    \centering
        \scalebox{.7}{
        \begin{tikzpicture}
            \node (a) at (0,0) {$\emptyset$};
            \node (b) at (0,3) {$A_1$};
            \node (c) at (0,6) {$\tilde{A_1}$};
            \node (d) at (0,9) {$G_2(a_1)$};
            \draw[-] (a)--(b) node[pos=0.5,midway, right]{$g_2$}--(c) node[pos=0.5,midway,right]{$m_1$}--(d) node[pos=0.5,midway, right]{$A_1$};
        \end{tikzpicture}
        }
    \caption{The Hasse diagram of $\overline{sub. reg. G_2}$ in the $G_2$ nilcone with each orbit specified by its Bala-Carter label. This is the same Hasse diagram for the Coulomb branch of $\mathcal Q_{\ref{eq_subregG2}}$.}
    \label{fig_HassesubregG2Coul}
\end{figure}

\paragraph{$G_2$ Example.}
First consider this theory at $n=3$, 
\begin{equation}
    \mathcal Q_{\ref{eq_subregG2}}=\raisebox{-.4\height}{\begin{tikzpicture}
        \node[gauge, label=below:$1$] (1) at (0,0){};
        \node[gauge, label=below:$2$] (2) at (1,0){};
        \node[gauge, label=below:$3$] (3) at (2,0) {};

        \draw[-] (1)--(2)--(3);
        \draw[-] (3) to[out=-45,in=45,looseness=8] (3);\label{eq_subregG2}
    \end{tikzpicture}}.
\end{equation}
The Coulomb branch of $\mathcal Q_{\ref{eq_subregG2}}$ is well known to be $\overline{sub. reg. G_2}$ \cite{Gaiotto:2012uq,Cremonesi:2014vla}. The Hasse diagram for this moduli space is given in \Figref{fig_HassesubregG2Coul}, which may be computed with quiver subtraction \cite{Cabrera:2018ann} or with the fusion and fission algorithm \cite{Bourget:2023dkj,Bourget:2024mgn}.

Lusztig-Spaltenstein duality \cite{Generic_singularities}, which acts on the special pieces of \Figref{fig_HassesubregG2Coul}, suggests that the Higgs branch is the S{\l}odowy slice to the minimal nilpotent orbit of $G_2$, i.e. $\mathcal S^{G_2}_{\mathcal N,A_1}$. Using the Higgs branch subtraction algorithm established in Section \ref{sec_DQDS}, the Hasse diagram of $\mathcal H\left(\mathcal Q_{\ref{eq_subregG2}}\right)$ is computed to be \Figref{fig_HassesubregG2a}, consistent with the Hasse diagram of $\mathcal S^{G_2}_{\mathcal N,A_1}$\cite{Generic_singularities} given in \Figref{fig_HassesubregG2b}. The top slice is identified as $G_2$ since this corresponds to the subtraction of an affine $D_4$ quiver with three decorated $\urm(1)$, as given in Table \ref{tab_foldedslices}. Additionally, the Namikawa-Weyl group for the Higgs branch is identified as $\mathcal W_{G_2}=\mathrm{Dih}_{6}$ of order 12 using the prescription in Section \ref{sec_NWG}. The Higgs branch Hilbert series of $\mathcal Q_{\ref{eq_subregG2}}$ is computed to be 
\begin{equation}
    \hs[\mathcal H(\mathcal Q_{\ref{eq_subregG2}})]=\pe\left[[2]_{\surm(2)}t^2+[3]_{\surm(2)}t^3-t^{12}\right]=\hs\left[\mathcal S^{G_2}_{\mathcal N,A_1}\right]
\end{equation} 
where the label $[n]_{G}$ specifies the character of the representation given by the Dynkin label. This confirms that the Higgs branch $\mathcal H(\mathcal Q_{\ref{eq_subregG2}})=\mathcal S^{G_2}_{\mathcal N,A_1}$.

\begin{figure}[h!]
    \centering
    \begin{subfigure}{0.49\textwidth}
    \centering
        \scalebox{.7}{
        \begin{tikzpicture}
        \node (a) at (0,0){$\begin{tikzpicture}
        \node[gauge, label=below:$1$] (1) at (0,0){};
        \node[gauge, label=below:$2$] (2) at (1,0){};
        \node[gauge, label=below:$3$] (3) at (2,0) {};
        \draw[-] (1)--(2)--(3);
        \draw[-] (3) to[out=-45,in=45,looseness=8] (3);
        \end{tikzpicture}$};
        
        \node (b) at (0,3){$\begin{tikzpicture}
        \node[gauge, label=below:$1$] (1) at (0,0){};
        \node[gauge, label=below:$2$] (2) at (1,0){};
        \node[gauge, label=below:$2$] (2r) at (2,0){};
        \node[gauge, label=left:$1$] (1t) at (1,1){};
        \draw[-] (1t) to[out=45,in=135,looseness=8] (1t);
        \draw[-] (1)--(2)--(2r) (2)--(1t);
        \draw[-] (2r) to[out=-45,in=45,looseness=8](2r){};
        \draw[blue] (1t) circle (0.2cm);
        \draw[blue] (2r) circle (0.2cm);
        \end{tikzpicture}$};
        
        \node (c) at (0,6){$\begin{tikzpicture}
        \node[gauge,label=below:$1$] (1) at (0,0){};
        \node[gauge, label=below:$2$] (2) at (1,0){};
        \node[gauge, label=above:$1$] (1r) at (2,0){};
        \node[gauge, label=above:$1$] (1t) at ({cos(45)+1},{sin(45)}){};
        \node[gauge, label=above:$1$] (1b) at ({cos(45)+1},{-sin(45)}){};
        \draw[-] (1r) to[out=45,in=315,looseness=8] (1r);
        \draw[-] (1t) to[out=45,in=315,looseness=8] (1t);
        \draw[-] (1b) to[out=45,in=315,looseness=8] (1b);
        \draw[-] (1)--(2)--(1r) (1t)--(2)--(1b);
        \draw[blue] (1t) circle (0.2cm);
        \draw[blue] (1r) circle (0.2cm);
        \draw[blue] (1b) circle (0.2cm);
        \end{tikzpicture}$};
        
        \node (d) at (0,9){$\begin{tikzpicture}
            \node[gauge,label=below:$1$] (1) at (0,0){};
            \draw[-] (1) to[out=45,in=135,looseness=8] node[pos=0.5,above]{$4$} (1);
        \end{tikzpicture}$};
        \draw[-] (a)--(b) node[pos=0.5,midway, right]{$m_1$}--(c) node[pos=0.5,midway,right]{$A_1$}--(d) node[pos=0.5,midway, right]{$G_2$};
    \end{tikzpicture}}
    \caption{}
    \label{fig_HassesubregG2a}
    \end{subfigure}
    \begin{subfigure}{0.49\textwidth}
        \centering
        \scalebox{.7}{
        \begin{tikzpicture}
            \node (a) at (0,0) {$A_1$};
            \node (b) at (0,3) {$\tilde{A_1}$};
            \node (c) at (0,6) {$G_2(a_1)$};
            \node (d) at (0,9) {$G_2$};
            \draw[-] (a)--(b) node[pos=0.5,midway, right]{$m_1$}--(c) node[pos=0.5,midway,right]{$A_1$}--(d) node[pos=0.5,midway, right]{$G_2$};
        \end{tikzpicture}
        }
        \caption{}
        \label{fig_HassesubregG2b}
    \end{subfigure}
    \caption{(\subref{fig_HassesubregG2a}): The Higgs branch Hasse diagram of $\mathcal Q_{\ref{eq_subregG2}}$. (\subref{fig_HassesubregG2b}): The Hasse diagram for the S{\l}odowy slice $\mathcal S^{G_2}_{\mathcal N,A_1}$ in the $G_2$ nilcone to the minimal orbit with orbits labelled by their Bala-Carter labels. This is the same Hasse diagram as in \Figref{fig_HassesubregG2a}.}
    \label{fig_HassesubregG2}
\end{figure}
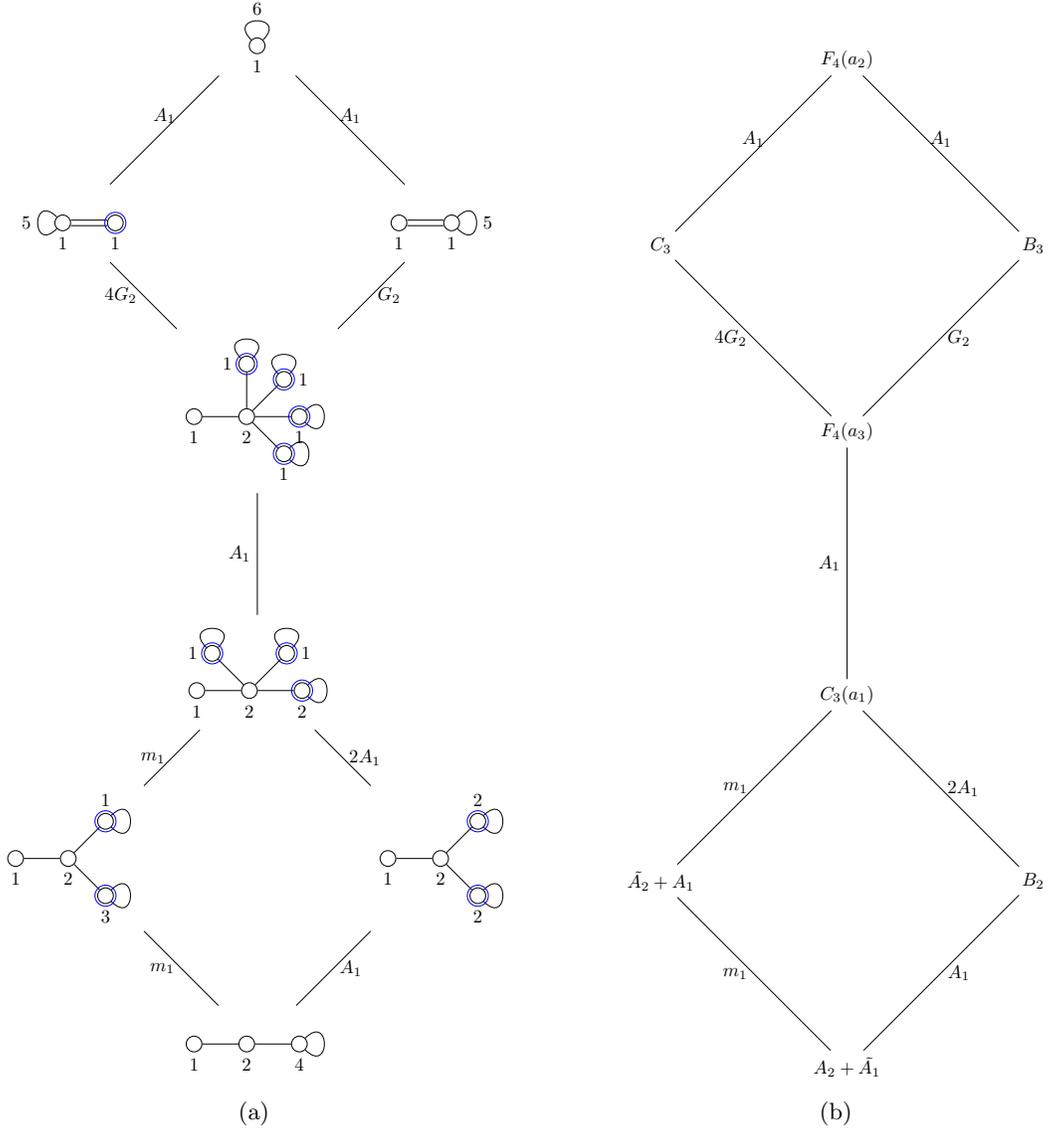
\begin{figure}[h]
    \centering
    \begin{subfigure}{0.49\textwidth}
        \centering
         \scalebox{.7}{
    \begin{tikzpicture}
        \node (124) at (0,0){$\begin{tikzpicture}
            \node[gauge, label=below:$1$] (1) at (0,0){};
            \node[gauge, label=below:$2$] (2) at (1,0){};
            \node[gauge, label=below:$4$] (4) at (2,0){};

            \draw (1) to (2) to (4) to[out=-45,in=45,looseness=8] (4);
        \end{tikzpicture}$};
        \node (1213) at ({-5*cos(45)},{5*sin(45)}) {$\begin{tikzpicture}
            \node[gauge, label=below:$1$] (1l) at (0,0){};
            \node[gauge, label=below:$2$] (2) at (1,0){};
            \node[gauge, label=above:$1$] (1t) at ({1+cos(45)},{sin(45)}){};
            \node[gauge, label=below:$3$] (3) at ({1+cos(45)},{-sin(45)}){};

            \draw (1l) to (2) to (3) to[out=-45,in=45,looseness=8] (3) (2) to (1t)to[out=-45,in=45,looseness=8] (1t);

            \draw[blue] (1t) circle (0.2cm);
            \draw[blue] (3) circle (0.2cm);
        \end{tikzpicture}$};
        \node (1222) at ({5*cos(45)},{5*sin(45)}){$\begin{tikzpicture}
            \node[gauge, label=below:$1$] (1l) at (0,0){};
            \node[gauge, label=below:$2$] (2) at (1,0){};
            \node[gauge, label=above:$2$] (2t) at ({1+cos(45)},{sin(45)}){};
            \node[gauge, label=below:$2$] (2b) at ({1+cos(45)},{-sin(45)}){};

            \draw (1l) to (2) to (2b) to[out=-45,in=45,looseness=8] (2b) (2) to (2t)to[out=-45,in=45,looseness=8] (2t);

            \draw[blue] (2t) circle (0.2cm);
            \draw[blue] (2b) circle (0.2cm);
        \end{tikzpicture}$};
        \node (12211) at ({0},{10*cos(45)}){$\begin{tikzpicture}
            \node[gauge, label=below:$1$] (1l) at (-1,0){};
            \node[gauge, label=below:$2$] (2) at (0,0){};
            \node[gauge, label=below:$2$] (2r) at (1,0){};
            \node[gauge, label=left:$1$] (1tl) at ({-cos(45)},{sin(45)}){};
            \node[gauge, label=right:$1$] (1tr) at ({cos(45)},{sin(45)}){};

            \draw (1l) to (2) to (2r) to[out=-45,in=45,looseness=8] (2r) (1tl) to[out=45,in=135,looseness=8] (1tl) to (2) to (1tr) to[out=45,in=135,looseness=8] (1tr);

            \draw[blue] (2r) circle (0.2cm);
            \draw[blue] (1tr) circle (0.2cm);
            \draw[blue] (1tl) circle (0.2cm);
        \end{tikzpicture}$};
        \node (121111) at ({0},{5+10*cos(45)}){$\begin{tikzpicture}
            \node[gauge, label=below:$2$] (2) at (0,0){};
            \node[gauge, label=below:$1$] (1l) at (-1,0){};
            \node[gauge, label=left:$1$] (1t) at (0,1){};
            \node[gauge, label=right:$1$] (1tr) at ({cos(45)},{sin(45)}){};
            \node[gauge, label=below:$1$] (1r) at (1,0){};
            \node[gauge, label=below:$1$] (1br) at ({cos(45)},{-sin(45)}){};
            \draw[-] (1t) to[out=45,in=135,looseness=8] (1t);
            \draw[-] (1tr) to[out=45,in=135,looseness=8] (1tr);
            \draw[-] (1r) to[out=45,in=315,looseness=8] (1r);
            \draw[-] (1br) to[out=45,in=315,looseness=8] (1br);
            
            \draw (1l) to (2) to (1t) (1tr) to (2)to (1r) (1br)to (2);

            \draw[blue] (1t) circle (0.2cm);
            \draw[blue] (1tr) circle (0.2cm);
            \draw[blue] (1r) circle (0.2cm);
            \draw[blue] (1br) circle (0.2cm);
        \end{tikzpicture}$};

        \node (11) at ({+5*sin(45)},{5+15*cos(45)}){$\begin{tikzpicture}
            \node[gauge, label=below:$1$] (l) at (0,0){};
            \node[gauge,label=below:$1$]  (r) at (1,0){};
            \draw[-] (r) to[out=45,in=315,looseness=8] node[pos=0.5,right]{$5$} (r);
            \draw[transform canvas={yshift=1.5pt}] (l)--(r);
            \draw[transform canvas={yshift=-1.5pt}] (l)--(r);
        \end{tikzpicture}$};
        \node (11d) at ({-5*sin(45)},{5+15*cos(45)}){$\begin{tikzpicture}
            \node[gauge, label=below:$1$] (l) at (0,0){};
            \node[gauge,label=below:$1$]  (r) at (1,0){};
            \draw[-] (l) to[out=135,in=225,looseness=8] node[pos=0.5,left]{$5$} (l);
            \draw[transform canvas={yshift=1.5pt}] (l)--(r);
            \draw[transform canvas={yshift=-1.5pt}] (l)--(r);
            \draw[blue] (r) circle (0.2cm);
        \end{tikzpicture}$};

        \node (1) at ({0},{5+20*cos(45)}){$\begin{tikzpicture}
            \node[gauge, label=below:$1$](1){};
            \draw[-] (1) to[out=45,in=135,looseness=8] node[pos=0.5,above]{$6$} (1);
        \end{tikzpicture}$};
        
        \draw[-] (124) -- (1213) node[pos=0.5,midway,left]{$m_1$}--(12211) node[pos=0.5,midway, left]{$m_1$}--(121111) node[pos=0.5,midway, left]{$A_1$}--(11d) node[pos=0.5,midway,left]{$4G_2$}--(1) node[pos=0.5,midway, above]{$A_1$} (124)--(1222) node[pos=0.5,midway, right]{$A_1$}--(12211)node[pos=0.5,midway,right]{$2A_1$} (121111)--(11) node[pos=0.5,midway, right]{$G_2$}--(1) node[pos=0.5,midway, above]{$A_1$};
    \end{tikzpicture}}
        \caption{}
        \label{fig_HasseF4a}
    \end{subfigure}
    \begin{subfigure}{0.49\textwidth}
    \centering
    \scalebox{.7}{
    \begin{tikzpicture}
        \node (124) at (0,0) {$A_2+\tilde{A_1}$};
        \node (1213) at ({-5*cos(45)},{5*sin(45)}) {$\tilde{A_2}+A_1$};
        \node (1222) at ({5*cos(45)},{5*sin(45)}) {$B_2$};
        \node (12211) at ({0},{10*cos(45)}) {$C_3(a_1)$};
        \node (121111) at ({0},{5+10*cos(45)}) {$F_4(a_3)$};
        \node (11) at ({+5*sin(45)},{5+15*cos(45)}) {$B_3$};
        \node (11d) at ({-5*sin(45)},{5+15*cos(45)}) {$C_3$};
        \node (1) at ({0},{5+20*cos(45)})  {$F_4(a_2)$};
         \draw[-] (124) -- (1213) node[pos=0.5,midway,left]{$m_1$}--(12211) node[pos=0.5,midway, left]{$m_1$}--(121111) node[pos=0.5,midway, left]{$A_1$}--(11d) node[pos=0.5,midway,left]{$4G_2$}--(1) node[pos=0.5,midway, above]{$A_1$} (124)--(1222) node[pos=0.5,midway, right]{$A_1$}--(12211)node[pos=0.5,midway,right]{$2A_1$} (121111)--(11) node[pos=0.5,midway, right]{$G_2$}--(1) node[pos=0.5,midway, above]{$A_1$};
    \end{tikzpicture}
    }
    \caption{}
    \label{fig_HasseF4b}
    \end{subfigure}
    \caption{(\subref{fig_HasseF4a}): The Higgs branch Hasse diagram for the quiver $\mathcal Q_{\ref{eq_F4}}$. (\subref{fig_HasseF4b}): The Hasse diagram for the S{\l}odowy intersection $\mathcal S^{F_4}_{F_4(a_2),A_2+\tilde A_1}$ in the $F_4$ nilcone with orbits labelled by their Bala-Carter labels. This is the same Hasse diagram as \Figref{fig_HasseF4a}.}
    \label{fig_HasseF4124}
\end{figure}

\FloatBarrier
\paragraph{$F_4$ Example.}
Next consider $\mathcal Q_{\ref{eq_12n+1loop}}$ at $n=4$
\begin{equation}
    \mathcal Q_{\ref{eq_F4}}=\raisebox{-.4\height}{\begin{tikzpicture}
        \node[gauge, label=below:$1$] (1) at (0,0){};
        \node[gauge, label=below:$2$] (2) at (1,0){};
        \node[gauge, label=below:$4$] (3) at (2,0) {};

        \draw[-] (1)--(2)--(3);
        \draw[-] (3) to[out=-45,in=45,looseness=8] (3);
    \end{tikzpicture}}\label{eq_F4}
\end{equation}
The Hasse diagram for the Higgs branch is shown in \Figref{fig_HasseF4a}, which matches the Hasse diagram of the S{\l}odowy intersection, $\mathcal S^{F_4}_{F_4(a_2),A_2+\tilde{A}_1}$, in the nilcone of $F_4$ \cite{Generic_singularities} explicitly shown in \Figref{fig_HasseF4b}. From the Higgs branch Hasse diagram, the Namikawa-Weyl group for the Higgs branch is identified as $\mathbb Z_2\times\mathbb Z_2$. Note that the quiver corresponding to $F_4(a_3)$ admits two different subtractions as given in Table \ref{tab_foldedslices}. One possibility is the subtraction of an affine $D_4$ with four decorated $\urm(1)$ nodes, this gives rise to the $G_2$ slice. The other possibility is the subtraction of affine $D_4$ with three decorated $\urm(1)$ with ${4\choose 3}=4$ different subtractions, giving rise to the $4G_2$ slice.

This identification of $\mathcal H\left(\mathcal Q_{\ref{eq_F4}}\right)=\mathcal S^{F_4}_{F_4(a_2),A_2+\tilde A_1}$ is also confirmed by the refined Hilbert series, where for brevity only the unrefined series is presented. The result matches with the calculation from localization for S{\l}odowy intersections \cite{Cabrera:2018ldc}.
\begin{align}
    \hs\left[\mathcal H(\mathcal Q_{\ref{eq_F4}})\right]&=\frac{1 + 2 t^4 + 2 t^5 + 3 t^6 + 3 t^8 + 4 t^9 + 6 t^{10} + 4 t^{11} + 
 3 t^{12} + 3 t^{14} + 2 t^{15} + 
 2 t^{16} + t^{20}}{(1 - t^2)^3  (1 - t^3)^4 (1 - t^4)^3}
 \\&= \hs\left[\mathcal S^{F_4}_{F_4(a_2),A_2+\tilde A_1}\right]. \notag
\end{align}

\begin{figure}[h]
\centering
\begin{tikzpicture}
\node (a) at (0,0) {$A_1+\tilde A_1$};
\node (b) at (0,3) {$A_2$};
\node (c) at (0,6) {$A_2+\tilde A_1$};
\node (d) at ({-3*cos(45)},{6+3*sin(45)}) {$\tilde A_2+A_1$};
\node (e) at ({+3*cos(45)},{6+3*sin(45)}) {$B_2$};
\node (f) at (0,{6+6*sin(45)}) {$C_3(a_1)$};
\node (g) at (0,{9+6*sin(45)}) {$F_4(a_3)$};

 \draw[-] (a) -- (b) node[pos=0.5,midway,left]{$A_1$}--(c) node[pos=0.5,midway, left]{$a_2^+$}--(d) node[pos=0.5,midway, left]{$m_1$}--(f) node[pos=0.5,midway, left]{$m_1$}--(g) node[pos=0.5,midway,left]{$A_1$};

 \draw[-] (c)--(e)node[pos=0.5,midway, right]{$A_1$}--(f)node[pos=0.5,midway, right]{$A_1$};
\end{tikzpicture}

\caption{The Hasse diagram for $\mathcal S^{F_4}_{F_4(a_3),A_1+\tilde A_1}$ in the $F_4$ nilcone with orbits labelled by their Bala-Carter labels. This Hasse diagram is the same as for the Coulomb branch of  $Q_{\ref{eq_F4}}$.}
\label{fig_F4HasseCoul}
\end{figure}
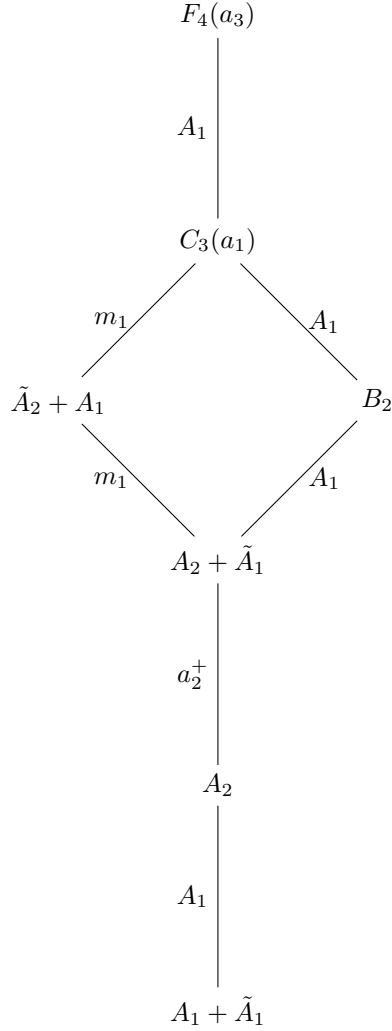

By Lusztig-Spaltenstein duality, the Coulomb branch of $\mathcal Q_{\ref{eq_F4}}$ also appears in the $F_4$ nilcone. To compute the Lusztig-Spaltenstein dual we note that $F_4(a_2)$ is a special orbit, hence there is no problem in computing its dual which is the special orbit $A_1+\tilde A_1$. On the other hand, the $A_2+\tilde A_1$ orbit is the lowest non special orbit in an $S_4$ special piece which is self dual. Hence the dual is the special orbit $F_4(a_3)$. Calculation of the refined Hilbert series for the Coulomb branch verifies that the Coulomb branch of $\mathcal Q_{\ref{eq_F4}}$ is the S{\l}odowy intersection $\mathcal S^{F_4}_{F_4(a_3),A_1+\tilde A_1}$. For brevity, only the unrefined Hilbert series is given below. \begin{align}
     \hs\left[\mathcal C(\mathcal Q_{\ref{eq_F4}})\right]&=\frac{\left(\begin{aligned}1 &+ t^2 + 4 t^3 + 5 t^4 + 6 t^5 + 12 t^6 + 14 t^7 + 19 t^8 + 
   20 t^9 + 20 t^{10} \\&+ 20 t^{11} + 19 t^{12} + 14 t^{13} + 12 t^{14} + 
   6 t^{15} + 5 t^{16} + 4 t^{17} + t^{18} + 
   t^{20}\end{aligned}\right)}{(1 - t^2)^5 (1 - t^3)^6(1 - t^4) }\\
   &=\hs\left[\mathcal S^{F_4}_{F_4(a_3),A_1+\tilde A_1}\right]. \notag 
\end{align}
The Hasse diagram for $\mathcal S^{F_4}_{F_4(a_3),A_1+\tilde A_1}$ is given in \Figref{fig_F4HasseCoul} which matches the Hasse diagram for $\mathcal C\left(\mathcal Q_{\ref{eq_F4}}\right)$.

\FloatBarrier
\paragraph{$E_8$ Example.}
Finally consider $\mathcal Q_{\ref{eq_12n+1loop}}$ at $n=5$
\begin{equation}
    \mathcal Q_{\ref{eq_125oneloop}}=\raisebox{-.4\height}{\begin{tikzpicture}
        \node[gauge, label=below:$1$] (1) at (0,0){};
        \node[gauge, label=below:$2$] (2) at (1,0){};
        \node[gauge, label=below:$5$] (3) at (2,0) {};

        \draw[-] (1)--(2)--(3);
        \draw[-] (3) to[out=-45,in=45,looseness=8] (3);
    \end{tikzpicture}}.
    \label{eq_125oneloop}
\end{equation}
The Hasse diagram for the Higgs branch is given in \Figref{fig_125oneloophassea}, which matches the Hasse diagram for the S{\l}odowy intersection $\mathcal S^{E_8}_{D_5+A_2,A_4+A_3}$ in the $E_8$ nilcone \cite{Generic_singularities}, explicitly shown in \Figref{fig_125oneloophasseb}. Note that the quiver corresponding to the leaf with Bala-Carter label $E_8(a_7)$ admits two possible subtractions as given in Table \ref{tab_foldedslices}. The first possible subtraction is of an affine $D_4$ quiver with four decorated $\urm(1)$ nodes, there are ${5\choose4}=5$ ways of subtracting it and so gives rise to the $5G_2$ slice. The other possibility is to subtract an affine $D_4$ quiver with three decorated $\urm(1)$ nodes, this time there are ${5\choose 3}=10$ ways of subtracting it and gives rise to the $10G_2$ slice. The Higgs branch Hasse diagram also allows the identification of the Higgs branch Namikawa-Weyl group as $\mathbb Z_2\times\mathbb Z_2$.
\newcommand{\EeiAfourAthree}{
\begin{tikzpicture}
\node (1) [gauge, label=below:$1$] at (0,0) {};
\node (2) [gauge, label=below:$2$] at (1,0) {};
\node (3) [gauge, label=below:$5$] at (2,0) {};
\draw[-] (3) to[out=45,in=315,looseness=8] (3);
\draw[-] (1)--(2)--(3);
\end{tikzpicture}
}

\newcommand{\EeiAfiveAone}{
\begin{tikzpicture}
\node (1) [gauge, label=below:$1$] at (0,0) {};
\node (2) [gauge, label=below:$2$] at (1,0) {};
\node (3) [gauge, label=left:$3$] at (0.5,0.866) {};
\node (4) [gauge, label=right:$2$] at (1.5,0.866) {};
\draw[-] (3) to[out=45,in=135,looseness=8] (3);
\draw[-] (4) to[out=45,in=135,looseness=8] (4);
\draw[-] (1)--(2)--(3);
\draw[-] (2)--(4);
\draw[blue] (3) circle (0.2cm);
\draw[blue] (4) circle (0.2cm);
\end{tikzpicture}
}

\newcommand{\EeiDfiveaoneAtwo}{
\begin{tikzpicture}
\node (1) [gauge, label=below:$1$] at (0,0) {};
\node (2) [gauge, label=below:$2$] at (1,0) {};
\node (3) [gauge, label=left:$4$] at (0.5,0.866) {};
\node (4) [gauge, label=right:$1$] at (1.5,0.866) {};
\draw[-] (3) to[out=45,in=135,looseness=8] (3);
\draw[-] (4) to[out=45,in=135,looseness=8] (4);
\draw[-] (1)--(2)--(3);
\draw[-] (2)--(4);
\draw[blue] (3) circle (0.2cm);
\draw[blue] (4) circle (0.2cm);
\end{tikzpicture}
}

\newcommand{\EeiEsixathreeaone}{
\begin{tikzpicture}
\node (1) [gauge, label=below:$1$] at (0,0) {};
\node (2) [gauge, label=below:$2$] at (1,0) {};
\node (3) [gauge, label=left:$3$] at (0.5,0.866) {};
\node (4) [gauge, label=right:$1$] at (1.5,0.866) {};
\node (5) [gauge, label=below:$1$] at (2,0) {};
\draw[-] (3) to[out=45,in=135,looseness=8] (3);
\draw[-] (4) to[out=45,in=135,looseness=8] (4);
\draw[-] (5) to[out=45,in=315,looseness=8] (5);
\draw[-] (1)--(2)--(3);
\draw[-] (4)--(2)--(5);
\draw[blue] (3) circle (0.2cm);
\draw[blue] (4) circle (0.2cm);
\draw[blue] (5) circle (0.2cm);
\end{tikzpicture}
}

\newcommand{\EeiDsixatwo}{
\begin{tikzpicture}
\node (1) [gauge, label=below:$1$] at (0,0) {};
\node (2) [gauge, label=below:$2$] at (1,0) {};
\node (3) [gauge, label=left:$2$] at (0.5,0.866) {};
\node (4) [gauge, label=right:$2$] at (1.5,0.866) {};
\node (5) [gauge, label=below:$1$] at (2,0) {};
\draw[-] (3) to[out=45,in=135,looseness=8] (3);
\draw[-] (4) to[out=45,in=135,looseness=8] (4);
\draw[-] (5) to[out=45,in=315,looseness=8] (5);
\draw[-] (1)--(2)--(3);
\draw[-] (4)--(2)--(5);
\draw[blue] (3) circle (0.2cm);
\draw[blue] (4) circle (0.2cm);
\draw[blue] (5) circle (0.2cm);
\end{tikzpicture}
}

\newcommand{\EeiEsevenafive}{
\begin{tikzpicture}
\node (1) [gauge, label=below:$1$] at (0,0) {};
\node (2) [gauge, label=below:$2$] at (1,0) {};
\node (3) [gauge, label=below:$1$] at (2,0) {};
\node (4) [gauge, label=left:$2$] at (0.1,0.866) {};
\node (5) [gauge, label=left:$1$] at (1,1) {};
\node (6) [gauge, label=right:$1$] at (1.9,0.866) {};
\draw[-] (3) to[out=45,in=315,looseness=8] (3);
\draw[-] (4) to[out=45,in=135,looseness=8] (4);
\draw[-] (5) to[out=45,in=135,looseness=8] (5);
\draw[-] (6) to[out=45,in=135,looseness=8] (6);
\draw[-] (1)--(2)--(3);
\draw[-] (5)--(2)--(4);
\draw[-] (2)--(6);
\draw[blue] (3) circle (0.2cm);
\draw[blue] (4) circle (0.2cm);
\draw[blue] (5) circle (0.2cm);
\draw[blue] (6) circle (0.2cm);
\end{tikzpicture}
}

\newcommand{\EeiDfiveAone}{
\begin{tikzpicture}
\node (1) [gauge, label=below:$1$] at (0,0) {};
\node (2) [gauge, label=below:$2$] at (1,0) {};
\draw[-] (1) to[out=225,in=135,looseness=8] node[pos=0.5,left]{$4$} (1);
\draw[-] (2) to[out=45,in=315,looseness=8] (2);
\draw[transform canvas={xshift=0pt, yshift=1.5pt}] (1)--(2);
\draw[transform canvas={xshift=0pt, yshift=-1.5pt}] (1)--(2);
\draw[blue] (2) circle (0.2cm);
\draw[red] (1) circle (0.2cm);
\end{tikzpicture}
}

\newcommand{\EeiEeightaseven}{
\begin{tikzpicture}
\node (1) [gauge, label=below:$1$] at (0,0) {};
\node (2) [gauge, label=below:$2$] at (1,0) {};
\node (3) [gauge, label=below:$1$] at (2,0) {};
\node (4) [gauge, label=left:$1$] at (0.5,0.866) {};
\node (5) [gauge, label=right:$1$] at (1.5,0.866) {};
\node (6) [gauge, label=left:$1$] at (0.5,-0.866) {};
\node (7) [gauge, label=right:$1$] at (1.5,-0.866) {};
\draw[-] (3) to[out=45,in=315,looseness=8] (3);
\draw[-] (4) to[out=45,in=135,looseness=8] (4);
\draw[-] (5) to[out=45,in=135,looseness=8] (5);
\draw[-] (6) to[out=225,in=315,looseness=8] (6);
\draw[-] (7) to[out=225,in=315,looseness=8] (7);
\draw[-] (1)--(2)--(3);
\draw[-] (4)--(2)--(5);
\draw[-] (6)--(2)--(7);
\draw[blue] (3) circle (0.2cm);
\draw[blue] (4) circle (0.2cm);
\draw[blue] (5) circle (0.2cm);
\draw[blue] (6) circle (0.2cm);
\draw[blue] (7) circle (0.2cm);
\end{tikzpicture}
}

\newcommand{\EeiDsixaone}{
\begin{tikzpicture}
\node (1) [gauge, label=below:$1$] at (0,0) {};
\node (2) [gauge, label=below:$1$] at (1,0) {};
\node (3) [gauge, label=left:$1$] at (0,1) {};
\draw[-] (1) to[out=225,in=135,looseness=8] node[pos=0.5,left]{$4$} (1);
\draw[-] (2) to[out=45,in=315,looseness=8] (2);
\draw[-] (3) to[out=45,in=135,looseness=8] (3);
\draw[transform canvas={xshift=0pt, yshift=1.5pt}] (1)--(2);
\draw[transform canvas={xshift=0pt, yshift=-1.5pt}] (1)--(2);
\draw[transform canvas={xshift=1.5pt, yshift=0pt}] (1)--(3);
\draw[transform canvas={xshift=-1.5pt, yshift=0pt}] (1)--(3);
\draw[red] (1) circle (0.2cm);
\draw[blue] (2) circle (0.2cm);
\draw[blue] (3) circle (0.2cm);
\end{tikzpicture}
}

\newcommand{\EeiAsix}{
\begin{tikzpicture}
\node (1) [gauge, label=below:$1$] at (0,0) {};
\node (2) [gauge, label=below:$1$] at (-1,0) {};
\node (3) [gauge, label=left:$1$] at (0,1) {};
\draw[-] (1) to[out=45,in=315,looseness=8] node[pos=0.5,right]{$5$} (1);
\draw[-] (3) to[out=45,in=135,looseness=8] (3);
\draw[transform canvas={xshift=0pt, yshift=1.5pt}] (1)--(2);
\draw[transform canvas={xshift=0pt, yshift=-1.5pt}] (1)--(2);
\draw[transform canvas={xshift=1.5pt, yshift=0pt}] (1)--(3);
\draw[transform canvas={xshift=-1.5pt, yshift=0pt}] (1)--(3);
\draw[red] (1) circle (0.2cm);
\draw[blue] (3) circle (0.2cm);
\end{tikzpicture}
}

\newcommand{\EeiEsevenafour}{
\begin{tikzpicture}
\node (1) [gauge, label=below:$1$] at (0,0) {};
\node (2) [gauge, label=below:$1$] at (1,0) {};
\draw[-] (1) to[out=225,in=135,looseness=8] node[pos=0.5,left]{$6$} (1);
\draw[-] (2) to[out=45,in=315,looseness=8] (2);
\draw[transform canvas={xshift=0pt, yshift=1.5pt}] (1)--(2);
\draw[transform canvas={xshift=0pt, yshift=-1.5pt}] (1)--(2);
\draw[blue] (1) circle (0.2cm);
\draw[blue] (2) circle (0.2cm);
\end{tikzpicture}
}

\newcommand{\EeiAsixAone}{
\begin{tikzpicture}
\node (1) [gauge, label=below:$1$] at (0,0) {};
\node (2) [gauge, label=below:$1$] at (1,0) {};
\draw[-] (2) to[out=45,in=315,looseness=8] node[pos=0.5,right]{$7$} (2);
\draw[transform canvas={xshift=0pt, yshift=1.5pt}] (1)--(2);
\draw[transform canvas={xshift=0pt, yshift=-1.5pt}] (1)--(2);
\draw[blue] (2) circle (0.2cm);
\end{tikzpicture}
}

\newcommand{\EeiDfiveAtwo}{
\begin{tikzpicture}
\node (1) [gauge, label=below:$1$] at (0,0) {};
\draw[-] (1) to[out=45,in=135,looseness=8] node[pos=0.5,above]{$8$} (1);
\draw[blue] (1) circle (0.2cm);
\end{tikzpicture}
}

\begin{figure}[h]
    \centering
    \begin{subfigure}{0.49\textwidth}
    \centering
    \scalebox{.4}{
    \begin{tikzpicture}
        \node (a) at (0,0) {\EeiAfourAthree};
        \node (b) at (-3,4) {\EeiAfiveAone};
        \node (c) at (3,4) {\EeiDfiveaoneAtwo};
        \node (d) at (-3,8) {\EeiEsixathreeaone};
        \node (e) at (3,8) {\EeiDsixatwo};
        \node (f) at (0,12) {\EeiEsevenafive};
        \node (g) at (-3,16) {\EeiDfiveAone};
        \node (h) at (3,16) {\EeiEeightaseven};
        \node (i) at (-3,20) {\EeiDsixaone};
        \node (j) at (3,20) {\EeiAsix};
        \node (k) at (-3,24) {\EeiEsevenafour};
        \node (l) at (3,24) {\EeiAsixAone};
        \node (m) at (0,28) {\EeiDfiveAtwo};
        \draw[-] (a)--(b) node[pos=0.5, left]{$m_1$};
        \draw[-] (a)--(c) node[pos=0.5, right]{$m_1$};
        \draw[-] (d)--(b) node[pos=0.5, right]{$A_1$};
        \draw[-] (d)--(c) node[pos=0.3, left]{$m_1$};
        \draw[-] (e)--(b) node[pos=0.3, right]{$m_1$};
        \draw[-] (e)--(c) node[pos=0.5, right]{$A_1$};
        \draw[-] (f)--(d) node[pos=0.5, left]{$m_1$};
        \draw[-] (f)--(e) node[pos=0.5, right]{$2A_1$};
        \draw[-] (f)--(g) node[pos=0.5, left]{$G_2$};
        \draw[-] (f)--(h) node[pos=0.5, right]{$A_1$};
        \draw[-] (i)--(g) node[pos=0.5, left]{$A_1$};
        \draw[-] (i)--(h) node[pos=0.5, left]{$10G_2$};
        \draw[-] (j)--(h) node[pos=0.5, right]{$5G_2$};
        \draw[-] (k)--(i) node[pos=0.5, left]{$2A_1$};
        \draw[-] (k)--(j) node[pos=0.5, left]{$A_1$};
        \draw[-] (l)--(j) node[pos=0.5, right]{$A_1$};
        \draw[-] (m)--(k) node[pos=0.5, left]{$A_1$};
        \draw[-] (m)--(l) node[pos=0.5, right]{$A_1$};
    \end{tikzpicture}
    }
    \caption{}
    \label{fig_125oneloophassea}
    \end{subfigure}
    \begin{subfigure}{0.49\textwidth}
    \centering
    \scalebox{.55}{
    \begin{tikzpicture}
        \node (a) at (0,0) {$A_4+A_3$};
        \node (b) at (-3,3) {$A_5+A_1$};
        \node (c) at (3,3) {$D_5(a_1)+A_2$};
        \node (d) at (-3,6) {$E_6(a_3)+a_1$};
        \node (e) at (3,6) {$D_6(a_2)$};
        \node (f) at (0,9) {$E_7(a_5)$};
        \node (g) at (-3,12) {$D_5+A_1$};
        \node (h) at (3,12) {$E_8(a_7)$};
        \node (i) at (-3,15) {$D_6(a_1)$};
        \node (j) at (3,15) {$A_6$};
        \node (k) at (-3,18) {$E_7(a_4)$};
        \node (l) at (3,18) {$A_6+A_1$};
        \node (m) at (0,21) {$D_5+A_2$};
        \draw[-] (a)--(b) node[pos=0.5, left]{$m_1$};
        \draw[-] (a)--(c) node[pos=0.5, right]{$m_1$};
        \draw[-] (d)--(b) node[pos=0.5, right]{$A_1$};
        \draw[-] (d)--(c) node[pos=0.3, left]{$m_1$};
        \draw[-] (e)--(b) node[pos=0.3, right]{$m_1$};
        \draw[-] (e)--(c) node[pos=0.5, right]{$A_1$};
        \draw[-] (f)--(d) node[pos=0.5, left]{$m_1$};
        \draw[-] (f)--(e) node[pos=0.5, right]{$2A_1$};
        \draw[-] (f)--(g) node[pos=0.5, left]{$G_2$};
        \draw[-] (f)--(h) node[pos=0.5, right]{$A_1$};
        \draw[-] (i)--(g) node[pos=0.5, left]{$A_1$};
        \draw[-] (i)--(h) node[pos=0.5, left]{$10G_2$};
        \draw[-] (j)--(h) node[pos=0.5, right]{$5G_2$};
        \draw[-] (k)--(i) node[pos=0.5, left]{$2A_1$};
        \draw[-] (k)--(j) node[pos=0.5, left]{$A_1$};
        \draw[-] (l)--(j) node[pos=0.5, right]{$A_1$};
        \draw[-] (m)--(k) node[pos=0.5, left]{$A_1$};
        \draw[-] (m)--(l) node[pos=0.5, right]{$A_1$};
    \end{tikzpicture}
    }
    \caption{}
    \label{fig_125oneloophasseb}
    \end{subfigure}
    \caption{(\subref{fig_125oneloophassea}): The Higgs branch Hasse diagram of $\mathcal Q_{\ref{eq_125oneloop}}$. (\subref{fig_125oneloophasseb}): The S{\l}odowy intersection $\mathcal S^{E_8}_{D_5+A_2,A_4+A_3}$ in the $E_8$ nilcone with orbits specified by their Bala-Carter labels. This is the same Hasse diagram as in \Figref{fig_125oneloophassea}.}
    \label{fig_125oneloophasse}
\end{figure}
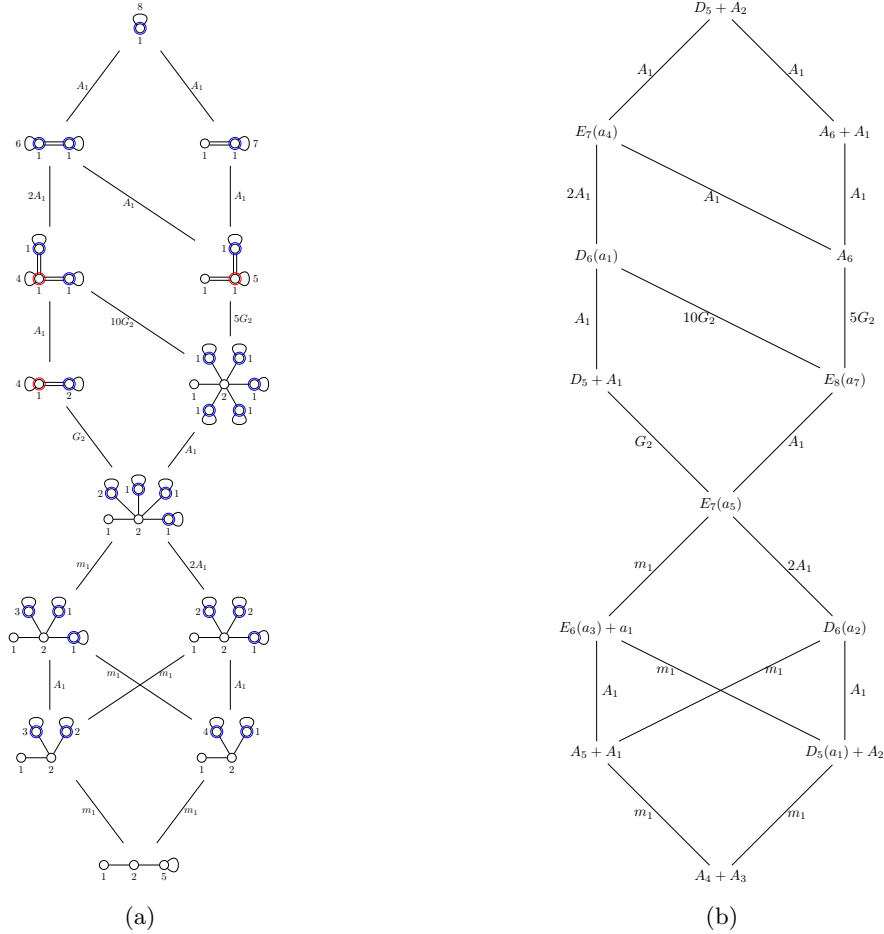

It is computationally challenging to confirm that the Hilbert series of the S{\l}odowy slice $\mathcal S^{E_8}_{D_5+A_2,A_4+A_3}$ matches that of $\mathcal H\left(\mathcal Q_{\ref{eq_125oneloop}}\right)$ which is \begin{equation}
\hs\left[\mathcal H\left(\mathcal Q_{\ref{eq_125oneloop}}\right)\right]=\frac{\left(\begin{aligned}1 &+ t^4 + 5 t^5 + 6 t^6 + 6 t^7 + 7 t^8 + 9 t^9 + 21 t^{10} + 
 30 t^{11} + 39 t^{12} + 39 t^{13} + 37 t^{14} + 45 t^{15} + 53 t^{16} \\&+ 
 65 t^{17} + 65 t^{18} + 53 t^{19} + 45 t^{20} + 37 t^{21} + 39 t^{22} + 
 39 t^{23} + 30 t^{24} + 21 t^{25} + 9 t^{26} + 7 t^{27} \\&+ 6 t^{28} + 6 t^{29} + 
 5 t^{30} + t^{31} + t^{35}\end{aligned}\right)}{(1 - t^2)^3 (1 - t^3)^4 (1 - t^4)^4  (1 - 
   t^5)^3}.
\end{equation}

\begin{figure}[h]
\centering
\scalebox{0.7}{
\begin{tikzpicture}
\node (1) at (0,0){$E_8(a_7)$};
\node (2) at (0,-3){$E_7(a_5)$};
\node (3) at (0,-6){$E_6(a_3)+A_1$};
\node (11) at (-3,-6){$D_6(a_2)$};
\node (4) at (-3,-9) {$D_5(a_1)+A_2$};
\node (12) at (0,-9) {$A_5+A_1$};
\node (6) at (-6,-12) {$D_4+A_2$};
\node (5) at (-3,-12) {$A_4+A_3$};
\node (7) at (3,-12) {$E_6(a_3)$};
\node (8) at (-6,-15) {$A_4+A_2+A_1$};
\node (9) at (-3,-15) {$D_5(a_1)+A_1$};
\node (13) at (0,-15) {$A_5$};
\node (10) at (-3,-18) {$A_4+A_2$};

\draw[-] (1)--(2)node[midway, right]{$A_1$}--(3) node[midway,right]{$m_1$}--(12) node[midway, right]{$A_1$}--(13) node[midway, right]{$g_2$}--(10)node[midway, right]{$A_1$};

\draw[-] (3)--(7)node[midway, right]{$g_2$}--(13)node[midway, below]{$A_1$};

\draw[-] (11)--(4)node[midway, left]{$A_1$}--(6)node[midway, above]{$a_2^+$}--(8)node[midway, left]{$A_1$}--(10)node[midway, below]{$A_1$};

\draw[-] (11)--(12)node[pos=0.25, left]{$m_1$};

\draw[-] (3)--(4)node[pos=0.25, right]{$m_1$}--(5)node[midway, right]{$m_1$}--(8)node[pos=0.25, below]{$\mathcal{Y}_{5}$};

\draw[-] (12)--(5)node[midway, right]{$m_1$};

\draw[-] (6)--(9)node[pos=0.25, above]{$A_1$}--(7)node[pos=0.75, above]{$A_1$};

\draw[-] (9)--(10)node[midway, right]{$A_1$};

\draw(11)--(2)node[midway, left]{$2A_1$};
\end{tikzpicture}}
\caption{The Hasse diagram for $\mathcal S^{E_8}_{E_8(a_7),A_4+A_2}$ in the $E_8$ nilcone with orbits specified by their Bala-Carter labels. This is the same Hasse diagram as for the Coulomb branch of $\mathcal Q_{\ref{eq_125oneloop}}$. The slice $\mathcal{Y}_{5}$ was introduced in \cite{Generic_singularities} (where it was called $\chi$) and further studied in \cite{bellamy2021newfamilyisolatedsymplectic, Bourget_2022dim6}.}
\label{fig_125oneloopCoul}
\end{figure}
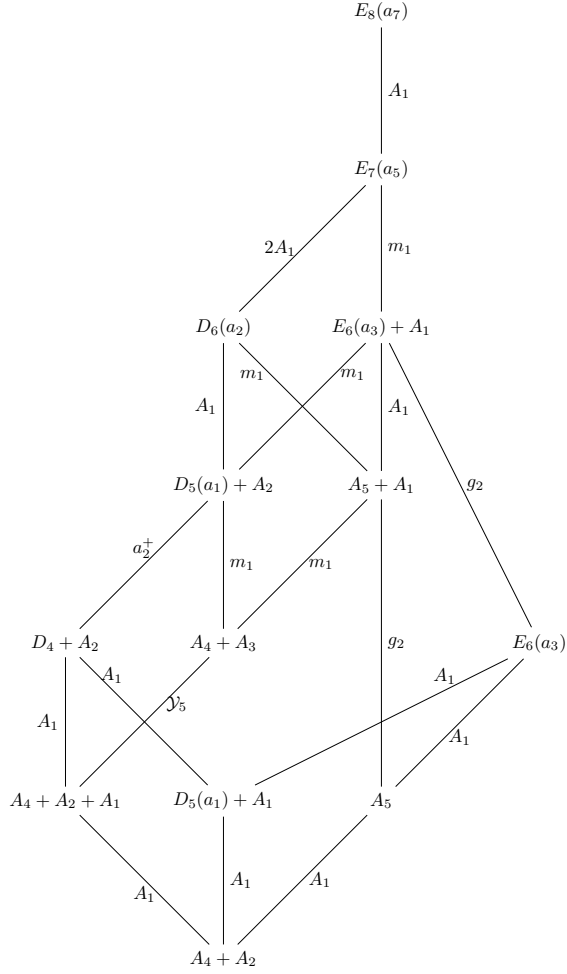
Lusztig-Spaltenstein duality suggests that the Coulomb branch is also in the $E_8$ nilcone, specifically being the S{\l}odowy intersection $\mathcal S^{E_8}_{E_8(a_7),A_4+A_2}$. The Hasse diagram for $\mathcal S^{E_8}_{E_8(a_7),A_4+A_2}$ is given in \Figref{fig_125oneloopCoul} and matches the Hasse diagram of the Coulomb branch of $\mathcal Q_{\ref{eq_125oneloop}}$.

Once again it is computationally challenging to confirm that the Hilbert series of the S{\l}odowy intersection $\mathcal S^{E_8}_{E_8(a_7),A_4+A_2}$ is the same as for $\mathcal C\left(\mathcal Q_{\ref{eq_125oneloop}}\right)$ which is \begin{equation}
\hs\left[\mathcal C\left(\mathcal Q_{\ref{eq_125oneloop}}\right)\right]=\frac{\left(\begin{aligned}1 &+ t^2 + 11 t^4 + 20 t^6 + 65 t^8 + 109 t^{10} + 221 t^{12} + 
   285 t^{14} + 401 t^{16} + 394 t^{18} \\&+ 401 t^{20} + 285 t^{22} + 221 t^{24} + 
   109 t^{26} + 65 t^{28} + 20 t^{30} + 11 t^{32} + t^{34} + 
   t^{36}\end{aligned}\right)}{(1 - t^2)^5 (1 - t^4)^7 (1 - t^6)^2}.
\end{equation}
\FloatBarrier
\subsection{A Second $F_4$ Example.}
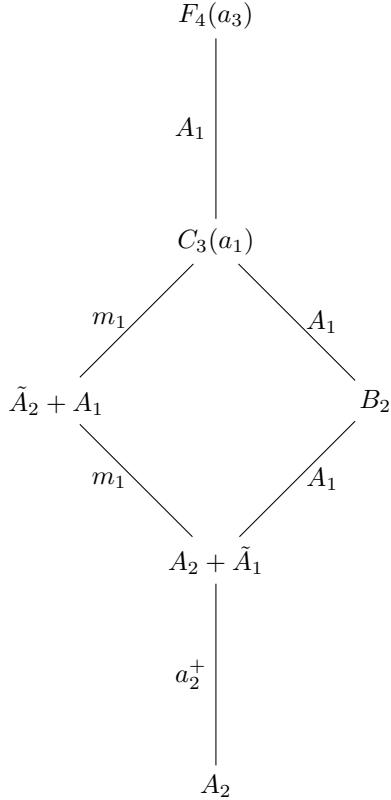
\begin{figure}[h!]
\centering
\begin{tikzpicture}
\node (b) at (0,3) {$A_2$};
\node (c) at (0,6) {$A_2+\tilde A_1$};
\node (d) at ({-3*cos(45)},{6+3*sin(45)}) {$\tilde A_2+A_1$};
\node (e) at ({+3*cos(45)},{6+3*sin(45)}) {$B_2$};
\node (f) at (0,{6+6*sin(45)}) {$C_3(a_1)$};
\node (g) at (0,{9+6*sin(45)}) {$F_4(a_3)$};

 \draw[-] (b) --(c) node[pos=0.5,midway, left]{$a_2^+$}--(d) node[pos=0.5,midway, left]{$m_1$}--(f) node[pos=0.5,midway, left]{$m_1$}--(g) node[pos=0.5,midway,left]{$A_1$};

 \draw[-] (c)--(e)node[pos=0.5,midway, right]{$A_1$}--(f)node[pos=0.5,midway, right]{$A_1$};
 
\end{tikzpicture}

\caption{The Hasse diagram for $\mathcal S^{F_4}_{F_4(a_3),A_2}$ in the $F_4$ nilcone with orbits specified by their Bala-Carter labels. This Hasse diagram is the same as for the Coulomb branch of $\mathcal Q_{\ref{eq_D4/S4}}$.}
\label{fig_D4/S4HasseCoul}
\end{figure}

Now consider the quiver $\mathcal Q_{\ref{eq_D4/S4}}$ which is similar to $\mathcal Q_{\ref{eq_F4}}$. \begin{equation}
    \mathcal Q_{\ref{eq_D4/S4}}= \raisebox{-.4\height}{ \begin{tikzpicture}
         \node[gauge, label=below:$2$] (2) at (0,0){};
            \node[gauge, label=below:$4$] (4) at (1,0){};

            \draw[-] (2) to (4) to[out=-45,in=45,looseness=8](4);
    \end{tikzpicture}} \label{eq_D4/S4}.
\end{equation}
The Coulomb branch of $\mathcal Q_{\ref{eq_D4/S4}}$ is known to be $d_4/S_4$ \cite{Bourget:2020bxh,Gledhill:2021cbe} but is also equivalently $\mathcal S^{F_4}_{F_4(a_3),A_2}$ \cite{Hanany:2023uzn}. The Hasse diagram for $\mathcal S^{F_4}_{F_4(a_3),A_2}$ is drawn in \Figref{fig_D4/S4HasseCoul} \cite{2023arXiv230807398F}, which matches the Hasse diagram of $\mathcal C\left(\mathcal Q_{\ref{eq_D4/S4}}\right)$. 

\begin{figure}[h]
    \centering
    \begin{subfigure}{0.49\textwidth}
    \scalebox{.7}{
    \begin{tikzpicture}
        \node (24) at (0,0){$\begin{tikzpicture}
            \node[gauge, label=below:$2$] (2) at (0,0){};
            \node[gauge, label=below:$4$] (4) at (1,0){};

            \draw[-] (2) to (4) to[out=-45,in=45,looseness=8](4);
        \end{tikzpicture}$};
        \node (22) at ({5*cos(45)},{5*sin(45)}) {$\begin{tikzpicture}
            \node[gauge, label=below:$2$] (2) at (0,0){};
            \node[gauge, label=above:$2$] (2t) at ({cos(45)},{sin(45)}){};
            \node[gauge, label=below:$2$] (2b) at ({cos(45)},{-sin(45)}){};
            \draw[-] (2t) to[out=-45,in=45,looseness=8] (2t) to (2) to (2b) to[out=-45,in=45,looseness=8] (2b);

             \draw[blue] (2t) circle (0.2cm);
            \draw[blue] (2b) circle (0.2cm);
        \end{tikzpicture}$};
        \node (123) at ({-5*cos(45)},{5*sin(45)}){$\begin{tikzpicture}
            \node[gauge, label=below:$2$] (2) at (0,0){};
            \node[gauge, label=above:$1$] (1) at ({cos(45)},{sin(45)}){};
            \node[gauge, label=below:$3$] (3) at ({cos(45)},{-sin(45)}){};
            \draw[-] (1) to[out=45, in=315, looseness=8] (1);
            \draw[-] (1) to (2) to (3) to[out=-45,in=45,looseness=8] (3);
            \draw[blue] (1) circle (0.2cm);
            \draw[blue] (3) circle (0.2cm);
        \end{tikzpicture}$};
        \node (112) at (0,{10*sin(45}){$\begin{tikzpicture}
            \node[gauge, label=below:$2$] (2) at (0,0){};
            \node[gauge, label=below:$2$] (2r) at (1,0){};
            \node[gauge, label=above:$1$] (1t) at (-{cos(45)},{sin(45)}){};
            \node[gauge, label=below:$1$] (1b) at (-{cos(45)},{-sin(45)}){};
            \draw[-] (1t) to[out=135, in=225, looseness=8] (1t);
            \draw[-] (1b) to[out=135, in=225, looseness=8] (1b);
            \draw[-] (1t) to (2) to (1b) (2) to (2r) to[out=-45,in=45,looseness=8](2r);

            \draw[blue] (1t) circle (0.2cm);
            \draw[blue] (2r) circle (0.2cm);
            \draw[blue] (1b) circle (0.2cm);
        \end{tikzpicture}$};
        \node (1111) at (0,{5+10*sin(45)}){$\begin{tikzpicture}
            \node[gauge, label=below:$2$] (2) at (0,0){};
            \node[gauge, label=right:$1$] (1tr) at ({cos(45)},{sin(45)}){};
            \node[gauge, label=left:$1$] (1tl) at ({-cos(45)},{sin(45)}){};
            \node[gauge, label=right:$1$] (1br) at ({cos(45)},{-sin(45)}){};
            \node[gauge, label=left:$1$] (1bl) at ({-cos(45)},{-sin(45)}){};
            \draw[-] (1tr) to[out=135, in=45, looseness=8] (1tr);
            \draw[-] (1tl) to[out=135, in=45, looseness=8] (1tl);
            \draw[-] (1br) to[out=315, in=225, looseness=8] (1br);
            \draw[-] (1bl) to[out=315, in=225, looseness=8] (1bl);
                        
            \draw[-] (1tr)--(2)--(1tl) (1br)--(2)--(1bl);

            \draw[blue] (1tl) circle (0.2cm);
            \draw[blue] (1bl) circle (0.2cm);
            \draw[blue] (1tr) circle (0.2cm);
            \draw[blue] (1br) circle (0.2cm);
            
        \end{tikzpicture}$};

        \node (0) at (0,{10+10*sin(45)}){$\begin{tikzpicture}
            \node[gauge, label=below:$1$] (1){};
            \draw[-] (1) to[out=45,in=135,looseness=8] node[pos=0.5,above]{$5$} (1);
        \end{tikzpicture}$};
        \draw[-] (22) -- (24) node[pos=0.5,midway,right]{$A_1$}--(123) node[pos=0.5,midway, left]{$m_1$}--(112) node[pos=0.5,midway, left]{$m_1$}--(1111) node[pos=0.5,midway,right]{$A_1$}--(0) node[pos=0.5,midway,right]{$G_2$};

        \draw[-](22)--(112)node[pos=0.5, midway, right]{$2A_1$};
    
    \end{tikzpicture}}
    \caption{}
    \label{fig_HasseD4/S4a}
    \end{subfigure}
    \begin{subfigure}{0.49\textwidth}
    \scalebox{.7}{
    \begin{tikzpicture}
        \node (24) at (0,0){$A_2+\tilde{A}_1$};
        \node (22) at ({5*cos(45)},{5*sin(45)}) {$B_2$};
        \node (123) at (-{5*cos(45)},{5*sin(45)}){$\tilde{A}_2+A_1$};
        \node (112) at (0,{10*sin(45}){$C_3(a_1)$};
        \node (1111) at (0,{5+10*sin(45)}){$F_4(a_3)$};

        \node (0) at (0,{10+10*sin(45)}){$B_3$};
        \draw[-] (22) -- (24) node[pos=0.5,midway,right]{$A_1$}--(123) node[pos=0.5,midway, left]{$m_1$}--(112) node[pos=0.5,midway, left]{$m_1$}--(1111) node[pos=0.5,midway,right]{$A_1$}--(0) node[pos=0.5,midway,right]{$G_2$};

        \draw[-](22)--(112)node[pos=0.5, midway, right]{$2A_1$};
    
    \end{tikzpicture}}
    \caption{}
    \label{fig_HasseD4/S4b}
    \end{subfigure}
    \caption{(\subref{fig_HasseD4/S4a}): The Higgs branch Hasse diagram of $\mathcal Q_{\ref{eq_D4/S4}}$. (\subref{fig_HasseD4/S4b}): The S{\l}odowy intersection $\mathcal S^{F_4}_{B_3,A_2+\tilde{A}_{1}}$ in the $F_4$ nilcone with orbits specified by their Bala-Carter labels. This is the same Hasse diagram as in \Figref{fig_HasseD4/S4a}.}
    \label{fig_HasseD4/S4}
\end{figure}
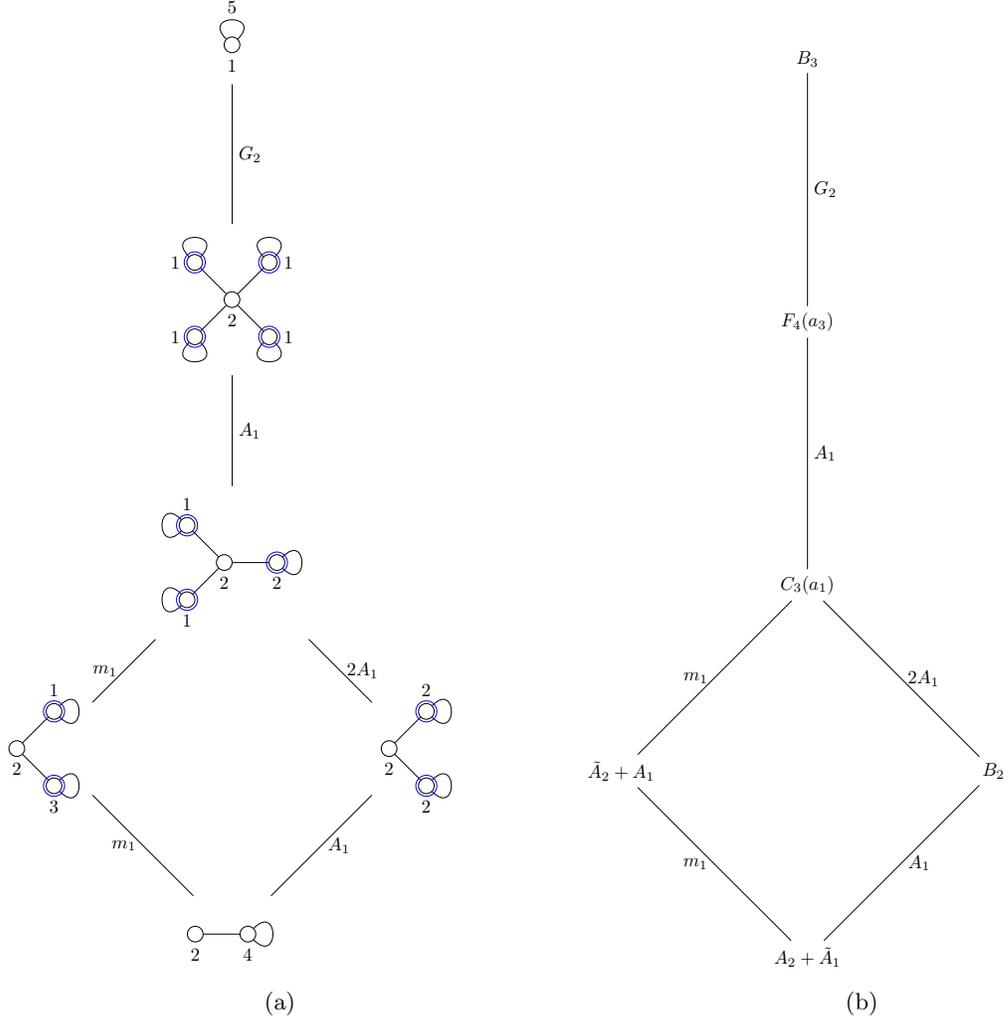

Lusztig-Spaltenstein duality suggests that the Higgs branch is again a slice in the $F_4$ nilcone. The Hasse diagram for the Higgs branch is computed using the rules introduced in this paper in \Figref{fig_HasseD4/S4a}, which matches the Hasse diagram for $\mathcal S^{F_4}_{B_3,A_2+\tilde{A}_1}$ explicitly given in \Figref{fig_HasseD4/S4b}. The Higgs branch Hilbert series is computed as
\begin{equation}
    \hs[\mathcal H(\mathcal Q_{\ref{eq_D4/S4}})]=(1 + t^4 + t^{10} + t^{14} - [3]_{\surm(2)}t^7 )\pe\left[[2]_{\surm(2)}t^2+[3]_{\surm(2)}t^3+[4]_{\surm(2)}t^4-t^4-t^{12}\right],
\end{equation}which is not a complete intersection, but does confirm that $\mathcal H(\mathcal Q_{\ref{eq_D4/S4}})=\mathcal S^{F_4}_{B_3,A_2+\tilde{A}_{1}}$. The Namikawa-Weyl group for the Higgs branch is the Weyl group of $G_2$, $\mathcal W_{G_2}=\mathrm{Dih}_{6}$ of order 12.


\begin{figure}
\centering
\begin{subfigure}{0.4\textwidth}
\centering
\scalebox{0.6}{\begin{tikzpicture}
\node (1) at (0,0){$F_4$};
\node (2) at (0,-3){$F_3(a_1)$};
\node (3) at (0,-6){$F_4(a_2)$};
\node (4) at (4,-9){$B_3$};
\node (5) at (-4,-9){$C_3$};
\node (6) at (0,-12){$F_4(a_3)$};
\node (7) at (0,-15){$C_3(a_1)$};
\node (8) at (-4,-18){$\tilde A_2+A_1$};
\node (9) at (4,-18){$B_2$};
\node (10) at (4,-19.5){$A_2+\tilde A_1$};
\node (11) at (-4,-21){$\tilde A_2$};
\node (12) at (4,-21){$A_2$};
\node (13) at (0,-24){$A_1+\tilde A_1$};
\node (14) at (0,-27){$\tilde A_1$};
\node (15) at (0,-30){$A_1$};
\node (16) at (0,-33){$0$};

\draw[-] (1)--(2)node[midway, left]{$F_4$}--(3)node[midway, left]{$C_3$}--(4)node[midway, above]{$A_1$}--(6)node[midway, below]{$G_2$}--(7)node[midway, left]{$A_1$}--(8)node[midway, above]{$m_1$}--(11)node[midway, left]{$g_2$}--(13)node[midway, below]{$A_1$}--(14)node[midway, left]{$a_3^+$}--(15)node[midway, left]{$c_3$}--(16)node[midway, left]{$f_4$};

\draw[-] (3)--(5)node[midway, above]{$A_1$}--(6)node[midway, below]{$4G_2$};

\draw[-] (7)--(9)node[midway, above]{$2A_1$}--(10)node[midway, right]{$A_1$}--(12)node[midway, right]{$a_2^+$}--(13)node[midway, below]{$A_1$};

\draw[-] (8)--(10)node[midway, above]{$m_1$};

  \draw[blue,rounded corners]
            (3.north west) -- (5.north west) -- (5.south west) -- (6.south west) -- (7.north west) -- (8.north west)--(8.south west)--(10.south west)--(10.south east)--(9.north east)--(7.north east)--(6.south east)--(4.south east)--(4.north east)--(3.north east)--cycle;
        
    \draw[red,rounded corners]
            (6.north east) -- (7.north east) -- (9.north east) -- (10.east) -- (12.south east) -- (13.south east) -- (13.south west)--(13.west)-- (12.north west) -- (10.south west) -- (8.south west) --(8.north west)-- (7.north west) -- (6.north west)--cycle;
 
\end{tikzpicture}}
\caption{}
\label{fig_F4Nilcone_a}
\end{subfigure}
\begin{subfigure}{0.4\textwidth}
\centering
\scalebox{0.6}{\begin{tikzpicture}
\node (1) at (0,0){$F_4$};
\node (2) at (0,-3){$F_3(a_1)$};
\node (3) at (0,-6){$F_4(a_2)$};
\node (4) at (4,-9){$B_3$};
\node (5) at (-4,-9){$C_3$};
\node (6) at (0,-12){$F_4(a_3)$};
\node (7) at (0,-15){$C_3(a_1)$};
\node (8) at (-4,-18){$\tilde A_2+A_1$};
\node (9) at (4,-18){$B_2$};
\node (10) at (4,-19.5){$A_2+\tilde A_1$};
\node (11) at (-4,-21){$\tilde A_2$};
\node (12) at (4,-21){$A_2$};
\node (13) at (0,-24){$A_1+\tilde A_1$};
\node (14) at (0,-27){$\tilde A_1$};
\node (15) at (0,-30){$A_1$};
\node (16) at (0,-33){$0$};

\draw[-] (1)--(2)node[midway, left]{$F_4$}--(3)node[midway, left]{$C_3$}--(4)node[midway, above]{$A_1$}--(6)node[midway, below]{$G_2$}--(7)node[midway, left]{$A_1$}--(8)node[midway, above]{$m_1$}--(11)node[midway, left]{$g_2$}--(13)node[midway, below]{$A_1$}--(14)node[midway, left]{$a_3^+$}--(15)node[midway, left]{$c_3$}--(16)node[midway, left]{$f_4$};

\draw[-] (3)--(5)node[midway, above]{$A_1$}--(6)node[midway, below]{$4G_2$};

\draw[-] (7)--(9)node[midway, above]{$2A_1$}--(10)node[midway, right]{$A_1$}--(12)node[midway, right]{$a_2^+$}--(13)node[midway, below]{$A_1$};

\draw[-] (8)--(10)node[midway, above]{$m_1$};

 \draw[green,rounded corners, radius=1.2mm](6.north east) -- (7.north east) -- (9.north east) -- (10.east) -- (12.south east) -- (12.south west)-- (10.south west) -- (8.south west) --(8.north west)-- (7.north west) -- (6.north west)--cycle;

\draw[orange,rounded corners](4.north west)--(6.north west) -- (7.north west) -- (8.north west) --(8.south west)--(10.south west)--(10.south east)--(9.north east)--(7.north east)--(6.south east)--(4.south east)--(4.north east)--cycle;
 
\end{tikzpicture}}
\caption{}
\label{fig_F4Nilcone_b}
\end{subfigure}
\caption{(\ref{fig_F4Nilcone_a}) The Hasse diagram of the $F_4$ nilcone where \textcolor{blue}{blue} is the intersection $\mathcal S^{F_4}_{F_4(a_2),A_2+\tilde{A}_1}=\mathcal H\left(\mathcal Q_{\ref{eq_F4}}\right)$ and \textcolor{red}{red} is the intersection $\mathcal S^{F_4}_{F_4(a_3),A_1+\tilde{A}_1}=\mathcal C\left(\mathcal Q_{\ref{eq_F4}}\right)$. (\ref{fig_F4Nilcone_b}) The Hasse diagram of the $F_4$ nilcone where \textcolor{orange}{orange} is the intersection $\mathcal S^{F_4}_{B_3,A_2+\tilde{A}_1}=\mathcal H\left(\mathcal Q_{\ref{eq_D4/S4}}\right)$ and \textcolor{green}{green} is the intersection $\mathcal S^{F_4}_{F_4(a_3),A_2}=\mathcal C\left(\mathcal Q_{\ref{eq_D4/S4}}\right)$.}
\label{fig_F4Nilcone}
\end{figure}
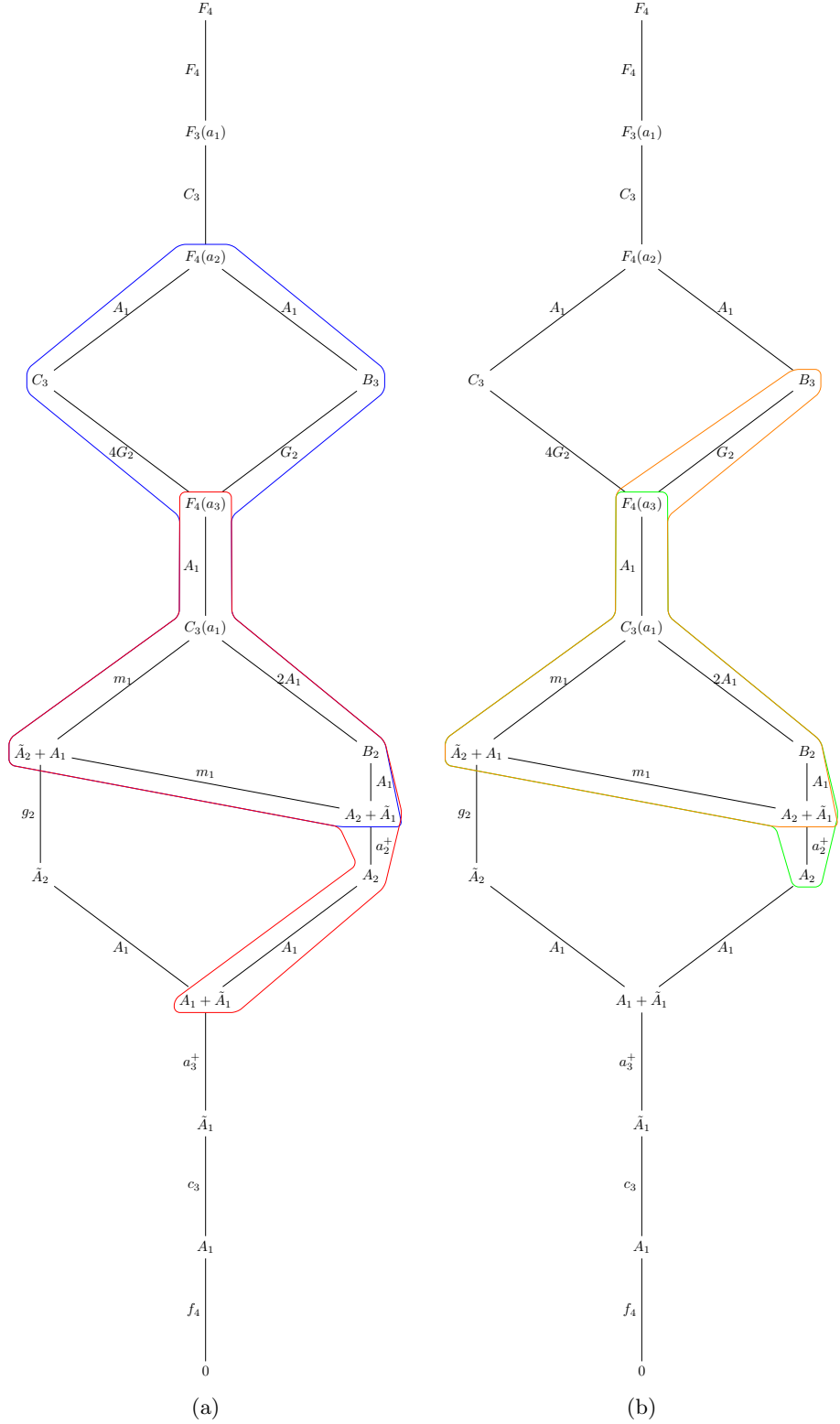

\paragraph{Comment.}
Note that the quiver $\mathcal Q_{\ref{eq_D4/S4}}$ is a subquiver of $\mathcal Q_{\ref{eq_F4}}$ - its Higgs and Coulomb branch Hasse diagrams are contained in those of $\mathcal Q_{\ref{eq_F4}}$. In \Figref{fig_F4Nilcone}, the Hasse diagram for the $F_4$ nilcone is given with the Higgs branch Hasse diagrams of $\mathcal Q_{\ref{eq_F4}}$ and $\mathcal Q_{\ref{eq_D4/S4}}$ outlined in \textcolor{blue}{blue} and \textcolor{orange}{orange} respectively. The Coulomb branch Hasse diagrams for these quivers are outlined in \textcolor{red}{red} and \textcolor{green}{green} respectively.

Below, the quivers $\mathcal Q_{\ref{eq_E7}}$ and $\mathcal Q_{\ref{eq_E72}}$ obey the same relationship. The Higgs branch Hasse diagram of $\mathcal Q_{\ref{eq_E72}}$ can be embedded into $\mathcal Q_{\ref{eq_E7}}$ explicitly, as apparent from \Figref{fig_HasseE72} and \Figref{fig_125E7}. 

In general, if a undecorated quiver $\mathcal{Q}_1$ is a subquiver of another undecorated quiver $\mathcal{Q}_2$, then the Higgs branch of $\mathcal{Q}_1$ is a leaf of the Higgs branch of $\mathcal{Q}_2$, and the Coulomb branch of $\mathcal{Q}_1$ is a slice of the Coulomb branch of $\mathcal{Q}_2$. 

\FloatBarrier

\subsection{A Second $E_8$ Example.}

\subsection{Example: $C_3$}
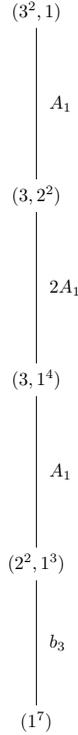
\begin{figure}[h!]
    \centering
    \scalebox{0.7}{
    \begin{tikzpicture}
    \node (a) at (0,0) {$(3^2,1)$};
    \node (b) at (0,-3.5) {$(3,2^2)$};
    \node (c) at (0,-7) {$(3,1^4)$};
    \node (d) at (0,-10.5) {$(2^2,1^3)$};
    \node (e) at (0,-13.5) {$(1^7)$};
    \draw[] (a)--(b) node[pos=0.5, label=right:$A_1$] {};

     \draw(b)--(c) node[pos=0.5, label=right:$2A_1$]{};
    
    \draw[](c) --(d) node[pos=0.5, label=right:$A_1$] {}--(e) node[pos=0.5, label=right:$b_3$] {};
    \end{tikzpicture}}
    \caption{The Hasse diagram for $\overline{\mathcal O}^{B_3}_{(3^2,1)}$ in the $B_3$ nilcone with orbits labelled by partitions of $7$. This Hasse diagram is the same as the Hasse diagram for the Coulomb branch of $\mathcal Q_{\ref{eq_B3Orb}}$.}
    \label{fig_HasseB3OrbitCoul}
    \end{figure}

Now consider the following quiver
\begin{equation}    \mathcal Q_{\ref{eq_B3Orb}}=\raisebox{-.4\height}{\begin{tikzpicture}
         \node[gauge, label=below:$1$] (1) at (0,0){};
            \node[gauge, label=below:$2$] (2) at (1,0){};
            \node[gauge, label=below:$3$] (3) at (2,0){};
            \node[gauge, label=below:$2$] (2r) at (3,0){};

            \draw[-] (1)--(2)--(3);
            \draw[transform canvas={yshift=1.5pt}] (2r)--(3);
            \draw[transform canvas={yshift=-1.5pt}] (2r)--(3);
    \end{tikzpicture}}\label{eq_B3Orb}.
\end{equation}
The Coulomb branch of $\mathcal Q_{\ref{eq_B3Orb}}$ is $\overline{\mathcal O}_{(3^2,1)}^{B_3}$ \cite{Gledhill:2021cbe,Hanany:2024fqf} which is a nilpotent orbit of height 4. This makes $\mathcal Q_{\ref{eq_B3Orb}}$ a rare example of a unitary quiver whose Coulomb branch is a height 4 nilpotent orbit for an algebra other than $A_k$. The Hasse diagram of $\overline{\mathcal O}_{(3^2,1)}^{B_3}$ is shown in \Figref{fig_HasseB3OrbitCoul} which matches the Hasse diagram of $\mathcal C\left(\mathcal Q_{\ref{eq_B3Orb}}\right)$. 

Lusztig-Spaltenstein duality suggests that the Higgs branch is the S{\l}odowy intersection $\mathcal S^{C_3}_{\mathcal N,(2^3)}$. The Hasse diagram for the Higgs branch is given in \Figref{fig_HasseB3Orbita}, which is consistent with the Hasse diagram of $\mathcal S^{C_3}_{\mathcal N,(2^3)}$ given in \Figref{fig_HasseB3Orbitb}. The computation of the Higgs branch Hilbert series proceeds as \begin{equation}
    \hs[\mathcal H(\mathcal Q_{\ref{eq_B3Orb}})]=\pe\left[[2]_{\surm(2)}t^2+[4]_{\surm(2)}t^4-t^8-t^{12}\right]=\hs\left[\mathcal S^{C_3}_{\mathcal N,(2^3)}\right],
\end{equation}which confirms that $\mathcal H(\mathcal Q_{\ref{eq_B3Orb}})=\mathcal S^{C_3}_{\mathcal N,(2^3)}$. The Higgs branch Namikawa-Weyl group is $\mathcal W_{C_3}=\mathbb Z_2^3\rtimes S_3$.
\begin{figure}[h!]
    \centering
    \begin{subfigure}{0.49\textwidth}
    \centering
    \scalebox{.7}{
    \begin{tikzpicture}
        \node (1232) at (0,0){$\begin{tikzpicture}
            \node[gauge, label=below:$1$] (1) at (0,0){};
            \node[gauge, label=below:$2$] (2) at (1,0){};
            \node[gauge, label=below:$3$] (3) at (2,0){};
            \node[gauge, label=below:$2$] (2r) at (3,0){};

            \draw[-] (1)--(2)--(3);
            \draw[transform canvas={yshift=1.5pt}] (2r)--(3);
            \draw[transform canvas={yshift=-1.5pt}] (2r)--(3);
        \end{tikzpicture}$};
        \node (12211) at ({-5*cos(45)},{5*sin(45)}){$\begin{tikzpicture}
            \node[gauge, label=below:$1$] (1) at (0,0){};
            \node[gauge, label=below:$2$] (2) at (1,0){};
            \node[gauge, label=below:$2$] (2r) at (2,0){};
            \node[gauge, label=below:$1$] (1r) at (3,0){};
            \node[gauge, label=right:$1$] (1t) at (1,1){};
            \draw[-] (1t) to[out=45,in=135,looseness=8] (1t);
            \draw[-] (1)--(2)--(2r) (1t)--(2);
            \draw[transform canvas={yshift=1.5pt}] (2r)--(1r);
            \draw[transform canvas={yshift=-1.5pt}] (2r)--(1r);

            \draw[blue] (1t) circle (0.2cm);
            \draw[blue] \convexpath{2r,1r} {0.2cm};
        \end{tikzpicture}$};
        \node (1221) at ({5*cos(45)},{5*sin(45)}){$\begin{tikzpicture}
            \node[gauge, label=below:$1$] (1) at (0,0){};
            \node[gauge, label=below:$2$] (2) at (1,0){};
            \node[gauge, label=above:$1$] (1t) at (1,1){};
            \node[gauge, label=below:$2$] (2r) at (2,0){};

            \draw (1) to (2) to (2r) to[out=-45,in=45,looseness=8] (2r) (2) to (1t);

            \draw[blue] (2r) circle (0.2cm);
        \end{tikzpicture}$};
        \node (12111) at (0,{10*sin(45)}){$\begin{tikzpicture}
            \node[gauge, label=below:$1$] (1) at (0,0){};
            \node[gauge, label=below:$2$] (2) at (1,0){};
            \node[gauge, label=left:$1$] (1T) at (1,1) {};
            \node[gauge, label=above:$1$] (1t) at ({1+cos(45)},{sin(45)}){};
            \node[gauge, label=below:$1$] (1b) at ({1+cos(45)},{-sin(45)}){};
            \draw (1)--(2)--(1t) (1T)--(2)--(1b);
            \draw[-] (1t) to[out=45,in=315,looseness=8](1t);
            \draw[-] (1b) to[out=45,in=315,looseness=8] (1b);
            \draw[blue] (1t) circle (0.2cm);
            \draw[blue] (1b) circle (0.2cm);
        \end{tikzpicture}$};
        \node (1) at (0,{5+10*sin(45)}){$\begin{tikzpicture}
            \node[gauge, label=below:$1$](1){};
            \draw[-] (1) to[out=45,in=135,looseness=8] node[pos=0.5,above]{$3$} (1);
        \end{tikzpicture}$};

        \draw[] (1232) -- (12211) node[pos=0.5, label=below:$A_1$]{} -- (12111) node[pos=0.5, label=above:$A_1$]{} -- (1) node[pos=0.5,label=right:$C_3$]{};

        \draw[] (1232)--(1221) node[pos=0.5,label=below:$A_1$]{}--(12111) node[pos=0.5,label=above:$A_1$]{};
        
    \end{tikzpicture}}
    \caption{}
    \label{fig_HasseB3Orbita}
    \end{subfigure}
    \begin{subfigure}{0.49\textwidth}
    \centering
    \scalebox{.7}{
    \begin{tikzpicture}
        \node (1232) at (0,0){$(2^3)$};
        \node (12211) at ({-5*cos(45)},{5*sin(45)}){$(4,1^2)$};
        \node (1221) at ({5*cos(45)},{5*sin(45)}){$(3^2)$};
        \node (12111) at (0,{10*sin(45)}){$(4,2)$};
        \node (1) at (0,{5+10*sin(45)}){$(6)$};

        \draw[] (1232) -- (12211) node[pos=0.5, label=below:$A_1$]{} -- (12111) node[pos=0.5, label=above:$A_1$]{} -- (1) node[pos=0.5,label=right:$C_3$]{};

        \draw[] (1232)--(1221) node[pos=0.5,label=below:$A_1$]{}--(12111) node[pos=0.5,label=above:$A_1$]{};
        
    \end{tikzpicture}
    }
    \caption{}
    \label{fig_HasseB3Orbitb}
    \end{subfigure}
    \caption{(\subref{fig_HasseB3Orbita}): The Higgs branch Hasse diagram of $\mathcal Q_{\ref{eq_B3Orb}}$. (\subref{fig_HasseB3Orbitb}): The S{\l}odowy intersection $\mathcal S^{C_3}_{\mathcal N,(2^3)}$ in the $C_3$ nilcone with orbits specified by partitions of $6$. This is the same Hasse diagram as in \Figref{fig_HasseB3Orbita}.}
    \label{fig_HasseB3Orbit}
\end{figure}
\FloatBarrier
\subsection{Example: $E_6$}
Consider now the theory given by $\mathcal Q_{\ref{eq_E6}}$.
\begin{equation}
    \mathcal Q_{\ref{eq_E6}}=\raisebox{-.4\height}{\begin{tikzpicture}
        \node[gauge, label=below:$1$] (1) at (0,0){};
        \node[gauge, label=below:$2$] (2) at (1,0){};
        \node[gauge, label=below:$3$] (3) at (2,0) {};
        \node[gauge, label=below:$1$] (4) at (3,0) {};

        \draw[-] (1)--(2)--(3)--(4);
        \draw[-] (3) to[out=45,in=135,looseness=8] (3);
    \end{tikzpicture}}\label{eq_E6}.
\end{equation}
The Hasse diagram for the Higgs branch of $\mathcal Q_{\ref{eq_E6}}$ is shown in \Figref{fig_E6a}, which matches the S{\l}odowy intersection $\mathcal S^{E_6}_{D_5(a_1),2A_2+A_1}$ in the $E_6$ nilcone shown in \Figref{fig_E6b}. The Higgs branch Namikawa-Weyl group is $S_3$. The unrefined Higgs branch Hilbert series is given below for brevity, however the refined Hilbert series is verified to match that of $\mathcal S^{E_6}_{D_5(a_1),2A_2+A_1}$.
\begin{align}
    \hs\left[\mathcal H(\mathcal Q_{\ref{eq_E6}})\right]=&\frac{\left(\begin{aligned}1 &+ 2 t^3 + t^4 + 4 t^5 + 7 t^6 + 4 t^7 + 8 t^8 + 10 t^9 + 6 t^{10} \\&+ 
 10 t^{11} + 8 t^{12} + 4 t^{13} + 7 t^{14} + 4 t^{15} + t^{16} + 
 2 t^{17} + t^{20}\end{aligned}\right)}{(1 - t^2)^3  (1 - t^3)^4 (1 - t^4)^3}\\
 &=\hs\left[\mathcal S^{E_6}_{D_5(a_1),2A_2+A_1}\right].
\end{align}

\newcommand{\EsitwoAtwoAone}{
\begin{tikzpicture}
\node (1) [gauge, label=below:$1$] at (0,0) {};
\node (2) [gauge, label=below:$2$] at (1,0) {};
\node (3) [gauge, label=below:$3$] at (2,0) {};
\node (4) [gauge, label=below:$1$] at (3,0) {};
\draw[-] (1)--(2)--(3)--(4);
\draw[-] (3) to[out=45,in=135,looseness=8] (3);
\end{tikzpicture}
}

\newcommand{\EsiAthreeAone}{
\begin{tikzpicture}
\node (1) [gauge, label=below:$1$] at (0,0) {};
\node (2) [gauge, label=below:$2$] at (1,0) {};
\node (3) [gauge, label=left:$2$] at (2,1) {};
\node (4) [gauge, label=left:$1$] at (2,-1) {};
\node (5) [gauge, label=below:$1$] at (3,0) {};
\draw[-] (1)--(2)--(3)--(5)--(4)--(2);
\draw[-] (3) to[out=45,in=135,looseness=8] (3);
\draw[-] (4) to[out=225,in=315,looseness=8] (4);
\draw[blue] (3) circle (0.2cm);
\draw[blue] (4) circle (0.2cm);
\end{tikzpicture}
}

\newcommand{\EsiDfouraone}{
\begin{tikzpicture}
\node (1) [gauge, label=below:$1$] at (0,0) {};
\node (2) [gauge, label=below:$2$] at (1,0) {};
\node (3) [gauge, label=below:$1$] at (2,0) {};
\node (4) [gauge, label=left:$1$] at (2,1) {};
\node (5) [gauge, label=left:$1$] at (2,-1) {};
\node (6) [gauge, label=below:$1$] at (3,0) {};
\draw[-] (1)--(2)--(3)--(6)--(4)--(2);
\draw[-] (2)--(5)--(6);
\draw[-] (3) to[out=45,in=135,looseness=8] (3);
\draw[-] (4) to[out=45,in=135,looseness=8] (4);
\draw[-] (5) to[out=225,in=315,looseness=8] (5);
\draw[blue] (3) circle (0.2cm);
\draw[blue] (4) circle (0.2cm);
\draw[blue] (5) circle (0.2cm);
\end{tikzpicture}
}

\newcommand{\EsiDfour}{
\begin{tikzpicture}
\node (1) [gauge, label=below:$1$] at (0,0) {};
\node (2) [gauge, label=below:$1$] at (1,0) {};
\draw[-] (1) to[out=225,in=135,looseness=8] node[pos=0.5,left]{$4$} (1);
\draw[transform canvas={xshift=0pt, yshift=2pt}] (1)--(2);
\draw[transform canvas={xshift=0pt, yshift=-2pt}] (1)--(2);
\draw[transform canvas={xshift=0pt, yshift=0pt}] (1)--(2);
\draw[red] (1) circle (0.2cm);
\end{tikzpicture}
}

\newcommand{\EsiAfour}{
\begin{tikzpicture}
\node (1) [gauge, label=below:$1$] at (0,0) {};
\node (2) [gauge, label=below:$1$] at (1,0) {};
\node (3) [gauge, label=above:$1$] at (0,1) {};
\node (4) [gauge, label=above:$1$] at (1,1) {};
\draw[-] (1)--(3)--(4)--(2);
\draw[transform canvas={xshift=0pt, yshift=1.5pt}] (1)--(2);
\draw[transform canvas={xshift=0pt, yshift=-1.5pt}] (1)--(2);
\draw[-] (1) to[out=225,in=135,looseness=8] (1);
\draw[-] (2) to[out=45,in=315,looseness=8] node[pos=0.5,right]{$3$} (2);
\draw[blue] (1) circle (0.2cm);
\draw[red] (2) circle (0.2cm);
\end{tikzpicture}
}

\newcommand{\EsiAfourAone}{
\begin{tikzpicture}
\node (1) [gauge, label=below:$1$] at (0,0) {};
\node (2) [gauge, label=below:$1$] at (1,0) {};
\node (3) [gauge, label=above:$1$] at (0.5,0.866) {};
\draw[-] (1)--(2)--(3)--(1);
\draw[-] (2) to[out=45,in=315,looseness=8] node[pos=0.5,right]{$5$} (2);
\draw[blue] (2) circle (0.2cm);
\end{tikzpicture}
}

\newcommand{\EsiDfiveaone}{
\begin{tikzpicture}
\node (1) [gauge, label=below:$1$] at (0,0) {};
\draw[-] (1) to[out=45,in=135,looseness=8] node[pos=0.5,above]{$6$} (1);
\draw[blue] (1) circle (0.2cm);
\end{tikzpicture}
}

\begin{figure}[H]
    \centering
    \begin{subfigure}{0.49\textwidth}
    \centering
    \scalebox{.5}{
    \begin{tikzpicture}
        \node (a) at (0,0) {\EsitwoAtwoAone};
        \node (b) at (0,5) {\EsiAthreeAone};
        \node (c) at (0,10) {\EsiDfouraone};
        \node (d) at (-3,15) {\EsiDfour};
        \node (e) at (3,15) {\EsiAfour};
        \node (f) at (3,20) {\EsiAfourAone};
        \node (g) at (0,25) {\EsiDfiveaone};
        \draw[-] (a)--(b) node[pos=0.5, left]{$m_1$};
        \draw[-] (b)--(c) node[pos=0.5, left]{$A_1$};
        \draw[-] (c)--(d) node[pos=0.5, right]{$G_2$};
        \draw[-] (c)--(e) node[pos=0.5, right]{$3C_2$};
        \draw[-] (e)--(f) node[pos=0.5, right]{$A_1$};
        \draw[-] (f)--(g) node[pos=0.5, right]{$A_2$};
        \draw[-] (d)--(g) node[pos=0.5, left]{$a_2$};
    \end{tikzpicture}
    }
    \caption{}
    \label{fig_E6a}
    \end{subfigure}
    \begin{subfigure}{0.49\textwidth}
    \centering
    \scalebox{.7}{
    \begin{tikzpicture}
        \node (a) at (0,0) {$2A_2+A_1$};
        \node (b) at (0,3.5) {$A_3+A_1$};
        \node (c) at (0,6.5) {$D_4(a_1)$};
        \node (d) at (-3,10) {$D_4$};
        \node (e) at (3,10) {$A_4$};
        \node (f) at (3,13.5) {$A_4+A_1$};
        \node (g) at (0,17.5) {$D_5(a_1)$};
        \draw[-] (a)--(b) node[pos=0.5, left]{$m_1$};
        \draw[-] (b)--(c) node[pos=0.5, left]{$A_1$};
        \draw[-] (c)--(d) node[pos=0.5, right]{$G_2$};
        \draw[-] (c)--(e) node[pos=0.5, right]{$3C_2$};
        \draw[-] (e)--(f) node[pos=0.5, right]{$A_1$};
        \draw[-] (f)--(g) node[pos=0.5, right]{$A_2$};
        \draw[-] (d)--(g) node[pos=0.5, left]{$a_2$};
    \end{tikzpicture}
    }
    \caption{}
    \label{fig_E6b}
    \end{subfigure}
    \caption{(\subref{fig_E6a}): The Higgs branch Hasse diagram of $\mathcal Q_{\ref{eq_E6}}$. (\subref{fig_E6b}): The Hasse diagram of $\mathcal S^{E_6}_{D_5(a_1),2A_2+A_1}$ in the $E_6$ nilcone with orbits specified by their Bala-Carter labels. This is the same Hasse diagram as in \Figref{fig_E6a}.}
    \label{fig_1231E6}
\end{figure}

The Coulomb branch of $\mathcal Q_{\ref{eq_E6}}$ is found to be $\mathcal S^{E_6}_{D_4(a_1),A_2+A_1}$ via a calculation of its refined Hilbert series. Again, only the unrefined series is presented here for brevity.\begin{equation}
    \hs\left[\mathcal C(\mathcal Q_{\ref{eq_E6}})\right]=\frac{(1-t)\left(\begin{aligned}1 &+ t + 4 t^2 + 7 t^3 + 17 t^4 + 26 t^5 + 46 t^6 + 62 t^7 + 89 t^8 + 
   105 t^9 + 125 t^{10} + 124 t^{11}\\ &+ 125 t^{12} + 105 t^{13} + 89 t^{14} + 
   62 t^{15} + 46 t^{16} + 26 t^{17} + 17 t^{18} + 7 t^{19} + 4 t^{20} + t^{21} + 
   t^{22}\end{aligned}\right)}{(1 - t^2)^6 (1 - t^3)^5(1 - t^4)^2 }
\end{equation}
This is as expected from Lusztig-Spaltenstein duality.
\begin{figure}[h!]
\centering
\begin{tikzpicture}
\node (1) at (0,0) {$D_4(a_1)$};
\node (2) at (0,-3) {$A_3+A_1$};
\node (3) at (-3,-6) {$A_3$};
\node (4) at (3,-6) {$2A_2+A_1$};
\node (5) at (-3,-9) {$A_2+2A_1$};
\node (6) at (3,-9) {$2A_2$};
\node (7) at (0,-12) {$A_2+A_1$};

\draw[-] (1)--(2)node[midway, right]{$A_1$}--(3)node[midway, left]{$b_2$}--(5)node[midway, left]{$A_1$}--(7)node[midway, left]{$a_2$};

\draw[-] (2)--(4)node[midway, right]{$m_1$}--(6)node[midway, right]{$g_2$}--(7)node[midway, right]{$A_2$};

\draw[-] (4)--(5)node[midway, above]{$h_{2,3}$};
\end{tikzpicture}
\caption{The Hasse diagram for $\mathcal S^{E_6}_{D_4(a_1),A_2+A_1}$ in the $E_6$ nilcone with orbits specified by their Bala-Carter labels. This is the same Hasse diagram as for the Coulomb branch of $\mathcal Q_{\ref{eq_E6}}$. More information regarding the slice $h_{2,3}$, which is called $\tau$ in \cite{Generic_singularities}, can be found in \cite{bourget2021branesquiversaffinegrassmannian}.}
\label{fig_E6Coul}
\end{figure}
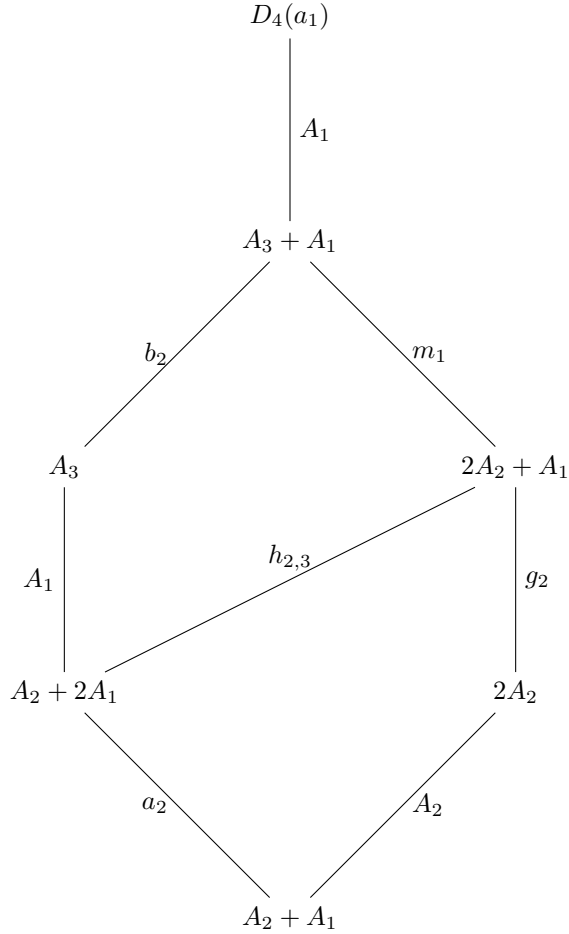

\FloatBarrier
\subsection{Example: $E_7$}
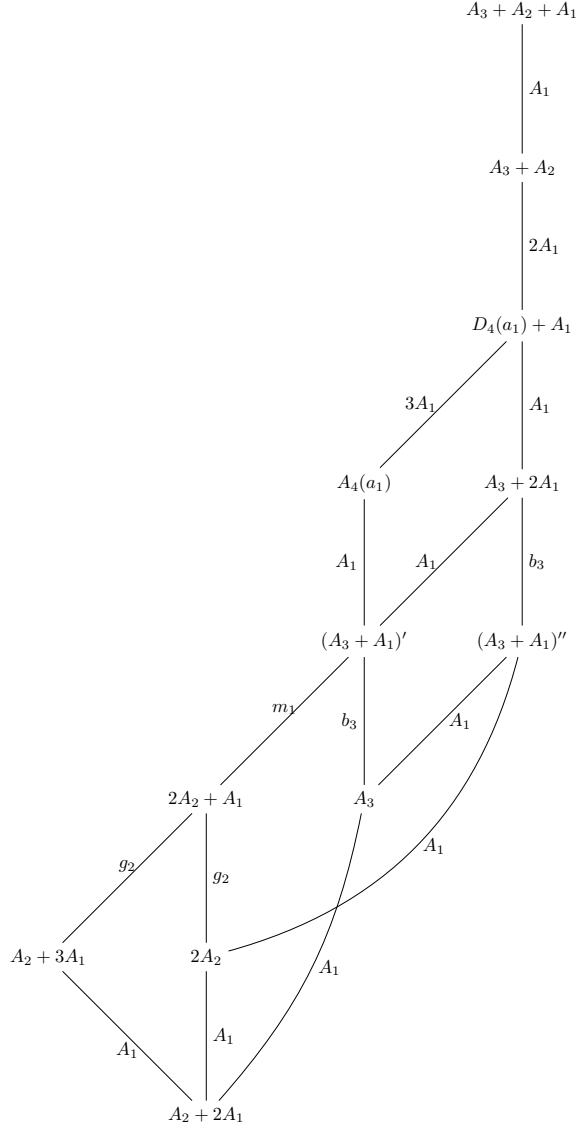
\begin{figure}[h]
\centering
\scalebox{0.7}{
\begin{tikzpicture}
\node (1) at (0,0){$A_3+A_2+A_1$};
\node (2) at (0,-3) {$A_3+A_2$};
\node (9) at (0,-6) {$D_4(a_1)+A_1$};
\node (3) at (0,-9) {$A_3+2A_1$};
\node (12) at ({-3},-9) {$A_4(a_1)$};
\node (10) at ({-3},-12) {$(A_3+A_1)'$};
\node (5) at (0,-12) {$(A_3+A_1)''$};
\node (4) at ({-6},-15) {$2A_2+A_1$};
\node (11) at ({-3},-15) {$A_3$};
\node (6) at ({-9},-18){$A_2+3A_1$};
\node (7) at ({-6},-18){$2A_2$};
\node (8) at ({-6},-21){$A_2+2A_1$};

\draw[-] (1)--(2) node[midway, right] {$A_1$};
\draw[-] (9)--(3)node[midway, right]{$A_1$}--(5) node[midway, right]{$b_3$};
\draw[-] (12)--(10)node[midway, left]{$A_1$}--(11)node[midway, left]{$b_3$};
\draw[-] (3)--(10) node[midway, left]{$A_1$};
\draw[-] (5)--(11) node[midway, right]{$A_1$};
\draw[-] (10)--(4)node[midway, above]{$m_1$}--(6)node[midway, above]{$g_2$}--(8)node[midway, below]{$A_1$};
\draw[-] (4)--(7)node[midway, right]{$g_2$}--(8)node[midway, right]{$A_1$};

\draw[-] (5) to[bend left=30] node[midway, right]{$A_1$} (7);
\draw[-] (11) to[bend left=15] node[midway, right]{$A_1$} (8);

\draw(9)--node[midway, right]{$2A_1$}(2);

\draw(9)--node[midway, left]{$3A_1$}(12);
\end{tikzpicture}}
\caption{The Hasse diagram of $\mathcal S^{E_7}_{A_3+A_2+A_1,A_2+2A_1}$ in the $E_7$ nilcone with orbits specified by their Bala-Carter labels. This is the same Hasse diagram as for the Coulomb branch of $\mathcal Q_{\ref{eq_E7}}$.}
\label{fig_E7Coul}
\end{figure}

Consider the theory given by $\mathcal Q_{\ref{eq_E7}}$.
\begin{equation}
    \mathcal Q_{\ref{eq_E7}}=\raisebox{-.4\height}{\begin{tikzpicture}
        \node[gauge, label=below:$1$] (1) at (0,0){};
        \node[gauge, label=below:$2$] (2) at (1,0){};
        \node[gauge, label=below:$4$] (3) at (2,0) {};
        \node[gauge, label=below:$3$] (4) at (3,0) {};

        \draw[-] (1)--(2)--(3);
        \draw[transform canvas={yshift=1.5pt}] (3)--(4);
        \draw[transform canvas={yshift=-1.5pt}] (3)--(4);
    \end{tikzpicture}}\label{eq_E7}.
\end{equation}
The Coulomb branch of $Q_{\ref{eq_E7}}$ is computed to be the S{\l}odowy intersection $\mathcal S^{E_7}_{A_3+A_2+A_1, A_2+2A_1}$ \cite{Gledhill:2021cbe,Hanany:2024fqf}. The Hasse diagram for $\mathcal S^{E_7}_{A_3+A_2+A_1, A_2+2A_1}$ is given in \Figref{fig_E7Coul} and matches the Hasse diagram for $\mathcal C\left(Q_{\ref{eq_E7}}\right)$. 

As predicted by Lusztig-Spaltenstein duality, the Higgs branch is the S{\l}odowy intersection $\mathcal S^{E_7}_{E_{7}(a_4), A_4+A_2}$ in the $E_7$ nilcone. The Hasse diagram for the Higgs branch is shown in \Figref{fig_125E7a}, which matches $\mathcal S^{E_7}_{E_{7}(a_4), A_4+A_2}$, as can be seen in \Figref{fig_125E7b}. The Higgs branch Namikawa-Weyl group is $\mathbb Z_2\times\mathbb Z_2\times \mathbb Z_2$. The identification of the moduli space is verified via the refined Higgs branch Hilbert series, presented in its unrefined form below. 
\begin{align}
    \hs\left[\mathcal H(\mathcal Q_{\ref{eq_E7}})\right]=&\frac{\left(\begin{aligned}1 &+ 2 t^4 + 6 t^6 + 8 t^8 + 12 t^{10} + 23 t^{12} + 27 t^{14} + 25 t^{16} \\&+ 
 27 t^{18} + 23 t^{20} + 12 t^{22} + 8 t^{24} + 6 t^{26} + 
 2 t^{28} + t^{32}\end{aligned}\right)}{(1 - t^2)^{3} (1 - t^4)^3 (1 - t^6)^4}\\
 &=\hs\left[\mathcal S^{E_7}_{E_7(a_4),A_4+A_2}\right].
\end{align}

\newcommand{\EseAfourAtwo}{
\begin{tikzpicture}
\node (1) [gauge, label=below:$1$] at (0,0) {};
\node (2) [gauge, label=below:$2$] at (1,0) {};
\node (3) [gauge, label=below:$4$] at (2,0) {};
\node (4) [gauge, label=below:$3$] at (3,0) {};
\draw[-] (1)--(2)--(3);
\draw[transform canvas={xshift=0pt, yshift=1.5pt}] (3)--(4);
\draw[transform canvas={xshift=0pt, yshift=-1.5pt}] (3)--(4);
\end{tikzpicture}  
}

\newcommand{\EseAfiveprime}{
\begin{tikzpicture}
\node (1) [gauge, label=below:$1$] at (0,0) {};
\node (2) [gauge, label=below:$2$] at (1,0) {};
\node (3) [gauge, label=below:$3$] at (2,0) {};
\node (4) [gauge, label=below:$2$] at (3,0) {};
\node (5) [gauge, label=left:$1$] at (1,1) {};
\draw[-] (5) to[out=45,in=135,looseness=8] (5);
\draw[-] (1)--(2)--(3);
\draw[-] (5)--(2);
\draw[transform canvas={xshift=0pt, yshift=1.5pt}] (3)--(4);
\draw[transform canvas={xshift=0pt, yshift=-1.5pt}] (3)--(4);
\draw[blue] (5) circle (0.2cm);
\end{tikzpicture}
}

\newcommand{\EseDfiveaone}{
\begin{tikzpicture}
\node (1) [gauge, label=below:$1$] at (0,0) {};
\node (2) [gauge, label=below:$2$] at (1,0) {};
\node (3) [gauge, label=below:$2$] at (2,0) {};
\node (4) [gauge, label=below:$1$] at (3,0) {};
\node (5) [gauge, label=left:$2$] at (1,1) {};
\draw[-] (5) to[out=45,in=135,looseness=8] (5);
\draw[-] (1)--(2)--(3);
\draw[-] (5)--(2);
\draw[transform canvas={xshift=0pt, yshift=1.5pt}] (3)--(4);
\draw[transform canvas={xshift=0pt, yshift=-1.5pt}] (3)--(4);
\draw[blue] (5) circle (0.2cm);
\end{tikzpicture}
}

\newcommand{\EseAfiveAone}{
\begin{tikzpicture}
\node (1) [gauge, label=below:$1$] at (0,0) {};
\node (2) [gauge, label=below:$2$] at (1,0) {};
\node (3) [gauge, label=below:$1$] at (2,0) {};
\node (5) [gauge, label=left:$3$] at (1,1) {};
\draw[-] (5) to[out=45,in=135,looseness=8] (5);
\draw[-] (1)--(2)--(3);
\draw[-] (5)--(2);
\draw[blue] (5) circle (0.2cm);
\end{tikzpicture}
}

\newcommand{\EseEsixathree}{
\begin{tikzpicture}
\node (1) [gauge, label=below:$1$] at (0,0) {};
\node (2) [gauge, label=below:$2$] at (1,0) {};
\node (3) [gauge, label=below:$2$] at (2,0) {};
\node (4) [gauge, label=below:$1$] at (3,0) {};
\node (5) [gauge, label=left:$1$] at (0.5,0.866) {};
\node (6) [gauge, label=right:$1$] at (1.5,0.866) {};
\draw[-] (5) to[out=45,in=135,looseness=8] (5);
\draw[-] (6) to[out=45,in=135,looseness=8] (6);
\draw[-] (1)--(2)--(3);
\draw[-] (5)--(2)--(6);
\draw[transform canvas={xshift=0pt, yshift=1.5pt}] (3)--(4);
\draw[transform canvas={xshift=0pt, yshift=-1.5pt}] (3)--(4);
\draw[blue] (5) circle (0.2cm);
\draw[blue] (6) circle (0.2cm);
\end{tikzpicture}
}

\newcommand{\EseDsixatwo}{
\begin{tikzpicture}
\node (1) [gauge, label=below:$1$] at (0,0) {};
\node (2) [gauge, label=below:$2$] at (1,0) {};
\node (3) [gauge, label=below:$1$] at (2,0) {};
\node (5) [gauge, label=left:$1$] at (0.5,0.866) {};
\node (6) [gauge, label=right:$2$] at (1.5,0.866) {};
\draw[-] (5) to[out=45,in=135,looseness=8] (5);
\draw[-] (6) to[out=45,in=135,looseness=8] (6);
\draw[-] (1)--(2)--(3);
\draw[-] (5)--(2)--(6);
\draw[blue] (5) circle (0.2cm);
\draw[blue] (6) circle (0.2cm);
\end{tikzpicture}
}

\newcommand{\EseDfive}{
\begin{tikzpicture}
\node (1) [gauge, label=below:$1$] at (0,0) {};
\node (2) [gauge, label=below:$1$] at (1,0) {};
\node (3) [gauge, label=right:$1$] at (1,1) {};
\draw[-] (3) to[out=45,in=135,looseness=8] node[pos=0.5,above]{$3$} (3);
\draw[transform canvas={xshift=0pt, yshift=1.5pt}] (1)--(2);
\draw[transform canvas={xshift=0pt, yshift=-1.5pt}] (1)--(2);
\draw[transform canvas={xshift=1.5pt, yshift=0pt}] (3)--(2);
\draw[transform canvas={xshift=-1.5pt, yshift=0pt}] (3)--(2);
\draw[blue] (3) circle (0.2cm);
\end{tikzpicture}
}

\newcommand{\EseEsevenafive}{
\begin{tikzpicture}
\node (1) [gauge, label=below:$1$] at (0,0) {};
\node (2) [gauge, label=below:$2$] at (1,0) {};
\node (3) [gauge, label=below:$1$] at (2,0) {};
\node (4) [gauge, label=left:$1$] at (0.1,0.866) {};
\node (5) [gauge, label=left:$1$] at (1,0.866) {};
\node (6) [gauge, label=right:$1$] at (1.9,0.866) {};
\draw[-] (4) to[out=45,in=135,looseness=8] (4);
\draw[-] (5) to[out=45,in=135,looseness=8] (5);
\draw[-] (6) to[out=45,in=135,looseness=8] (6);
\draw[-] (1)--(2)--(3);
\draw[-] (4)--(2)--(5);
\draw[-] (6)--(2);
\draw[blue] (4) circle (0.2cm);
\draw[blue] (5) circle (0.2cm);
\draw[blue] (6) circle (0.2cm);
\end{tikzpicture}
}

\newcommand{\EseDsixaone}{
\begin{tikzpicture}
\node (1) [gauge, label=below:$1$] at (0,0) {};
\node (2) [gauge, label=below:$1$] at (1,0) {};
\draw[-] (1) to[out=225,in=135,looseness=8] (1);
\draw[-] (2) to[out=45,in=315,looseness=8] node[pos=0.5,right]{$3$} (2);
\draw[transform canvas={xshift=0pt, yshift=1.5pt}] (1)--(2);
\draw[transform canvas={xshift=0pt, yshift=-1.5pt}] (1)--(2);
\draw[red] (1) circle (0.2cm);
\draw[blue] (2) circle (0.2cm);
\end{tikzpicture}
}

\newcommand{\EseDfiveAone}{
\begin{tikzpicture}
\node (1) [gauge, label=below:$1$] at (0,0) {};
\node (2) [gauge, label=below:$1$] at (1,0) {};
\draw[-] (2) to[out=45,in=315,looseness=8] node[pos=0.5,right]{$4$} (2);
\draw[transform canvas={xshift=0pt, yshift=1.5pt}] (1)--(2);
\draw[transform canvas={xshift=0pt, yshift=-1.5pt}] (1)--(2);
\draw[blue] (2) circle (0.2cm);
\end{tikzpicture}
}

\newcommand{\EseAsix}{
\begin{tikzpicture}
\node (1) [gauge, label=below:$1$] at (0,0) {};
\node (2) [gauge, label=below:$1$] at (1,0) {};
\draw[-] (1) to[out=225,in=135,looseness=8] node[pos=0.5,left]{$4$} (1);
\draw[transform canvas={xshift=0pt, yshift=1.5pt}] (1)--(2);
\draw[transform canvas={xshift=0pt, yshift=-1.5pt}] (1)--(2);
\draw[blue] (1) circle (0.2cm);
\end{tikzpicture}
}

\newcommand{\EseEsevenafour}{
\begin{tikzpicture}
\node (1) [gauge, label=below:$1$] at (0,0) {};
\draw[-] (1) to[out=45,in=135,looseness=8] node[pos=0.5,above]{$5$} (1);
\draw[blue] (1) circle (0.2cm);
\end{tikzpicture}
}

\begin{figure}[h]
    \centering
    \begin{subfigure}{0.49\textwidth}
    \centering
    \scalebox{.4}{
    \begin{tikzpicture}
        \node (a) at (0,0) {\EseAfourAtwo};
        \node (b) at (-5,5) {\EseAfiveprime};
        \node (c) at (0,5) {\EseDfiveaone};
        \node (d) at (5,5) {\EseAfiveAone};
        \node (e) at (-2.5,10) {\EseEsixathree};
        \node (f) at (2.5,10) {\EseDsixatwo};
        \node (g) at (-2.5,15) {\EseDfive};
        \node (h) at (2.5,15) {\EseEsevenafive};
        \node (i) at (-5,20) {\EseDsixaone};
        \node (j) at (0,20) {\EseDfiveAone};
        \node (k) at (5,20) {\EseAsix};
        \node (l) at (0,25) {\EseEsevenafour};
        \draw[-] (a)--(b) node[pos=0.5, left]{$A_1$};
        \draw[-] (a)--(c) node[pos=0.5, left]{$A_1$};
        \draw[-] (a)--(d) node[pos=0.5, right]{$A_1$};
        \draw[-] (b)--(e) node[pos=0.5, left]{$A_1$};
        \draw[-] (b)--(f) node[pos=0.8, left]{$A_1$};
        \draw[-] (c)--(e) node[pos=0.8, left]{$A_1$};
        \draw[-] (c)--(f) node[pos=0.5, left]{$A_1$};
        \draw[-] (d)--(f) node[pos=0.5, right]{$m_1$};
        \draw[-] (e)--(g) node[pos=0.5, left]{$C_3$};
        \draw[-] (e)--(h) node[pos=0.5, left]{$A_1$};
        \draw[-] (f)--(h) node[pos=0.5, right]{$A_1$};
        \draw[-] (g)--(i) node[pos=0.5, left]{$A_1$};
        \draw[-] (g)--(j) node[pos=0.8, left]{$A_1$};
        \draw[-] (h)--(i) node[pos=0.6, left]{$3C_3$};
        \draw[-] (h)--(j) node[pos=0.5, left]{$G_2$};
        \draw[-] (h)--(k) node[pos=0.5, right]{$G_2$};
        \draw[-] (i)--(l) node[pos=0.5, left]{$A_1$};
        \draw[-] (j)--(l) node[pos=0.5, left]{$A_1$};
        \draw[-] (k)--(l) node[pos=0.5, right]{$A_1$};
    \end{tikzpicture}
    }
    \caption{}
    \label{fig_125E7a}
    \end{subfigure}
    \begin{subfigure}{0.49\textwidth}
    \centering
    \scalebox{.6}{
    \begin{tikzpicture}
        \node (a) at (0,0) {$A_4+A_2$};
        \node (b) at (-3,3) {$A_5^\prime$};
        \node (c) at (0,3) {$D_5(a_1)+A_1$};
        \node (d) at (3,3) {$A_5+A_1$};
        \node (e) at (-1.5,6.5) {$E_6(a_3)$};
        \node (f) at (1.5,6.5) {$D_6(a_2)$};
        \node (g) at (-1.5,10) {$D_5$};
        \node (h) at (1.5,10) {$E_7(a_5)$};
        \node (i) at (-3,13) {$D_6(a_1)$};
        \node (j) at (0,13) {$D_5+A_1$};
        \node (k) at (3,13) {$A_6$};
        \node (l) at (0,16) {$E_7(a_4)$};
        \draw[-] (a)--(b) node[pos=0.5, left]{$A_1$};
        \draw[-] (a)--(c) node[pos=0.5, left]{$A_1$};
        \draw[-] (a)--(d) node[pos=0.5, right]{$A_1$};
        \draw[-] (b)--(e) node[pos=0.5, left]{$A_1$};
        \draw[-] (b)--(f) node[pos=0.8, left]{$A_1$};
        \draw[-] (c)--(e) node[pos=0.8, left]{$A_1$};
        \draw[-] (c)--(f) node[pos=0.5, left]{$A_1$};
        \draw[-] (d)--(f) node[pos=0.5, right]{$m_1$};
        \draw[-] (e)--(g) node[pos=0.5, left]{$C_3$};
        \draw[-] (e)--(h) node[pos=0.5, left]{$A_1$};
        \draw[-] (f)--(h) node[pos=0.5, right]{$A_1$};
        \draw[-] (g)--(i) node[pos=0.5, left]{$A_1$};
        \draw[-] (g)--(j) node[pos=0.8, left]{$A_1$};
        \draw[-] (h)--(i) node[pos=0.6, left]{$3C_3$};
        \draw[-] (h)--(j) node[pos=0.5, left]{$G_2$};
        \draw[-] (h)--(k) node[pos=0.5, right]{$G_2$};
        \draw[-] (i)--(l) node[pos=0.5, left]{$A_1$};
        \draw[-] (j)--(l) node[pos=0.5, left]{$A_1$};
        \draw[-] (k)--(l) node[pos=0.5, right]{$A_1$};
    \end{tikzpicture}
    }
    \caption{}
    \label{fig_125E7b}
    \end{subfigure}
    \caption{(\subref{fig_125E7a}): The Higgs branch Hasse diagram of $\mathcal Q_{\ref{eq_E7}}$. (\subref{fig_125E7b}): The Hasse diagram of $\mathcal S^{E_7}_{E_7(a_4),A_4+A_2}$ in the $E_7$ nilcone with orbits specified by their Bala-Carter labels. This is the same Hasse diagram as \Figref{fig_125E7a}.}
    \label{fig_125E7}
\end{figure}

The quivers for the slice between $A_5^\prime$ and $E_7(a_4)$, and for the slice between $D_5(a_1)+A_1$ and $E_7(a_4)$ have the same Hasse diagram, although their moduli spaces are not identical since they have different Hilbert series. 

\paragraph{Example: $E_7$ Again}
Now consider the following theory:
\begin{equation}
     \mathcal Q_{\ref{eq_E72}}=\raisebox{-.4\height}{\begin{tikzpicture}
        \node[gauge, label=below:$2$] (2) at (1,0){};
        \node[gauge, label=below:$4$] (3) at (2,0) {};
        \node[gauge, label=below:$3$] (4) at (3,0) {};

        \draw[-] (2)--(3);
        \draw[transform canvas={yshift=1.5pt}] (3)--(4);
        \draw[transform canvas={yshift=-1.5pt}] (3)--(4);
    \end{tikzpicture}}\label{eq_E72}.
\end{equation}

The Coulomb branch Hilbert series of \Quiver{eq_E72} identifies the moduli space as $\mathcal C(\text{\Quiver{eq_E72}}) = {\mathcal S}^{E_7}_{A_3+A_2+A_1,A_2+3A_1}$ \cite{Hanany:2024fqf}. The Hasse diagram of ${\mathcal S}^{E_7}_{A_3+A_2+A_1,A_2+3A_1}$ is given in \Figref{fig_D4/S4HasseCoul} and matches the Hasse diagram for ${\mathcal C}(\text{\Quiver{eq_E72}})$.

\begin{figure}[h]
    \centering
    \scalebox{0.7}{
    \begin{tikzpicture}
    \node (a) at (0,0) {$A_3+A_2+A_1$};
    \node (b) at (0,-3.5) {$A_3+A_2$};
    \node (c) at (0,-7) {$D_4(a_1)+A_1$};
    \node (d) at (-3.5,-10.5) {$A_3+2A_1$};
    \node (e) at (0,-14) {$(A_3+A_1)'$};
    \node (f) at (0,-17.5){$2A_2+A_1'$};
    \node (g) at (0,-21){$A_2+3A_1$};
    \node (h) at (0,-10.5){$D_4(a_1)$};

    \draw[] (a)--(b) node[pos=0.5, label=right:$A_1$] {};
    
   \draw(b)--(c) node[pos=0.5, label=right:$2A_1$]{};
    
    \draw(c)--(h) node[pos=0.5, label=right:$3A_1$]{};
    \draw[](c)--(h)--(e) node[pos=0.5, label=right:$A_1$]{};

    \draw[](c)--(d) node[pos=0.5, label=above:$A_1$]{}--(e) node[pos=0.5, label=right:$A_1$] {}--(f) node[pos=0.5,label=right:$m_1$] {}--(g) node[pos=0.5, label=right:$g_2$]{};
    \end{tikzpicture}}
    \caption{The Hasse diagram of $\mathcal S^{E_7}_{A_3+A_2+A_1,A_2+3A_1}$ in the $E_7$ nilcone with orbits specified by their Bala-Carter labels. This is the same Hasse diagram as for the Coulomb branch of $\mathcal Q_{\ref{eq_D4/S4}}$.}
    \label{fig_HasseE72Coul}
    \end{figure}

The Higgs branch Hasse diagram of $\mathcal Q_{\ref{eq_E72}}$ is computed as \Figref{fig_HasseE72a}. This Hasse diagram matches that of $\mathcal S^{E_7}_{A_6,A_4+A_2}$ shown explicitly in \Figref{fig_HasseE72b}. The Namikawa-Weyl group of the Higgs branch is identified as $\mathcal W_{G_2}=\mathrm{Dih}_{6}$ of order 12. The unrefined Higgs branch Hilbert series is given below, which (when refined) agrees with that for $\mathcal S^{E_7}_{A_6,A_4+A_2}$.\begin{equation}
    \hs\left[\mathcal H(\mathcal Q_{\ref{eq_E72}}\right]=\frac{(1-t^{12}) (1 + 3 t^2 + 6 t^4 + 13 t^6 + 24 t^8 + 29 t^{10} + 
     29 t^{12} + 29 t^{14} + 24 t^{16} + 13 t^{18} + 6 t^{20} + 3 t^{22} + 
     t^{24})}{(1 - t^4)^5 (1 - t^6)^4}. 
\end{equation}

\begin{figure}[h]
    \centering
    \begin{subfigure}{0.49\textwidth}
    \scalebox{.7}{
    \begin{tikzpicture}
        \node (243) at (0,-2){$\begin{tikzpicture}
            \node[gauge, label=below:$2$] (1) at (0,0){};
            \node[gauge, label=below:$4$] (2) at (1,0){};
            \node[gauge, label=below:$3$] (3) at (2,0){};
            \draw (1) to (2);
            \draw[transform canvas={yshift=1.5pt}] (2)--(3);
            \draw[transform canvas={yshift=-1.5pt}] (2)--(3);
    \end{tikzpicture}$};
        
        \node (321) at ({-6*cos(45)},{5*sin(45)}) {$\begin{tikzpicture}
            \node[gauge, label=below:$2$] (1) at (0,0){};
            \node[gauge, label=below:$1$] (2) at (1,0){};
            \node[gauge, label=left:$3$] (3) at (0,1){};

            \draw (2) to (1) to (3) to[out=45,in=135,looseness=8] (3);
    \end{tikzpicture}$};
        
        \node (2221) at ({6*cos(45)},{5*sin(45)}){$\begin{tikzpicture}
            \node[gauge, label=left:$2$] (1) at (0,1){};
            \node[gauge, label=below:$2$] (2) at (0,0){};
            \node[gauge, label=below:$2$] (3) at (1,0){};
            \node[gauge, label=below:$1$] (4) at (2,0){};

            \draw (3) to (2) to (1) to [out=45,in=135,looseness=8] (1);
            \draw[transform canvas={yshift=1.5pt}] (3)--(4);
            \draw[transform canvas={yshift=-1.5pt}] (3)--(4);
            \draw[blue] (1) circle (0.2cm);
            \draw[blue] \convexpath{3,4} {0.2cm};

    \end{tikzpicture}$};

        \node (1232) at (0,{5*sin(45)}){$\begin{tikzpicture}
            \node[gauge, label=left:$1$] (1) at (0,1){};
            \node[gauge, label=below:$2$] (2) at (0,0){};
            \node[gauge, label=below:$3$] (3) at (1,0){};
            \node[gauge, label=below:$2$] (4) at (2,0){};

            \draw (3) to (2) to (1) to [out=45,in=135,looseness=8] (1) ;
            \draw[transform canvas={yshift=1.5pt}] (3)--(4);
            \draw[transform canvas={yshift=-1.5pt}] (3)--(4);
            \draw[blue] (1) circle (0.2cm);
            \draw[blue] \convexpath{3,4} {0.2cm};
    \end{tikzpicture}$};
        
        \node (2211) at ({-3},9){$\begin{tikzpicture}
            \node[gauge, label=below:$2$] (1) at (0,0){};
            \node[gauge, label=below:$1$] (2) at (1,0){};
            \node[gauge, label=left:$2$] (3) at ({-cos(45)},{sin(45)}){};
            \node[gauge, label=right:$1$] (4) at ({cos(45)},{sin(45)}){};
            \draw (2) to (1) to (3) to [out=45,in=135,looseness=8] (3) to (1) to (4) to  [out=45,in=135,looseness=8] (4);
            \draw[blue] (3) circle (0.2cm);
            \draw[blue] (4) circle (0.2cm);
        \end{tikzpicture}$};

        \node (22111) at (3,9){$\begin{tikzpicture}
            \node[gauge, label=right:$1$] (1) at (0.5,1){};
            \node[gauge, label=below:$2$] (2) at (0,0){};
            \node[gauge, label=below:$2$] (3) at (1,0){};
            \node[gauge, label=below:$1$] (4) at (2,0){};
            \node[gauge, label=left:$1$] (5) at (-0.5,1){};

            \draw (3) to (2) to (1) to [out=45,in=135,looseness=8] (1) to (2) to (5) to [out=45,in=135,looseness=8] (5);
            \draw[transform canvas={yshift=1.5pt}] (3)--(4);
            \draw[transform canvas={yshift=-1.5pt}] (3)--(4);
            \draw[blue] (1) circle (0.2cm);
            \draw[blue] (5) circle (0.2cm);
            \draw[blue] \convexpath{3,4} {0.2cm};
    \end{tikzpicture}$};
        
        \node (21111) at (0,15){$\begin{tikzpicture}
            \node[gauge, label=below:$2$] (2) at (0,0){};
            \node[gauge, label=below:$1$] (1l) at (-1,0){};
            \node[gauge, label=left:$1$] (1t) at (0,1){};
            \node[gauge, label=above:$1$] (1tr) at ({cos(45)},{sin(45)}){};
            \node[gauge, label=below:$1$] (1r) at (1,0){};

            \draw[-] (1t) to[out=45,in=135,looseness=8] (1t);
            \draw[-] (1tr) to[out=45,in=315,looseness=8] (1tr);
            \draw[-] (1r) to[out=45,in=315,looseness=8] (1r);
            \draw (1l) to (2) to (1t) (1tr) to (2)to (1r) to (2);

            \draw[blue] (1t) circle (0.2cm);
            \draw[blue] (1tr) circle (0.2cm);
            \draw[blue] (1r) circle (0.2cm);
    \end{tikzpicture}$};

        \node (1) at (0,18){$\begin{tikzpicture}
            \node[gauge, label=below:$1$] (1) at (0,0){};
            \draw[-] (1) to[out=45,in=135,looseness=8] node[pos=0.5,above]{$4$} (1);
    \end{tikzpicture}$};

    \draw[-] (243) -- (321) node[pos=0.5, midway, left]{$A_1$};

        \draw[-] (321) -- (2211) node[pos=0.5, midway, left]{$m_1$};

        \draw[-] (2211) -- (21111) node[pos=0.5, midway, left]{$A_1$};

        \draw[-] (21111) -- (1) node[pos=0.5,midway, left]{$G_2$};

        \draw[-] (243) -- (1232) node[pos=0.5,midway, right]{$A_1$};

        \draw[-] (243) -- (2221) node[pos=0.5,midway, right]{$A_1$};

        \draw[-] (1232) -- (2211) node[pos=0.5,midway, left]{$A_1$};

        \draw[-] (1232) -- (22111) node[pos=0.5,midway, right]{$A_1$};

        \draw[-] (2221) -- (2211) node[pos=0.5,midway, left]{$A_1$};

        \draw[-] (2221) -- (22111) node[pos=0.5,midway, right]{$A_1$};

        \draw[-] (22111) -- (21111) node[pos=0.5,midway, right]{$A_1$};

         \draw[-] (2221) --(2211) node[pos=0.5,midway,left]{$A_1$};
        \draw[-] (1232) -- (2211) node[pos=0.5,midway,left]{$A_1$};
    \end{tikzpicture}}
    \caption{}
    \label{fig_HasseE72a}
    \end{subfigure}
    \begin{subfigure}{0.49\textwidth}
    \centering
    \scalebox{0.7}{\begin{tikzpicture}

     \node (243) at (0,-2){$A_4+A_2$};
        
        \node (321) at ({-6*cos(45)},{5*sin(45)}) {$A_5+A_1$};
        
        \node (2221) at ({6*cos(45)},{5*sin(45)}){$D_5(a_1)+A_1$};

        \node (1232) at (0,{5*sin(45)}){$A_5'$};
        
        \node (2211) at ({-3},9){$D_6(a_2)$};

        \node (22111) at (3,9){$E_6(a_3)$};
        
        \node (21111) at (0,15){$E_7(a_5)$};

        \node (1) at (0,18){$A_6$};
        
        \draw[-] (243) -- (321) node[pos=0.5, midway, left]{$A_1$};

        \draw[-] (321) -- (2211) node[pos=0.5, midway, left]{$m_1$};

        \draw[-] (2211) -- (21111) node[pos=0.5, midway, left]{$A_1$};

        \draw[-] (21111) -- (1) node[pos=0.5,midway, left]{$G_2$};

        \draw[-] (243) -- (1232) node[pos=0.5,midway, right]{$A_1$};

        \draw[-] (243) -- (2221) node[pos=0.5,midway, right]{$A_1$};

        \draw[-] (1232) -- (2211) node[pos=0.5,midway, left]{$A_1$};

        \draw[-] (1232) -- (22111) node[pos=0.5,midway, right]{$A_1$};

        \draw[-] (2221) -- (2211) node[pos=0.5,midway, left]{$A_1$};

        \draw[-] (2221) -- (22111) node[pos=0.5,midway, right]{$A_1$};

        \draw[-] (22111) -- (21111) node[pos=0.5,midway, right]{$A_1$};

         \draw[-] (2221) --(2211) node[pos=0.5,midway,left]{$A_1$};
        \draw[-] (1232) -- (2211) node[pos=0.5,midway,left]{$A_1$};
    
    \end{tikzpicture}}
    \caption{}
    \label{fig_HasseE72b}
    \end{subfigure}
    \caption{(\subref{fig_HasseE72a}): The Higgs branch Hasse diagram of $\mathcal Q_{\ref{eq_E72}}$. (\subref{fig_HasseE72b}): The Hasse diagram of $\mathcal S^{E_7}_{A_6,A_4+A_2}$ in the $E_7$ nilcone with orbits specified by their Bala-Carter labels. This is the same Hasse diagram as in \Figref{fig_HasseE72a}.}
    \label{fig_HasseE72}
\end{figure}
 
\subsection{A Family of Quivers with Multiple Adjoint Loops}

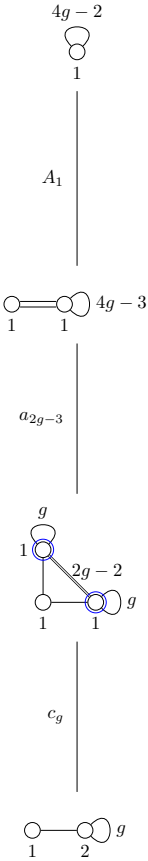
\begin{figure}
    \centering
    \scalebox{.7}{
    \begin{tikzpicture}
        \node (1a) at (0,0){$\begin{tikzpicture}
            \node[gauge, label=below:$1$] (1) at (0,0) {};
            \node[gauge, label=below:$2$] (2) at (1,0){};
            \draw (1) to (2)to[out=-45,in=45,looseness=8]node[pos=0.5,right]{$g$}(2);
            \end{tikzpicture}$};
        \node (2a) at (0,5){$\begin{tikzpicture}
            \node[gauge, label=below:$1$] (1) at (0,0) {};
            \node[gauge, label=below:$1$] (2) at (1,0){};
            \node[gauge, label=left:$1$] (3) at (0,1){};
            \draw (1) to (2)to[out=-45,in=45,looseness=8]node[pos=0.5,right]{$g$}(2);
            \draw (1) to (3) to[out=45,in=135,looseness=8] node[pos=0.5,above]{$g$} (3);
            \draw[double] (2)--(3)node[pos=0.6,right]{$2g-2$};
            \draw[blue] (2) circle (0.2cm);
            \draw[blue] (3) circle (0.2cm);
            \end{tikzpicture}$};
        \node (3a) at (0,10){$\begin{tikzpicture}
            \node (1) [gauge, label=below:$1$] at (0,0) {};
            \node (2) [gauge, label=below:$1$] at (1,0) {};
            \draw[-] (2) to[out=45,in=315,looseness=8] node[pos=0.5,right]{$4g-3$}(2);
            \draw[transform canvas={xshift=0pt, yshift=1.5pt}] (1)--(2);
            \draw[transform canvas={xshift=0pt, yshift=-1.5pt}] (1)--(2);
            \end{tikzpicture}$};
        \node (4a) at (0,15){$\begin{tikzpicture}
            \node (1) [gauge, label=below:$1$] at (0,0) {};
            \draw[-] (1) to[out=45,in=135,looseness=8] node[pos=0.5,above]{$4g-2$} (1);
            \end{tikzpicture}$};
        \draw[] (1a) -- (2a) node[pos=0.5, label=left:$c_g$]{} -- (3a) node[pos=0.5, label=left:$a^+_{2g-3}$]{} -- (4a) node[pos=0.5,label=left:$A_1$]{};
    \end{tikzpicture}
    }
    \caption{The Higgs branch Hasse diagram of the one-parameter family of quivers $\mathcal Q_{\ref{eq_CnFam}}$.}
    \label{fig_12g}
\end{figure}

The one-parameter family of theories labelled by the number of adjoint hypermultiplet loops $g$ given in $\mathcal Q_{\ref{eq_CnFam}}$ also arise in the nilcones of simple Lie algebras. 
\begin{equation}
   \mathcal Q_{\ref{eq_CnFam}}= \raisebox{-.4\height}{\begin{tikzpicture}
        \node[gauge, label=below:$1$] (1) at (0,0) {};
        \node[gauge, label=below:$2$] (2) at (1,0){};
        \draw (1) to (2)to[out=-45,in=45,looseness=8]node[pos=0.5,right]{$g$}(2);
    \end{tikzpicture}}\label{eq_CnFam}
\end{equation}
The Higgs branch Hasse diagram for $\mathcal Q_{\ref{eq_CnFam}}$ is shown in \Figref{fig_12g}. The Namikawa Weyl group for the Higgs branch is $\mathbb Z_2$ in all cases. It is currently not known how to compute the Coulomb branch Hasse diagram for quivers with multiple loops.

Nevertheless, Hilbert series computations may still be performed to identify the moduli spaces of vacua. The identification may be supported additionally with the use of the Lusztig-Spaltenstein duality for slices in classical and exceptional nilcones, as has been extensively used in the previous examples. In particular slices in $\sprm(k)$ nilcones are Spaltenstein dual to slices in $\sorm(2k+1)$ nilcones.

However, some slices in the nilcones of $\sprm(k)$ algebras enjoy a more exotic duality to other slices in the same $\sprm(k)$ nilcone, under the metaplectic Lusztig-Spaltenstein duality \cite{Moeglin2002,2014arXiv1412.8742J,2020arXiv201016089B}. This type of duality also has a realisation in string theory from Type IIB brane systems with two $\tilde{O}^+$ orientifold planes \cite{Hanany:2001iy}.
\paragraph{Sp to SO duality.}
In particular, for $g\geq 2$ the Higgs branch is a S{\l}odowy slice in the $C_{g+1}$ nilcone 
\begin{equation}
    \mathcal H\left(\mathcal Q_{\ref{eq_CnFam}}\right)=\mathrm{Norm}\left[\mathcal S^{C_{g+1}}_{\left(3^2,1^{2g-4}\right),\left(2,1^{2g}\right)}\right]
\end{equation} 
where `$\mathrm{Norm}$' denotes the normalisation of the singularity.

In fact, for $g\geq 4$ the nilpotent orbit closure $\overline{\mathcal O}^{C_{g+1}}_{(3^2,1^{2g-4})}$ is non-normal \cite{Hanany:2016gbz} as is the S{\l}odowy intersection from this orbit to the $\overline{\mathcal O}^{C_{g+1}}_{(2,1^{2g})}$.
It is convenient to convert the Hilbert series for these one parameter family of moduli spaces to a Highest Weight Generating funciton (HWG) \cite{Hanany:2014dia} which is given as
\begin{equation}
    \hwg\left[\mathcal S^{C_{g+1}}_{\left(3^2,1^{2g-4}\right),\left(2,1^{2g}\right)}\right]=\begin{cases}
    \pe[\mu_1^2t^2+\mu_1t^3+\mu_2^2t^4+\mu_2t^4+\mu_1\mu_2t^5-\mu_1^2\mu_2^2t^{10}]&g=2\\\pe[\mu_1^2t^2+\mu_1t^3+\mu_3t^3+\mu_2^2t^4+\mu_2t^4+\mu_1\mu_2t^5-\mu_1^2\mu_2^2t^{10}]&g=3\\\begin{aligned}&(1 + \mu_1 \mu_2 t^5 + \mu_1 \mu_3 t^6 + \mu_2 \mu_3 t^7 + \mu_1 \mu_2 \mu_3 t^8 - 
 \mu_1 \mu_2 \mu_3 t^{10})\\&\times\pe[\mu_1^2t^2+\mu_1t^3+\mu_2^2t^4+\mu_3^2t^6+\mu_2t^4]\end{aligned}&g\geq4
    \end{cases},
\end{equation}
for the S{\l}odowy intersection and also for its normalisation \begin{equation}
    \hwg\left[\mathrm{Norm}\left[\mathcal S^{C_{g+1}}_{\left(3^2,1^{2g-4}\right),\left(2,1^{2g}\right)}\right]\right]=\begin{cases} \pe[\mu_1^2t^2+\mu_1t^3+\mu_2^2t^4+\mu_2t^4+\mu_1\mu_2t^5-\mu_1^2\mu_2^2t^{10}],&g=2.\\\pe[\mu_1^2t^2+\mu_1t^3+\mu_3t^3+\mu_2^2t^4+\mu_2t^4+\mu_1\mu_2t^5-\mu_1^2\mu_2^2t^{10}],&g\geq 3.
    \end{cases}
\end{equation}

The $\mu_i$ are highest weight fugacities for $\sprm(g)$.

Lusztig-Spaltenstein duality predicts that the Coulomb branch is a slice in $B_{g+1}$. Indeed, $\mathcal C\left(\mathcal Q_{\ref{eq_CnFam}}\right)=\mathcal S^{B_{g+1}}_{\left(2g+1,1^2\right),\left(2g-1,2^2\right)}$. The HWG is \begin{equation}
    \hwg\left[\mathcal S^{B_{g+1}}_{\left(2g+1,1^2\right),\left(2g-1,2^2\right)}\right]=\pe[\mu^2t^2+t^4+\mu t^{2g-1}+\mu t^{2g+1}-\mu^2t^{4g+2}],\quad g\geq 2,
\end{equation}where $\mu$ is a highest weight fugacity for $\surm(2)\simeq \sprm(1)$.

Note the collapse of the partitions for the transposition for the duality. The above results were computed for up to $g=5$.

\paragraph{Sp to Sp duality.}
\label{sec_CnFamSp2Sp}
The same moduli spaces may be thought of in a different way which brings to light a metaplectic Luszztig-Spaltenstein duality between the Coulomb and Higgs branches. The Coulomb branch of $\mathcal Q_{\ref{eq_CnFam}}$ may also be identified as the following slice in the nilcone of $C_g$ \cite{Finkelberg:2020cb},
\begin{equation}\mathcal C\left(\mathcal Q_{\ref{eq_CnFam}}\right)=\mathcal S^{C_{g}}_{\left(2g\right),\left(2g-2,1^2\right)}.\end{equation}

The Higgs branch may be also be identified as the double cover of the following nilpotent orbit closure in the nilcone of $C_g$, 
\begin{equation}\mathcal H\left(\mathcal Q_{\ref{eq_CnFam}}\right)/S_2=\overline{\mathcal O}^{C_{g}}_{(4,1^{2g-4})}.\end{equation}

This duality agrees with the metaplectic Lusztig-Spaltenstein duality defined in \cite{2020arXiv201016089B}. The double cover relation can be checked by comparing the volumes of the moduli spaces through the Hilbert series, \begin{equation}\Vol\left[ \mathcal H\left(\mathcal Q_{\ref{eq_CnFam}}\right) \right]/\Vol\left[ \overline{\mathcal O}^{C_{g}}_{(4,1^{2g-4})} \right]=2,\end{equation} with explicit examples for $g=2,3,4$ shown in Table \ref{tab_spspdual}.

The HWG for $\overline{O}^{C_g}_{(4,1^{2g-4})}$ is \begin{equation}
\hwg\left[\overline{O}^{C_g}_{(4,1^{2g-4})}\right]=\begin{cases}\pe\left[\mu_1^2t^2+\mu_2^2t^4+\mu_1^2t^6+\mu_2t^4+\mu_1^2\mu_2t^8-\mu_1^4\mu_2^2t^{16}\right]&g=2\\\left(1+\mu_1\mu_3t^6+\mu_1^2\mu_2t^8+\mu_1\mu_2\mu_3t^8\right)\pe\left[\mu_1^2t^2+\mu_2^2t^4+\mu_2t^4+\mu_1^2t^6+\mu_3^2t^6\right]&g\geq 3\end{cases}.
\end{equation}

The $S_2$ action may be seen at the level of the HWG by the following assignment of $S_2$ irreps to the terms in the HWG of $\mathcal H\left(\mathcal Q_{\ref{eq_CnFam}}\right)$ then performing a group average \cite{Bourget:2020bxh},
\begin{equation}
    \hwg\left[\mathcal H\left(\mathcal Q_{\ref{eq_CnFam}}\right)/S_2\right]=\begin{cases} \pe[\mu_1^2t^2+\epsilon\mu_1t^3+\mu_2^2t^4+\mu_2t^4+\epsilon\mu_1\mu_2t^5-\mu_1^2\mu_2^2t^{10}],&g=2.\\\pe[\mu_1^2t^2+\epsilon\mu_1t^3+\epsilon\mu_3t^3+\mu_2^2t^4+\mu_2t^4+\epsilon\mu_1\mu_2t^5-\mu_1^2\mu_2^2t^{10}],&g\geq 3,
    \end{cases}
\end{equation}where $\epsilon$ denotes the signed representation of $S_2$ and the trivial representation is not explicitly written.

\begin{table}[H]
\centering \ra{1.5}
\begin{tabular}{ccc}
\toprule
 Quiver/Label & HS & Volume \\ \midrule 
 \raisebox{-.4 \height}{\begin{tikzpicture}
        \node[gauge, label=below:$1$] (1) at (0,0) {};
        \node[gauge, label=below:$2$] (2) at (1,0){};
        \draw (1) to (2)to[out=-45,in=45,looseness=8]node[pos=0.5,right]{$2$}(2);
    \end{tikzpicture}} & $\frac{(1 + t^2) (1 + t^2 + 4 t^3 + t^4 + t^6)}{(1 - t^2)^8}$ & $\frac{1}{16}$ \\
    $\overline{\mathcal O}^{C_{2}}_{(4)}$ & $\frac{(1 + t^2) (1 + t^2 + t^4 + t^6)}{(1 - t^2)^8}$ & $\frac{1}{32}$ \\ 
    \midrule
        \raisebox{-.4 \height}{\begin{tikzpicture}
        \node[gauge, label=below:$1$] (1) at (0,0) {};
        \node[gauge, label=below:$2$] (2) at (1,0){};
        \draw (1) to (2)to[out=-45,in=45,looseness=8]node[pos=0.5,right]{$3$}(2);
    \end{tikzpicture}} & $\frac{(1 + t^2) (1 + 6 t^2 + 20 t^3 + 21 t^4 + 36 t^5 + 56 t^6 + 36 t^7 + 21 t^8 + 20 t^9 + 6 t^{10} + t^{12})}{(1 - t^2)^{14} }$ & $\frac{7}{256}$ \\
     $\overline{\mathcal O}^{C_{3}}_{(4,1^{2})}$ & $\frac{(1 + t^2) (1 + 6 t^2 + 21 t^4 + 56 t^6 + 21 t^8 + 6 t^{10} + t^{12})}{(1 - t^2)^{14}}$ & $\frac{7}{512}$ \\ 
     \midrule
        \raisebox{-.4 \height}{\begin{tikzpicture}
        \node[gauge, label=below:$1$] (1) at (0,0) {};
        \node[gauge, label=below:$2$] (2) at (1,0){};
        \draw (1) to (2)to[out=-45,in=45,looseness=8]node[pos=0.5,right]{$4$}(2);
    \end{tikzpicture}} & $\frac{(1 + t^2) (1 + 15 t^2 + 56 t^3 + 120 t^4 + 336 t^5 + 680 t^6 + 960 t^7 + 
 1296 t^8 + 1520 t^9 + \; \cdots \; + t^{18})}{(1 - t^2)^{20}}$ & $\frac{33}{2048}$ \\
    $\overline{\mathcal O}^{C_{4}}_{(4,1^{4})}$ & $\frac{(1 + t^2) (1 + 15 t^2 + 120 t^4 + 680 t^6 + 1296 t^8 + 1296 t^{10} + 680 t^{12} +  120 t^{14} + 15 t^{16} + t^{18})}{(1 - t^2)^{20}}$ & $\frac{33}{4096}$ \\ \bottomrule
    \end{tabular}
\caption{The Hilbert series and volume of the Higgs branch of $\mathcal Q_{\ref{eq_CnFam}}$ and the corresponding orbit closure for $g=2,3,4$. The Hilbert series of the orbit closures are firstly computed in \cite{Hanany:2016gbz}. The $\cdots$ denotes a palindromic series.}
\label{tab_spspdual}
\end{table}

Note that in this family, the Coulomb branch and Higgs branch have a different number of leaves.
\subsection{Example: $E_8$ with Multiloops}
\begin{figure}[h]
\centering
        \begin{tikzpicture}
                \node (a) at (0,0){$E_8(b_6)$};
                \node (b) at (0,3){$D_7(a_1)$};
                \node (c) at (0,6){$D_8(a_6)$};

                 \draw[-] (a)--(b) node[pos=0.5, left]{$\mu$}--(c) node[pos=0.5,left]{$C_2$};

        \end{tikzpicture}
        \caption{The Hasse diagram of $\mathcal S^{E_8}_{E_8(a_6),E_8(b_6)}$ in the $E_8$ nilcone with orbits specified by their Bala-Carter labels. Where $\mu$ is a non-normal slice with normalisation $A_3$ as explained in \cite{Generic_singularities,Hanany_2023uzn}.}
        \label{fig_E8_multiloopsCoul}
\end{figure}
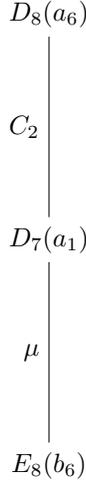

The final example arising as a slice in a nilcone is that of $\mathcal Q_{\ref{eq_E8_multiloops}}$.
\begin{equation}
    \mathcal Q_{\ref{eq_E8_multiloops}}=\raisebox{-.4\height}{\begin{tikzpicture}
        \node[gauge, label=below:$3$] (1) at (0,0){};

        \draw[-] (1) to[out=135,in=45,looseness=8] node[pos=0.5,above]{2} (1);
    \end{tikzpicture}}
    \label{eq_E8_multiloops}
\end{equation}
As first shown in \cite{Hanany:2023uzn}, the Coulomb branch of $\mathcal Q_{\ref{eq_E8_multiloops}}$ is $a_2/S_4$. The Coulomb branch Hilbert series is computed as 
\begin{equation}
\hs\left[\mathcal C\left(\mathcal Q_{\ref{eq_E8_multiloops}}\right)\right]=\frac{1+2t^6+2t^8+t^{14}}{(1-t^4)^3(1-t^6)}.
\end{equation}
This moduli space is particularly interesting as it has been shown in \cite{2023arXiv230807398F} that $a_2/S_4$ appears as a slice in the nilcone of $E_8$, specifically as $\mathcal S^{E_8}_{E_8(a_6),E_8(b_6)}$. The Hasse diagram for this slice is given in \Figref{fig_E8_multiloopsCoul}. It is currently not possible to make a comparison with the Coulomb branch Hasse diagram of $\mathcal Q_{\ref{eq_E8_multiloops}}$ since it is not known how to compute the Coulomb branch Hasse diagram for quivers with multiple adjoint hypermultiplet loops.

Since the Coulomb branch of $\mathcal Q_{\ref{eq_E8_multiloops}}$ appears as a slice in $E_8$ one should expect that the Higgs branch is related to the Spaltenstein dual slice in the nilcone of $E_8$ up to some factor of $\mathbb Z_2$, which may be acting as a quotient or as a cover on this slice. This factor of $\mathbb Z_2$ arises due to $\mathcal Q_{\ref{eq_E8_multiloops}}$ having two adjoint hypermultiplet loops, which is the same as seen for $\mathcal Q_{\ref{eq_CnFam}}$ studied in Section \ref{sec_CnFamSp2Sp}.

In the example at hand, the Higgs branch is the double cover of the S{\l}odowy intersection $\mathcal S^{E_8}_{D_4(a_1) + A_2,2A_2 + 2A_1}$, verified via the refined Hilbert series. For brevity, only the unrefined Hilbert series are presented 
\begin{align}
\hs\left[\mathcal H\left(\mathcal Q_{\ref{eq_E8_multiloops}}\right)\right]&=\frac{\left(\begin{aligned}1 &+ t^2 + 13 t^3 + 20 t^4 + 29 t^5 + 96 t^6 + 160 t^7 + 188 t^8 + 
   304 t^9 + 420 t^{10} + 388 t^{11} + 388 t^{12} \\&+ 420 t^{13} + 304 t^{14} + 
   188 t^{15} + 160 t^{16} + 96 t^{17} + 29 t^{18} + 20 t^{19} + 13 t^{20} + 
   t^{21} + t^{23}\end{aligned}\right)}{(1 - t^2)^9 (1 - t^3)^7}\\
   \hs\left[\mathcal S^{E_8}_{E_8(a_6),E_8(b_6)}\right]&=\frac{\left(\begin{aligned}1 &+ 5 t^2 + 13 t^3 + 
   25 t^4 + 61 t^5 + 131 t^6 + 229 t^7 + 413 t^8 + 655 t^9 + 
   973 t^{10} + 1381 t^{11} + 1785 t^{12} \\&+ 2157 t^{13} + 2490 t^{14} + 
   2641 t^{15} + 2641 t^{16} + 2490 t^{17} + 2157 t^{18} + 1785 t^{19} + 
   1381 t^{20} \\&+ 973 t^{21} + 655 t^{22} + 413 t^{23} + 229 t^{24} + 131 t^{25} + 
   61 t^{26} + 25 t^{27} + 13 t^{28} + 5 t^{29} + t^{31}\end{aligned}\right)}{(1 - t^2)^5  (1 - t^3)^7 (1 - t^4)^4},
\end{align}from which it is clear to see that \begin{equation}\frac{\Vol\left[\mathcal H\left(\mathcal Q_{\ref{eq_E8_multiloops}}\right)\right]}{\Vol\left[\mathcal S^{E_8}_{E_8(a_6),E_8(b_6)}\right]}=\lim_{t\rightarrow1}\frac{\hs\left[\mathcal H\left(\mathcal Q_{\ref{eq_E8_multiloops}}\right)\right]}{\hs\left[\mathcal \mathcal S^{E_8}_{E_8(a_6),E_8(b_6)}\right]}=2,\end{equation} indicating that $\mathcal H\left(\mathcal Q_{\ref{eq_E8_multiloops}}\right)$ is the double cover of $\mathcal S^{E_8}_{D_4(a_1) + A_2,2A_2 + 2A_1}$.

The Higgs branch Hasse diagram is computed as \Figref{fig_E8_multiloopsa}.
As predicted by Lusztig-Spaltenstein duality, the Coulomb branch is the S{\l}odowy intersection $\mathcal S^{E_8}_{E_8(a_6),E_8(b_6)}$, with corresponding Hasse diagram given in \Figref{fig_E8_multiloopsb}. The Higgs branch Namikawa-Weyl group is trivial as there is no codimension 1 slice. The top slice of \Figref{fig_E8_multiloopsa} is $a_3=d_3$ and under a $\mathbb Z_2$ quotient this top slice breaks into two, $A_1$ then $b_2$ as seen in \Figref{fig_E8_multiloopsb}. This phenomenon has been observed in the study of $6d\;\mathcal N=(1,0)$ Higgs branches using magnetic quivers, in particular those theories that have an F-theory description involving the collapse of a $(-2)$-curve \cite{Hanany:2018vph,Hanany:2018cgo,Cabrera:2019izd,Cabrera:2019dob,Hanany:2022itc}.
\newcommand{\threetwo}{
\begin{tikzpicture}
\node (1) [gauge, label=below:$3$] at (0,0) {};
\draw[-] (1) to[out=45,in=135,looseness=8] node[pos=0.5,above]{$2$} (1);
\end{tikzpicture}
}

\newcommand{\onetwo}{
\begin{tikzpicture}
\node (1) [gauge, label=below:$1$] at (0,0) {};
\node (2) [gauge, label=below:$2$] at (1,0) {};
\draw[transform canvas={xshift=0pt, yshift=1.5pt}] (1)--(2);
\draw[transform canvas={xshift=0pt, yshift=-1.5pt}] (1)--(2);
\draw[-] (1) to[out=225,in=135,looseness=8] node[pos=0.5,left]{$2$} (1);
\draw[-] (2) to[out=45,in=315,looseness=8] node[pos=0.5,right]{$2$} (2);
\draw[blue] (1) circle (0.2cm);
\draw[blue] (2) circle (0.2cm);
\end{tikzpicture}
}

\newcommand{\oneoneone}{
\begin{tikzpicture}
\node (1) [gauge, label=below:$1$] at (0,0) {};
\node (2) [gauge, label=below:$1$] at (1,0) {};
\node (3) [gauge, label=left:$1$] at (0.5,0.866) {};
\draw[transform canvas={xshift=0pt, yshift=1.5pt}] (1)--(2);
\draw[transform canvas={xshift=0pt, yshift=-1.5pt}] (1)--(2);
\draw[transform canvas={xshift=1.5*0.866pt, yshift=-1.5*0.5pt}] (1)--(3);
\draw[transform canvas={xshift=-1.5*0.866pt, yshift=1.5*0.5pt}] (1)--(3);
\draw[transform canvas={xshift=1.5*0.866pt, yshift=1.5*0.5pt}] (3)--(2);
\draw[transform canvas={xshift=-1.5*0.866pt, yshift=-1.5*0.5pt}] (3)--(2);
\draw[-] (1) to[out=225,in=135,looseness=8] node[pos=0.5,left]{$2$} (1);
\draw[-] (2) to[out=45,in=315,looseness=8] node[pos=0.5,right]{$2$} (2);
\draw[-] (3) to[out=45,in=135,looseness=8] node[pos=0.5,above]{$2$} (3);
\draw[blue] (1) circle (0.2cm);
\draw[blue] (2) circle (0.2cm);
\draw[blue] (3) circle (0.2cm);
\end{tikzpicture}
}

\newcommand{\oneone}{
\begin{tikzpicture}
\node (1) [gauge, label=below:$1$] at (0,0) {};
\node (2) [gauge, label=below:$1$] at (1,0) {};
\draw[transform canvas={xshift=0pt, yshift=1pt}] (1)--(2);
\draw[transform canvas={xshift=0pt, yshift=3pt}] (1)--(2);
\draw[transform canvas={xshift=0pt, yshift=-1pt}] (1)--(2);
\draw[transform canvas={xshift=0pt, yshift=-3pt}] (1)--(2);
\draw[-] (1) to[out=225,in=135,looseness=8] node[pos=0.5,left]{$5$} (1);
\draw[-] (2) to[out=45,in=315,looseness=8] node[pos=0.5,right]{$2$} (2);
\draw[blue] (1) circle (0.2cm);
\draw[blue] (2) circle (0.2cm);
\end{tikzpicture}
}

\newcommand{\one}{
\begin{tikzpicture}
\node (1) [gauge, label=below:$1$] at (0,0) {};
\draw[-] (1) to[out=45,in=135,looseness=8] node[pos=0.5,above]{$10$} (1);
\draw[blue] (1) circle (0.2cm);
\end{tikzpicture}
}

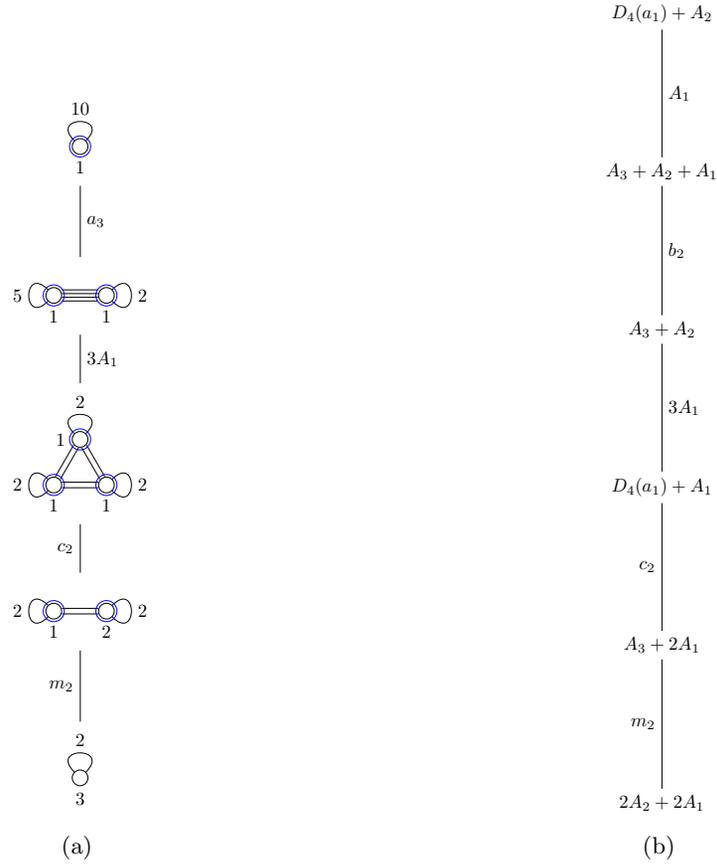
\begin{figure}[H]
    \centering
    \begin{subfigure}{0.49\textwidth}
        \centering
        \scalebox{.7}{
    \begin{tikzpicture}
        \node (a) at (0,0) {\threetwo};
        \node (b) at (0,3) {\onetwo};
        \node (c) at (0,6) {\oneoneone};
        \node (d) at (0,9) {\oneone};
        \node (e) at (0,12) {\one};
        \draw[-] (a)--(b) node[pos=0.5, left]{$m_2$};
        \draw[-] (b)--(c) node[pos=0.5, left]{$c_2$};
        \draw[-] (c)--(d) node[pos=0.5, right]{$3A_1$};
        \draw[-] (d)--(e) node[pos=0.5, right]{$a_3$};
        \end{tikzpicture}}
        \caption{}
        \label{fig_E8_multiloopsa}
    \end{subfigure}
    \begin{subfigure}{0.49\textwidth}
    \centering
    \scalebox{0.7}{
    \begin{tikzpicture}
         \node (a) at (0,0) {$2A_2+2A_1$};
        \node (b) at (0,3) {$A_3+2A_1$};
        \node (c) at (0,6) {$D_4(a_1)+A_1$};
        \node (d) at (0,9) {$A_3+A_2$};
        \node (e) at (0,12) {$A_3+A_2+A_1$};
        \node (f) at (0,15) {$D_4(a_1)+A_2$};
        \draw[-] (a)--(b) node[pos=0.5, left]{$m_2$}--(c) node[pos=0.5, left]{$c_2$}--(d) node[pos=0.5, right]{$3A_1$}--(e) node[pos=0.5, right]{$b_2$}--(f) node[pos=0.5, right]{$A_1$};
    \end{tikzpicture}}
        \caption{}
        \label{fig_E8_multiloopsb}
    \end{subfigure}
    \caption{(\subref{fig_E8_multiloopsa}): Higgs branch Hasse diagram of \eqref{eq_E8_multiloops}, which equals the double cover of $\mathcal S^{E_8}_{D_4(a_1) + A_2,2A_2 + 2A_1}$. (\subref{fig_E8_multiloopsb}): Hasse diagram of  $\mathcal S^{E_8}_{D_4(a_1) + A_2,2A_2 + 2A_1}$ with orbits specified by their Bala-Carter labels.}
\end{figure}

\begin{table}[H]
\centering \ra{1.5}
\scalebox{.85}{
\begin{tabular}{cccc}
\toprule
 Quiver & Coulomb Branch &Higgs Branch & Checked with HS? \\ \midrule 
 \multirow{4}{*}{\raisebox{-.5 \height}{\begin{tikzpicture}
        \node[gauge, label=below:$1$] (0) at (-1,0) {};
        \node[gauge, label=below:$2$] (1) at (0,0) {};
        \node[gauge, label=below:$3$] (2) at (1,0){};
        \draw (0)--(1) to (2)to[out=-45,in=45,looseness=8]node[pos=0.5,right]{}(2);
    \end{tikzpicture}}} &$\overline{sub. reg. G_2}$ & $\mathcal S^{G_2}_{\mathcal N,A_1}$& \checkmark \\
    &$\mathcal S^{F_4}_{F_4(a_3),\tilde A_2}$ & $\mathcal S^{F_4}_{B_2,\tilde{A}_2+A_1}$& \checkmark \\
    &$\mathcal S^{E_6}_{D_4(a_1),2A_2}$ & $\mathcal S^{E_6}_{D_4,2A_2+A_1}$& \checkmark\\
    &$\mathcal S^{E_7}_{D_4(a_1),A_2+3A_1}$ & $\mathcal S^{E_7}_{A_6,A_5+A_1}$& \checkmark\\
    &$\mathcal S^{E_7}_{D_4(a_1),2A_2}$ & $\mathcal S^{E_7}_{D_5+A_1,A_5+A_1}$& \checkmark\\
    &$\mathcal S^{E_7}_{E_7(a_5),A_5''}$ & $\mathcal S^{E_7}_{D_4,2A_2+A_1}$& \checkmark\\
    &$\mathcal S^{E_8}_{E_8(b_5),E_6}$ & $\mathcal S^{E_8}_{D_4,2A_2+A_1}$& \\
     &  &  \\ 
    \midrule
        \raisebox{-.5 \height}{\begin{tikzpicture}
        \node[gauge, label=below:$1$] (0) at (-1,0) {};
        \node[gauge, label=below:$2$] (1) at (0,0) {};
        \node[gauge, label=below:$4$] (2) at (1,0){};
        \draw (0)--(1) to (2)to[out=-45,in=45,looseness=8]node[pos=0.5,right]{}(2);
    \end{tikzpicture}} & $\mathcal S^{F_4}_{F_4(a_3),A_1+\tilde{A}_1}$& $\mathcal S^{F_4}_{F_4(a_2),A_2+\tilde A_1}$ & \checkmark \\
      & &  \\ 
    \midrule
        \raisebox{-.5 \height}{\begin{tikzpicture}
        \node[gauge, label=below:$1$] (0) at (-1,0) {};
        \node[gauge, label=below:$2$] (1) at (0,0) {};
        \node[gauge, label=below:$5$] (2) at (1,0){};
        \draw (0)--(1) to (2)to[out=-45,in=45,looseness=8]node[pos=0.5,right]{}(2);
    \end{tikzpicture}} & $\mathcal S^{E_8}_{E_8(a_7),A_4+A_2}$& $\mathcal S^{E_8}_{D_5+A_2,A_4+A_3}$ &  \\
      & &  \\ 
     \midrule
        \raisebox{-.5 \height}{\begin{tikzpicture}
        \node[gauge, label=below:$2$] (1) at (0,0) {};
        \node[gauge, label=below:$4$] (2) at (1,0){};
        \draw (1) to (2)to[out=-45,in=45,looseness=8]node[pos=0.5,right]{}(2);
    \end{tikzpicture}} & $\mathcal S^{F_4}_{F_4(a_3),A_2}$& $\mathcal S^{F_4}_{B_3,A_2+\tilde A_1}$ & \checkmark \\
     &  & \\ 
    \midrule
        \raisebox{-.5 \height}{\begin{tikzpicture}
        \node[gauge, label=below:$2$] (1) at (0,0) {};
        \node[gauge, label=below:$5$] (2) at (1,0){};
        \draw (1) to (2)to[out=-45,in=45,looseness=8]node[pos=0.5,right]{}(2);
    \end{tikzpicture}} & $\mathcal S^{E_8}_{E_8(a_7),A_4+A_2+A_1}$ & $\mathcal S^{E_8}_{A_6+A_1,A_4+A_3}$ &  \\
         &  & \\ 
    \midrule
        \raisebox{-.5\height}{\begin{tikzpicture}
         \node[gauge, label=below:$1$] (1) at (0,0){};
            \node[gauge, label=below:$2$] (2) at (1,0){};
            \node[gauge, label=below:$3$] (3) at (2,0){};
            \node[gauge, label=below:$2$] (2r) at (3,0){};

            \draw[-] (1)--(2)--(3);
            \draw[transform canvas={yshift=1.5pt}] (2r)--(3);
            \draw[transform canvas={yshift=-1.5pt}] (2r)--(3);
    \end{tikzpicture}} & $\overline{\mathcal O}_{(3^2,1)}^{B_3}$& $\mathcal S^{C_3}_{\mathcal N,(2^3)}$ & \checkmark \\
         &  & \\ 
        \midrule
       \raisebox{-.4\height}{\begin{tikzpicture}
        \node[gauge, label=below:$1$] (1) at (0,0){};
        \node[gauge, label=below:$2$] (2) at (1,0){};
        \node[gauge, label=below:$3$] (3) at (2,0) {};
        \node[gauge, label=below:$1$] (4) at (3,0) {};

        \draw[-] (1)--(2)--(3)--(4);
        \draw[-] (3) to[out=45,in=135,looseness=8] (3);
    \end{tikzpicture}} & $\mathcal S^{E_6}_{D_4(a_1),A_2+A_1}$ & $\mathcal S^{E_6}_{D_5(a_1),2A_2+A_1}$ & \checkmark \\
     &  & \\ 
     \midrule
        \raisebox{-.5\height}{\begin{tikzpicture}
        \node[gauge, label=below:$1$] (1) at (0,0){};
        \node[gauge, label=below:$2$] (2) at (1,0){};
        \node[gauge, label=below:$4$] (3) at (2,0) {};
        \node[gauge, label=below:$3$] (4) at (3,0) {};

        \draw[-] (1)--(2)--(3);
        \draw[transform canvas={yshift=1.5pt}] (3)--(4);
        \draw[transform canvas={yshift=-1.5pt}] (3)--(4);
    \end{tikzpicture}} & $\mathcal S^{E_7}_{A_3+A_2+A_1, A_2+2A_1}$& $\mathcal S^{E_7}_{E_7(a_4),A_4+A_2}$ & \checkmark \\
         &  & \\ 
         \midrule
        \raisebox{-.5\height}{\begin{tikzpicture}
        \node[gauge, label=below:$2$] (2) at (1,0){};
        \node[gauge, label=below:$4$] (3) at (2,0) {};
        \node[gauge, label=below:$3$] (4) at (3,0) {};

        \draw[-] (2)--(3);
        \draw[transform canvas={yshift=1.5pt}] (3)--(4);
        \draw[transform canvas={yshift=-1.5pt}] (3)--(4);
    \end{tikzpicture}} & $\mathcal S^{E_7}_{A_3+A_2+A_1,A_2+3A_1}$& $\mathcal S^{E_7}_{A_6,A_4+A_2}$ & \checkmark \\
         &  & \\ 
        \midrule
        \raisebox{-.4\height}{\begin{tikzpicture}
        \node[gauge, label=below:$1$] (1) at (0,0) {};
        \node[gauge, label=below:$2$] (2) at (1,0){};
        \draw (1) to (2)to[out=-45,in=45,looseness=8]node[pos=0.5,right]{$g$}(2);
    \end{tikzpicture}} & $\mathrm{Norm}\left[\mathcal S^{C_{g+1}}_{\left(3^2,1^{2g-4}\right),\left(2,1^{2g}\right)}\right]$& $\mathcal S^{B_{g+1}}_{\left(2g+1,1^2\right),\left(2g-1,2^2\right)}$ & \checkmark \\
    &$\mathcal S^{C_{g}}_{\left(2g\right),\left(2g-2,1^2\right)}$ &$S_2$ cover of $\overline{\mathcal O}^{C_{g}}_{(4,1^{2g-4})}$ & \checkmark \\
         &  & \\ 
    \midrule
        \raisebox{-.4\height}{\begin{tikzpicture}
        \node[gauge, label=below:$3$] (1) at (0,0){};

        \draw[-] (1) to[out=135,in=45,looseness=8] node[pos=0.5,above]{2} (1);
    \end{tikzpicture}} & $\mathcal S^{E_8}_{E_8(a_6),E_8(b_6)}$ & $S_2$ cover of $\mathcal S^{E_8}_{D_4(a_1) + A_2,2A_2 + 2A_1}$ & \checkmark \\
    \bottomrule
\end{tabular}}
\caption{Summary of unitary quivers studied here whose moduli spaces of vacua are slices in nilcones. The slices in $E_8$ remains unchecked due to the computational complexity of the Hilbert series.}
\label{}
\end{table}
\section{Outlook}
\label{sec_outlook}
The algorithm presented in this work successfully determines the Higgs branch Hasse diagram for a wide variety of unitary 3d $\mathcal{N}=4$ quiver gauge theories, and the techniques admit extensions to various families of more exotic quivers. For instance, it would be interesting to augment the algorithm in this paper such that it includes non-simply-laced unitary theories, alongside developing a similar algorithm for orthosymplectic quivers. Given the successes of the Coulomb and Higgs branch subtraction algorithms used to determine the moduli space for theories with 8 supercharges, it may be time to turn to theories with less supersymmetry where the moduli spaces of vacua are less constrained and much more difficult to study. Further work might even consider an algorithmic description of the Higgs branch Hasse diagram for 4d $\mathcal{N}=1$ quivers. Other extensions include a subtraction algorithm for the Higgs branch of a theory with non-general values of FI deformations.
\\\par A slightly different point of view may consider the implications of the monodromies present in this work to theories in six-dimensions; for instance, it would be interesting to consider whether such global data can be seen from F-theory constructions and whether there is a similar global structure around the tensor branch. 
\\\par Slightly more in keeping with the focus of this paper, it would be interesting to sharpen current understanding of the effect of changing $\urm(n)$ gauge nodes to $\surm(n)$ on the Higgs branch. Previous work on this matter includes \cite{Bourget:2021jwo}, wherein the $\surm(n)$ nodes open up a $d_k$ transition.
\\\par

The question of whether a slice can be defined globally also remains unsettled. As already mentioned, a neighbourhood of any point in a given symplectic singularity decomposes as a product of a neighbourhood of the leaf through the point and a neighbourhood of the origin of the slice. In some cases the slice is not just defined in a neighbourhood, but as a subset of the whole symplectic singularity. S{\l}odowy intersections and slices in affine Grassmanians are such examples. Globally a slice within the S{\l}odowy intersections can be defined as the intersection of a transverse slice and an orbit. In the case of affine Grassmanian slices, a slice in the Hasse diagram is defined again as the intersection of some (infinite dimensional) transverse space and some orbit. These are two examples where a slice can be globally defined. It thus remains a question as to the conditions under which such a global definition is available for an arbitrary symplectic singularity.
\\\par The decoration on Higgs and Coulomb branch can be used to generalise the notion of ``special" and ``non-special" to the symplectic singularity outside of nilcone and predict the symplectic duality of the slices. This will be discussed in the future papers.
\\\par Further, the field theory description of the decorated quiver remains unexplored.

\paragraph{Acknowledgements.}
We gratefully acknowledge discussions with Michael Finkelberg, Rudolph Kalveks, Paul Levy, Hiraku Nakajima, and  Jasper van de Kreeke. In particular, We thank Gwyn Bellamy for the examples he provided in UNIST Workshop on Symplectic Singularities 2023, which inspired this work, and for the valuable discussions through email. We thank Travis Schedler for pointing out the three rules introduced in the present paper are equivalent to the three types of minimal imaginary roots that will appear in \cite{Travis_toappear}. We thank Ben Webster for his early insights into the similarity between the Higgsing process and the Coulomb branch algorithm. The work of SB, AH, GK, CL, and DL is partially supported by STFC Consolidated Grants ST/T000791/1 and ST/X000575/1. The work of SB is supported by the STFC DTP research studentship grant ST/Y509231/1. The work of GK is supported by STFC DTP research studentship grant ST/X508433/1. The work of MS is supported by Austrian Science Fund (FWF), START project STA 73-N. MS also acknowledges support from the Faculty of Physics, University of Vienna.

\appendix
\section{Higgs Branch Subtraction Cheat Sheet}
\label{sec:cheat_sheet}
It merits repeating that the local rules, introduced in Section~\ref{sec_RHBQS}, gives partial information regarding the slice associated to a given subtraction. This is of course a result of a monodromy and is a global effect. In section \ref{sec_NWG}, the question of monodromy (which on the quiver is drawn as a decoration) was addressed at length, alongside a prescription for applying decorations to quivers on the Higgs branch Hasse diagram.

The steps below give a quick summary of the the Higgs branch subtraction algorithm. Generic examples may contain further complexities (the likes of which are described in the main text) and hence this cheat sheet is only intended to indicate the standard approach.
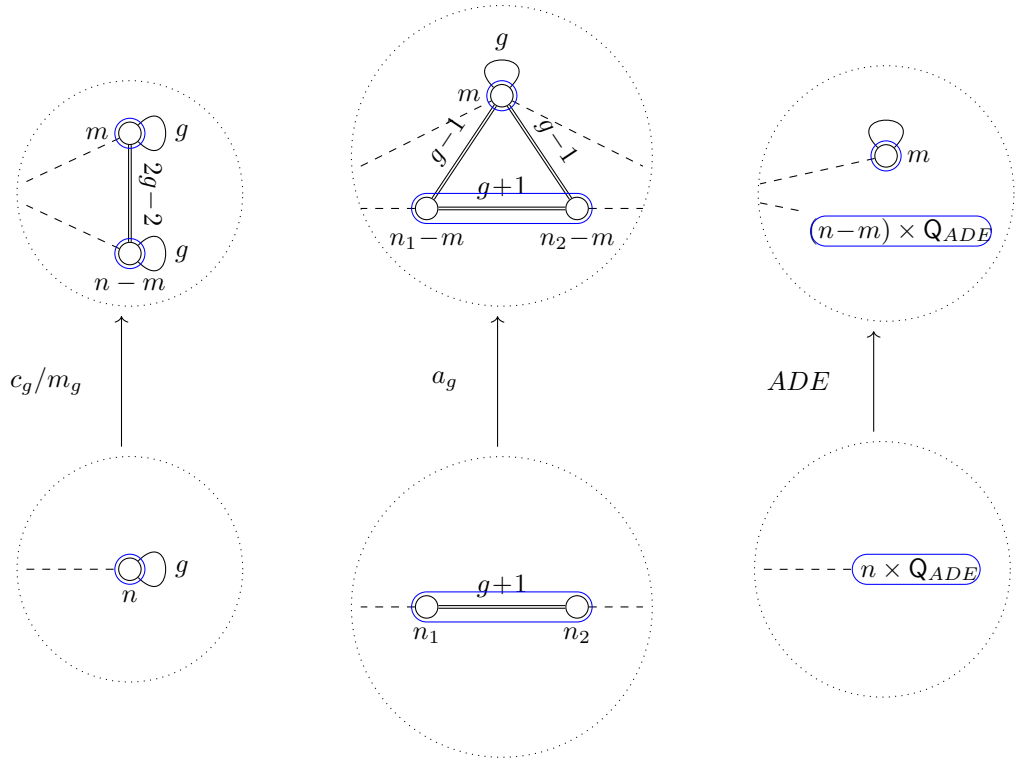
\begin{figure}[H]
    \centering
    \begin{tikzpicture}
        \node (aa) at (-5,0) {      \begin{tikzpicture}
        \node (a) at (0,0) {$\raisebox{-.5 \height}{ \begin{tikzpicture}
            \node (1u) at (-0.5,0.1) {};
            \node (1d) at (-0.5,-0.1) {};
            \draw[dotted] (1,0) circle (1.5cm);
            \node[gauge,label=below:{$n-m$}] (2) at (1,-0.8) {};
            \node[gauge,label=left:{$m$}] (3) at (1,0.8) {};
            \draw (2) to [in=45,out=-45,looseness=8]node[pos=0.5,right]{$g$} (2);
            \draw (3) to [in=45,out=-45,looseness=8]node[pos=0.5,right]{$g$} (3);
            \draw[double] (3) to node[pos=0.5,above,sloped]{$2g\!-\!2$} (2);
            \draw[dashed] (3)--(1u) (2)--(1d);
            \draw[blue] (2) circle (0.2cm);
            \draw[blue] (3) circle (0.2cm);
        \end{tikzpicture}}$};
        \node at (-1,-2.5) {$c_g/m_g$};
        \node (b) at (0,-5) {$\raisebox{-.5 \height}{ \begin{tikzpicture}
            \node (1) at (-0.5,0) {};
            \draw[dotted] (1,0) circle (1.5cm);
            \node[gauge,label=below:{$n$}] (2) at (1,0) {};
            \draw (2) to [in=45,out=-45,looseness=8]node[pos=0.5,right]{$g$} (2);
            \draw[dashed] (1)--(2);
            \draw[blue] (2) circle (0.2cm);
        \end{tikzpicture}}$};
        \draw[->] (b)--(a);
    \end{tikzpicture}  };
        \node (bb) at (0,0) {  \begin{tikzpicture}
        \node (a) at (0,0) {$\raisebox{-.5 \height}{ \begin{tikzpicture}
            \node (1l) at (0,0) {};
            \node (1r) at (4,0) {};
            \node (1lu) at (0,0.5) {};
            \node (1ru) at (4,0.5) {};
            \draw[dotted] (2,0.7) circle (2cm);
            \node[gauge,label=below:{$n_1\!-\!m$}] (2) at (1,0) {};
            \node[gauge,label=below:{$n_2\!-\!m$}] (3) at (3,0) {};
            \node[gauge,label=left:{$m$}] (4) at (2,1.5) {};
            \draw[double] (2)--(3)node[pos=0.5,above,sloped]{$g\!+\!1$};
            \draw[double] (2)--(4)node[pos=0.5,above,sloped]{$g\!-\!1$};
            \draw[double] (3)--(4)node[pos=0.5,above,sloped]{$g\!-\!1$};
            \draw (4) to [in=45,out=135,looseness=8]node[pos=0.5,above,sloped]{$g$}(4);
            \draw[dashed] (2)--(1l) (3)--(1r) (1lu)--(4)--(1ru);
            \draw[blue] (4) circle (0.2cm);
            \draw[blue] \convexpath{2,3} {0.2cm};
        \end{tikzpicture}}$};
        \node at (-0.7,-3) {$a_g$};
        \node (b) at (0,-6) {$\raisebox{-.5 \height}{ \begin{tikzpicture}
            \node (1l) at (0,0) {};
            \node (1r) at (4,0) {};
            \draw[dotted] (2,0) circle (2cm);
            \node[gauge,label=below:{$n_1$}] (2) at (1,0) {};
            \node[gauge,label=below:{$n_2$}] (3) at (3,0) {};
            \draw[dashed] (2)--(1l) (3)--(1r);
            \draw[double] (2)--(3)node[pos=0.5,above,sloped]{$g\!+\!1$};
            \draw[blue] \convexpath{2,3} {0.2cm};
        \end{tikzpicture}}$};
        \draw[->] (b)--(a);
\end{tikzpicture} };
        \node (cc) at (5,0) {      \begin{tikzpicture}
        \node (a) at (0,0) {$\raisebox{-.5 \height}{ \begin{tikzpicture}
            \node (1u) at (-2,0.6) {};
            \node (1d) at (-2,0.4) {};
            \draw[dotted] (-0.2,0.5) circle (1.7cm);
            \node (2) at (0,0) {$(n\!-\!m)\times\mathsf{Q}_{ADE}$};
            \node[gauge,label=right:{$m$}] (3) at (-0.2,1) {};
            \draw (3) to [out=45,in=135,looseness=8] (3);
            \draw[dashed] (2)--(1d) (3)--(1u);
            \draw[blue] (3) circle (0.2cm);
            \node (aux1) at (-1,0) {};
            \node (aux2) at (1,0) {};
            \draw[blue] \convexpath{aux1,aux2} {0.2cm};
        \end{tikzpicture}}$};
        \node at (-1,-2.5) {$ADE$};
        \node (b) at (0,-5) {$\raisebox{-.5 \height}{ \begin{tikzpicture}
            \node (1) at (-0.2,0) {};
            \draw[dotted] (1.5,0) circle (1.7cm);
            \node (2) at (2,0) {$n\times\mathsf{Q}_{ADE}$};
            \draw[dashed] (1)--(2);
            \node (aux1) at (1.3,0) {};
            \node (aux2) at (2.6,0) {};
            \draw[blue] \convexpath{aux1,aux2} {0.2cm};
        \end{tikzpicture}}$};
        \draw[->] (b)--(a);
    \end{tikzpicture}  };
    \end{tikzpicture}
    \caption{A reproduction of \Figref{fig_decorationinhert}. Recall that the decorations on the bottom row of quivers, although formally trivial, are applied before the subtraction step such that the subtraction products ``inherit'' the decoration from the initial nodes.}
    \label{fig:cheat_sheet}
\end{figure}
\begin{enumerate}
    \item Scan the quiver for a sub-quiver that admits a Higgsing via Rules \hyperref[fig_ag]{1}, \hyperref[fig_ag]{2} or \hyperref[fig_ADE]{3}. If such a sub-quiver exists, and is not already entirely decorated, apply a decoration.
    \item Using \Figref{fig:cheat_sheet}, apply the appropriate rule to the decorated sub-quiver, whose subtraction products will inherit the decoration as shown.
\end{enumerate}
The process described above is the most consistent way to think about the Higgs branch subtraction process. Of course, given the identification of \Figref{fig_decorationcreation}, the decorations on the bottom row of \Figref{fig:cheat_sheet} can strictly be omitted --- their inclusion serves only as a reminder that decorations are considered to be ``inherited'' under subtraction. 
\section{One Example to Rule Them All}
\label{appendix_A}
The example in \Figref{fig_allrules} uses all the features of the subtraction algorithm introduced in this paper, including Rules \hyperref[fig_cgmg]{1}, \hyperref[fig_ag]{2}, and \hyperref[fig_ADE]{3}, multiple loops,  decoration of nodes, merging of decorations, unions of cones, and non-trivial monodromy. The example is included to indicate the level of complexity that often arises in the course of Higgs branch quiver subtraction. The following will go through the transitions step-by-step. Note that since the algorithm proceeds from the bottom up, `the $n$th row' counts from the bottom.
\begin{itemize}
\item In the first step, either the $\urm(3)$ node or the $\urm(2)$ node can be split using \hyperref[fig_cgmg]{Rule 1}, giving $m_1$ and $c_1$ transitions respectively. The associated monodromies are tracked via the introduction of red and blue decorations.\\
\item For the quiver with blue decoration in the second row, the only transition that is permitted is to break the node of rank 3 into two nodes of ranks 1 and 2 under \hyperref[fig_cgmg]{Rule 1}. Since the node of rank 3 is initially undecorated, a red decoration is introduced on the products, as seen in the third row.
\item The quiver with a red decoration on the second row sees two transitions. The $c_1$ transition simply comes from splitting the decorated node of rank 2 into two nodes of rank 1 under \hyperref[fig_cgmg]{Rule 1}. Since the node undergoing the splitting is initially decorated, no new decoration is imposed. The $A_1$ transition arises from the subtraction of a subdiagram of affine $A_1$ under \hyperref[fig_ag]{Rule 2} or \hyperref[fig_ADE]{Rule 3}. 
\\
\item The third row contains three quivers. The quiver with green decoration admits only one transition, which under \hyperref[fig_cgmg]{Rule 1} splits the node of rank 2 into two nodes of rank 1 (inheriting both the decoration and loops). The quiver in the middle of the third row similarly admits a transition under \hyperref[fig_cgmg]{Rule 1}, splitting the node of rank 2 into two nodes of rank 1. In this case, since the node of rank 2 is initially undecorated, a new decoration is picked up. The quiver in the third row with both red and blue decorations has two possible transitions, the first of which (under \hyperref[fig_cgmg]{Rule 1}) has the same result as the transition just mentioned. The other possible transition uses \hyperref[fig_ADE]{Rule 3} and the blue and red decorations merge to form a green decoration.\\
\item The leftmost quiver in the fourth row has six possible subtractions of an affine $A_1$ quiver with one node decorated red and the other blue under \hyperref[fig_ADE]{Rule 3}. This multiplicity is reflected in the slice $6A_1$. The rightmost quiver in the fourth row has two transitions, the first of which (giving a slice $c_1$) comes from the application of \hyperref[fig_ag]{Rule 2} to the node of rank 2. The other transition results from applying \hyperref[fig_ADE]{Rule 3} to affine $A_1$ sub-quiver spanned by the nodes with blue and green decorations.\\
\item The fifth row contains two quivers, the leftmost of which sees three \hyperref[fig_ADE]{Rule 3} transitions. The rightmost quiver undergoes a transition under \hyperref[fig_cgmg]{Rule 1} to split the node of rank 2 into two nodes of rank 1, each inheriting the original red decoration.
\item The sixth row contains three quivers, the leftmost of which sees the \hyperref[fig_ADE]{Rule 3} transitions $2A_1$ and $a_{3}^{+}$. The former arises with multiplicity 2 since the two green nodes are decorated under the same decoration, while the latter is a subtraction of the affine $A_3$ quiver with loops defined by the green decorated nodes. The central quiver has three \hyperref[fig_ADE]{Rule 3} transitions, with the $a_3$ slice arising from an affine $A_3$ subtraction along the two rank 1 nodes with four bifundamental hypers between them. The rightmost quiver has only one $a_3$ transition of multiplicity 2, arising from an affine $A_3$ subtraction on each of the two decorated nodes.
\item The seventh row contains two quivers, each of which see a single $a_5$ transition under \hyperref[fig_ADE]{Rule 3}.
\item The eighth row contains a single quiver which sees one $a_3$ transition under \hyperref[fig_ADE]{Rule 3} on account of the four hypermultiplets in the bifundamental representation of the two rank 1 gauge nodes.

\end{itemize}
\newcommand{\threetwotwo}{
\begin{tikzpicture}
\node (1) [gauge, label=below:$3$] at (0,0) {};
\node (2) [gauge, label=below:$2$] at (1,0) {};
\draw[-] (1) to[out=225,in=135,looseness=8] (1);
\draw[-] (2) to[out=45,in=315,looseness=8] (2);
\draw[transform canvas={xshift=0pt, yshift=1.5pt}] (1)--(2);
\draw[transform canvas={xshift=0pt, yshift=-1.5pt}] (1)--(2);
\end{tikzpicture}
}

\newcommand{\twotwoone}{
\begin{tikzpicture}
\node (1) [gauge, label=below:$2$] at (0,0) {};
\node (2) [gauge, label=below:$2$] at (1,0) {};
\node (3) [gauge, label=right:$1$] at (1,1) {};
\draw[-] (1) to[out=225,in=135,looseness=8] (1);
\draw[-] (2) to[out=45,in=315,looseness=8] (2);
\draw[-] (3) to[out=45,in=135,looseness=8] (3);
\draw[transform canvas={xshift=0pt, yshift=1.5pt}] (1)--(2);
\draw[transform canvas={xshift=0pt, yshift=-1.5pt}] (1)--(2);
\draw[transform canvas={xshift=1.5pt, yshift=0pt}] (3)--(2);
\draw[transform canvas={xshift=-1.5pt, yshift=0pt}] (3)--(2);
\draw[red] (1) circle (0.2cm);
\draw[red] (3) circle (0.2cm);
\end{tikzpicture}
}

\newcommand{\threeoneone}{
\begin{tikzpicture}
\node (1) [gauge, label=below:$3$] at (0,0) {};
\node (2) [gauge, label=below:$1$] at (1,0) {};
\node (3) [gauge, label=right:$1$] at (0,1) {};
\draw[-] (1) to[out=225,in=135,looseness=8] (1);
\draw[-] (2) to[out=45,in=315,looseness=8] (2);
\draw[-] (3) to[out=45,in=135,looseness=8] (3);
\draw[transform canvas={xshift=0pt, yshift=1.5pt}] (1)--(2);
\draw[transform canvas={xshift=0pt, yshift=-1.5pt}] (1)--(2);
\draw[transform canvas={xshift=1.5pt, yshift=0pt}] (3)--(1);
\draw[transform canvas={xshift=-1.5pt, yshift=0pt}] (3)--(1);
\draw[blue] (2) circle (0.2cm);
\draw[blue] (3) circle (0.2cm);
\end{tikzpicture}
}

\newcommand{\onetwoloop}{
\begin{tikzpicture}
\node (1) [gauge, label=below:$1$] at (0,0) {};
\node (2) [gauge, label=below:$2$] at (1,0) {};
\draw[-] (2) to[out=45,in=315,looseness=8] node[pos=0.5,right]{$3$} (2);
\draw[-] (1) to[out=225,in=135,looseness=8] (1);
\draw[transform canvas={xshift=0pt, yshift=1.5pt}] (1)--(2);
\draw[transform canvas={xshift=0pt, yshift=-1.5pt}] (1)--(2);
\draw[greed] (2) circle (0.2cm);
\end{tikzpicture}
}

\newcommand{\twooneoneone}{
\begin{tikzpicture}
\node (1) [gauge, label=below:$1$] at (0,0) {};
\node (2) [gauge, label=above right:$2$] at (1,0) {};
\node (3) [gauge, label=below:$1$] at (2,0) {};
\node (4) [gauge, label=right:$1$] at (1,1) {};
\draw[-] (1) to[out=225,in=135,looseness=8] (1);
\draw[-] (2) to[out=225,in=315,looseness=8] (2);
\draw[-] (3) to[out=45,in=315,looseness=8] (3);
\draw[-] (4) to[out=45,in=135,looseness=8] (4);
\draw[transform canvas={xshift=0pt, yshift=1.5pt}] (1)--(2);
\draw[transform canvas={xshift=0pt, yshift=-1.5pt}] (1)--(2);
\draw[transform canvas={xshift=0pt, yshift=1.5pt}] (3)--(2);
\draw[transform canvas={xshift=0pt, yshift=-1.5pt}] (3)--(2);
\draw[transform canvas={xshift=1.5pt, yshift=0pt}] (4)--(2);
\draw[transform canvas={xshift=-1.5pt, yshift=0pt}] (4)--(2);
\draw[red] (1) circle (0.2cm);
\draw[red] (3) circle (0.2cm);
\draw[red] (4) circle (0.2cm);

\node (2l) at (0.9,0) {};
\node (2r) at (1.1,0) {};
\end{tikzpicture}
}

\newcommand{\twothreeones}{
\begin{tikzpicture}
\node (1) [gauge, label=below:$2$] at (0,0) {};
\node (2) [gauge, label=below:$1$] at (1,0) {};
\node (3) [gauge, label=above:$1$] at (0,1) {};
\node (4) [gauge, label=above:$1$] at (1,1) {};
\draw[-] (1) to[out=225,in=135,looseness=8] (1);
\draw[-] (2) to[out=45,in=315,looseness=8] (2);
\draw[-] (3) to[out=225,in=135,looseness=8] (3);
\draw[-] (4) to[out=45,in=315,looseness=8] (4);
\draw[transform canvas={xshift=0pt, yshift=1.5pt}] (1)--(2);
\draw[transform canvas={xshift=0pt, yshift=-1.5pt}] (1)--(2);
\draw[transform canvas={xshift=0pt, yshift=1.5pt}] (3)--(4);
\draw[transform canvas={xshift=0pt, yshift=-1.5pt}] (3)--(4);
\draw[transform canvas={xshift=1.5pt, yshift=0pt}] (1)--(3);
\draw[transform canvas={xshift=-1.5pt, yshift=0pt}] (1)--(3);
\draw[transform canvas={xshift=1.5pt, yshift=0pt}] (4)--(2);
\draw[transform canvas={xshift=-1.5pt, yshift=0pt}] (4)--(2);
\draw[red] (1) circle (0.2cm);
\draw[blue] (2) circle (0.2cm);
\draw[blue] (3) circle (0.2cm);
\draw[red] (4) circle (0.2cm);
\end{tikzpicture}
}

\newcommand{\fiveones}{
\begin{tikzpicture}
\node (1) [gauge, label=below:$1$] at (0,0) {};
\node (2) [gauge, label=below:$1$] at (1,0) {};
\node (3) [gauge, label=below:$1$] at (2,0) {};
\node (4) [gauge, label=right:$1$] at (1,1) {};
\node (5) [gauge, label=right:$1$] at (1,-1) {};
\draw[-] (1) to[out=225,in=135,looseness=8] (1);
\draw[-] (2) to[out=45,in=135,looseness=8] (2);
\draw[-] (3) to[out=45,in=315,looseness=8] (3);
\draw[-] (4) to[out=45,in=135,looseness=8] (4);
\draw[-] (5) to[out=225,in=315,looseness=8] (5);
\draw[transform canvas={xshift=0pt, yshift=1.5pt}] (1)--(2);
\draw[transform canvas={xshift=0pt, yshift=-1.5pt}] (1)--(2);
\draw[transform canvas={xshift=0pt, yshift=1.5pt}] (3)--(2);
\draw[transform canvas={xshift=0pt, yshift=-1.5pt}] (3)--(2);
\draw[transform canvas={xshift=-1.5*0.707pt, yshift=1.5*0.707pt}] (1)--(4);
\draw[transform canvas={xshift=1.5*0.707pt, yshift=-1.5*0.707pt}] (1)--(4);
\draw[transform canvas={xshift=1.5*0.707pt, yshift=1.5*0.707pt}] (4)--(3);
\draw[transform canvas={xshift=-1.5*0.707pt, yshift=-1.5*0.707pt}] (4)--(3);
\draw[transform canvas={xshift=1.5*0.707pt, yshift=1.5*0.707pt}] (5)--(1);
\draw[transform canvas={xshift=-1.5*0.707pt, yshift=-1.5*0.707pt}] (5)--(1);
\draw[transform canvas={xshift=-1.5*0.707pt, yshift=1.5*0.707pt}] (5)--(3);
\draw[transform canvas={xshift=1.5*0.707pt, yshift=-1.5*0.707pt}] (5)--(3);
\draw[blue] (1) circle (0.2cm);
\draw[red] (2) circle (0.2cm);
\draw[blue] (3) circle (0.2cm);
\draw[red] (4) circle (0.2cm);
\draw[red] (5) circle (0.2cm);
\end{tikzpicture}
}

\newcommand{\twooneone}{
\begin{tikzpicture}
\node (1) [gauge, label=below:$2$] at (0,0) {};
\node (2) [gauge, label=below:$1$] at (1,0) {};
\node (3) [gauge, label=right:$1$] at (0,1) {};
\draw[-] (1) to[out=225,in=135,looseness=8] (1);
\draw[-] (2) to[out=45,in=315,looseness=8] (2);
\draw[-] (3) to[out=45,in=135,looseness=8] node[pos=0.5,above]{$3$} (3);
\draw[transform canvas={xshift=0pt, yshift=1.5pt}] (1)--(2);
\draw[transform canvas={xshift=0pt, yshift=-1.5pt}] (1)--(2);
\draw[transform canvas={xshift=1.5pt, yshift=0pt}] (3)--(1);
\draw[transform canvas={xshift=-1.5pt, yshift=0pt}] (3)--(1);
\draw[transform canvas={xshift=1.5*0.707pt, yshift=1.5*0.707pt}] (2)--(3);
\draw[transform canvas={xshift=-1.5*0.707pt, yshift=-1.5*0.707pt}] (2)--(3);
\draw[red] (1) circle (0.2cm);
\draw[blue] (2) circle (0.2cm);
\draw[greed] (3) circle (0.2cm);
\end{tikzpicture}
}

\newcommand{\oneoneoneone}{
\begin{tikzpicture}
\node (1) [gauge, label=below:$1$] at (0,0) {};
\node (2) [gauge, label=below:$1$] at (2,0) {};
\node (3) [gauge, label=right:$1$] at (1,1.732/3) {};
\node (4) [gauge, label=right:$1$] at (1,-1.732/3) {};
\draw[-] (1) to[out=225,in=135,looseness=8] (1);
\draw[-] (2) to[out=45,in=315,looseness=8] (2);
\draw[-] (3) to[out=45,in=135,looseness=8] (3);
\draw[-] (4) to[out=225,in=315,looseness=8] node[pos=0.5,below]{$3$} (4);
\draw[transform canvas={xshift=1.5pt, yshift=0pt}] (3)--(4);
\draw[transform canvas={xshift=-1.5pt, yshift=0pt}] (3)--(4);
\draw[transform canvas={xshift=1.5*0.5pt, yshift=-1.5*0.866pt}] (3)--(1);
\draw[transform canvas={xshift=-1.5*0.5pt, yshift=1.5*0.866pt}] (3)--(1);
\draw[transform canvas={xshift=1.5*0.5pt, yshift=-1.5*0.866pt}] (2)--(4);
\draw[transform canvas={xshift=-1.5*0.5pt, yshift=1.5*0.866pt}] (2)--(4);
\draw[transform canvas={xshift=1.5*0.5pt, yshift=1.5*0.866pt}] (3)--(2);
\draw[transform canvas={xshift=-1.5*0.5pt, yshift=-1.5*0.866pt}] (3)--(2);
\draw[transform canvas={xshift=1.5*0.5pt, yshift=1.5*0.866pt}] (1)--(4);
\draw[transform canvas={xshift=-1.5*0.5pt, yshift=-1.5*0.866pt}] (1)--(4);
\draw[red] (1) circle (0.2cm);
\draw[red] (2) circle (0.2cm);
\draw[blue] (3) circle (0.2cm);
\draw[greed] (4) circle (0.2cm);
\end{tikzpicture}
}

\newcommand{\twoonewithfour}{
\begin{tikzpicture}
\node (1) [gauge, label=below:$2$] at (0,0) {};
\node (2) [gauge, label=below:$1$] at (1,0) {};
\draw[-] (1) to[out=225,in=135,looseness=8] (1);
\draw[-] (2) to[out=45,in=315,looseness=8] node[pos=0.5,right]{$5$} (2);
\draw[double] (2)--(1)node[pos=0.5,above]{$4$};
\draw[red] (1) circle (0.2cm);
\end{tikzpicture}
}

\newcommand{\oneoneonewithfour}{
\begin{tikzpicture}
\node (1) [gauge, label=below:$1$] at (0,0) {};
\node (2) [gauge, label=above:$1$] at (1,1.732/3) {};
\node (3) [gauge, label=below:$1$] at (1,-1.732/3) {};
\draw[-] (1) to[out=135,in=225,looseness=8] (1);
\draw[-] (2) to[out=45,in=315,looseness=8] node[pos=0.5,right] {$3$} (2);
\draw[-] (3) to[out=45,in=315,looseness=8] node[pos=0.5,right] {$3$} (3);
\draw[double] (2)--(3)node[pos=0.5,right]{$4$};
\draw[transform canvas={xshift=1.5*0.5pt, yshift=-1.5*0.866pt}] (2)--(1);
\draw[transform canvas={xshift=-1.5*0.5pt, yshift=1.5*0.866pt}] (2)--(1);
\draw[transform canvas={xshift=1.5*0.5pt, yshift=1.5*0.866pt}] (3)--(1);
\draw[transform canvas={xshift=-1.5*0.5pt, yshift=-1.5*0.866pt}] (3)--(1);
\draw[greed] (2) circle (0.2cm);
\draw[greed] (3) circle (0.2cm);
\end{tikzpicture}
}

\newcommand{\oneoneonewithfiveloops}{
\begin{tikzpicture}
\node (1) [gauge, label=below:$1$] at (0,0) {};
\node (2) [gauge, label=above:$1$] at (1,1.732/3) {};
\node (3) [gauge, label=below:$1$] at (1,-1.732/3) {};
\draw[-] (1) to[out=135,in=225,looseness=8] (1);
\draw[-] (2) to[out=45,in=315,looseness=8] (2);
\draw[-] (3) to[out=45,in=315,looseness=8] node[pos=0.5,right] {$5$} (3);
\draw[double] (2)--(3)node[pos=0.5,right]{$4$};
\draw[transform canvas={xshift=1.5*0.5pt, yshift=-1.5*0.866pt}] (2)--(1);
\draw[transform canvas={xshift=-1.5*0.5pt, yshift=1.5*0.866pt}] (2)--(1);
\draw[transform canvas={xshift=1.5*0.5pt, yshift=1.5*0.866pt}] (3)--(1);
\draw[transform canvas={xshift=-1.5*0.5pt, yshift=-1.5*0.866pt}] (3)--(1);
\end{tikzpicture}
}

\newcommand{\twoathree}{
\begin{tikzpicture}
\node (1) [gauge, label=below:$1$] at (0,0) {};
\node (2) [gauge, label=below:$1$] at (1,0) {};
\node (3) [gauge, label=below:$1$] at (2,0) {};
\draw[-] (1) to[out=135,in=225,looseness=8] (1);
\draw[-] (3) to[out=45,in=315,looseness=8] (3);
\draw[-] (2) to[out=45,in=135,looseness=8] node[pos=0.5,above] {$5$} (2);
\draw[double] (2)--(1)node[pos=0.5,above]{$4$};
\draw[double] (2)--(3)node[pos=0.5,above]{$4$};
\draw[red] (1) circle (0.2cm);
\draw[red] (3) circle (0.2cm);
\end{tikzpicture}
}

\newcommand{\oneonesixfivethree}{
\begin{tikzpicture}
\node (1) [gauge, label=below:$1$] at (0,0) {};
\node (2) [gauge, label=below:$1$] at (1,0) {};
\draw[double] (2)--(1)node[pos=0.5,above]{$6$};
\draw[-] (1) to[out=225,in=135,looseness=8] node[pos=0.5,left] {$5$} (1);
\draw[-] (2) to[out=45,in=315,looseness=8] node[pos=0.5,right] {$3$} (2);
\end{tikzpicture}
}

\newcommand{\oneonesixsevenone}{
\begin{tikzpicture}
\node (1) [gauge, label=below:$1$] at (0,0) {};
\node (2) [gauge, label=below:$1$] at (1,0) {};
\draw[double] (2)--(1)node[pos=0.5,below]{$6$};
\draw[-] (1) to[out=225,in=135,looseness=8] node[pos=0.5,left] {$7$} (1);
\draw[-] (2) to[out=45,in=315,looseness=8] (2);
\end{tikzpicture}
}

\newcommand{\oneonefournineone}{
\begin{tikzpicture}
\node (1) [gauge, label=below:$1$] at (0,0) {};
\node (2) [gauge, label=below:$1$] at (1,0) {};
\draw[double] (2)--(1)node[pos=0.5,above]{$4$};
\draw[-] (1) to[out=225,in=135,looseness=8] node[pos=0.5,left] {$9$} (1);
\draw[-] (2) to[out=45,in=315,looseness=8] (2);
\end{tikzpicture}
}

\newcommand{\onewithnineloops}{
\begin{tikzpicture}
\node (1) [gauge, label=below:$1$] at (0,0) {};
\draw[-] (1) to[out=45,in=135,looseness=8] node[pos=0.5,above]{$13$} (1);
\end{tikzpicture}
}

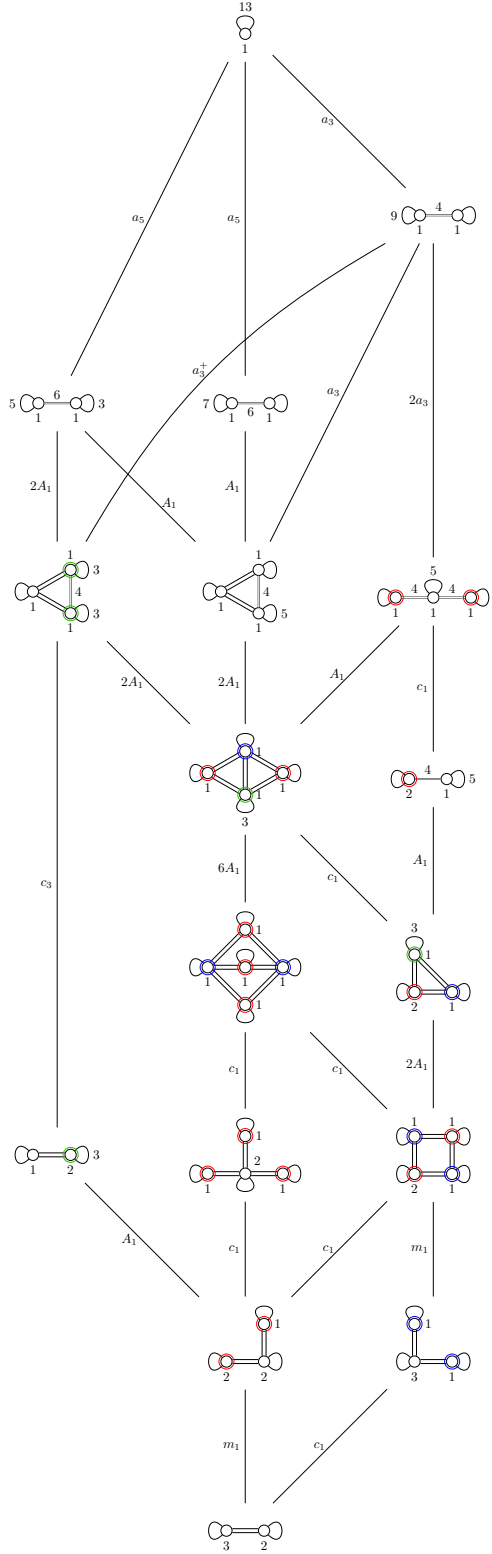
\begin{figure}[H]
    \centering
    \scalebox{.5}{
    \begin{tikzpicture}
        \node (a) at (0,0) {\threetwotwo};
        \node (b) at (0,5) {\twotwoone};
        \node (c) at (5,5) {\threeoneone};
        \node (d) at (-5,10) {\onetwoloop};
        \node (e) at (0,10) {\twooneoneone};
        \node (f) at (5,10) {\twothreeones};
        \node (g) at (0,15) {\fiveones};
        \node (h) at (5,15) {\twooneone};
        \node (i) at (0,20) {\oneoneoneone};
        \node (j) at (5,20) {\twoonewithfour};
        \node (k) at (-5,25) {\oneoneonewithfour};
        \node (l) at (0,25) {\oneoneonewithfiveloops};
        \node (m) at (5,25) {\twoathree};
        \node (n) at (-5,30) {\oneonesixfivethree};
        \node (o) at (0,30) {\oneonesixsevenone};
        \node (p) at (5,35) {\oneonefournineone};
        \node (q) at (0,40) {\onewithnineloops};
        \draw[-] (a)--(b) node[pos=0.5, left]{$m_1$};
        \draw[-] (a)--(c) node[pos=0.5, left]{$c_1$};
        \draw[-] (d)--(b) node[pos=0.5, left]{$A_1$};
        \draw[-] (e)--(b) node[pos=0.5, left]{$c_1$};
        \draw[-] (f)--(b) node[pos=0.5, left]{$c_1$};
        \draw[-] (c)--(f) node[pos=0.5, left]{$m_1$};
        \draw[-] (d)--(k) node[pos=0.5, left]{$c_3$};
        \draw[-] (g)--(f) node[pos=0.5, left]{$c_1$};
        \draw[-] (g)--(e) node[pos=0.5, left]{$c_1$};
        \draw[-] (h)--(f) node[pos=0.5, left]{$2A_1$};
        \draw[-] (g)--(i) node[pos=0.5, left]{$6A_1$};
        \draw[-] (h)--(i) node[pos=0.5, left]{$c_1$};
        \draw[-] (h)--(j) node[pos=0.5, left]{$A_1$};
        \draw[-] (i)--(k) node[pos=0.5, left]{$2A_1$};
        \draw[-] (i)--(l) node[pos=0.5, left]{$2A_1$};
        \draw[-] (i)--(m) node[pos=0.5, left]{$A_1$};
        \draw[-] (j)--(m) node[pos=0.5, left]{$c_1$};
        \draw[-] (k)--(n) node[pos=0.5, left]{$2A_1$};
         \draw[-] (k) to[bend left=15] node[pos=0.5, left]{$a_3^+$} (p);
        \draw[-] (l)--(n) node[pos=0.35, right]{$A_1$};
        \draw[-] (l)--(p) node[pos=0.5, left]{$a_3$};
        \draw[-] (l)--(o) node[pos=0.5, left]{$A_1$};
        \draw[-] (m)--(p) node[pos=0.5, left]{$2a_3$};
        \draw[-] (n)--(q) node[pos=0.5, left]{$a_5$};
        \draw[-] (o)--(q) node[pos=0.5, left]{$a_5$};
        \draw[-] (p)--(q) node[pos=0.5, left]{$a_3$};
    \end{tikzpicture}}
    \caption{The red decoration comes from the $\urm(3)$ node with a loop, the blue decoration comes from the $\urm(2)$ node with a loop, the green decoration comes from the $2$ copies of $\mathsf{Q}_{A_1}$ subquiver, or equivalently, it can be thought of as the result of merging the red and blue decorations. $a_3^+$ is $a_3$ with an $S_2$ monodromy, coming from the two identically decorated nodes.}
    \label{fig_allrules}
\end{figure}

\section{The Non-normal Varieties $m_g$}
\label{sec_mg}
The non-normal variety $m_1$ of quaternionic dimension $1$ is the simplest example of a non-normal variety that is both symplectic and holomorphic \cite{vinberg1972class}. The simplest definition of $m_1$ starts with the ring of functions of $\mathbb C^2$ with the two generators removed. The Hilbert series and Highest Weight Generating function describe this process\begin{align}
    \HS(m_1)&=\frac{1}{(1-a t)(1-t/a)}-(a+1/a)t \ , \\
    \hwg(m_1)&=\frac{1}{1-\mu t}-\mu t=\sum_{1 \neq n\in \mathbb{Z}_{\geq0}} (\mu t)^n=\pe[\mu^2t^2+\mu^3t^3-\mu^6t^6]\ ,
\end{align}
with $a$ and $\mu$ the fugacity and highest weight fugacity respectively of the $\sprm(1)$ symmetry acting on $m_1$ .

The ring of functions of $m_1$ can now be specified, using coordinates $x$ and $y$ of $\mathbb C^2$ as:
\begin{equation}
\mathcal{R}_{m_1}=\C[x^2,xy,y^2,x^3,x^2y,xy^2,y^3],
    \label{eq_ringm}
\end{equation}
with $m_1=\text{Spec}\ \mathcal{R}_{m_1}$.

The variety $m_1$ appears as the fixed locus of $S_k\times S_{n-k}$ in $\C^{2n}/S_n$ when $k\neq n-k$. To see this, take the coordinates of $\C^{2n}$ as $(x_1,y_1,\cdots,x_n,y_n)$ and remove the free $\C^2$ by setting $\sum_ix_i=\sum_iy_i=0$. The invariant subspace fixed by the action of $S_k\times S_{n-k}$ is (up to conjugation),
\begin{equation}
V=(\underbrace{-(n-k) x, -(n-k) y, \cdots, -(n-k) x, -(n-k) y}_{k}, \underbrace{k x, k y, \cdots, k x, k y}_{n-k}).
\end{equation}
The invariant local functions on $V$ under $S_n$ give the ring of functions of $m_1$ i.e, $V/S_n=m_1$. 

The above construction of $m_1$ may be generalised easily to any dimension. Namely taking the ring of functions of $\mathbb C^l,\;l\in \N^+$ and removing the $l$ generators. However, the holomorphic-symplectic property of the variety is preserved if only if $l$ is even. 

The construction of the $g$ quaternionic-dimensional non-normal variety $m_g$ proceeds by taking $l=2g,\;g\in \N^+$. This is, once again, easily seen through the Hilbert series and Highest Weight Generating function \begin{align}
    \HS(m_g)&=\pe\left[[1,0,\cdots,0]_{\sprm(g)}t\right]-[1,0,\cdots,0]_{\sprm(g)}t\ , \\
    \hwg(m_g)&=\pe[\mu_1^2t^2+\mu_1^3t^3-\mu_1^6t^6]\ ,
\end{align}
where the Dynkin label $[1,0,\cdots,0]_{\sprm(g)}$ is shorthand for the character of the fundamental representation of $\sprm(g)$ and $\mu_1$ the highest weight fugacity for $\sprm(g)$.

The coordinate ring is specified as:
\begin{equation}
\mathcal{R}_{m_g}=\C[x_1,\dots,x_g,y_1,\dots,y_g]\backslash \{x_1,\dots,x_g,y_1,\dots,y_g\}.
    \label{eq_ringmg}
\end{equation}
Similarly as in \eqref{eq_ringm}, the ring \eqref{eq_ringmg} is generated by generators at degree $2$ and $3$. There are $g(2g+1)$ generators at degree $2$ and $\frac{2g(g+1)(2g+1)}{3}$ generators at degree $3$. 

\bibliographystyle{JHEP}
\bibliography{bibli.bib}

\providecommand{\href}[2]{#2}\begingroup\raggedright\begin{thebibliography}{10}

\bibitem{englert1964broken}
F.~Englert and R.~Brout, \emph{{Broken Symmetry and the Mass of Gauge Vector Mesons}}, \href{http://dx.doi.org/10.1103/PhysRevLett.13.321}{\emph{Phys. Rev. Lett.} {\bfseries 13} (1964) 321--323}.

\bibitem{higgs1964broken}
P.~W. Higgs, \emph{{Broken Symmetries and the Masses of Gauge Bosons}}, \href{http://dx.doi.org/10.1103/PhysRevLett.13.508}{\emph{Phys. Rev. Lett.} {\bfseries 13} (1964) 508--509}.

\bibitem{guralnik1964global}
G.~S. Guralnik, C.~R. Hagen and T.~W.~B. Kibble, \emph{{Global Conservation Laws and Massless Particles}}, \href{http://dx.doi.org/10.1103/PhysRevLett.13.585}{\emph{Phys. Rev. Lett.} {\bfseries 13} (1964) 585--587}.

\bibitem{kibble1967symmetry}
T.~W.~B. Kibble, \emph{{Symmetry breaking in nonAbelian gauge theories}}, \href{http://dx.doi.org/10.1103/PhysRev.155.1554}{\emph{Phys. Rev.} {\bfseries 155} (1967) 1554--1561}.

\bibitem{Cremonesi:2015lsa}
S.~Cremonesi, G.~Ferlito, A.~Hanany and N.~Mekareeya, \emph{{Instanton Operators and the Higgs Branch at Infinite Coupling}}, \href{http://dx.doi.org/10.1007/JHEP04(2017)042}{\emph{JHEP} {\bfseries 04} (2017) 042}, [\href{https://arxiv.org/abs/1505.06302}{{\ttfamily 1505.06302}}].

\bibitem{Ferlito:2017xdq}
G.~Ferlito, A.~Hanany, N.~Mekareeya and G.~Zafrir, \emph{{3d Coulomb branch and 5d Higgs branch at infinite coupling}}, \href{http://dx.doi.org/10.1007/JHEP07(2018)061}{\emph{JHEP} {\bfseries 07} (2018) 061}, [\href{https://arxiv.org/abs/1712.06604}{{\ttfamily 1712.06604}}].

\bibitem{Cabrera:2018jxt}
S.~Cabrera, A.~Hanany and F.~Yagi, \emph{{Tropical Geometry and Five Dimensional Higgs Branches at Infinite Coupling}}, \href{http://dx.doi.org/10.1007/JHEP01(2019)068}{\emph{JHEP} {\bfseries 01} (2019) 068}, [\href{https://arxiv.org/abs/1810.01379}{{\ttfamily 1810.01379}}].

\bibitem{Cabrera:2019izd}
S.~Cabrera, A.~Hanany and M.~Sperling, \emph{{Magnetic quivers, Higgs branches, and 6d $N$=(1,0) theories}}, \href{http://dx.doi.org/10.1007/JHEP06(2019)071}{\emph{JHEP} {\bfseries 06} (2019) 071}, [\href{https://arxiv.org/abs/1904.12293}{{\ttfamily 1904.12293}}].

\bibitem{Bourget:2019rtl}
A.~Bourget, S.~Cabrera, J.~F. Grimminger, A.~Hanany and Z.~Zhong, \emph{{Brane Webs and Magnetic Quivers for SQCD}}, \href{http://dx.doi.org/10.1007/JHEP03(2020)176}{\emph{JHEP} {\bfseries 03} (2020) 176}, [\href{https://arxiv.org/abs/1909.00667}{{\ttfamily 1909.00667}}].

\bibitem{Cabrera:2019dob}
S.~Cabrera, A.~Hanany and M.~Sperling, \emph{{Magnetic quivers, Higgs branches, and 6d $ \mathcal{N} $ = (1, 0) theories \textemdash{} orthogonal and symplectic gauge groups}}, \href{http://dx.doi.org/10.1007/JHEP02(2020)184}{\emph{JHEP} {\bfseries 02} (2020) 184}, [\href{https://arxiv.org/abs/1912.02773}{{\ttfamily 1912.02773}}].

\bibitem{Aharony:1997bx}
O.~Aharony, A.~Hanany, K.~A. Intriligator, N.~Seiberg and M.~J. Strassler, \emph{{Aspects of N=2 supersymmetric gauge theories in three-dimensions}}, \href{http://dx.doi.org/10.1016/S0550-3213(97)00323-4}{\emph{Nucl. Phys. B} {\bfseries 499} (1997) 67--99}, [\href{https://arxiv.org/abs/hep-th/9703110}{{\ttfamily hep-th/9703110}}].

\bibitem{Borokhov:2002ib}
V.~Borokhov, A.~Kapustin and X.-k. Wu, \emph{{Topological disorder operators in three-dimensional conformal field theory}}, \href{http://dx.doi.org/10.1088/1126-6708/2002/11/049}{\emph{JHEP} {\bfseries 11} (2002) 049}, [\href{https://arxiv.org/abs/hep-th/0206054}{{\ttfamily hep-th/0206054}}].

\bibitem{Borokhov:2002cg}
V.~Borokhov, A.~Kapustin and X.-k. Wu, \emph{{Monopole operators and mirror symmetry in three-dimensions}}, \href{http://dx.doi.org/10.1088/1126-6708/2002/12/044}{\emph{JHEP} {\bfseries 12} (2002) 044}, [\href{https://arxiv.org/abs/hep-th/0207074}{{\ttfamily hep-th/0207074}}].

\bibitem{Gaiotto:2008ak}
D.~Gaiotto and E.~Witten, \emph{{S-Duality of Boundary Conditions In N=4 Super Yang-Mills Theory}}, \href{http://dx.doi.org/10.4310/ATMP.2009.v13.n3.a5}{\emph{Adv. Theor. Math. Phys.} {\bfseries 13} (2009) 721--896}, [\href{https://arxiv.org/abs/0807.3720}{{\ttfamily 0807.3720}}].

\bibitem{Bashkirov:2010hj}
D.~Bashkirov, \emph{{Examples of global symmetry enhancement by monopole operators}},  \href{https://arxiv.org/abs/1009.3477}{{\ttfamily 1009.3477}}.

\bibitem{Cremonesi:2013lqa}
S.~Cremonesi, A.~Hanany and A.~Zaffaroni, \emph{{Monopole operators and Hilbert series of Coulomb branches of $3d$ $\mathcal{N} = 4$ gauge theories}}, \href{http://dx.doi.org/10.1007/JHEP01(2014)005}{\emph{JHEP} {\bfseries 01} (2014) 005}, [\href{https://arxiv.org/abs/1309.2657}{{\ttfamily 1309.2657}}].

\bibitem{Cremonesi:2014kwa}
S.~Cremonesi, A.~Hanany, N.~Mekareeya and A.~Zaffaroni, \emph{{Coulomb branch Hilbert series and Hall-Littlewood polynomials}}, \href{http://dx.doi.org/10.1007/JHEP09(2014)178}{\emph{JHEP} {\bfseries 09} (2014) 178}, [\href{https://arxiv.org/abs/1403.0585}{{\ttfamily 1403.0585}}].

\bibitem{Cremonesi:2014xha}
S.~Cremonesi, G.~Ferlito, A.~Hanany and N.~Mekareeya, \emph{{Coulomb Branch and The Moduli Space of Instantons}}, \href{http://dx.doi.org/10.1007/JHEP12(2014)103}{\emph{JHEP} {\bfseries 12} (2014) 103}, [\href{https://arxiv.org/abs/1408.6835}{{\ttfamily 1408.6835}}].

\bibitem{Cremonesi:2014uva}
S.~Cremonesi, A.~Hanany, N.~Mekareeya and A.~Zaffaroni, \emph{{T$_{\rho}^{\sigma}$ (G) theories and their Hilbert series}}, \href{http://dx.doi.org/10.1007/JHEP01(2015)150}{\emph{JHEP} {\bfseries 01} (2015) 150}, [\href{https://arxiv.org/abs/1410.1548}{{\ttfamily 1410.1548}}].

\bibitem{Bullimore:2015lsa}
M.~Bullimore, T.~Dimofte and D.~Gaiotto, \emph{{The Coulomb Branch of 3d ${\mathcal{N}= 4}$ Theories}}, \href{http://dx.doi.org/10.1007/s00220-017-2903-0}{\emph{Commun. Math. Phys.} {\bfseries 354} (2017) 671--751}, [\href{https://arxiv.org/abs/1503.04817}{{\ttfamily 1503.04817}}].

\bibitem{Nakajima:2015txa}
H.~Nakajima, \emph{{Towards a mathematical definition of Coulomb branches of $3$-dimensional $\mathcal{N}=4$ gauge theories, I}}, \href{http://dx.doi.org/10.4310/ATMP.2016.v20.n3.a4}{\emph{Adv. Theor. Math. Phys.} {\bfseries 20} (2016) 595--669}, [\href{https://arxiv.org/abs/1503.03676}{{\ttfamily 1503.03676}}].

\bibitem{Braverman:2016wma}
A.~Braverman, M.~Finkelberg and H.~Nakajima, \emph{{Towards a mathematical definition of Coulomb branches of $3$-dimensional $\mathcal{N} = 4$ gauge theories, II}}, \href{http://dx.doi.org/10.4310/ATMP.2018.v22.n5.a1}{\emph{Adv. Theor. Math. Phys.} {\bfseries 22} (2018) 1071--1147}, [\href{https://arxiv.org/abs/1601.03586}{{\ttfamily 1601.03586}}].

\bibitem{Cabrera:2018ann}
S.~Cabrera and A.~Hanany, \emph{{Quiver Subtractions}}, \href{http://dx.doi.org/10.1007/JHEP09(2018)008}{\emph{JHEP} {\bfseries 09} (2018) 008}, [\href{https://arxiv.org/abs/1803.11205}{{\ttfamily 1803.11205}}].

\bibitem{Bourget:2019aer}
A.~Bourget, S.~Cabrera, J.~F. Grimminger, A.~Hanany, M.~Sperling, A.~Zajac et~al., \emph{{The Higgs mechanism \textemdash{} Hasse diagrams for symplectic singularities}}, \href{http://dx.doi.org/10.1007/JHEP01(2020)157}{\emph{JHEP} {\bfseries 01} (2020) 157}, [\href{https://arxiv.org/abs/1908.04245}{{\ttfamily 1908.04245}}].

\bibitem{Bourget:2023dkj}
A.~Bourget, M.~Sperling and Z.~Zhong, \emph{{Decay and Fission of Magnetic Quivers}},  \href{https://arxiv.org/abs/2312.05304}{{\ttfamily 2312.05304}}.

\bibitem{Bourget:2024mgn}
A.~Bourget, M.~Sperling and Z.~Zhong, \emph{{Higgs branch RG-flows via Decay and Fission}},  \href{https://arxiv.org/abs/2401.08757}{{\ttfamily 2401.08757}}.

\bibitem{kaledin2006symplectic}
D.~Kaledin, \emph{Symplectic singularities from the poisson point of view}, .

\bibitem{Nakajima:1994nid}
H.~Nakajima, \emph{{Instantons on ALE spaces, quiver varieties, and Kac-Moody algebras}}, \href{http://dx.doi.org/10.1215/S0012-7094-94-07613-8}{\emph{Duke Math. J.} {\bfseries 76} (1994) 365--416}.

\bibitem{Nakajima:1998}
H.~Nakajima, \emph{{Quiver varieties and Kac-Moody algebras}}, \href{http://dx.doi.org/10.1215/S0012-7094-98-09120-7}{\emph{Duke Mathematical Journal} {\bfseries 91} (1998) 515 -- 560}.

\bibitem{Crawley-Boevey1}
W.~Crawley-boevey, \emph{{Geometry of the Moment Map for Representations of Quivers}}, \href{http://dx.doi.org/10.1023/A:1017558904030}{\emph{Compositio Mathematica} {\bfseries 126} (05, 2000) }.

\bibitem{Crawley-Boevey2}
W.~Crawley-Boevey, \emph{Normality of marsden-weinstein reductions for representations of quivers},  \href{https://arxiv.org/abs/math/0105247}{{\ttfamily math/0105247}}.

\bibitem{Travis_toappear}
T.~Schedler and G.~Bellamy, \emph{Minimal degenerations of quiver varieties}, {\emph{work in progress} }.

\bibitem{Gaiotto:2013bwa}
D.~Gaiotto and P.~Koroteev, \emph{{On Three Dimensional Quiver Gauge Theories and Integrability}}, \href{http://dx.doi.org/10.1007/JHEP05(2013)126}{\emph{JHEP} {\bfseries 05} (2013) 126}, [\href{https://arxiv.org/abs/1304.0779}{{\ttfamily 1304.0779}}].

\bibitem{Cabrera:2016vvv}
S.~Cabrera and A.~Hanany, \emph{{Branes and the Kraft-Procesi Transition}}, \href{http://dx.doi.org/10.1007/JHEP11(2016)175}{\emph{JHEP} {\bfseries 11} (2016) 175}, [\href{https://arxiv.org/abs/1609.07798}{{\ttfamily 1609.07798}}].

\bibitem{Gu:2022dac}
J.~Gu, Y.~Jiang and M.~Sperling, \emph{{Rational $Q$-systems, Higgsing and mirror symmetry}}, \href{http://dx.doi.org/10.21468/SciPostPhys.14.3.034}{\emph{SciPost Phys.} {\bfseries 14} (2023) 034}, [\href{https://arxiv.org/abs/2208.10047}{{\ttfamily 2208.10047}}].

\bibitem{Hayashi:2018bkd}
H.~Hayashi, S.-S. Kim, K.~Lee and F.~Yagi, \emph{{5-brane webs for 5d $ \mathcal{N} $ = 1 G$_{2}$ gauge theories}}, \href{http://dx.doi.org/10.1007/JHEP03(2018)125}{\emph{JHEP} {\bfseries 03} (2018) 125}, [\href{https://arxiv.org/abs/1801.03916}{{\ttfamily 1801.03916}}].

\bibitem{Hayashi:2018lyv}
H.~Hayashi, S.-S. Kim, K.~Lee and F.~Yagi, \emph{{Dualities and 5-brane webs for 5d rank 2 SCFTs}}, \href{http://dx.doi.org/10.1007/JHEP12(2018)016}{\emph{JHEP} {\bfseries 12} (2018) 016}, [\href{https://arxiv.org/abs/1806.10569}{{\ttfamily 1806.10569}}].

\bibitem{Generic_singularities}
B.~{Fu}, D.~{Juteau}, P.~{Levy} and E.~{Sommers}, \emph{{Generic singularities of nilpotent orbit closures}}, \href{http://dx.doi.org/10.48550/arXiv.1502.05770}{\emph{arXiv e-prints} (Feb., 2015) arXiv:1502.05770}, [\href{https://arxiv.org/abs/1502.05770}{{\ttfamily 1502.05770}}].

\bibitem{losev2024unipotent}
I.~Losev, L.~Mason-Brown and D.~Matvieievskyi, \emph{Unipotent ideals and harish-chandra bimodules},  2024.

\bibitem{Namikawa_2011}
Y.~Namikawa, \emph{Poisson deformations of affine symplectic varieties}, \href{http://dx.doi.org/10.1215/00127094-2010-066}{\emph{Duke Mathematical Journal} {\bfseries 156} (Jan., 2011) }.

\bibitem{Namikawa_2010}
Y.~Namikawa, \emph{Poisson deformations of affine symplectic varieties, ii}, \href{http://dx.doi.org/10.1215/0023608x-2010-012}{\emph{Kyoto Journal of Mathematics} {\bfseries 50} (Jan., 2010) }.

\bibitem{Englert:1964}
F.~Englert and R.~Brout, \emph{{Broken Symmetry and the Mass of Gauge Vector Mesons}}, \href{http://dx.doi.org/10.1103/PhysRevLett.13.321}{\emph{Phys. Rev. Lett.} {\bfseries 13} (1964) 321--323}.

\bibitem{Higgs:1964}
P.~W. Higgs, \emph{{Broken Symmetries and the Masses of Gauge Bosons}}, \href{http://dx.doi.org/10.1103/PhysRevLett.13.508}{\emph{Phys. Rev. Lett.} {\bfseries 13} (1964) 508--509}.

\bibitem{Guralnik:1964}
G.~S. Guralnik, C.~R. Hagen and T.~W.~B. Kibble, \emph{{Global Conservation Laws and Massless Particles}}, \href{http://dx.doi.org/10.1103/PhysRevLett.13.585}{\emph{Phys. Rev. Lett.} {\bfseries 13} (1964) 585--587}.

\bibitem{Kibble:1967}
T.~W.~B. Kibble, \emph{Symmetry breaking in non-abelian gauge theories}, \href{http://dx.doi.org/10.1103/PhysRev.155.1554}{\emph{Phys. Rev.} {\bfseries 155} (Mar, 1967) 1554--1561}.

\bibitem{Hanany:2023uzn}
A.~Hanany, G.~Kumaran, C.~Li, D.~Liu and M.~Sperling, \emph{{Actions on the quiver -- Discrete quotients on the Coulomb branch}},  \href{https://arxiv.org/abs/2311.02773}{{\ttfamily 2311.02773}}.

\bibitem{Assel:2017jgo}
B.~Assel and S.~Cremonesi, \emph{{The Infrared Physics of Bad Theories}}, \href{http://dx.doi.org/10.21468/SciPostPhys.3.3.024}{\emph{SciPost Phys.} {\bfseries 3} (2017) 024}, [\href{https://arxiv.org/abs/1707.03403}{{\ttfamily 1707.03403}}].

\bibitem{de_Boer_1997}
J.~de~Boer, K.~Hori, H.~Ooguri, Y.~Oz and Z.~Yin, \emph{Mirror symmetry in three-dimensional gauge theories, and d-brane moduli spaces}, \href{http://dx.doi.org/10.1016/s0550-3213(97)00115-6}{\emph{Nuclear Physics B} {\bfseries 493} (may, 1997) 148--176}.

\bibitem{Cremonesi_2014}
S.~Cremonesi, G.~Ferlito, A.~Hanany and N.~Mekareeya, \emph{Coulomb branch and the moduli space of instantons}, \href{http://dx.doi.org/10.1007/jhep12(2014)103}{\emph{Journal of High Energy Physics} {\bfseries 2014} (Dec., 2014) }.

\bibitem{Slodowy1980SimpleSA}
P.~Slodowy, \emph{Simple singularities and simple algebraic groups},  1980.

\bibitem{OALiegroup}
E.~B.~V. Arkadij L.~Onishchik, \emph{Lie groups and algebraic groups},  1990.
\newblock \href{http://dx.doi.org/10.1007/978-3-642-74334-4}{DOI}.

\bibitem{Bourget_2022instanton}
A.~Bourget, J.~F. Grimminger, A.~Hanany and Z.~Zhong, \emph{The hasse diagram of the moduli space of instantons}, \href{http://dx.doi.org/10.1007/jhep08(2022)283}{\emph{Journal of High Energy Physics} {\bfseries 2022} (aug, 2022) }.

\bibitem{Bourget_2022dim6}
A.~Bourget and J.~F. Grimminger, \emph{Fibrations and hasse diagrams for 6d {SCFTs}}, \href{http://dx.doi.org/10.1007/jhep12(2022)159}{\emph{Journal of High Energy Physics} {\bfseries 2022} (dec, 2022) }.

\bibitem{Bourget:2020bxh}
A.~Bourget, A.~Hanany and D.~Miketa, \emph{{Quiver origami: discrete gauging and folding}}, \href{http://dx.doi.org/10.1007/JHEP01(2021)086}{\emph{JHEP} {\bfseries 01} (2021) 086}, [\href{https://arxiv.org/abs/2005.05273}{{\ttfamily 2005.05273}}].

\bibitem{Braden:2014iea}
T.~Braden, A.~Licata, N.~Proudfoot and B.~Webster, \emph{{Quantizations of conical symplectic resolutions II: category $\mathcal O$ and symplectic duality}}, {\emph{Asterisque} {\bfseries 384} (2016) 75--179}, [\href{https://arxiv.org/abs/1407.0964}{{\ttfamily 1407.0964}}].

\bibitem{wu2023namikawaweyl}
Y.~Wu, \emph{Namikawa-weyl groups of affinizations of smooth nakajima quiver varieties},  2023.

\bibitem{Gadde:2011uv}
A.~Gadde, L.~Rastelli, S.~S. Razamat and W.~Yan, \emph{{Gauge Theories and Macdonald Polynomials}}, \href{http://dx.doi.org/10.1007/s00220-012-1607-8}{\emph{Commun. Math. Phys.} {\bfseries 319} (2013) 147--193}, [\href{https://arxiv.org/abs/1110.3740}{{\ttfamily 1110.3740}}].

\bibitem{Hanany:2017ooe}
A.~Hanany and R.~Kalveks, \emph{{Quiver Theories and Formulae for Nilpotent Orbits of Exceptional Algebras}}, \href{http://dx.doi.org/10.1007/JHEP11(2017)126}{\emph{JHEP} {\bfseries 11} (2017) 126}, [\href{https://arxiv.org/abs/1709.05818}{{\ttfamily 1709.05818}}].

\bibitem{Cabrera:2018ldc}
S.~Cabrera, A.~Hanany and R.~Kalveks, \emph{{Quiver Theories and Formulae for Slodowy Slices of Classical Algebras}}, \href{http://dx.doi.org/10.1016/j.nuclphysb.2018.12.022}{\emph{Nucl. Phys. B} {\bfseries 939} (2019) 308--357}, [\href{https://arxiv.org/abs/1807.02521}{{\ttfamily 1807.02521}}].

\bibitem{Hanany:2019tji}
A.~Hanany and R.~Kalveks, \emph{{Quiver Theories and Hilbert Series of Classical Slodowy Intersections}}, \href{http://dx.doi.org/10.1016/j.nuclphysb.2020.114939}{\emph{Nucl. Phys. B} {\bfseries 952} (2020) 114939}, [\href{https://arxiv.org/abs/1909.12793}{{\ttfamily 1909.12793}}].

\bibitem{lusztig1979class}
G.~Lusztig, \emph{A class of irreducible representations of a weyl group},  in \emph{Indagationes Mathematicae (Proceedings)}, vol.~82, pp.~323--335, North-Holland, 1979.

\bibitem{lusztig1982class}
G.~Lusztig, \emph{A class of irreducible representations of a weyl group. ii},  in \emph{Indagationes Mathematicae (Proceedings)}, vol.~85, pp.~219--226, Elsevier, 1982.

\bibitem{spaltenstein2006classes}
N.~Spaltenstein, \emph{Classes unipotentes et sous-groupes de Borel}, vol.~946.
\newblock Springer, 2006.

\bibitem{juteau2023minimal}
D.~Juteau, P.~Levy and E.~Sommers, \emph{Minimal special degenerations and duality}, {\emph{arXiv preprint arXiv:2310.00521} (2023) }.

\bibitem{Gaiotto:2009we}
D.~Gaiotto, \emph{{N=2 dualities}}, \href{http://dx.doi.org/10.1007/JHEP08(2012)034}{\emph{JHEP} {\bfseries 08} (2012) 034}, [\href{https://arxiv.org/abs/0904.2715}{{\ttfamily 0904.2715}}].

\bibitem{Benini:2010uu}
F.~Benini, Y.~Tachikawa and D.~Xie, \emph{{Mirrors of 3d Sicilian theories}}, \href{http://dx.doi.org/10.1007/JHEP09(2010)063}{\emph{JHEP} {\bfseries 09} (2010) 063}, [\href{https://arxiv.org/abs/1007.0992}{{\ttfamily 1007.0992}}].

\bibitem{Gaiotto:2012uq}
D.~Gaiotto and S.~S. Razamat, \emph{{Exceptional Indices}}, \href{http://dx.doi.org/10.1007/JHEP05(2012)145}{\emph{JHEP} {\bfseries 05} (2012) 145}, [\href{https://arxiv.org/abs/1203.5517}{{\ttfamily 1203.5517}}].

\bibitem{Cremonesi:2014vla}
S.~Cremonesi, A.~Hanany, N.~Mekareeya and A.~Zaffaroni, \emph{{Coulomb branch Hilbert series and Three Dimensional Sicilian Theories}}, \href{http://dx.doi.org/10.1007/JHEP09(2014)185}{\emph{JHEP} {\bfseries 09} (2014) 185}, [\href{https://arxiv.org/abs/1403.2384}{{\ttfamily 1403.2384}}].

\bibitem{bellamy2021newfamilyisolatedsymplectic}
G.~Bellamy, C.~Bonnafé, B.~Fu, D.~Juteau, P.~Levy and E.~Sommers, \emph{A new family of isolated symplectic singularities with trivial local fundamental group},  2021.

\bibitem{Gledhill:2021cbe}
K.~Gledhill and A.~Hanany, \emph{{Coulomb branch global symmetry and quiver addition}}, \href{http://dx.doi.org/10.1007/JHEP12(2021)127}{\emph{JHEP} {\bfseries 12} (2021) 127}, [\href{https://arxiv.org/abs/2109.07237}{{\ttfamily 2109.07237}}].

\bibitem{2023arXiv230807398F}
B.~{Fu}, D.~{Juteau}, P.~{Levy} and E.~{Sommers}, \emph{{Local geometry of special pieces of nilpotent orbits}}, \href{http://dx.doi.org/10.48550/arXiv.2308.07398}{\emph{arXiv e-prints} (Aug., 2023) arXiv:2308.07398}, [\href{https://arxiv.org/abs/2308.07398}{{\ttfamily 2308.07398}}].

\bibitem{Hanany:2024fqf}
A.~Hanany, R.~Kalveks and G.~Kumaran, \emph{{Quiver Polymerisation}},  \href{https://arxiv.org/abs/2406.11561}{{\ttfamily 2406.11561}}.

\bibitem{bourget2021branesquiversaffinegrassmannian}
A.~Bourget, J.~F. Grimminger, A.~Hanany, M.~Sperling and Z.~Zhong, \emph{Branes, quivers, and the affine grassmannian},  2021.

\bibitem{Moeglin2002}
C.~Moeglin, \emph{Sur la classification des séries discrètes des groupes classiques p-adiques: paramètres de langlands et exhaustivité}, {\emph{Journal of the European Mathematical Society} {\bfseries 004} (2002) 143--200}.

\bibitem{2014arXiv1412.8742J}
D.~{Jiang}, B.~{Liu} and G.~{Savin}, \emph{{Raising nilpotent orbits in wave-front sets}}, \href{http://dx.doi.org/10.48550/arXiv.1412.8742}{\emph{arXiv e-prints} (Dec., 2014) arXiv:1412.8742}, [\href{https://arxiv.org/abs/1412.8742}{{\ttfamily 1412.8742}}].

\bibitem{2020arXiv201016089B}
D.~{Barbasch}, J.-J. {Ma}, B.~{Sun} and C.-B. {Zhu}, \emph{{On the notion of metaplectic Barbasch-Vogan duality}}, \href{http://dx.doi.org/10.48550/arXiv.2010.16089}{\emph{arXiv e-prints} (Oct., 2020) arXiv:2010.16089}, [\href{https://arxiv.org/abs/2010.16089}{{\ttfamily 2010.16089}}].

\bibitem{Hanany:2001iy}
A.~Hanany and J.~Troost, \emph{{Orientifold planes, affine algebras and magnetic monopoles}}, \href{http://dx.doi.org/10.1088/1126-6708/2001/08/021}{\emph{JHEP} {\bfseries 08} (2001) 021}, [\href{https://arxiv.org/abs/hep-th/0107153}{{\ttfamily hep-th/0107153}}].

\bibitem{Hanany:2016gbz}
A.~Hanany and R.~Kalveks, \emph{{Quiver Theories for Moduli Spaces of Classical Group Nilpotent Orbits}}, \href{http://dx.doi.org/10.1007/JHEP06(2016)130}{\emph{JHEP} {\bfseries 06} (2016) 130}, [\href{https://arxiv.org/abs/1601.04020}{{\ttfamily 1601.04020}}].

\bibitem{Hanany:2014dia}
A.~Hanany and R.~Kalveks, \emph{{Highest Weight Generating Functions for Hilbert Series}}, \href{http://dx.doi.org/10.1007/JHEP10(2014)152}{\emph{JHEP} {\bfseries 10} (2014) 152}, [\href{https://arxiv.org/abs/1408.4690}{{\ttfamily 1408.4690}}].

\bibitem{Finkelberg:2020cb}
M.~Finkelberg and E.~Goncharov, \emph{{Coulomb branch of a multiloop quiver gauge theory}}, \href{http://dx.doi.org/10.1134/S0016266319040014}{\emph{Functional Analysis and Its Applications} {\bfseries 53} (2019) 241--249}, [\href{https://arxiv.org/abs/1903.05822}{{\ttfamily 1903.05822}}].

\bibitem{Hanany_2023uzn}
A.~Hanany, G.~Kumaran, C.~Li, D.~Liu and M.~Sperling, \emph{{Actions on the quiver -- Discrete quotients on the Coulomb branch}},  \href{https://arxiv.org/abs/2311.02773}{{\ttfamily 2311.02773}}.

\bibitem{Hanany:2018vph}
A.~Hanany and G.~Zafrir, \emph{{Discrete Gauging in Six Dimensions}}, \href{http://dx.doi.org/10.1007/JHEP07(2018)168}{\emph{JHEP} {\bfseries 07} (2018) 168}, [\href{https://arxiv.org/abs/1804.08857}{{\ttfamily 1804.08857}}].

\bibitem{Hanany:2018cgo}
A.~Hanany and M.~Sperling, \emph{{Discrete quotients of 3-dimensional $ \mathcal{N}=4 $ Coulomb branches via the cycle index}}, \href{http://dx.doi.org/10.1007/JHEP08(2018)157}{\emph{JHEP} {\bfseries 08} (2018) 157}, [\href{https://arxiv.org/abs/1807.02784}{{\ttfamily 1807.02784}}].

\bibitem{Hanany:2022itc}
A.~Hanany and M.~Sperling, \emph{{Magnetic quivers and negatively charged branes}}, \href{http://dx.doi.org/10.1007/JHEP11(2022)010}{\emph{JHEP} {\bfseries 11} (2022) 010}, [\href{https://arxiv.org/abs/2208.07270}{{\ttfamily 2208.07270}}].

\bibitem{Bourget:2021jwo}
A.~Bourget, J.~F. Grimminger, A.~Hanany, R.~Kalveks and Z.~Zhong, \emph{{Higgs branches of U/SU quivers via brane locking}}, \href{http://dx.doi.org/10.1007/JHEP08(2022)061}{\emph{JHEP} {\bfseries 08} (2022) 061}, [\href{https://arxiv.org/abs/2111.04745}{{\ttfamily 2111.04745}}].

\bibitem{vinberg1972class}
{\`E}.~B. Vinberg and V.~L. Popov, \emph{On a class of quasihomogeneous affine varieties}, {\emph{Mathematics of the USSR-Izvestiya} {\bfseries 6} (1972) 743}.

\end{thebibliography}\endgroup

\end{document}